\documentclass[printer]{aa}
	
	\usepackage{footnote}
	\makesavenoteenv{tabular}
	\makesavenoteenv{table}
	
	\usepackage{hyperref}

	\usepackage{booktabs}

	\usepackage{afterpage}%
	\usepackage{float}%
	\usepackage[flushleft]{threeparttable}
	\usepackage{tikz,pgfplots}
	\usepackage[capitalize]{cleveref}
	
	\usepackage{longtable}
	
	\usepackage[arrows=pgf]{mhchem}
	
	\pgfplotsset{compat=newest}
	
	\usepackage{csvsimple}
	\usepackage{lscape}

\RequirePackage{txfonts}
\RequirePackage{microtype}
\RequirePackage{graphicx}
\RequirePackage{booktabs}
\RequirePackage{amsmath,amssymb}
\RequirePackage{cleveref}
\RequirePackage{pdfpages}
\RequirePackage{textgreek}

\newcommand*{\rangeto}{\textnormal{\hskip.1em\relax--\hskip.1em\relax}}
\RequirePackage[separate-uncertainty=true,range-units=single,range-phrase=\rangeto,forbid-literal-units=true]{siunitx}

\let\sun\odot

\newcommand*{\solarmass}{\si{\solarmass}}
\newcommand*{\solarluminosity}{\si{\solarluminosity}}
\DeclareSIUnit\lightspeed{$c$}
\DeclareSIUnit\rydberg{Ry}
\DeclareSIUnit\erg{erg}
\DeclareSIUnit\magnitude{mag}
\DeclareSIUnit\jansky{Jy}
\DeclareSIUnit\gauss{G}
\DeclareSIUnit\h{$h$}
\DeclareSIUnit\hseven{$h$_7}
\DeclareSIUnit\parsec{pc}
\DeclareSIUnit\au{AU}
\DeclareSIUnit\year{yr}
\DeclareSIUnit\solarluminosity{\ensuremath{L_\sun}}
\DeclareSIUnit\solarmass{\ensuremath{M_\sun}}
\DeclareSIUnit\jupitermass{\ensuremath{M_\mathrm{J}}}
\DeclareSIUnit\solarmassinenergy{\ensuremath{M_\sun|c^2}}
\DeclareSIUnit\solarradius{\ensuremath{R_\sun}}

\DeclareSIUnit\arcsecond{as}
\DeclareSIUnit\clight{\ensuremath c}
\DeclareRobustCommand\SIfrac{\@ifstar\astrosym@SIfracplain\astrosym@SIfracparen}

\newcommand{\kms}{\ensuremath{\mathrm{km~s}^{-1}}}

\newcommand{\erg}{\ensuremath{\mathrm{erg}}}
\newcommand{\persec}{\ensuremath{\mathrm{s}^{-1}}}

\newcommand{\msun}{\ensuremath{\mathrm{M}_\odot}}

\newcommand{\lsun}{\ensuremath{\mathrm{L}_\odot}}
\newcommand{\mstar}{\ensuremath{M_*}}
\newcommand{\lstar}{\ensuremath{L_*}}

\newcommand{\mdisk}{\ensuremath{M_\mathrm{disk}}}

\newcommand{\zetacr}{\ensuremath{\zeta_\mathrm{CR}}}
\newcommand{\au}{\ensuremath{\mathrm{AU}}}
\newcommand{\av}{\ensuremath{\mathrm{A}_V}}
\newcommand{\micron}{\ensuremath{\mu\mathrm{m}}}
\newcommand{\teff}{\ensuremath{T_\mathrm{eff}}}
\newcommand{\kelvin}{\ensuremath{\mathrm{K}}}
\newcommand{\fuv}{\ensuremath{f_\mathrm{UV}}}
\newcommand{\luv}{\ensuremath{L_\mathrm{UV}}}
\newcommand{\puv}{\ensuremath{p_\mathrm{UV}}}

\newcommand{\lxr}{\ensuremath{L_\mathrm{X-ray}}}
\newcommand{\rtaper}{\ensuremath{R_\mathrm{taper}}}
\newcommand{\rin}{\ensuremath{R_\mathrm{in}}}
\newcommand{\rout}{\ensuremath{R_\mathrm{out}}}
\newcommand{\amin}{\ensuremath{a_\mathrm{min}}}
\newcommand{\amax}{\ensuremath{a_\mathrm{max}}}
\newcommand{\apow}{\ensuremath{a_\mathrm{pow}}}

\newcommand{\tgas}{\ensuremath{T_\mathrm{gas}}}
\newcommand{\tdust}{\ensuremath{T_\mathrm{dust}}}

\newcommand{\jwst}{\mbox{\textit{JWST}}}
\newcommand{\spitzer}{\mbox{\textit{Spitzer}}}
\newcommand{\flits}{\texttt{FLiTs}}
\newcommand{\prodimo}{\texttt{ProDiMo}}

\newcommand{\herschel}{\mbox{\textit{Herschel}}}

\newcommand{\elt}{\mbox{ELT}}

\newcommand{\fline}{\ensuremath{F_\mathrm{line}}}

\newcommand{\mr}[1]{\ensuremath{\mathrm{#1}}}  %

	\usepackage{placeins}
	\usepackage[arrows=pgf]{mhchem}
	
	\newcommand{\cem}[1]{\mbox{\ensuremath{\ce{#1}}}}

\begin{document}

		\title{The infrared line-emitting regions of T Tauri protoplanetary disks}

		\author{A.J. Greenwood$^1$
			\and
			I. Kamp$^1$ \and L.B.F.M. Waters$^{2}$ \and P. Woitke$^3$ \and W.-F. Thi$^4$
		}
		
		\institute{$^1$ Kapteyn Astronomical Institute, University of Groningen, Postbus 800, 9700 AV Groningen, The Netherlands\\
					             \email{kamp@astro.rug.nl} \\
			$^2$ SRON Netherlands Institute for Space Research, Sorbonnelaan 2, 3584 CA Utrecht, The Netherlands \\
			$^3$ SUPA, School of Physics \& Astronomy, University of St. Andrews, North Haugh, St. Andrews KY16 9SS, UK \\
			$^4$ Max Planck Institute for Extraterrestrial Physics, Gie\textbeta{}enbachstra\textbeta{}e 1, 85741 Garching, Germany}

\abstract{
		Mid-infrared molecular line emission detected with the \spitzer{} Space Telescope is often interpreted using slab models. However, we need to understand the mid-infrared line emission in 2D disk models, such that we gain information about from where the lines are being emitted and under which conditions, such that we gain information about number densities, temperatures, and optical depths in both the radial and vertical directions. In this paper, we introduce a series of $2$D thermochemical models of a prototypical T~Tauri protoplanetary disk, in order to examine how sensitive the line-emitting regions are to changes in the UV and X-ray fluxes, the disk flaring angle, dust settling, and the dust-to-gas ratio. These all affect the heating of the inner disk, and thus can affect the mid-infrared spectral lines.
		
		Using the \prodimo{} and \flits{} codes, we produce a series of $2$D thermochemical disk models. We find that there is often a significant difference between the gas and dust temperatures in the line emitting regions, and we illustrate that the size of the line emitting regions is relatively robust against changes in the stellar and disk parameters (namely, the UV and X-ray fluxes, the flaring angle, and dust settling). These results demonstrate the potential for localized variations in the line-emitting region to greatly affect the resulting spectra and line fluxes, and the necessity of allowing for such variations in our models.
}
		\maketitle
		
		\section{Introduction}

		In the past decade, beginning with the \spitzer{} space telescope, we have begun to observe mid-infrared (mid-IR) molecular lines in protoplanetary disks, and to gain a better understanding of the chemistry in the mid-IR line-emitting regions.	
Throughout this era,  most modelling efforts of \spitzer{} spectra have been limited to local thermodynamic equilibrium (LTE) slab models. Such models require many assumptions because the level populations are assumed only to depend on the gas temperature. In non-LTE scenarios, radiation and collisional processes are also accounted for in calculating level populations. In both LTE and non-LTE scenarios, determining the column densities of individual species from spectra is unreliable because the disks are optically thick in the mid-IR continuum. Non-LTE effects are present in mid-IR spectral lines, and have been found to affect line fluxes by a factor of a few in HCN \citep{Bruderer:2015iw} and \cem{CO2} \citep{2017AA...601A..36B}. Perhaps the more important result of non-LTE models is that mid-IR lines can be excited out to about $10~\au$ in T~Tauri disks, resulting in a larger line-emitting area than is typically assumed in slab models  \citep{Bruderer:2015iw}.
		
The most significant disadvantage of a slab model is that it does not account for effects such as temperature and opacity gradients across the line-emitting region, and is independent of any calculations or assumptions about the spatial extent of the line-emitting region. In thermochemical models that include the effects of flaring and dust settling, we also see significant differences in optical depths and gas and dust temperatures across the line-emitting region. A slab model cannot properly account for important factors such as dust settling, disk flaring, and differences in species abundance and gas temperature as functions of both radius and height above the midplane.  Being able to match the near- and mid-IR lines of a $2$D (or $3$D) model to an observed spectrum would be great progress towards truly understanding the inner few AU of protoplanetary disks, which is a step that requires more advanced modelling techniques to achieve.
		
		Constraining our models will become significantly easier in the near future. The \textit{James Webb Space Telescope} (\jwst{}) and the Extremely Large Telescope (\elt{}) will provide significant gains in sensitivity and resolution, with spectral sensitivity 100 times greater than \spitzer{} \citep{Brandl:2014ea, Glasse:2015cg}. Additionally, the diffraction-limited imaging capabilities of the METIS Integral Field Unit on the \elt{} can resolve AU-scale structure in the closest protoplanetary disks  \citep{Brandl:2014ea}, allowing us to directly observe the kinematics and spatial distribution of gas species in nearby disks.
		
		These improved observational capabilities demand a more complex approach to understanding the data, such as using high-resolution $2$D models. We can construct thermochemical models of T~Tauri stars and fit them to ALMA and \herschel{} observations \citep{Woitke:2016gp}, but it is still difficult to model their near- and mid-IR spectra, which trace the regions of terrestrial planet formation. The focus of this paper is the combination of two modelling tools: first using \prodimo{}  \citep{Woitke:2009jf,Kamp:2010ek,Aresu:2011cm} to model the inner disk, and then using the line-tracing code \flits{}  \citep{Woitke:2018ux} to calculate high-resolution infrared (IR) spectra of a small series of disk models. We use the resulting spectra to show how certain disk parameters can affect the mid-IR lines of a series of molecules in different ways.
		
		By combining \flits{} and \prodimo{} we can produce high-resolution, multi-molecule IR spectra of a protoplanetary disk model in $2$D. The main goal of this paper is to show the properties of the \cem{CO2}, \cem{C2H2}, \cem{HCN}, \cem{H2O}, \cem{OH}, and \cem{NH3} line-emitting regions of T~Tauri disks: how large the line-emitting regions are, and how the properties such as the gas temperature vary across them. We emphasize the relevance of thermal de-coupling between the gas and dust for mid-IR lines. This decoupling hinges on our understanding of the gas heating and cooling processes, and the latter have not -- in the context of $2$D thermochemical disk models -- been explored very much in the literature. This paper serves further to demonstrate the capabilities of the \flits{} and \prodimo{} codes combined, with respect to their ability to analyse data from upcoming observatories such as \jwst{} and \elt{}.
		
		Other recent research has also produced spectra from $2$D disk models \citep{2017AA...601A..36B}, but these models assume that $\tgas = \tdust$. Because we allow the gas and dust temperatures to vary independently, the gas temperature in our models is determined not only by the radiative transfer of the dust, but also by the gas chemistry. It is this extra step which we suggest allows for more realistic gas temperature structures that may be better able to explain why some molecules are very bright in some disks yet absent in others.

		\section{FLiTs}

		\flits{} \emph{(Fast Line Tracer)} is a new code
		(described in detail by \citealt{Woitke:2018ux}) which can quickly (and accurately, insofar as the input model is correct) compute molecular lines in the IR.   A $2$D thermochemical disk model  is used as an input, in order to fix the structure of the disk. \citet{Woitke:2018ux} show that using \flits{} on a standard T Tauri disk model from \prodimo{}  produces spectra that are very similar to observations of disks such as TW Hya and RW Aur.
		The input data include at each grid point the dust and gas temperatures, number densities of species, and level populations (which can be in non-LTE). Currently this  input is a \prodimo{} disk model, and \flits{}  is already configured to read the data structures written by \prodimo{}.

\begin{figure}
			\centering
			\includegraphics[width=80mm]{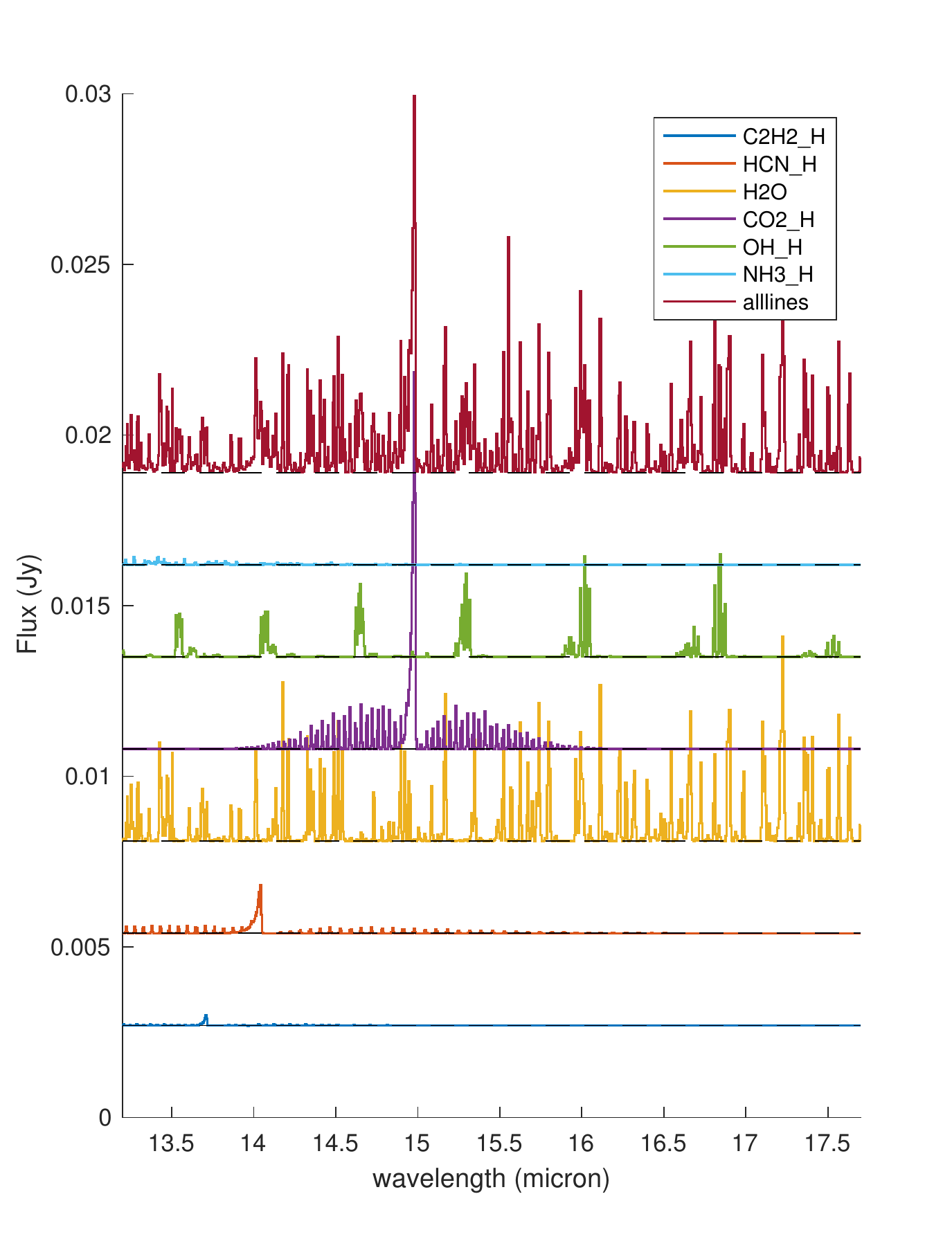}
			\caption{\flits{} spectra of the standard T~Tauri disk model, at a spectral resolution of $R=2\,800$. Each individual spectrum has been vertically offset by an arbitrary amount (the continuum levels of each spectrum are indicated by  horizontal dashed lines). The top spectrum ``alllines'' includes all of the other plotted species together, as well as many other species such as \cem{CH4} and \cem{Fe} which do not greatly affect the spectra seen here and can -- until better observatories such as \jwst{} might detect them -- be disregarded. The ``\_H'' in the legend refers to the fact that the ro-vibrational spectroscopic data are taken from the HITRAN database.}
			\label{fig:TT_flitsspec}
		\end{figure}

		\flits{} can then compute the near- and mid-IR lines of many molecules at once, producing a single output spectrum with potentially many thousands of blended lines.
		The result is a high-resolution (e.g. $1~\kms$)  spectrum that contains many molecules, with line blends computed self-consistently\footnote{Meaning that the effects of opacity overlap are accounted for when lines overlap and shield each other from radiative pumping.}. \Cref{fig:TT_flitsspec} shows an example of a \flits{} spectrum, from our standard T Tauri model, where the full spectrum including all species listed in \cref{tab:speclist} is compared against the individual contributions of \cem{C2H2}, \cem{HCN}, \cem{H2O}, \cem{CO2}, \cem{OH}, and \cem{NH3}.

		\section{Disk models: a standard T Tauri model}
		
		The disk models are based upon a previously-established, ``standard'' T~Tauri disk model, computed with the $2$D thermochemical disk modelling code \prodimo{}. The ``standard'' T~Tauri model in this paper is the ``TT\_LU'' large-chemical-network model introduced by  \citet{Woitke:2016gp} and further described by \citet{Kamp:2017bx}. \Cref{tab:stdmodelparams} describes the parameters of this model.
		
\begin{figure}
			\centering
			\includegraphics[width=0.47\textwidth,page=1]{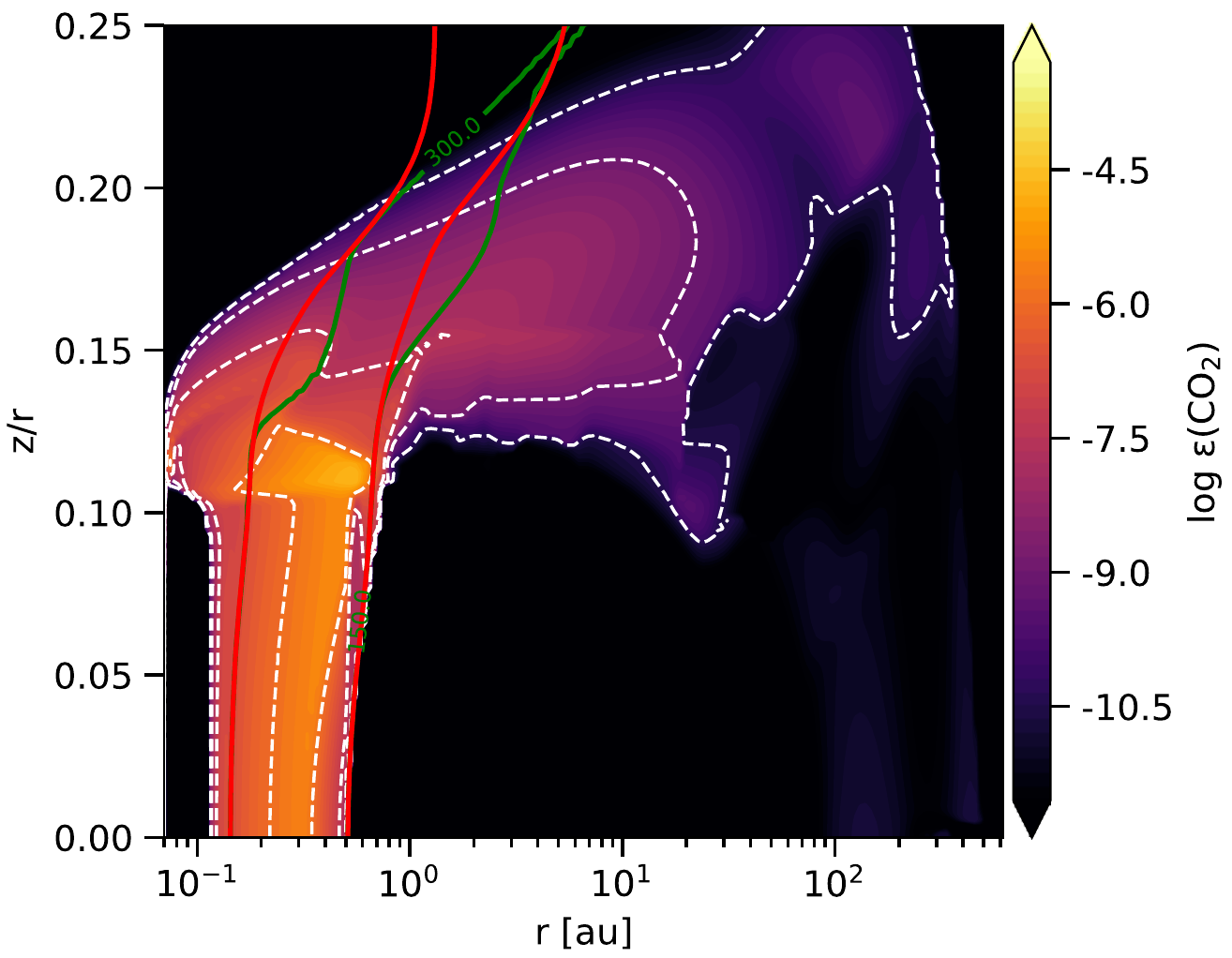} \\
			\includegraphics[width=0.47\textwidth,page=2]{./figures/TT_plots.pdf}
			\caption{Top panel: \cem{CO2}  abundance (relative to hydrogen) of the standard T~Tauri model. The white dashed contour lines trace the \cem{CO2} abundance at the levels labelled on the colour-bar, the solid green contour lines trace where $\tgas=150$ and $300~\kelvin$, while the red lines trace where $\tdust=150$ and $300~\kelvin$. Bottom panel: gas temperature of the same model. The thick, solid white contour lines indicate the level of visual extinction, at $\av=1$, $5$, and $10$. The dashed contour lines correspond to the temperature labels on the colour bar (i.e. they are at $10~\kelvin$, $20~\kelvin$, $40~\kelvin$, and so on).}
			\label{fig:stdmodel}
		\end{figure}

		Although we focus on the mid-IR-emitting regions, the \prodimo{} model is computed to an outer radius of $600~\au$.  The outer disk does not directly affect the inner disk, and at high inclinations the flared outer disk can absorb radiation from the inner disk along the line of sight. This way, we account for this effect and also provide the opportunity for analysis of the sub-mm regions of the exact same model.   \Cref{fig:stdmodel} shows the gas temperature and \cem{CO2} abundance of the standard disk model, to give a general idea of the results we get from \prodimo{}. The \cem{CO2} snow line occurs at about $0.4~\au$ in the mid-plane, while in the upper layers, \cem{CO2} gas extends in significant abundances out to about $10~\au$. Near the mid-plane, the depletion of gas-phase \cem{CO2} is caused by the appearance of water ice, which forms at around $\tgas=150~\kelvin$. In this optically-thick region near the mid-plane, the gas temperature is determined by the dust temperature. In the upper layers where \cem{CO2} occurs at larger radii, the gas becomes warmer than the dust, mostly due to chemical heating. This can be seen in \cref{fig:stdmodel}, where the gas temperature contours depart from the dust temperature contours.
		The results that are of most interest to us in this paper are the locations where certain heating and cooling processes are dominant, and the inequality between the gas and dust temperatures: these are the factors which we find can significantly affect the mid-IR spectra, and which have not previously been addressed in the literature.

		\begin{table*}

				\caption{Fundamental parameters of the standard ``TT highres'' T~Tauri disk model, based on parameters from \cite{Woitke:2016gp}. Parameter definitions are further explained by \cite{Woitke:2009jf}.  References are as follows: 1: \cite{Woitke:2016gp}, 2: \cite{Dorschner:1995wq}, 3: \cite{Zubko:1996fn}, 4: \cite{Dubrulle:1995jn}, 5: \cite{Draine:1978ec}.  }
				\label{tab:stdmodelparams}
				\medskip
						 \centering
				\small
				\begin{tabular}{l l r }
					Symbol & Quantity (units) & Parameter value \\ \hline
					$\mstar$ & Stellar mass (\msun) & $0.7$ \\
					$\lstar$ & Stellar luminosity (\lsun) & $1.0$  \\
					\teff & Effective temperature (\kelvin) & $4000$\\
					$\fuv$ & UV excess ($\luv / \lstar$) & $0.01$  \\
					$\puv$ & UV power law exponent & $1.3$\vspace{0.13em} \\
					$\lxr$ & \parbox[t]{7cm}{X-ray luminosity (\erg~\persec, bremsstrahlung continuum) $^1$} & $10^{30}$\vspace{0.13em} \\
					$\zetacr$ & \parbox[t]{7cm}{Cosmic ray \cem{H2} ionization rate ($\persec$)} & $1.7 \times 10^{-17}$\vspace{0.13em}  \\
					$\mdisk$ & Disk mass ($\times 10^{-4}~\msun$) & $100$\\
					${\rho_\mathrm{d}} / {\rho_\mathrm{g}}$ & Dust-to-gas ratio & $0.01$  \\
					$\rin$ & Inner disc radius (\au) & $0.07$\vspace{0.13em} \\
					$\rout$ & Outer disc radius (\au) & $600$\vspace{0.13em} \\
					$\rtaper$ & Tapering-off radius (\au) & $100$ \\
					$H_0$ & Scale height at $100~\au$ (\au) & $10$  \\
					$\beta$ & \parbox[t]{7cm}{Flaring power index \mbox{$H(r)=H_0 \left( {r} / {r_0} \right)^\beta $}} & $1.15$\vspace{0.13em} \\
					$N$ & Number of grid points & $240 \times 180$\vspace{0.13em} \\
					$\apow$ & \parbox[t]{7cm}{Dust size distribution \mbox{$f(a)\propto{}a^{-\apow}$}} & $3.5$\vspace{0.13em} \\
					& \multicolumn{2}{l}{\parbox[t]{7cm}{Dust grain mixture: $60\%$ amorphous \cem{Mg_{0.7} Fe_{0.3} Si O_3} silicates $^2$, $15\%$ amorphous carbon $^3$, $25\%$ vacuum for porosity~$^\dag$}}\vspace{0.13em} \\
					$\amin$ & Min. dust grain size (\micron) & $0.05$\\
					$\amax$ & Max. dust grain size (\micron) & $3000$ \\
					$i$ & Inclination angle ($^\circ$) & $45$\vspace{0.13em} \\
					$\alpha$ & \parbox[t]{7cm}{Turbulent viscosity, for Dubrulle settling of dust grains $^4$}\vspace{0.13em} & $0.01$ \\
					$\chi_{\mathrm{ISM}}$ & \parbox[t]{7cm}{Strength of incident UV w.r.t. ISM field $^5$} & 1 \\
					\midrule
				\end{tabular} \\
				\raggedright
			\dag: {The dust is a distribution of hollow spheres, where the maximum fractional volume filled by the central void is $0.8$ \citep{Min:2005uy,Min:2016hr}.}
		\end{table*}

We define the line-emitting region as the area from which $70\%$ of the flux originates, in both the radial and vertical directions. The lower limit, $\mr{x15}$, is defined so that $85\%$ of the total line flux of that spectral line is emitted at radii greater than $\mr{x15}$. The upper limit, $\mr{x85}$, is the opposite -- only $15\%$ of the total line flux of that spectral line is emitted at a radius greater than $\mr{x85}$. 
At each radius, $\mr{z15}$ and $\mr{z85}$ define the heights in the disk above which $15\%$ and $85\%$ of the flux is emitted respectively. The properties of the line-emitting region for a given species (such as \tgas)  are averaged and weighted with the volume density of that species across the line-emitting region.

The calculations of the line-emitting area are done by calculating the line flux at each grid point in the model using an escape probability method and the effects of radial optical depth are not accounted for.
Wherever we mention the flux of an individual spectral line, these fluxes are calculated using a vertical escape probability method \citep{Woitke:2009jf}. These line fluxes are accurate only for a face-on disk, because no detailed radiative transfer is taken into account.  In contrast, where we discuss \flits{} spectra, these have been calculated for an inclined disk and account for both the radial and vertical optical depth. These spectra are also convolved to an instrumental resolution, so that we measure the peak flux of a complex of lines and not the integrated flux of a single line.

To define a sample of lines that we are investigating, \cref{tab:transitions} details exactly the molecular lines chosen for analysis, one for each species. For ease of comparison, where possible, we analyze spectroscopic lines that have previously been analyzed in other literature. Whenever an individual molecular line is referenced in this paper, it refers to the line in this table.
		
\begin{figure}
			\centering
			\includegraphics[width=0.47\textwidth]{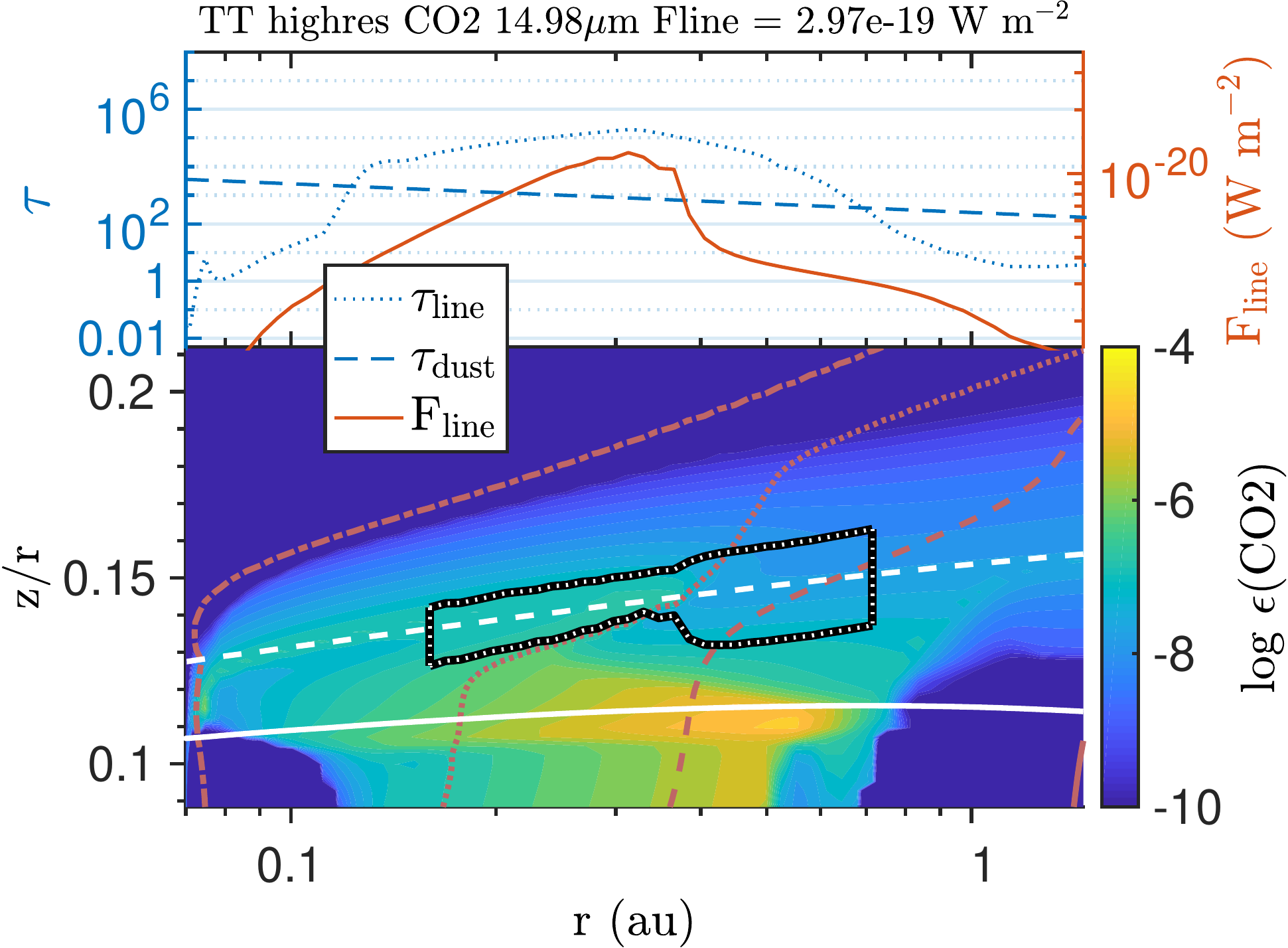}
			\includegraphics[width=0.47\textwidth]{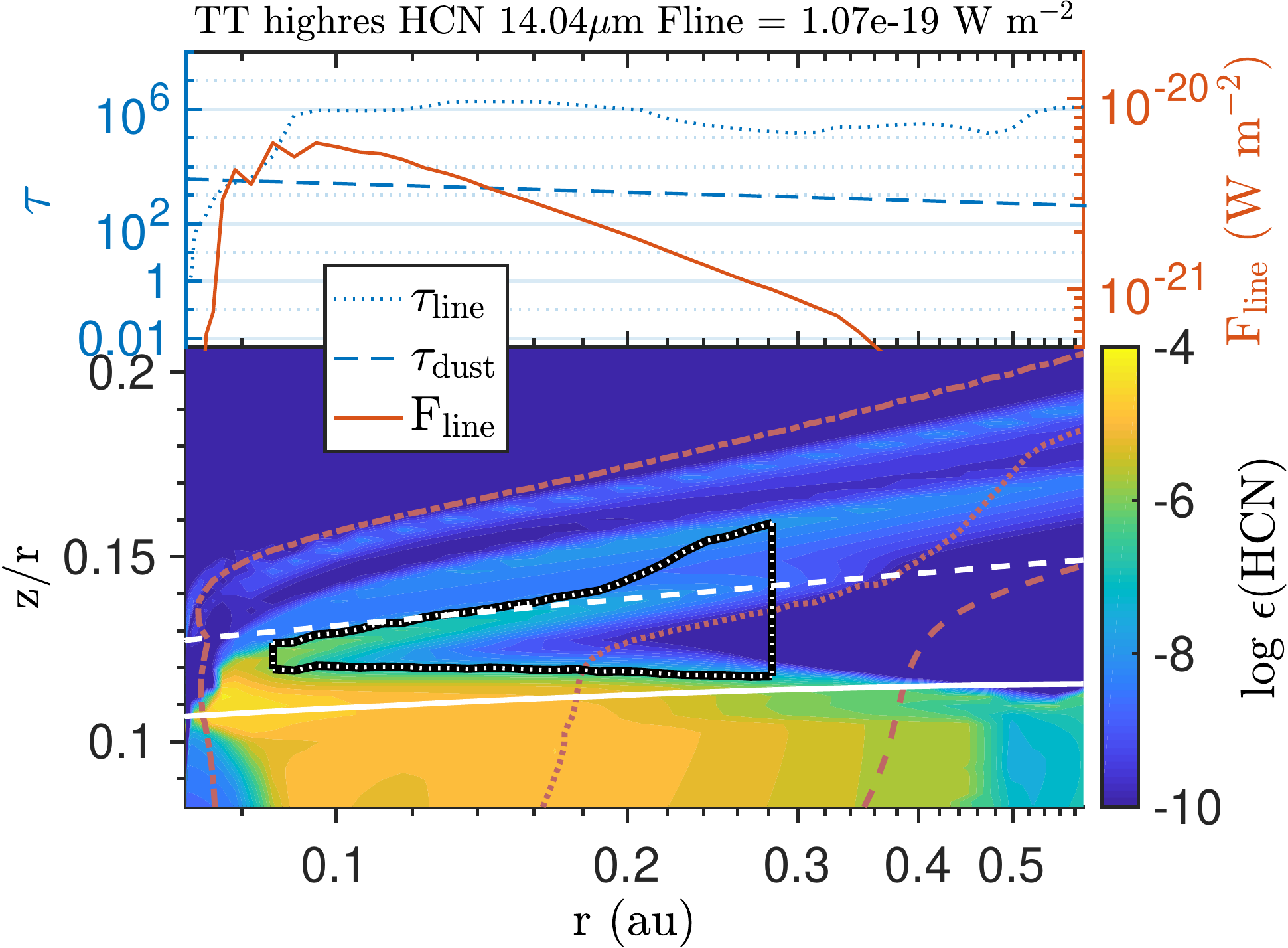}
			\includegraphics[width=0.47\textwidth]{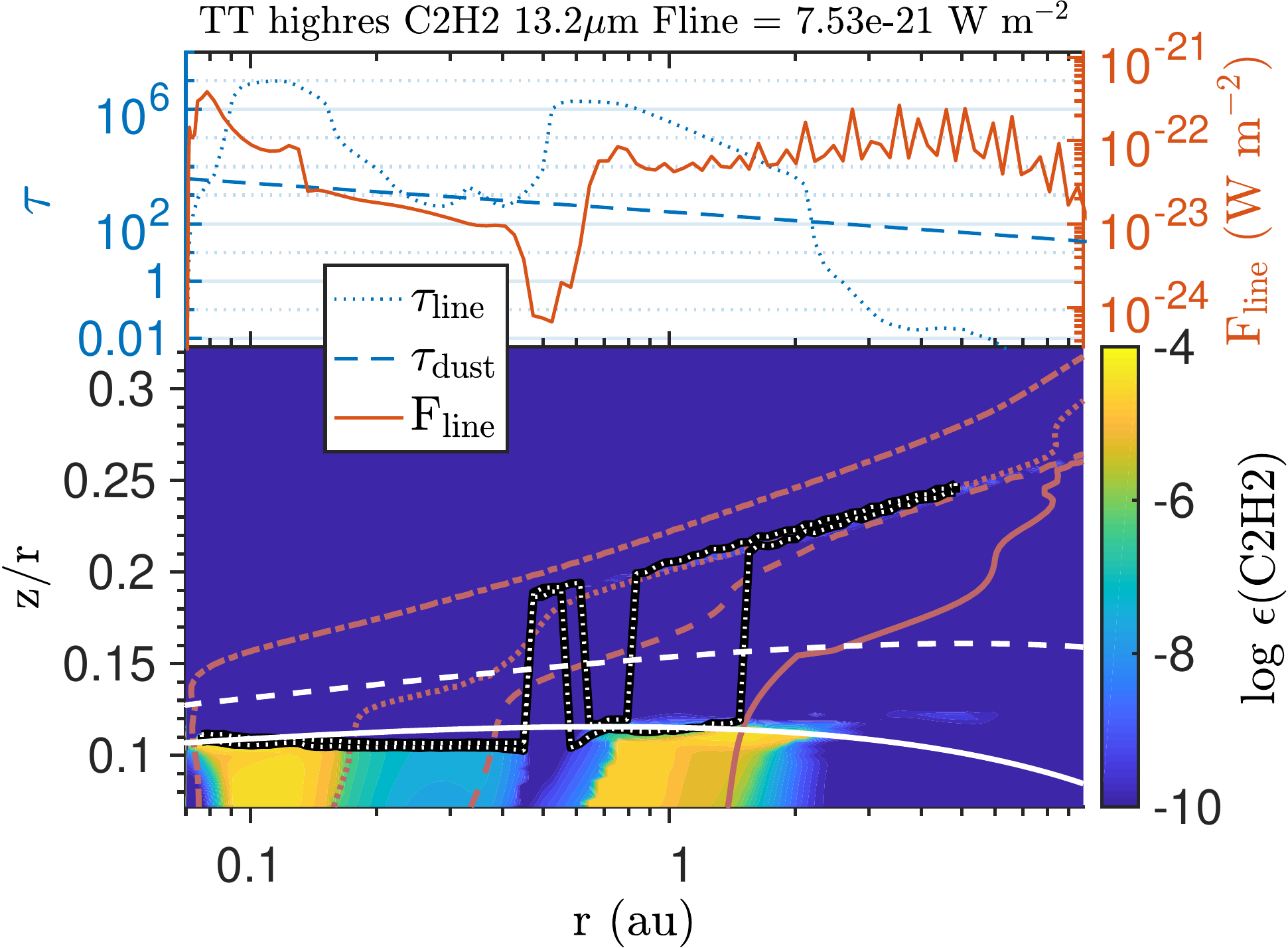}
			
			\caption{Line-emitting regions of \cem{CO2} (top), \cem{HCN} (middle), and \cem{C2H2} (bottom). On the upper panels of each sub-plot, the $20~\micron$ dust continuum and gas line optical depths and the line flux are plotted. The optical depths and line fluxes are calculated using the vertical escape probability.  The noisy appearance of the \cem{C2H2} line flux is an artefact of the low vertical resolution in comparison to the very thin line-emitting region. On the lower panels, the plotted colour map is the molecular abundance (relative to the total hydrogen abundance $\langle \cem{H} \rangle$), with contour increments every $0.5$ dex. The over-plotted contours are the main line-emitting area (black-and-white line), the visual extinction (dashed white line at $\av = 1$, and solid white line at $\av = 1$), and gas temperature. The gas temperature is plotted in red at $100$, $200$, $300$, and $1000~\kelvin$ (solid, dashed, dotted, and dash-dotted lines respectively).}
			\label{fig:TT_lineemittingregions}
		\end{figure}

		\begin{table*}
			\begin{center}
				\caption{Emission line of each species chosen for analysis, including upper level energies  $E_\mathrm{up}$ and the Einstein $A$ coefficient (giving the rate of spontaneous emission).
					The description of the ro-vibrational lines of \cem{CO2}, \cem{C2H2}, \cem{HCN}, and \cem{NH3} is an abbreviated form of that described in \citet{Jacquemart:2003bc,Rothman:2005cj}, where $v_j$  are the normal mode vibrational quantum numbers, $l_j$ are the vibrational angular momentum quantum numbers, and $l$ is the absolute value of the sum of $l_j$. The final entry, for example $R11e$, denotes that it is an $R$-branch transition, the lower-state rotational energy level is $11$, and $e$ or $f$ denotes the symmetry for $l$-type doubling.}
				\medskip
				\centering
				\begin{tabular}{l l l l l l}
					Species & $\lambda~(\micron)$ & Transition & $E_\mathrm{up}~(K)$ & $A~(\mathrm{s}^{-1})$ & Reference \\  \midrule 
					\cem{CO2} & $14.98299$ & $v_1v_2l_2v_3r = 01101 \rightarrow 00001$, $Q6e$ &  $983.85$ & $1.527$ & \cite{2017AA...601A..36B} \\
					\cem{C2H2} & $13.20393$ & $v_1v_2v_3v_4v_5l\pm = 000011 \rightarrow 000000$,  $R11e$ & $1313.1$ & $3.509$ & \citet{Woitke:2018ux} \\
					\cem{HCN} & $14.03930$ & $v_1v_2l_2v_3 = 0110 \rightarrow 0000$, $Q6e$ & $1114.1$ & $2.028$	& \cite{Bruderer:2015iw} \\
					\cem{o-H2O} & $17.75408$ & $J'=6 \rightarrow J''=5$ & $1278.5$ & $0.002869$ & \cite{Notsu:2017jc} \\
					\cem{NH3} & $10.33756$ & $v_1v_2v_3v_4 = 0100 \rightarrow 0000$, $J'=3 \rightarrow J''=3$ & $1515.3$ & $11.57$ &  \\
					\cem{OH} & $20.11506$ &$J'=13.5 \rightarrow J''=12.5$  & $5527.2$ & $50.47$ & \citet{Woitke:2018ux} \\ 
				\end{tabular}
				\label{tab:transitions}
			\end{center}
		\end{table*}
		\Cref{fig:TT_lineemittingregions} shows the line-emitting region for \cem{CO2} at $14.98~\micron$. The bold, black box traces the line-emitting area, as calculated above. This figure shows that most of the \cem{CO2} emission comes from a region where there is a significant gradient in the gas temperature. The situation is similar for \cem{C2H2} and \cem{HCN}, except that \cref{fig:TT_lineemittingregions} shows that the mid-IR lines from these molecules tend to come from slightly smaller radii in our T~Tauri disk model. As already noted by \citet{Woitke:2018ux}, each of the molecules that are commonly observed in mid-IR Spitzer spectra have line-emitting regions that trace different regions of the disk. For example, the line-emitting region of \cem{CO2} is located at a radius of around $0.1 - 1~\au$, the line-emitting region of \cem{HCN} is at around $0.08 - 0.3~\au$, and the line-emitting region of \cem{C2H2} is at around $0.07 - 5~\au$. Because the spatial location of the line emitting regions also differs significantly between molecules, the average gas temperature at which each molecule emits can also change significantly.
		
\begin{figure}
			\centering
		\includegraphics[width=0.47\textwidth,page=2]{./figures/C2H2_H_ratio.pdf} \\
				\includegraphics[width=0.47\textwidth,page=1]{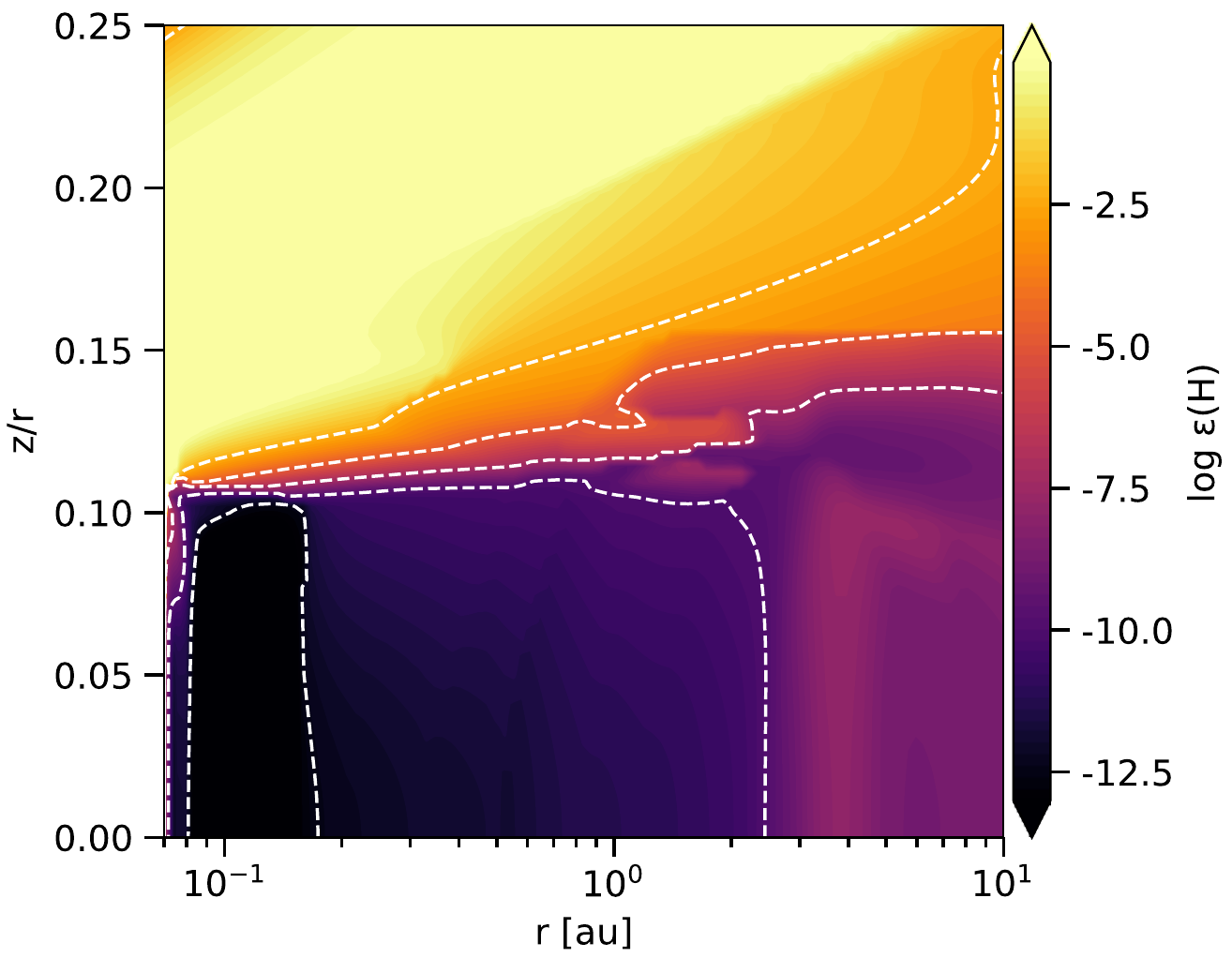}
			\caption{{Top:} ratio between the \cem{C2H2} and \cem{H} abundances for the inner $10~\au$ of our standard T~Tauri model. {Bottom:} \cem{H} abundance of the same model (the hydrogen abundance is the ratio between the number density of monatomic hydrogen and the total number density of hydrogen across all species).}
			\label{fig:stdTT_hydrogenabund}
		\end{figure}

		The emission of \cem{C2H2} is somewhat complicated by the fact that there are two distinct layers of \cem{C2H2}, resulting in two layers of line emission. An unmistakeably similar structure is also seen in models by \cite{Walsh:2015jr}.
		\Cref{fig:stdTT_hydrogenabund} compares the abundances of \cem{H} and \cem{C2H2} in the disk: below about $z/r=0.1$, there is very little atomic hydrogen and few free electrons, and the \cem{C2H2} abundances are high.   The formation of \cem{C2H2} in this lower layer is dominated by neutral-neutral and ion-neutral chemistry, through reactions such as \cem{H2 + C2H \rightarrow C2H2 + H}. Atomic hydrogen does not survive in this region, as it very rapidly forms \cem{H2} on the surfaces of dust grains: it plays no significant part in the formation of \cem{C2H2} here.
		The abundance of \cem{C2H2} in the upper layer is relatively low, but this region of the disk is optically thin and so its contribution to the total line flux is significant. Atomic hydrogen is abundant, and the dominant formation mechanism is \cem{H + C2H3 \rightarrow C2H2 + H2}. The two separate layers of \cem{C2H2} emission appear to result from a dichotomy in the formation pathways, possibly attributable in part to the H/\cem{H2} transition. Such separate layers also occur in models by \cite{Agundez:2018wz}, who compare the disk chemistry of T~Tauri and Herbig stars. Notably, \cem{HCN} also has a gap in its abundance similar to \cem{C2H2}, however, this gap is less significant and is generally outside of the line-emitting region.
		
		\section{Permutations to the standard model}
		
To explore the effects of disk geometry and the radiation environment on the mid-IR spectral lines, we have modified several parameters of the standard model: the dust-to-gas ratio, the flaring angle, the amount of settling, the UV flux, and the X-ray flux. These parameters are significant drivers of heating in the upper layers of the disk: it is this heating which drives the gas temperature to become higher than the dust temperature, which can then greatly affect the fluxes and line-emitting areas of the  mid-IR lines.
		
		\Cref{tab:modelseries} describes how each model in this series differs from the standard model. We have also produced a series of models with a gas-to-dust ratio of $1000$, hereafter called the ``lessdust'' models. The only change with respect to the models in  \cref{tab:modelseries} is that the mass of dust in the disk has been decreased. The total disk mass remains the same. In cases where we have changed the flaring angle of the disk, we have also modified the reference radius and scale height such that the height of the disk at the inner edge is the same in each model, irrespective of the flaring angle: in these cases, the reference radius is $0.07~\au$ and the scale height at this radius is $0.0024~\au$. 
		
		\begin{table}
			\caption{Description of the series of \prodimo{} models (all at an inclination of $45^\circ$).}
			\medskip
			\centering
			\small
			\begin{tabular}{l l c}
				Model name & Description \\ \midrule
				TT highres &  Standard T~Tauri model (STT) \\
				UV low & STT with $1\times 10^{-3}~\lstar$ UV excess ($10\%$ of STT)\\
				UV high & STT with $2\times 10^{-2}~\lstar$ UV excess \\
				Xray low & STT with $1\times 10^{29}~\mathrm{erg~s}^{-1}$ X-ray luminosity \\
				Xray high & STT with $1\times 10^{31}~\mathrm{erg~s}^{-1}$ X-ray luminosity \\
				Flaring low & STT with flaring index $\beta = 1.05$ \\
				Flaring high & STT with flaring index $\beta= 1.25$ \\
				Turbulence low & STT with turbulence $\alpha = 10^{-3}$ \\
				Turbulence high & STT with turbulence $\alpha = 10^{-1}$ \\
			\end{tabular}
			\label{tab:modelseries}
		\end{table}
		
		\subsection{Changes in the line-emitting area}

		Knowing the basic properties of the line-emitting regions, and how they differ between species, is an important step towards understanding the spectra that we observe. This has previously been done by fitting synthetic slab model spectra to \spitzer{} observations \citep{Carr:2011hl}. In this paper we show the line-emitting regions of a series of $2$D disk models computed by ProDiMo. This approach gives both radial and vertical information about the line emission. We demonstrate how sensitive the line-emitting regions are to changes in the disk geometry and radiation environment, and quantify the differences between $\tgas$ and $\tdust$ in line-emitting regions.
		
		\Cref{fig:CO2_lineemission,fig:HCN_lineemission,fig:C2H2_lineemission,fig:H2O_lineemission,fig:NH3_lineemission,fig:OH_lineemission} show how the line-emitting regions of our standard model series respond to the parameter changes listed in \cref{tab:modelseries}, while \cref{fig:CO2_lessdust_lineemission,fig:HCN_lessdust_lineemission,fig:C2H2_lessdust_lineemission,fig:H2O_lessdust_lineemission,fig:NH3_lessdust_lineemission,fig:OH_lessdust_lineemission} show the same but for the ``lessdust'' configuration. We stress that the \cem{C2H2} and \cem{NH3} line emission results for the $g/d=100$ scenario have been omitted because the line fluxes for these molecules are very low. Numerical noise makes the calculated properties of the line-emitting area for these species unreliable. 	
		\Cref{fig:LER_TT_highres,fig:LER_UV_high,fig:LER_UV_low,fig:LER_Xray_high,fig:LER_Xray_low,fig:LER_flaring_high,fig:LER_flaring_low_scaleheightfix_run5,fig:LER_settling_high,fig:LER_settling_low} and \cref{fig:LER_TT_highres_lessdust,fig:LER_UV_high_lessdust,fig:LER_UV_low_lessdust,fig:LER_Xray_high_lessdust,fig:LER_Xray_low_lessdust,fig:LER_flaring_high_lessdust,fig:LER_flaring_low_scaleheightfix_run5_lessdust,fig:LER_settling_high_lessdust,fig:LER_settling_low_lessdust} show each of the line-emitting regions in more detail, along with the continuum and line optical depths and vertically-summed line flux, for the standard and lessdust model series respectively.
		
		\cem{C2H2} has two distinct layers of emission: one at around $z/r=0.1$ and another at around $z/r=0.2$. The lower layer is optically thick, and the upper layer is optically thin. \cem{HCN} has a somewhat weaker upper layer of emission, mostly visible in the ``lessdust'' models. \cem{CO2} has the least complicated line-emitting area, which in every case is relatively rectangular and the shape and size of its line-emitting region is more robust against changes in the disk parameters than the other species.
		
		For all three molecules, the most significant dependencies are on the flaring angle and the dust mass. Increasing the flaring angle also increases the radius and height of the line-emitting region, thus also increasing the emitting area as a whole. Increasing the UV flux has a similar effect. Decreasing the dust mass also shows the same effects, and because of reduced optical depths the boundaries of the line-emitting region in the $1000{:}1$ scenario become more sensitive to the other parameters.

		\subsection{Sensitivity to gas and dust temperatures}
		
		We observe a significant difference between the gas and dust temperatures in the \cem{CO2} line-emitting regions. For most models with a gas-to-dust ratio of $100$, the gas in the line-emitting region is about $50~\kelvin$ warmer than the dust. For models with a gas-to-dust ratio of $1000$, this temperature difference increases to about $200~\kelvin$ (see  \cref{fig:CO2_lessdust_lineemission}). The reason for this is that the disk model with less dust has significantly lower continuum optical depths (at $20~\micron$) in the line-emitting regions, allowing the radiation-driven chemical heating of gas to be more effective and thus the dust and gas temperatures decouple further. At continuum optical depths $\av > 2$, the assumption that $\tgas = \tdust$ remains valid.
		
		This decoupling between the gas and dust temperatures is observed once again when changing the UV fluxes: by decreasing the UV flux by a factor of $10$ in the ``UV low'' model, the gas and dust temperatures are effectively equal in the line-emitting regions. In the $1000{:}1$ low-UV scenario the difference between gas and dust temperatures in the \cem{CO2} line-emitting region is $65~\kelvin$, as opposed to almost $200~\kelvin$ in the standard $1000{:}1$ model. This shows that the decoupling of gas and dust temperatures in this region is primarily a result of UV-driven heating processes such as photoelectric heating, heating by photo-dissociation of \cem{H2}, and chemical heating through exothermic reactions triggered by a UV photon.
		
		The other trend visible in the \cem{CO2} temperature results is that both models with high and low degrees of flaring have $\tgas-\tdust$ values that are lower than the standard model. This can be explained in the ``flaring low'' model by the fact that the line-emitting area is located closer towards both the inner rim and the mid-plane of the disk.  The greater continuum optical depth of the line-emitting region causes the gas and dust temperatures to couple together more tightly.  By comparison, although the line-emitting region in the ``flaring high'' model sits at larger radii and further above the mid-plane, the optical depth in the line-emitting region is also slightly greater than in the standard model.

		\cem{HCN} and \cem{C2H2} display much more variable sensitivities. For the models with a gas-to-dust ratio of $100$, we see the same trends as for \cem{CO2} when changing the UV radiation, indicating that the UV flux is again very important when determining the temperature of the line-emitting area. Likewise with \cem{CO2}, the gas and dust temperatures couple more tightly when both increasing and decreasing the flaring angle.  For interpreting observations, the sensitivities of \cem{C2H2} and \cem{HCN} may prove useful: on the other hand, correctly interpreting those sensitivities requires a more accurate model of the underlying disk structure and chemistry.
		
		The situation changes significantly in the ``lessdust'' scenario. Here, all models show strongly coupled gas and dust temperatures in the \cem{HCN} and \cem{C2H2} line-emitting regions. We show that the reason for this is that the lines are emitted from lower heights in the disk (in terms of $z/r$) than the models with $g/d=100{:}1$. Although increasing the gas-to-dust ratio also increases the temperature difference between \tgas{} and \tdust{} in the upper disk, the molecules also respond to this difference and change their line-emitting areas. The outcome is that \tgas{} and \tdust{} are more tightly coupled.

		\subsection{Spectra from the model series}
		
		The previous discussions rely on vertical escape probability line flux calculations. Because of the strictly vertical nature of these calculations, they do not accurately represent an inclined disk. The changes seen in the spectral flux densities in the \flits{} spectra in response to varying disk parameters are qualitatively similar to the changes in escape probability line fluxes, and such calculations are useful to determine from where in the disk a line originates. However, in order to calculate a full synthetic mid-IR spectrum of an inclined disk it is necessary to use a different technique. We use \flits{} to calculate \cem{CO2}, \cem{HCN}, \cem{C2H2}, \cem{OH}, \cem{H2O}, and \cem{NH3} spectra of our model series, where our disks are inclined at an angle of $45^\circ$. \Cref{fig:all_spectra,fig:all_spectra_lessdust} show select regions of the \flits{} spectra of individual molecules for each model, for a disk inclined at $45^\circ$ and convolved to a \jwst{}-like resolution of $R=2\,800$.
		
		\cem{CO2} is remarkably robust across the parameter changes, except for when the gas-to-dust ratio is increased to $1000$. In this case, the \cem{CO2} fluxes increase by about a factor of 20. \cem{H2O} also responds similarly.
		
		Some other trends we can observe are as follows. For both the $100{:}1$ and $1000{:}1$ cases, and for every species, the line fluxes increase in the high UV model by factors of a few to ten as compared to the low UV model. Interestingly, the \cem{C2H2} flux density increases even further for the UV low model, however the flux density of every other species decreases. We see no significant differences when changing the X-ray fluxes for the $100{:}1$ cases, except for OH which is fairly sensitive to X-rays. The \cem{C2H2} flux density stays almost constant. When the dust-to-gas ratio is increased to $1000{:}1$, the \cem{C2H2} flux density increases by a factor of 10 between the low and high X-ray cases, while the \cem{OH} flux density increases by a factor of 5.
		
When increasing the flaring angle, we almost universally see increases in the flux densities by a factor of a few to ten (compared to the low flaring model), depending on the species. The exceptions to this are in the $100{:}1$ models, where the HCN flux density decreases slightly and \cem{NH3} turns into absorption. These changes are likely attributable to temperature gradients in the line-emitting regions, because the escape-probability fluxes of individual lines  increase as we would expect. Similarly, we see decreased line fluxes when decreasing the flaring indices.  Finally, changing the turbulence parameter (which affects dust settling) makes no significant differences to the \flits{} spectra. Although settling only has a minor effect on the dust in these line-emitting regions, in a forthcoming paper we examine the effects of dust evolution and find that the spectra can indeed be greatly affected by the dust.
		
		\subsection{Absorption lines}
		\label{sec:flitspaper_absorptionlines}

		Although there are only a few known cases of disks with absorption features in their mid-IR molecular lines, it is a known phenomenon that has no definite explanation. Absorption lines are visible when gas absorbs background continuum radiation. The geometry of these systems remains unclear: it is sometimes argued that absorption lines are a sign of a highly-inclined disk. However, if the outer disk is flared then we find when varying the disk inclination, the outer disk will very rapidly occlude the mid-IR emitting regions and then produce no lines at all. 
		
		DG Tau B is one such case, where \cem{CO2} absorption lines have been detected \citep{Kruger:2011iw}. \citet{Eisloffel:1998cg} find that the jet of DG Tau B is likely highly inclined $i>65^\circ$, while  \citet{Kruger:2011iw} report that SED fitting (including envelope accretion) is unable to constrain its inclination. Thus, assuming the jets are perpendicular to the disk, an inclination of $i<35^\circ$ is likely. Another known disk with absorption features is GV Tau N, which has strong near-IR silicate absorption features that are indicative of a high inclination, as well as \cem{C2H2}, \cem{HCN}, and \cem{CO2} mid-IR absorption lines  \citep{Doppmann:2008bx,2013AA...551A.118B}. Finally, IRS 46 (also known as GY 274 or YLW 16b) has very strong \cem{C2H2}, \cem{HCN}, and \cem{CO2} absorption lines. The inclination of IRS 46 has been fitted as $75^\circ$  \citep{Lahuis:2006dg}.  Our models suggest another possible explanation: that in some disks, a cloud of cooler gas can sit above warmer gas in the line-emitting region, thus resulting in the absorption lines we see while not requiring a high inclination.
		
		\Cref{fig:all_spectra_inc0} shows the spectra of our model series, at a gas-to-dust ratio of $100{:}1$, for a face-on disk (that is, the inclination in \flits{} has been set to $i=0^\circ$). For all models except the ``Xray low'' model, there is a significant component of absorption in \cem{C2H2}. Note also that  \cref{fig:C2H2_lineemission} shows that for each model except ``Xray low'', the gas temperature of the line-emitting region is lower than the dust temperature.
		For \cem{HCN} and \cem{NH3}, only the ``UV low'', ``flaring low'', and ``flaring high'' models have an absorption component to the spectra. \Cref{fig:HCN_lineemission} shows that all three of these models have $\tgas < \tdust$ in the \cem{HCN} line-emitting region. However, \cref{fig:NH3_lineemission} does show $\tgas > \tdust$ in the line-emitting region of the ``flaring low'' and ``flaring high'' models. 
		Although there is a very small component of absorption visible in \cem{CO2} for the ``flaring low'' model, in general the other species are fully in emission.
		
		Usually, we expect the gas in the IR-emitting surface layers of the disk to be warmer than the dust. However, around $\av=1$ the gas temperature undershoots the dust temperature \citep{Kamp:2004cy}. We can trace the absorption lines that we see back to this temperature gradient across the line-emitting regions. This gradient occurs such that a cloud of colder gas is located above the warmer gas, and is visible in both the gas temperature isotherms in  \cref{fig:tgasundershoot} and the line-emitting regions plotted in 
		\cref{fig:LER_TT_highres,fig:LER_flaring_high}.
		We suggest that the cause of the gas temperature undershoot is efficient cooling of the gas through mid-IR lines:  disk densities are too low for the gas and dust temperatures to strongly couple together through collisions \citep{Kamp:2004cy}. This scenario is supported by the ``lessdust'' model, which has lower dust densities and thus an even greater region of cold gas in the upper layers of the inner disk. Enabled by the relatively optically-thin environment, the mid-IR molecular lines are an efficient cooling mechanism. The line-emitting regions of these species, particularly \cem{CO2}, \cem{HCN}, \cem{NH3}, and \cem{H2O}, are always co-spatial with the gas temperature undershoot (see \cref{fig:LER_TT_highres,fig:LER_UV_high,fig:LER_UV_low,fig:LER_Xray_high,fig:LER_Xray_low,fig:LER_flaring_high,fig:LER_flaring_low_scaleheightfix_run5,fig:LER_settling_high,fig:LER_settling_low,fig:LER_TT_highres_lessdust,fig:LER_UV_high_lessdust,fig:LER_UV_low_lessdust,fig:LER_Xray_high_lessdust,fig:LER_Xray_low_lessdust,fig:LER_flaring_high_lessdust,fig:LER_flaring_low_scaleheightfix_run5_lessdust,fig:LER_settling_high_lessdust,fig:LER_settling_low_lessdust}).
		Absorption by colder gas in the upper layers of the line-emitting region is very likely the cause of absorption lines in these models. This effect could also apply to some of the other known cases of mid-IR absorption lines in T~Tauri disks.

		\subsection{Dust settling}
		
		\citet{Salyk:2011jz} and \citet{Pontoppidan:2010gw} suggest that mid-IR line fluxes are likely to be stronger in disks with high levels of dust settling. High levels of settling can be inferred by observing the spectral index of the disk between $13~\micron$ and $30~\micron$ \citep{KesslerSilacci:2006gf,Furlan:2006bb}. However, we see only small differences in the spectra and line-emitting regions when increasing or decreasing the Dubrulle settling coefficient by a factor of 10.  The Dubrulle 
		settling does not appear to have a significant influence on the distribution of sub-micron dust grains
		in the planet-forming regions of protoplanetary disks, where gas densities are high (however, the ``turbulence low'' models do have a steeper SED slope between $13~\micron$ and $30~\micron$). \Cref{fig:settling_comparison} shows the effect of changing the amount of settling; note particularly that the $\av=1$ and $\av=10$ contours scarcely move. Sub-micron dust grains are the main carriers of opacity in the mid-IR, thus it follows that the effects of such settling on the mid-IR lines are minimal. \cite{Antonellini:2017bu} find similar results, that settling does not generally affect the mid-IR water lines. However, a simple settling description is not sufficient for describing the full variety of dust distributions possible in a disk: in a forthcoming paper, we examine the effects of dust evolution on the mid-IR lines, where much more dramatic changes in the dust size distribution can have an equally dramatic effect on the mid-IR lines.

		\section{Conclusions}
		
		We have created a series of T~Tauri disk models, in order to examine how changes in the radiation environment affect the mid-IR spectral lines.  The UV and X-ray fluxes, the flaring angle, the level of dust settling, and the dust-to-gas ratio may vary significantly between different T~Tauri disks. In this paper, we analyze how the line-emitting regions and spectra of \cem{C2H2}, \cem{HCN}, \cem{NH3}, \cem{OH}, \cem{H2O}, and \cem{CO2} change in response to these parameters.
		
		We find that the gas and dust temperatures can vary significantly across the spatial extent of a line-emitting region, and it is important to calculate these temperatures separately and not to assume that they are coupled or that there is a fixed temperature difference that fits every case. The separation between $\tgas$ and $\tdust$ can be $200~\mathrm{K}$ or more, depending on the model and the species' line-emitting region. 
		
		Likewise, the line-emitting areas are not stable across the parameter space. This area can change significantly, particularly in response to the UV flux, the dust-to-gas ratio, and to the flaring angle.
		
		The radial regions and vertical layers differ not only by molecule, but also by line: each of the thousands of molecular lines emits from a slightly different region of the disk. These effects cannot be 
		captured properly by slab-models, but require a full $2$D modelling approach. 
		
		The mid-IR line-emitting regions can span a large range of radii, heights, and gas and dust temperatures.  The gas and dust temperatures can also vary strongly within the line-emitting region, and a temperature inversion can exist where cold gas sits on top of warmer gas due to efficient line cooling. This result highlights how important it is that future modelling efforts capture the vertical structure of the disk, because the temperature structure of the line-emitting region can have a significant effect on the resulting fluxes.
		
		These models demonstrate the difficulties in interpreting observations by calculating or assuming a line-emitting area.  Although the escape probability fluxes of an individual line may respond significantly to changes in disk parameters, the flux densities measured when using \flits{} to calculate the spectrum of an inclined disk are generally less responsive, thus small changes to disk parameters have little effect on the spectrum.  The ability of \elt{} to spatially resolve some disks in the near- to mid-IR will prove extremely useful in breaking these degeneracies. With further large-scale modelling efforts, it may be possible to develop an approach to parametrize these results for use in $1$D models, which would be significantly faster to compute and could feasibly observational data to a wide range of model parameters.

		\section{Acknowledgements}
		
		We would like to thank the Center for Information Technology of the University of Groningen for their support and for providing access to the Peregrine high performance computing cluster, and Ilaria Pascucci for the discussions that led to the research in this paper.

		\bibliographystyle{aa}
		\bibliography{mylocalbib,fullbib_local}

		\FloatBarrier
		\appendix
		
		\section{Supplementary data}

		In \cref{tab:speclist}, we list all of the species included in the \flits{} models. This is not an exhaustive list of all of the species in the \prodimo{} model, but only of the subset for which level populations are calculated.  \Cref{fig:CO2_lineemission,fig:HCN_lineemission,fig:C2H2_lineemission,fig:H2O_lineemission,fig:NH3_lineemission,fig:OH_lineemission,fig:CO2_lessdust_lineemission,fig:HCN_lessdust_lineemission,fig:C2H2_lessdust_lineemission,fig:H2O_lessdust_lineemission,fig:NH3_lessdust_lineemission,fig:OH_lessdust_lineemission} show how the line-emitting regions respond to the parameter changes listed in \cref{tab:modelseries} (the latter figures refer to the ``lessdust'' configuration).  In  \cref{fig:LER_TT_highres,fig:LER_UV_high,fig:LER_UV_low,fig:LER_Xray_high,fig:LER_Xray_low,fig:LER_flaring_high,fig:LER_flaring_low_scaleheightfix_run5,fig:LER_settling_high,fig:LER_settling_low,fig:LER_TT_highres_lessdust,fig:LER_UV_high_lessdust,fig:LER_UV_low_lessdust,fig:LER_Xray_high_lessdust,fig:LER_Xray_low_lessdust,fig:LER_flaring_high_lessdust,fig:LER_flaring_low_scaleheightfix_run5_lessdust,fig:LER_settling_high_lessdust,fig:LER_settling_low_lessdust},  we plot the line-emitting regions of each line that we analyse in the paper, for every model.

		\begin{table*}
			\caption{A list of species included in the \flits{} models. For non-LTE molecules, the level populations are computed using escape probabilities \citep{Woitke:2009jf}. The transition data for the LTE species come from the HITRAN database (denoted by "\_H"). Selection rules are used to limit the number of ro-vibrational lines selected from HITRAN, for computational reasons  \citep{Woitke:2018ux}.}
			\centering
			\small \medskip
			\begin{tabular}{l c >{$}r<{$} >{$}r<{$} |  l c >{$}r<{$} >{$}r<{$}}
				Species  & treatment & \mathrm{\#levels} & \mathrm{\#lines}  & Species  & treatment & \mathrm{\#levels} & \mathrm{\#lines} \\ \midrule
				\cem{C+      }   &   non-LTE       &    18  &    57    & 						\cem{H       }   &   non-LTE       &    25  &    75   \\
				\cem{O       }   &   non-LTE       &     3  &     3    & 						\cem{HNC     }   &   non-LTE       &    26  &    25   \\
				\cem{CO      }   &   non-LTE       &    41  &    40    & 						\cem{o-NH3   }   &   non-LTE       &    22  &    24   \\
				\cem{O       }   &   non-LTE       &    91  &   647    & 						\cem{p-NH3   }   &   non-LTE       &    24  &    28   \\
				\cem{C       }   &   non-LTE       &    59  &   117    & 						\cem{Ar+     }   &   non-LTE       &     2  &     1   \\
				\cem{Mg+     }   &   non-LTE       &     8  &    12    & 						\cem{Ar++    }   &   non-LTE       &     5  &     9   \\
				\cem{Fe+     }   &   non-LTE       &   120  &   956    & 						\cem{O++     }   &   non-LTE       &     6  &    11   \\
				\cem{Si+     }   &   non-LTE       &    15  &    35    &					        \cem{O+      }   &   non-LTE       &     5  &    10   \\
				\cem{S+      }   &   non-LTE       &     5  &     9    & 						\cem{S++     }   &   non-LTE       &     5  &     9   \\
				\cem{o-H2    }   &   non-LTE       &    80  &   803    & 						\cem{Ne++    }   &   non-LTE       &     5  &     9   \\
				\cem{p-H2    }   &   non-LTE       &    80  &   736    & 						\cem{N++     }   &   non-LTE       &     2  &     1   \\
				\cem{CO      }   &   non-LTE       &   360  &  2699    & 						\cem{C^18O    }   &   non-LTE       &    41  &    40   \\
				\cem{o-H2O   }   &   non-LTE       &   411  &  4248    & 						\cem{H2CO\_H }    & LTE         & 3134 & 1567 \\
				\cem{p-H2O   }   &   non-LTE       &   413  &  3942    & 						\cem{CO2\_H  }    &  LTE       &    252  &   126   \\
				\cem{^13CO    }   &   non-LTE       &    41  &    40    & 						\cem{C2H2\_H }    &  LTE       &   1992  &   996   \\
				\cem{OH      }   &   non-LTE       &    20  &    50    & 						\cem{HCN\_H  }    &  LTE       &    252  &   126   \\
				\cem{SiO     }   &   non-LTE       &    41  &    40    & 						\cem{CH4\_H  }    &  LTE       &    430  &   215   \\
				\cem{NO      }   &   non-LTE       &    80  &   139    & 						\cem{NH3\_H  }    &  LTE       &   5932  &  2966   \\
				\cem{S       }   &   non-LTE       &     3  &     3    & 						\cem{OH\_H   }    &  LTE       &   2528  &  1264   \\
				\cem{CS      }   &   non-LTE       &    31  &    30    & 		 				\cem{p-H2CO  }   &   non-LTE       &    41  &   107   \\
				\cem{HCN     }   &   non-LTE       &    30  &    29    & 	                        		\cem{ N2H+   }    &   non-LTE       &    31  &    30   \\
				\cem{CN      }   &   non-LTE       &    41  &    59    & 			                        \cem{C2H     }   &   non-LTE       &   102  &   245   \\
				\cem{HCO+    }   &   non-LTE       &    31  &    30    & 			                        \cem{CO+     }   &   non-LTE       &     9  &    11   \\
				\cem{CH+     }   &   non-LTE       &    16  &    15    & 				                \cem{OH+     }   &   non-LTE       &    49  &   152   \\
				\cem{N+      }   &   non-LTE       &    23  &    86    & 				                \cem{O2      }   &   non-LTE       &    48  &    77   \\
				\cem{OH-hfs  }   &   non-LTE       &    24  &    95    & 				                \cem{o-H2S   }   &   non-LTE       &    45  &   139   \\
				\cem{o-H2CO  }   &   non-LTE       &    40  &   104    & 				                \cem{p-H2S   }   &   non-LTE       &    45  &   140   \\
				\cem{Ne+     }   &   non-LTE       &     3  &     3    & 				                \cem{HCS+    }   &   non-LTE       &    31  &    30   \\
				\cem{SO      }   &   non-LTE       &    91  &   301    & 				                \cem{E-CH3O  }   &   non-LTE       &   256  &  2324   \\
				\cem{SO2     }   &   non-LTE       &   198  &   855    & 				                \cem{A-CH3O  }   &   non-LTE       &   256  &  1853   \\
				\cem{OCS     }   &   non-LTE       &    99  &    98    & 				                \cem{C^17O    }   &   non-LTE       &    41  &    40   \\
				\cem{o-H3O+  }   &   non-LTE       &     9  &     8    & 				                \cem{O2\_H   }    &  LTE       &      0  &     0   \\
				\cem{p-H3O+  }   &   non-LTE       &    14  &    17    & 				                \cem{NO\_H   }    &  LTE       &    372  &   186   \\
				\cem{CH3OH\_H}   & LTE & 28570 & 14285 &                                                             \cem{CS\_H   }   & LTE & 14 & 7 \\
				\cem{SO2\_H  }   & LTE & 41590 & 20795 &                                                             \cem{p-C3H2  }   & non-LTE & 48 & 154 \\
				\cem{o-C3H2  }   & non-LTE & 47 & 156 \\

			\end{tabular}
			\label{tab:speclist}
		\end{table*}

		\begin{figure*}
			\centering
			\includegraphics[width=0.48\textwidth]{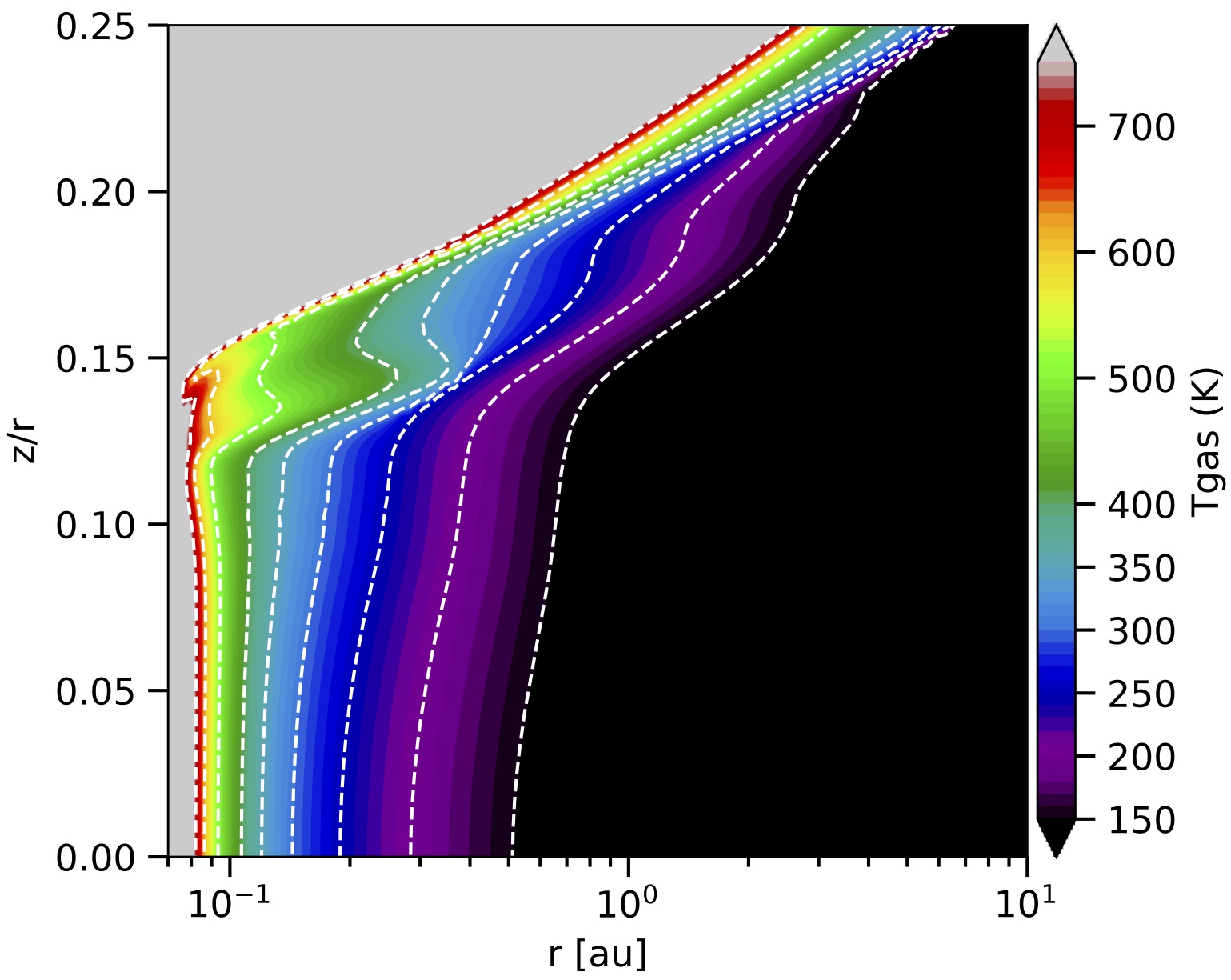} \quad
			\includegraphics[width=0.48\textwidth]{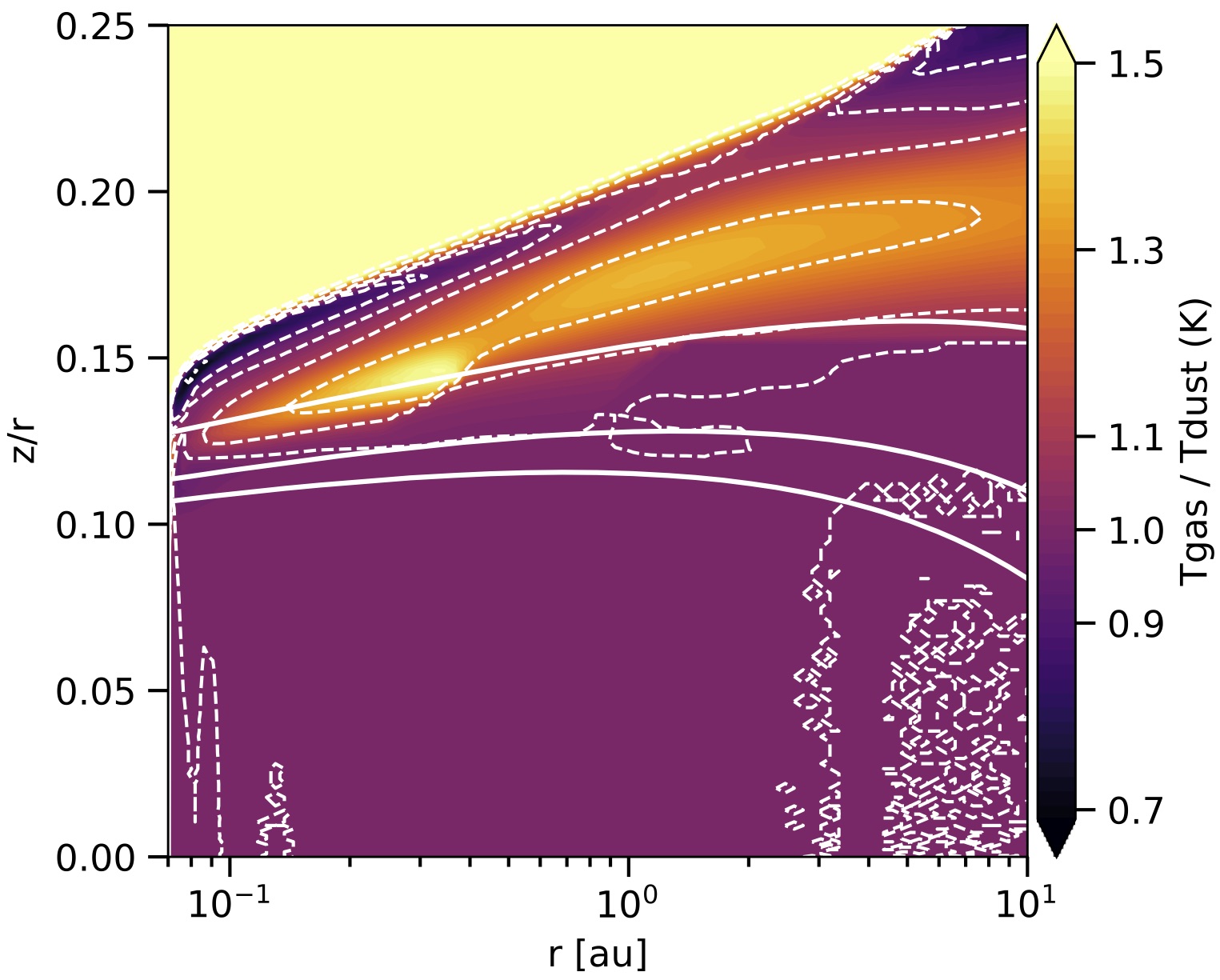} \\
			\includegraphics[width=0.48\textwidth]{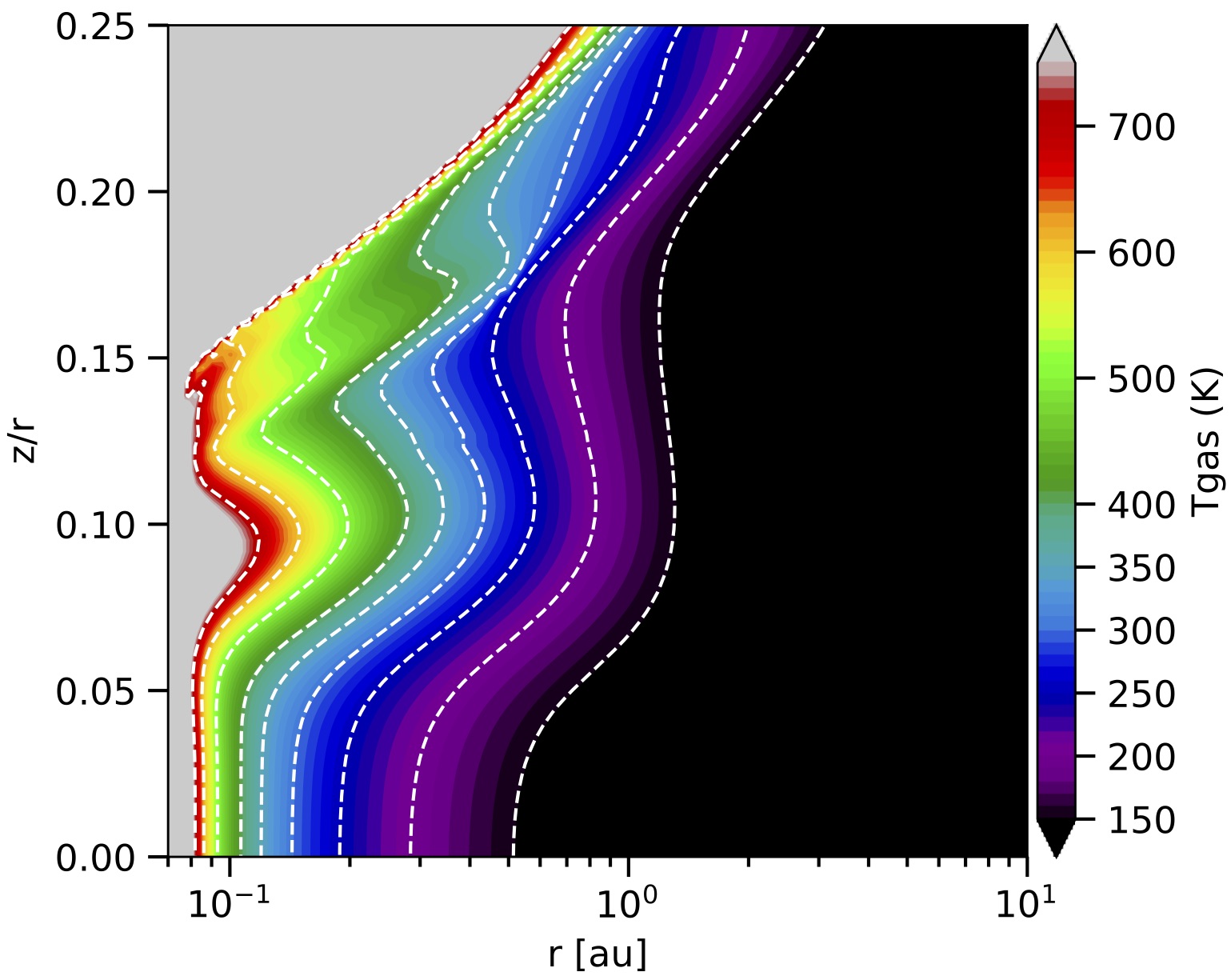} \quad
			\includegraphics[width=0.48\textwidth]{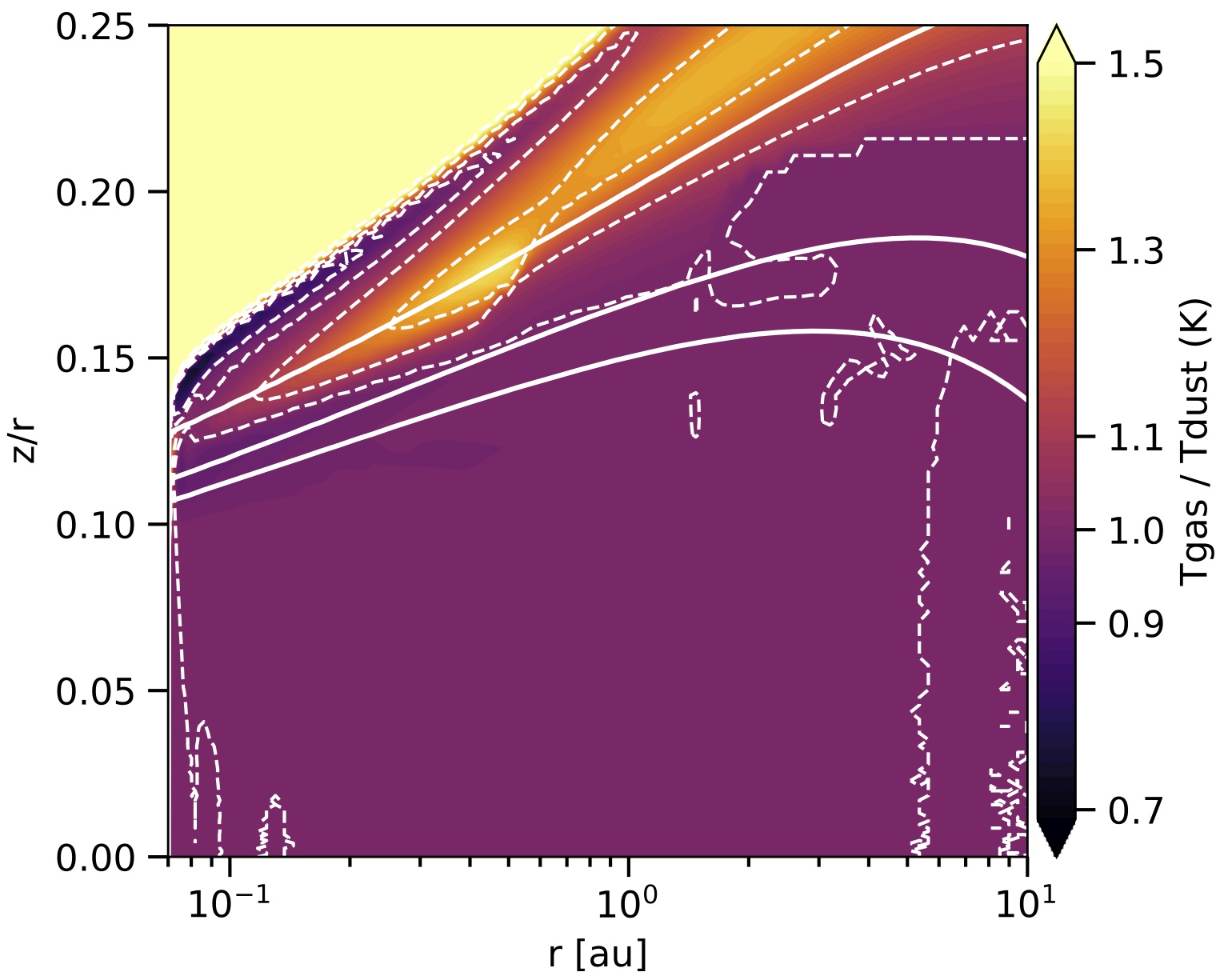} \\
			\includegraphics[width=0.48\textwidth]{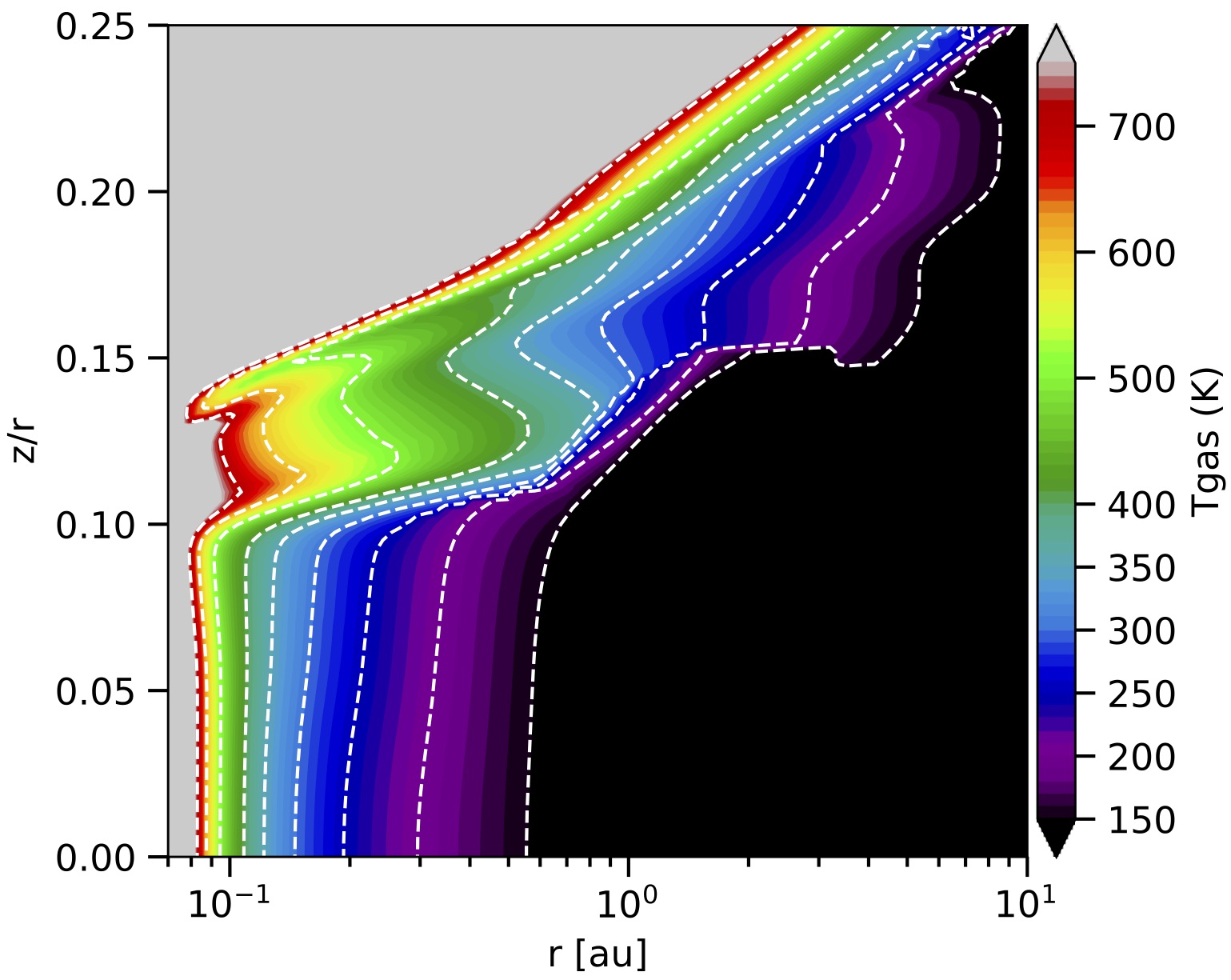} \quad
			\includegraphics[width=0.48\textwidth]{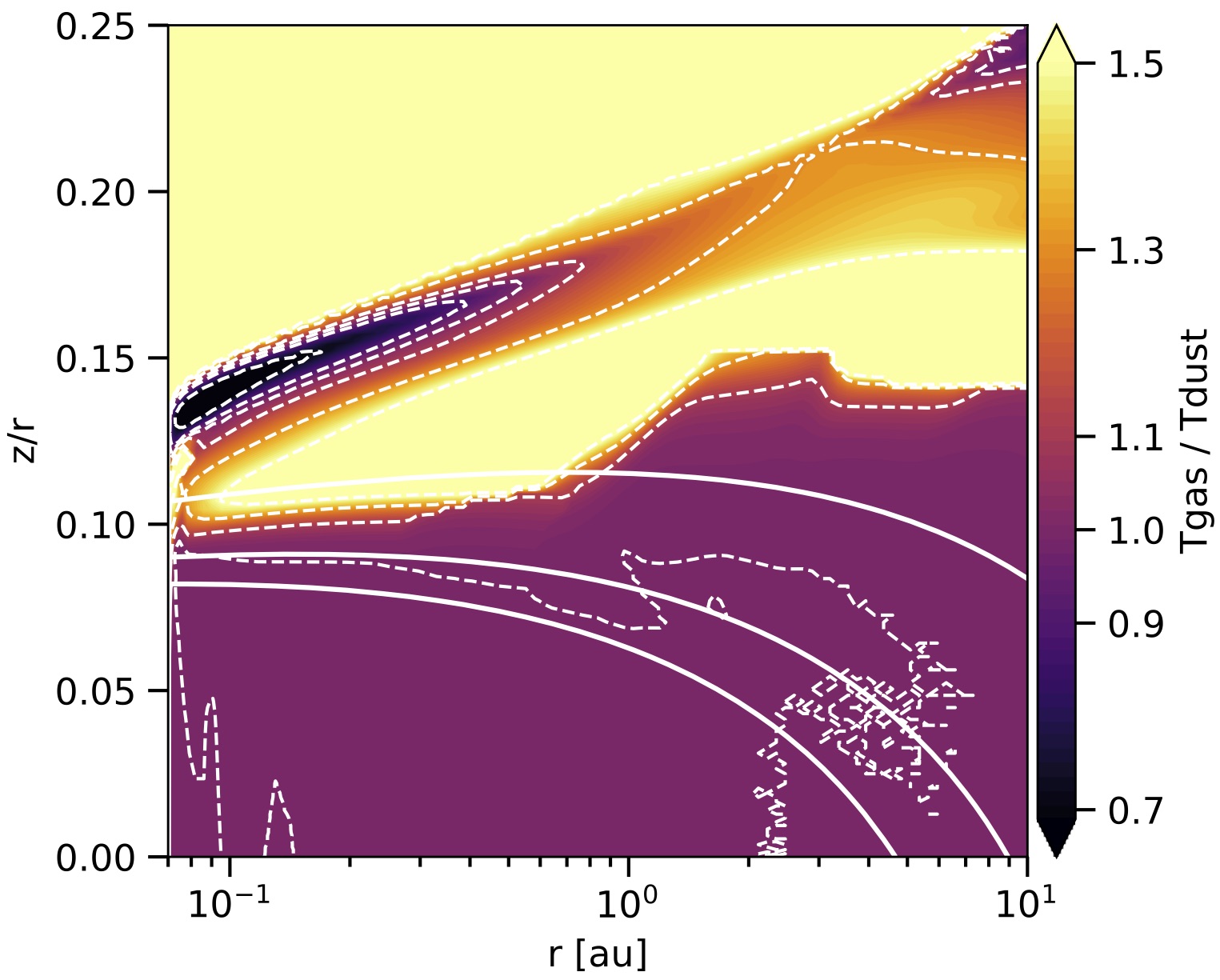} \\
			\caption{Upper left panel: gas temperature of the standard T~Tauri model. Upper right panel: ratio of gas temperature to dust temperature.  Middle two panels: same, but for the ``flaring high'' model. Bottom: ``lessdust'' model.  The colour scheme in the left-hand panels is made to exaggerate small differences. The dashed white contour lines correspond to the temperature labels on the colour bar, in the same manner as  \cref{fig:stdmodel}.  The solid white contours indicate vertical optical depths $\av=1$, $5$, and $10$.}
			\label{fig:tgasundershoot}
		\end{figure*}

		\begin{figure*}
			\centering
			\includegraphics[width=\textwidth]{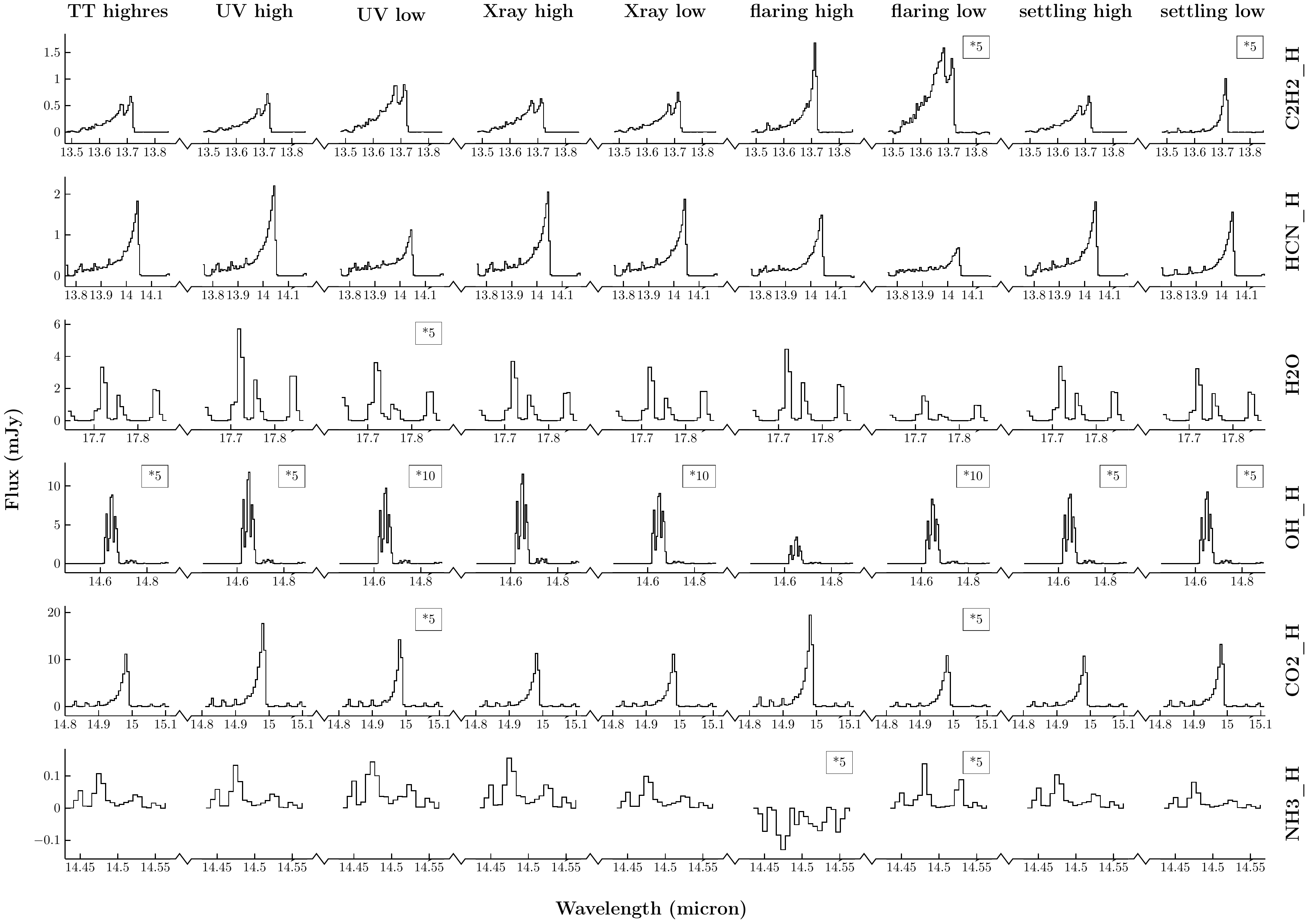}
			\caption{\flits{} spectra with a gas-to-dust ratio of $100{:}1$ at $R=2\,800$. Across each row, the indicated molecule's spectrum is plotted for every model (with constant flux and wavelength axes). As indicated, some spectra have been multiplied by a factor of $5$, $10$, or $50$.}
			\label{fig:all_spectra}
		\end{figure*}

		\begin{figure*}
			\centering
			\includegraphics[width=\textwidth]{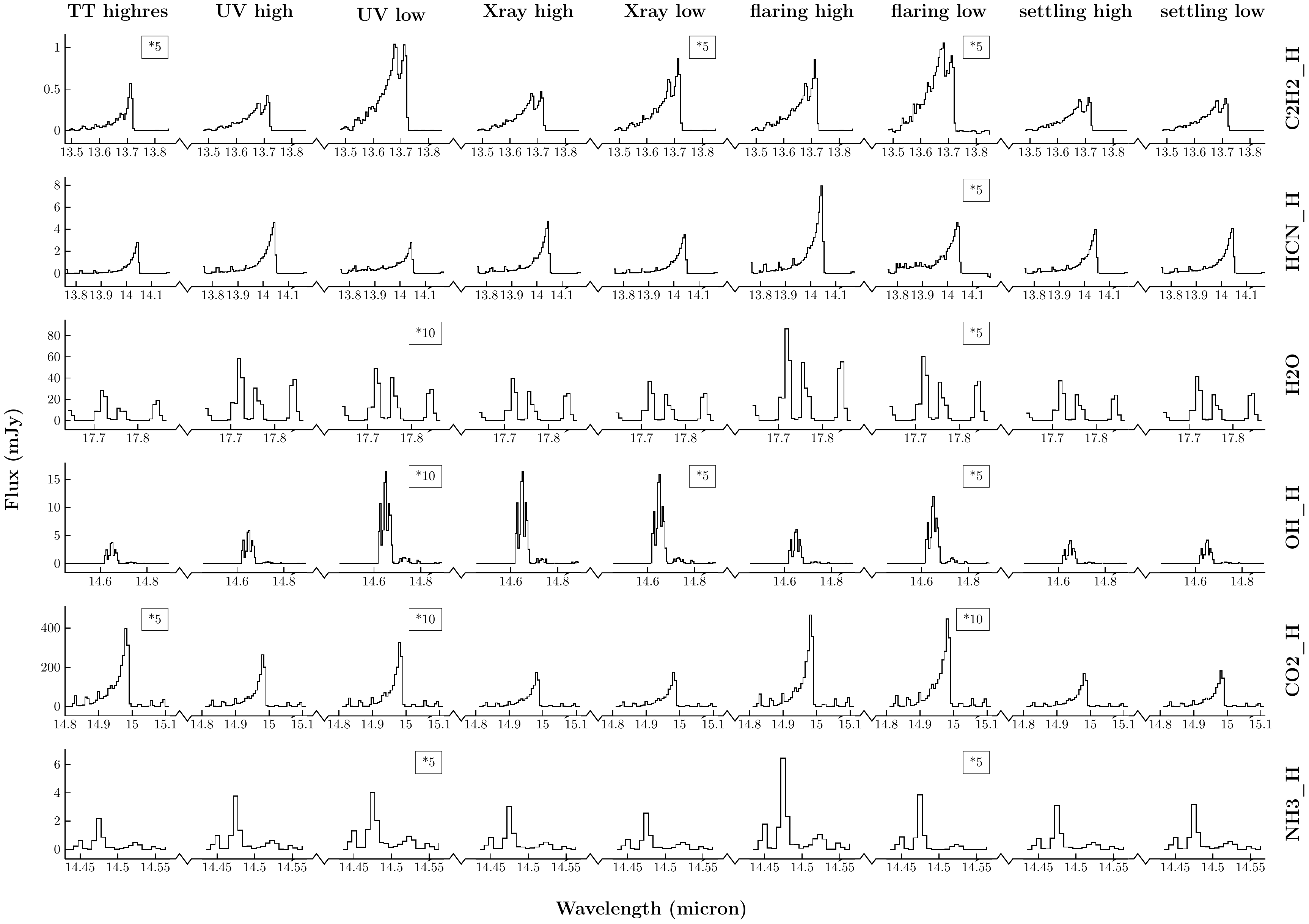}
			\caption{\flits{} spectra with a gas-to-dust ratio of $1000{:}1$ at $R=2\,800$. Across each row, the indicated molecule's spectrum is plotted for every model (with constant flux and wavelength axes). As indicated, some spectra have been multiplied by a factor of $5$, $10$, or $50$.}
			\label{fig:all_spectra_lessdust}
		\end{figure*}

		\begin{figure*}
			\centering
			\includegraphics[width=\textwidth]{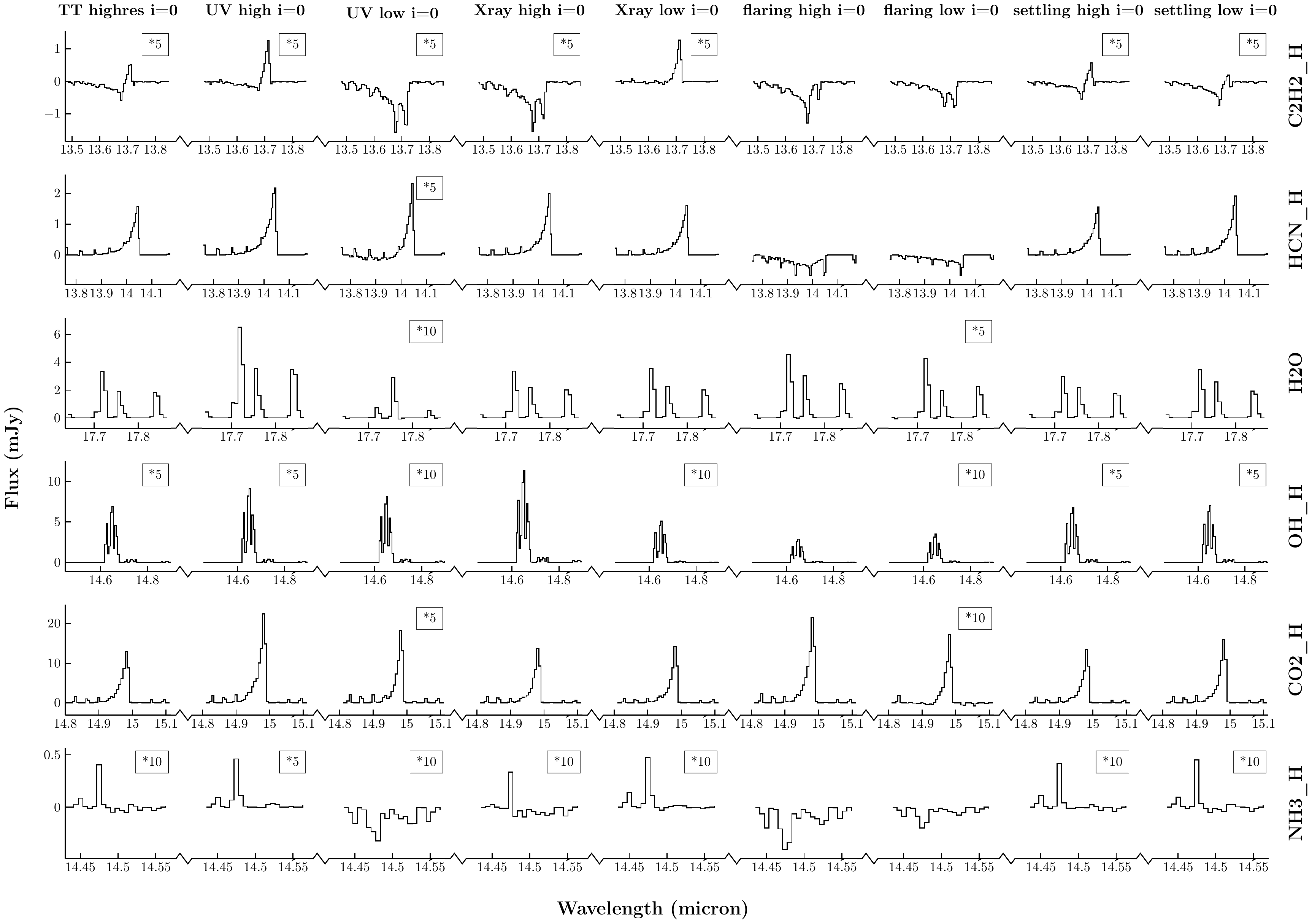}
			\caption{\flits{} spectra with a gas-to-dust ratio of $100{:}1$ at $R=2\,800$ and inclination $i=0$. The other details are the same as \cref{fig:all_spectra}. }
			\label{fig:all_spectra_inc0}
		\end{figure*}

		\begin{figure}
			\centering
			\includegraphics[width=0.47\textwidth,page=1]{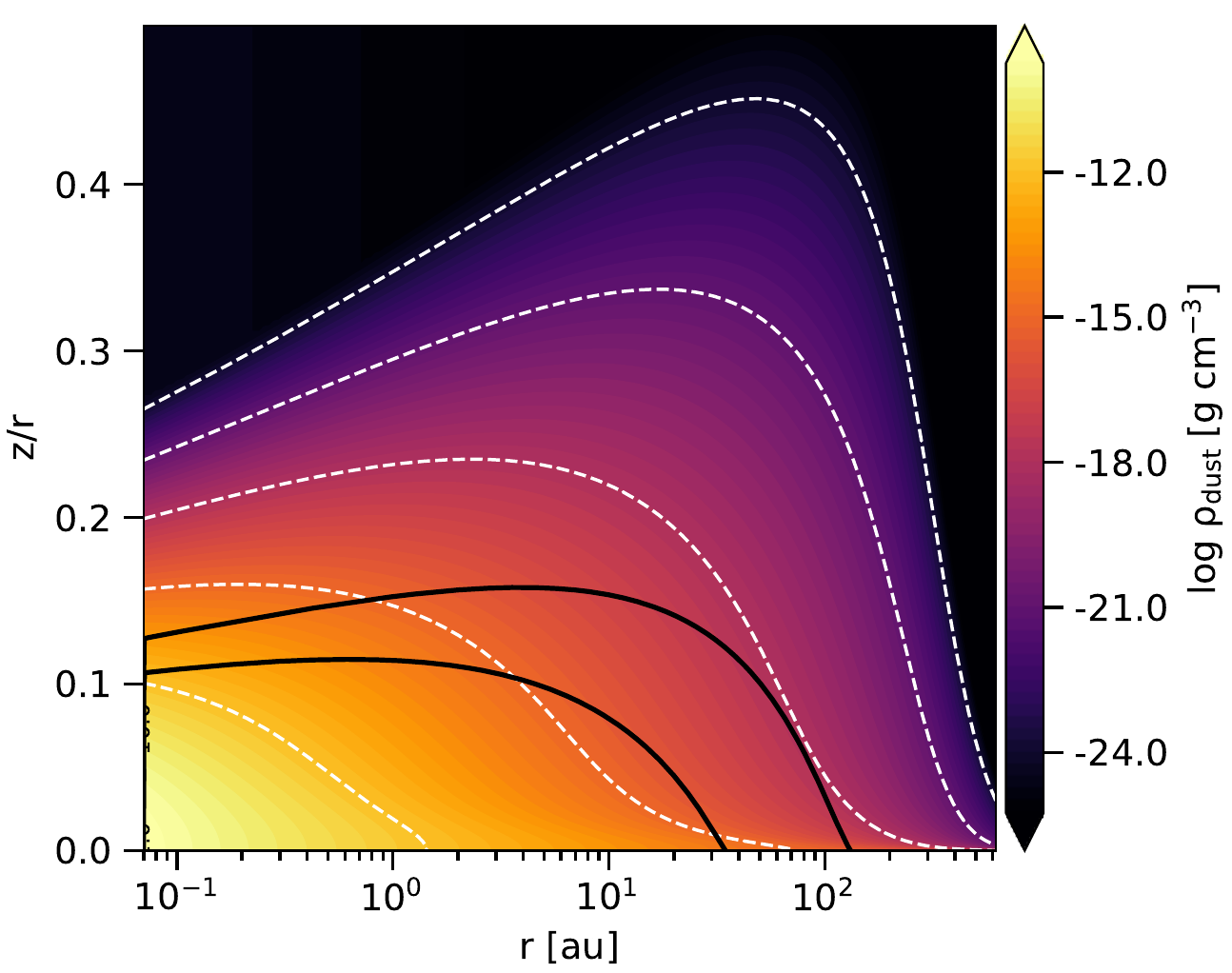}
			\includegraphics[width=0.47\textwidth,page=2]{./figures/settle_plots.pdf}
			\includegraphics[width=0.47\textwidth,page=3]{./figures/settle_plots.pdf}
			\caption{Dust surface density of the ``turbulence low'' ($\alpha= 10^{-3}$) model (top), the standard T~Tauri ($\alpha= 10^{-2}$) model (middle), and ``turbulence high'' ($\alpha= 10^{-1}$) model (bottom). The black contour lines indicate the $\av=1$ and $\av=10$ lines.}
			\label{fig:settling_comparison}
		\end{figure}

		\begin{figure*}
			\centering \large \cem{CO2($14.98$)} \par \medskip
			\includegraphics[width=\textwidth]{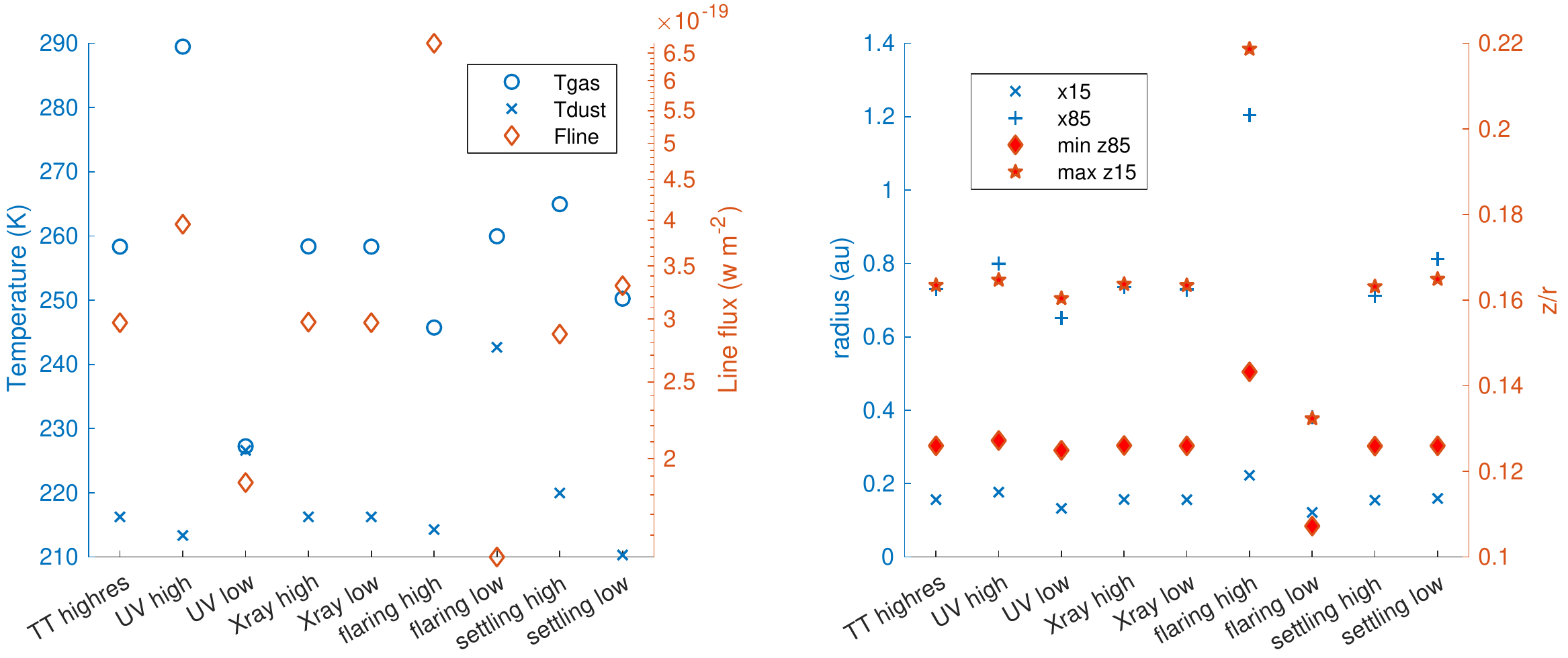}
			\caption{Properties of the line-emitting area of the \cem{CO2} line at $14.98299~\micron$, calculated using the escape probability method for our series of models with a gas-to-dust ratio of $100{:}1$. In the left-hand panel are the average gas and dust temperatures, weighted by the mass of each grid point (left-hand $y$-axis), and the difference between them (on the right-hand $y$-axis). On the right-hand panel are the inner (``\mr{x15}'') and outer (``\mr{x85}'') radial boundaries of the line-emitting region (on the left-hand $y-axis$), and the vertical boundaries of the line-emitting region (on the right-hand $y$-axis). If the line-emitting region is shaped like a box, as in  \cref{fig:TT_lineemittingregions}, then ``min \mr{z85}'' is the lower-left corner, and ``max \mr{z15}'' is the upper-right corner.}
			\label{fig:CO2_lineemission}
		\end{figure*}
		\begin{figure*}
			\centering \large \cem{HCN($14.04$)} \par \medskip
			\includegraphics[width=\textwidth]{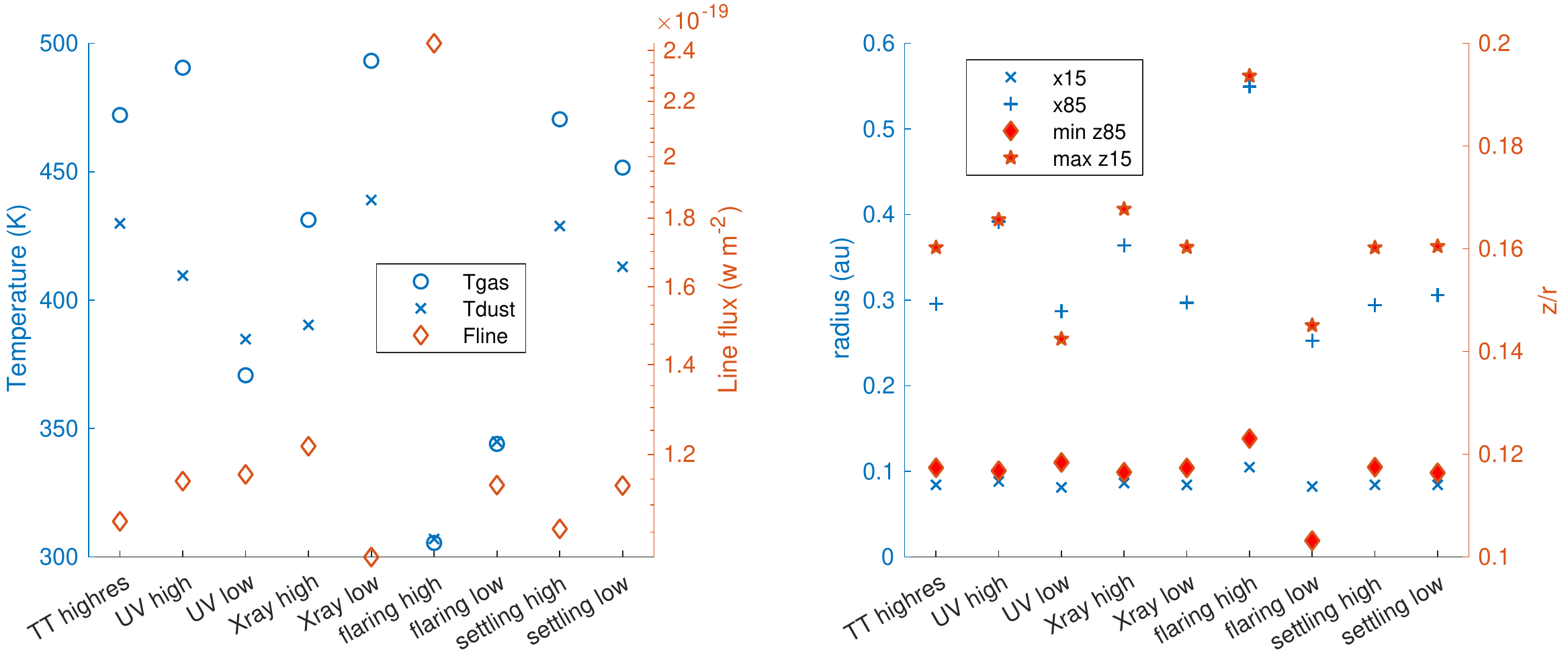}
			\caption{Properties of the line-emitting area of the \cem{HCN} line at $14.03930~\micron$, calculated using the escape probability method for our series of models with a gas-to-dust ratio of $100{:}1$. The description of each sub-figure is the same as \cref{fig:CO2_lineemission}.}
			\label{fig:HCN_lineemission}
		\end{figure*}
		
		\begin{figure*}
			\centering \large \cem{C2H2($13.20$)} \par \medskip
			\includegraphics[width=\textwidth]{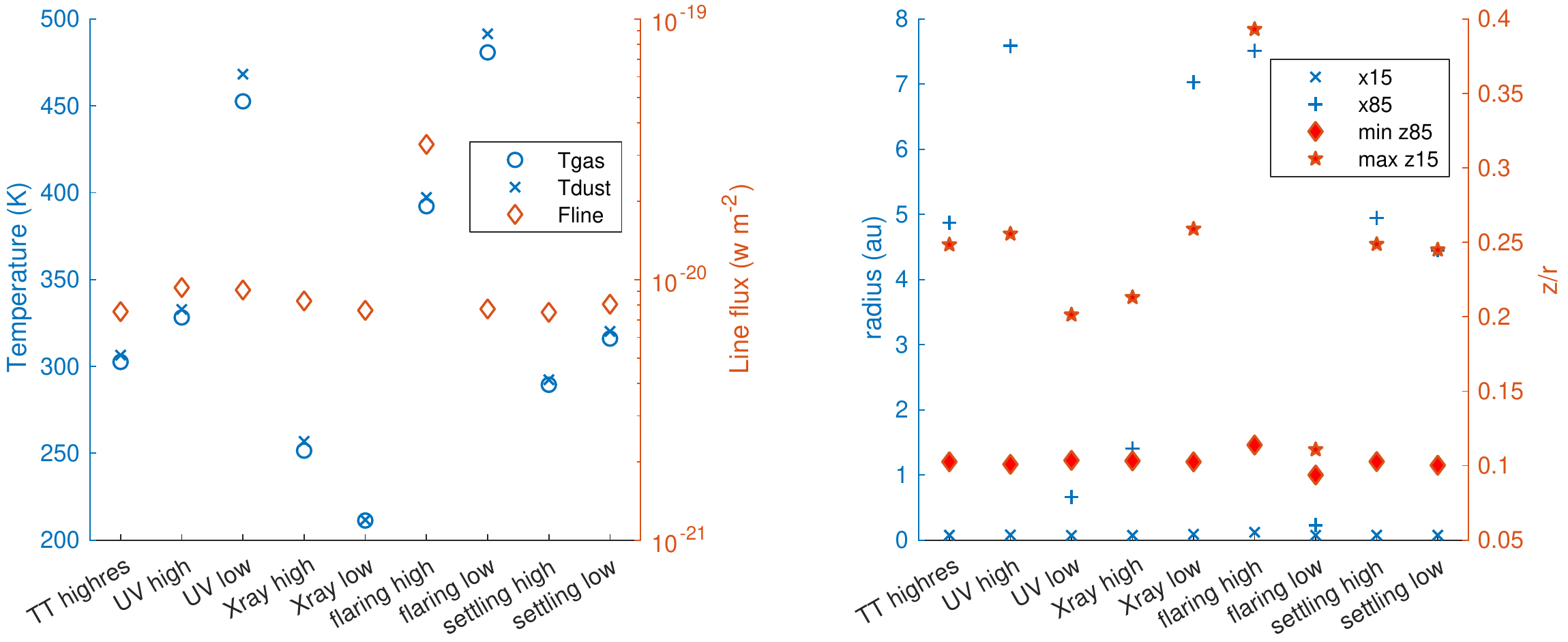}
			\caption{Properties of the line-emitting area of the \cem{C2H2} line at $13.20393~\micron$, calculated using the escape probability method for our series of models with a gas-to-dust ratio of $100{:}1$. The description of each sub-figure is the same as \cref{fig:CO2_lineemission}.}
			\label{fig:C2H2_lineemission}
		\end{figure*}

		\begin{figure*}
			\centering \large \cem{H2O($17.75$)} \par \medskip
			\includegraphics[width=\textwidth]{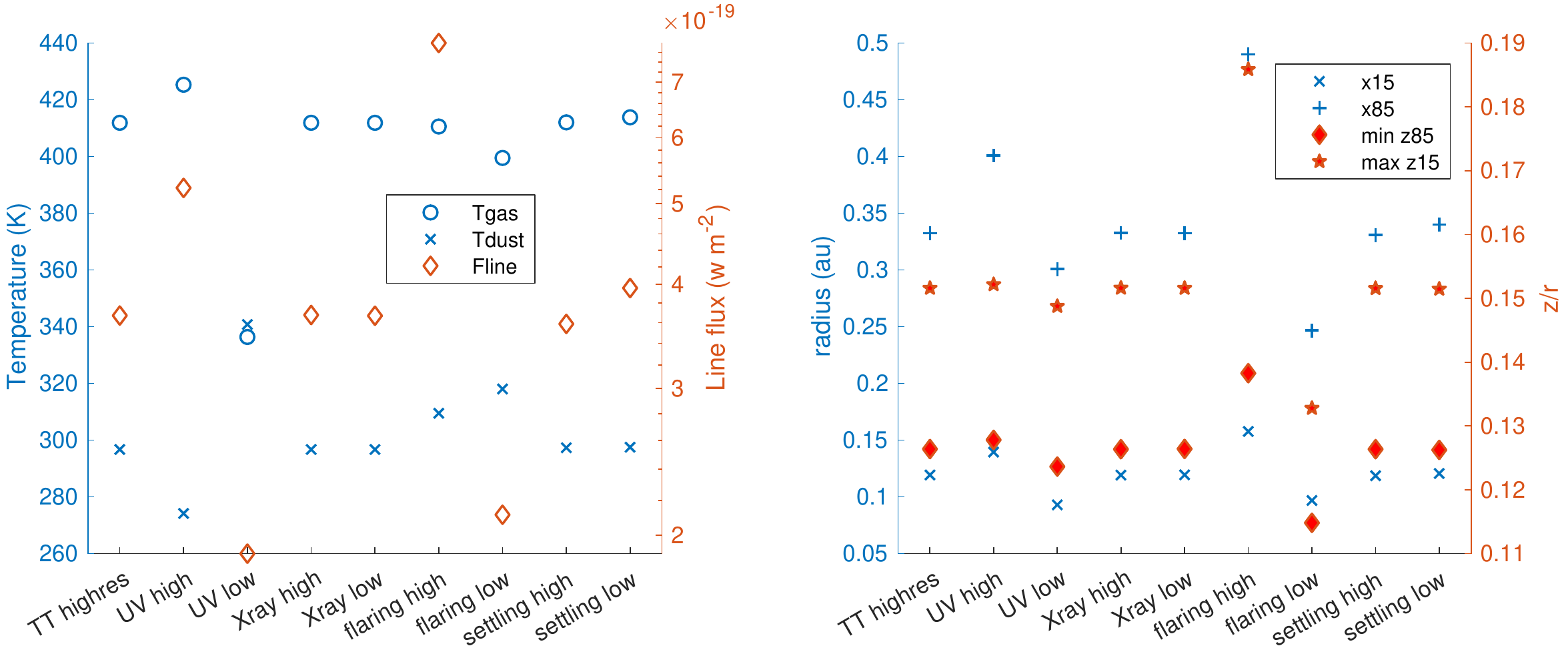}
			\caption{Properties of the line-emitting area of the \cem{H2O} line at $17.7541~\micron$, calculated using the escape probability method for our series of models with a gas-to-dust ratio of $100{:}1$.. The description of each sub-figure is the same as  \cref{fig:CO2_lineemission}.}
			\label{fig:H2O_lineemission}		
		\end{figure*}
		
		\begin{figure*}
			\centering \large \cem{NH3($10.34$)} \par \medskip
			\includegraphics[width=\textwidth]{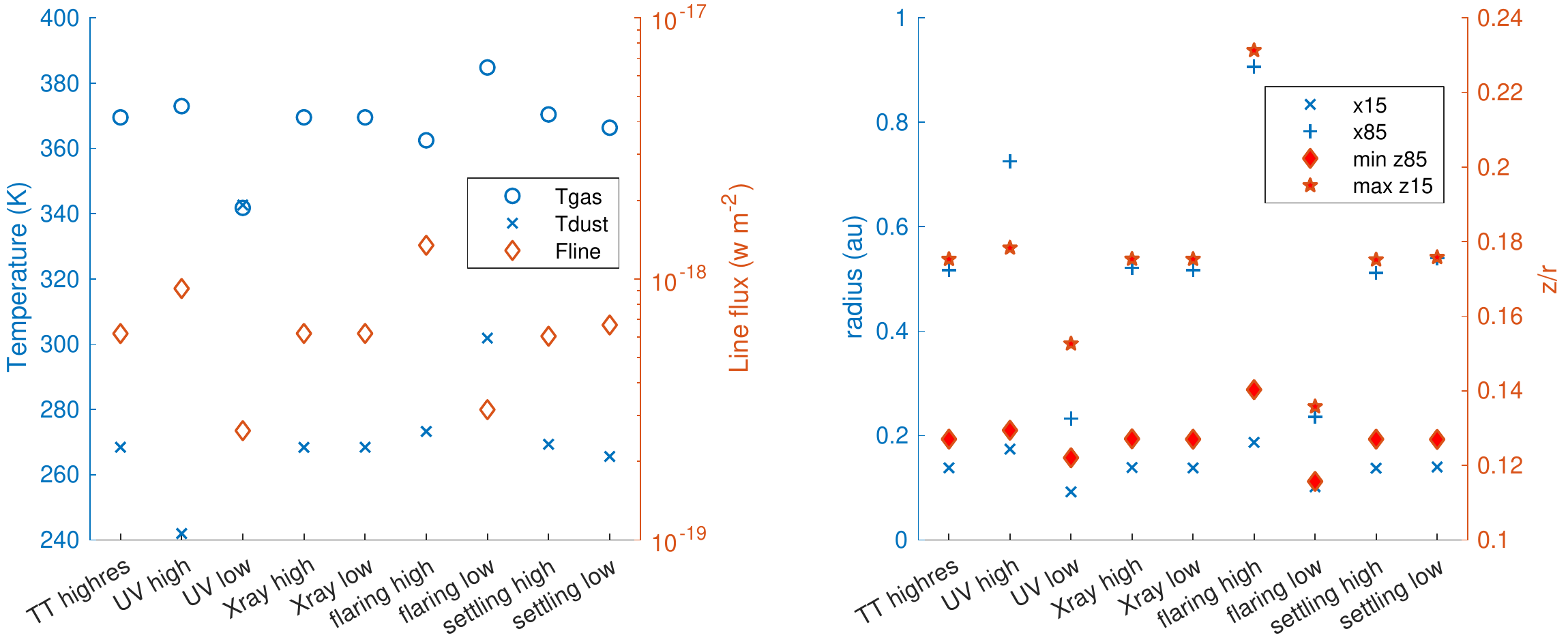}
			\caption{Properties of the line-emitting area of the \cem{NH3} line at $10.3376~\micron$, calculated using the escape probability method for our series of models with a gas-to-dust ratio of $100{:}1$. The description of each sub-figure is the same as  \cref{fig:CO2_lineemission}.}
			\label{fig:NH3_lineemission}		
		\end{figure*}
		
		\begin{figure*}
			\centering \large \cem{OH($20.12$)} \par \medskip
			\includegraphics[width=\textwidth]{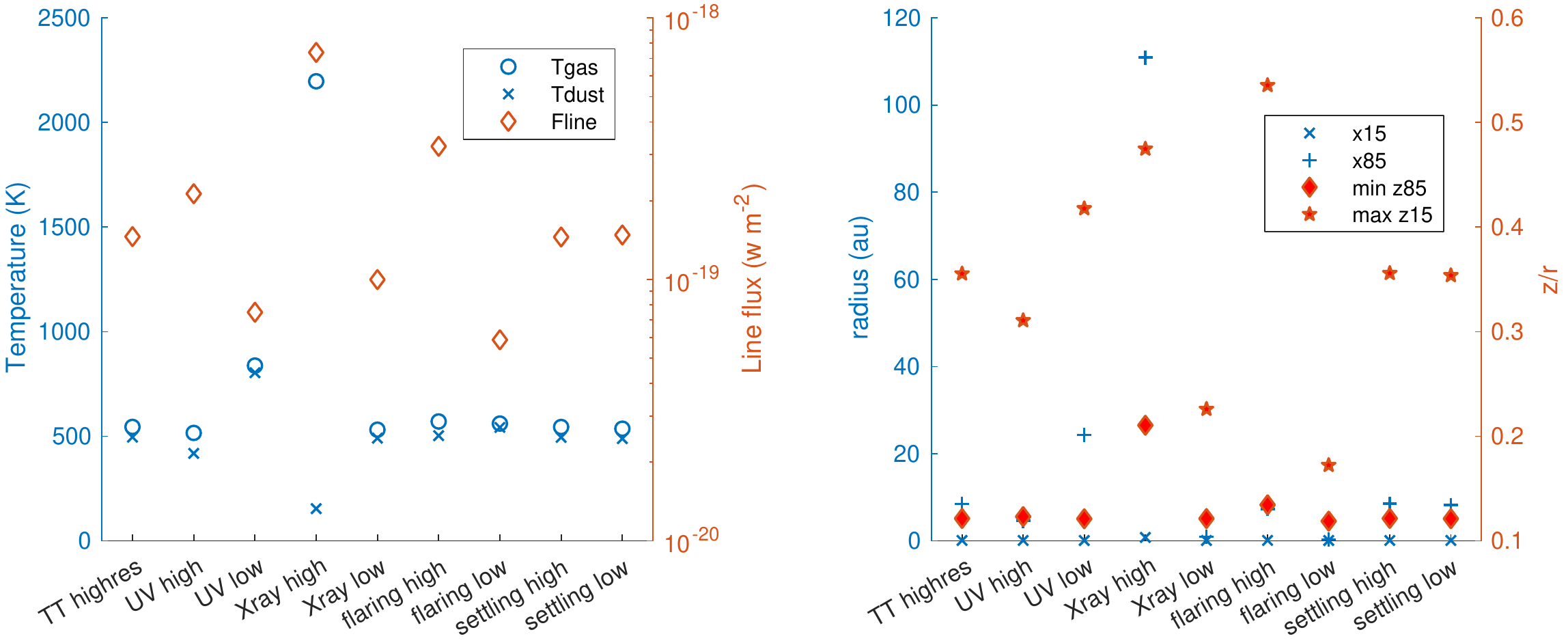}
			\caption{Properties of the line-emitting area of the \cem{OH} line at $20.1151~\micron$, calculated using the escape probability method for our series of models with a gas-to-dust ratio of $100{:}1$. The description of each sub-figure is the same as  \cref{fig:CO2_lineemission}. {Note: a few of the \tdust{} points are obscured behind the \tgas{} and \fline{} data points}.}
			\label{fig:OH_lineemission}		
		\end{figure*}

		\begin{figure*}
			\centering \large \cem{CO2($14.98$)} lessdust \par \medskip
			\includegraphics[width=\textwidth]{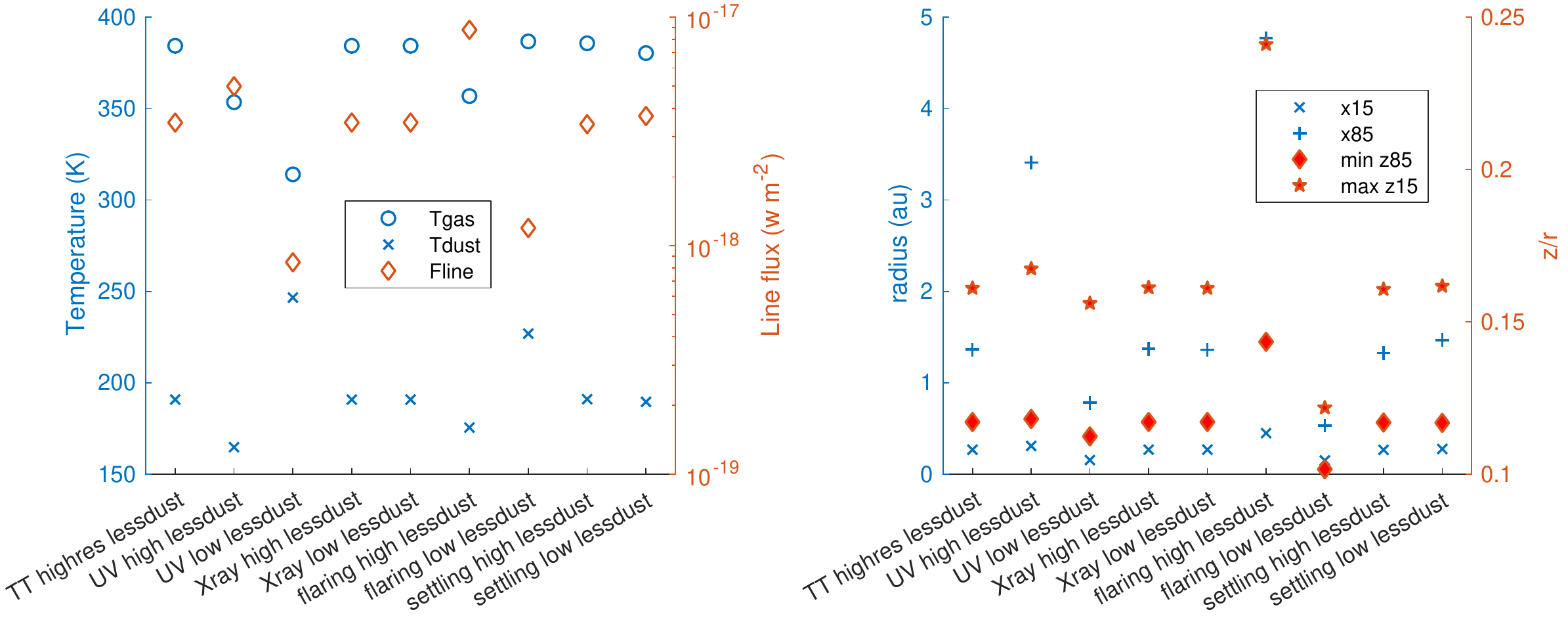}
			\caption{Properties of the line-emitting area of the \cem{CO2} line at $14.98299~\micron$, calculated using the escape probability method for our series of models with a gas-to-dust ratio of $1000{:}1$. The description of each sub-figure is the same as  \cref{fig:CO2_lineemission}.}
			\label{fig:CO2_lessdust_lineemission}
		\end{figure*}
		\begin{figure*}
			\centering \large \cem{HCN($14.04$)} lessdust \par \medskip
			\includegraphics[width=\textwidth]{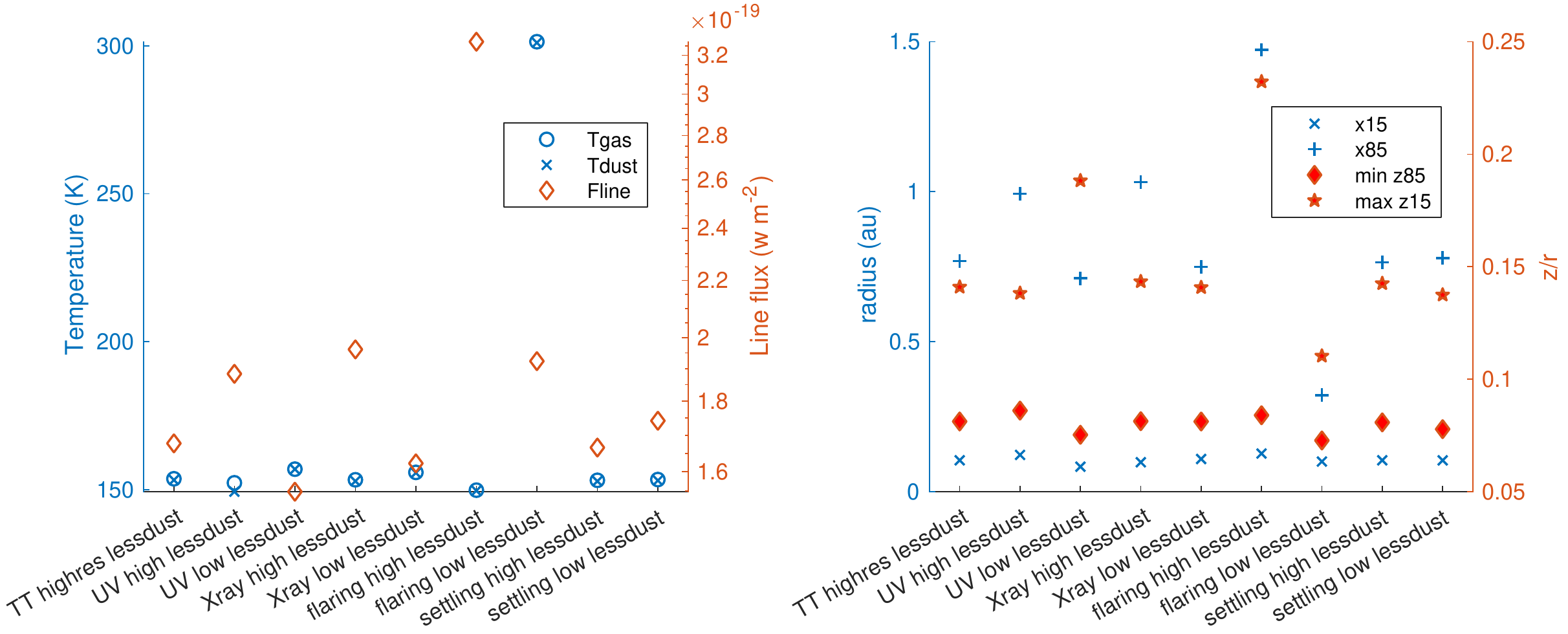}
			\caption{Properties of the line-emitting area of the \cem{HCN} line at $14.03930~\micron$, calculated using the escape probability method for our series of models with a gas-to-dust ratio of $1000{:}1$. The description of each sub-figure is the same as \cref{fig:CO2_lineemission}.}
			\label{fig:HCN_lessdust_lineemission}
		\end{figure*}
		\begin{figure*}
			\centering \large \cem{C2H2($13.20$)} lessdust \par \medskip
			\includegraphics[width=\textwidth]{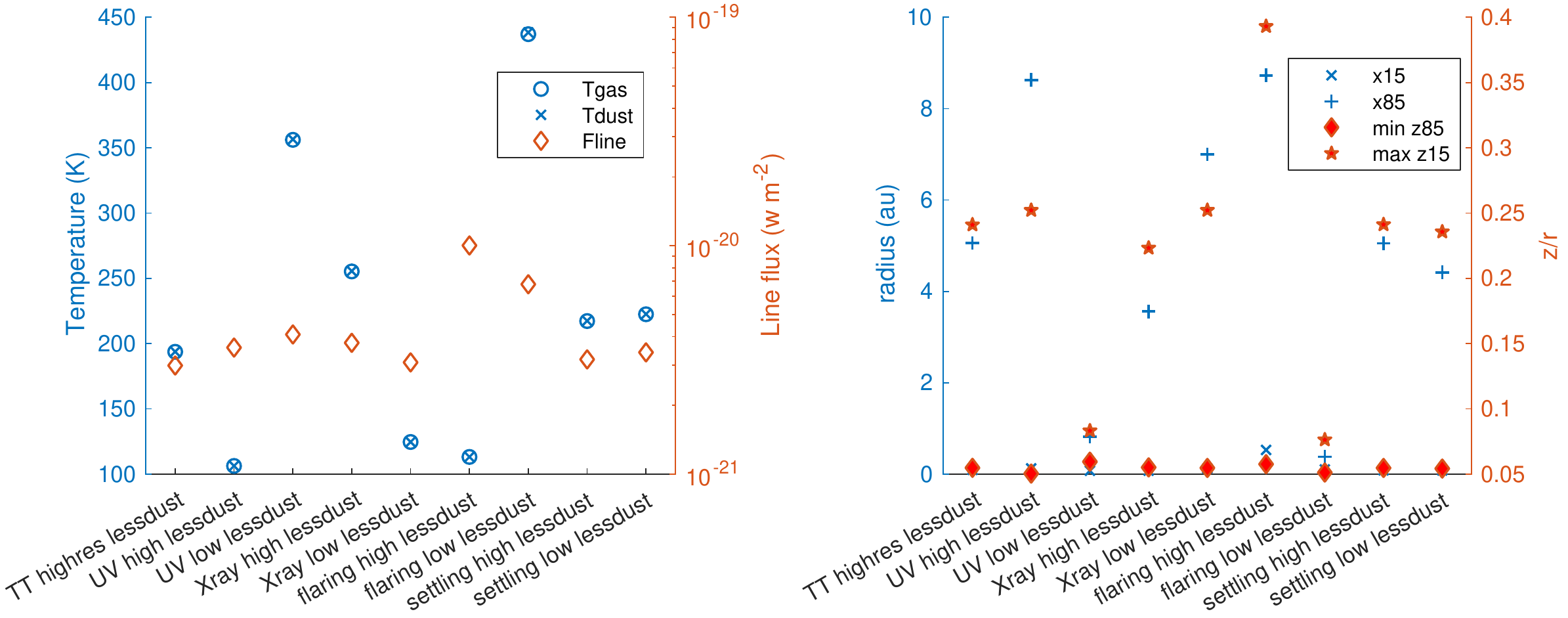}
			\caption{Properties of the line-emitting area of the \cem{C2H2} line at $13.20393~\micron$, calculated using the escape probability method for our series of models with a gas-to-dust ratio of $1000{:}1$. The description of each sub-figure is the same as  \cref{fig:CO2_lineemission}.}
			\label{fig:C2H2_lessdust_lineemission}
		\end{figure*}
		
		\begin{figure*}
			\centering \large \cem{H2O($17.75$)} lessdust \par \medskip
			\includegraphics[width=\textwidth]{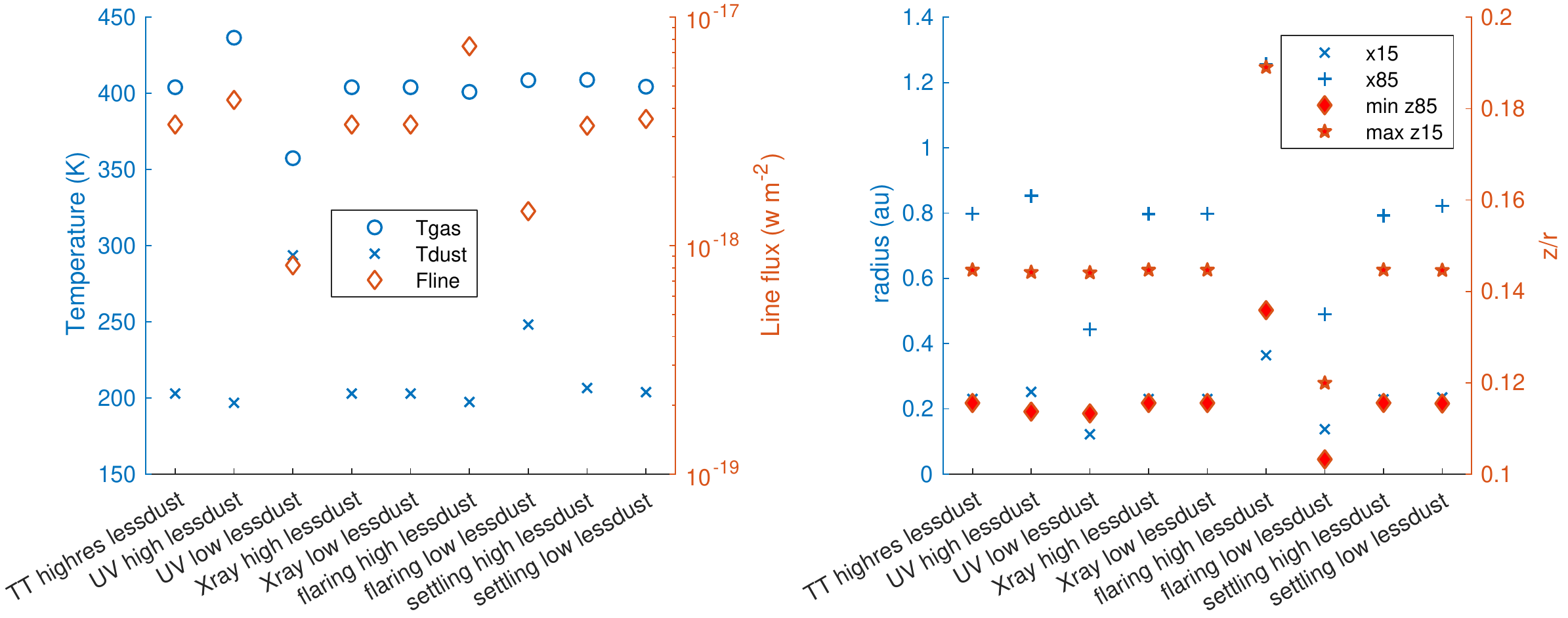}
			\caption{Properties of the line-emitting area of the \cem{H2O} line at $17.7541~\micron$, calculated using the escape probability method for our series of models with a gas-to-dust ratio of $1000{:}1$.. The description of each sub-figure is the same as \cref{fig:CO2_lineemission}.}
			\label{fig:H2O_lessdust_lineemission}		
		\end{figure*}
		\begin{figure*}
			\centering \large \cem{NH3($10.34$)} lessdust \par \medskip
			\includegraphics[width=\textwidth]{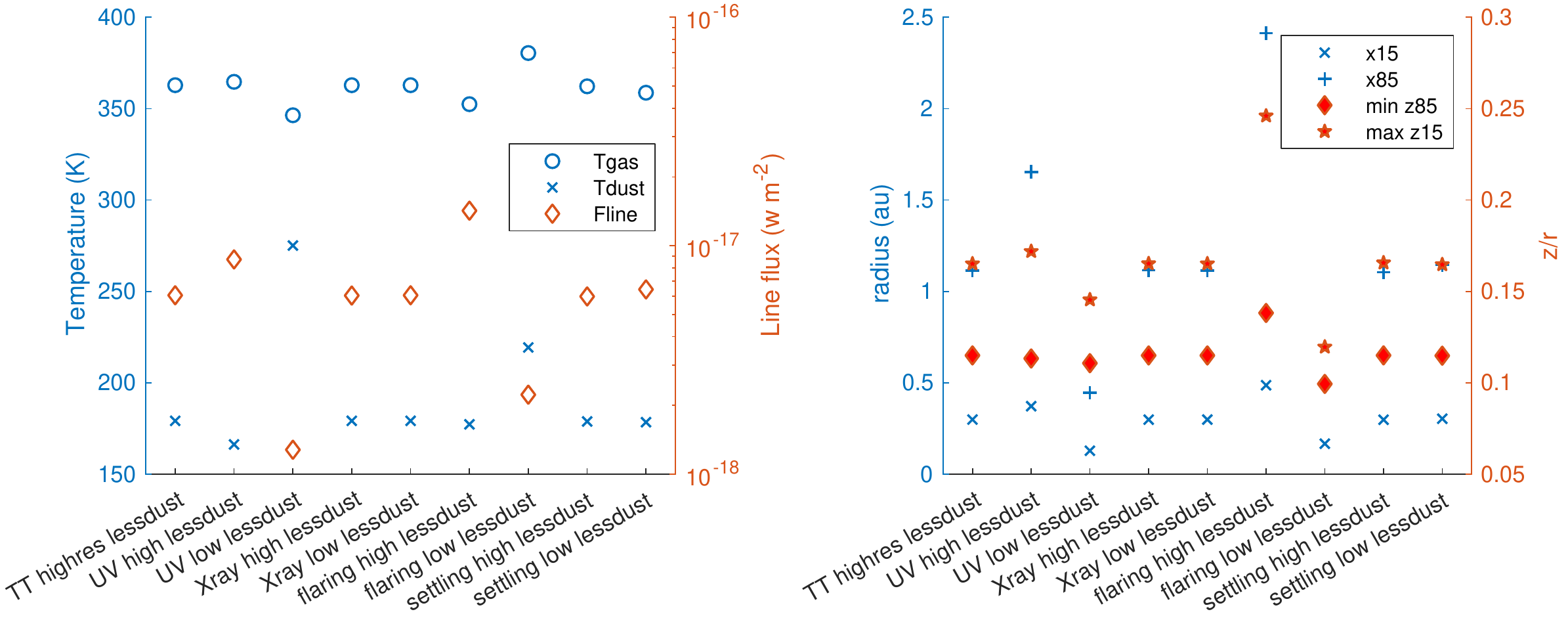}
			\caption{Properties of the line-emitting area of the \cem{NH3} line at $10.3376~\micron$, calculated using the escape probability method for our series of models with a gas-to-dust ratio of $1000{:}1$. The description of each sub-figure is the same as  \cref{fig:CO2_lineemission}.}
			\label{fig:NH3_lessdust_lineemission}		
		\end{figure*}
		\begin{figure*}
			\centering \large \cem{OH($20.12$)} lessdust \par \medskip
			\includegraphics[width=\textwidth]{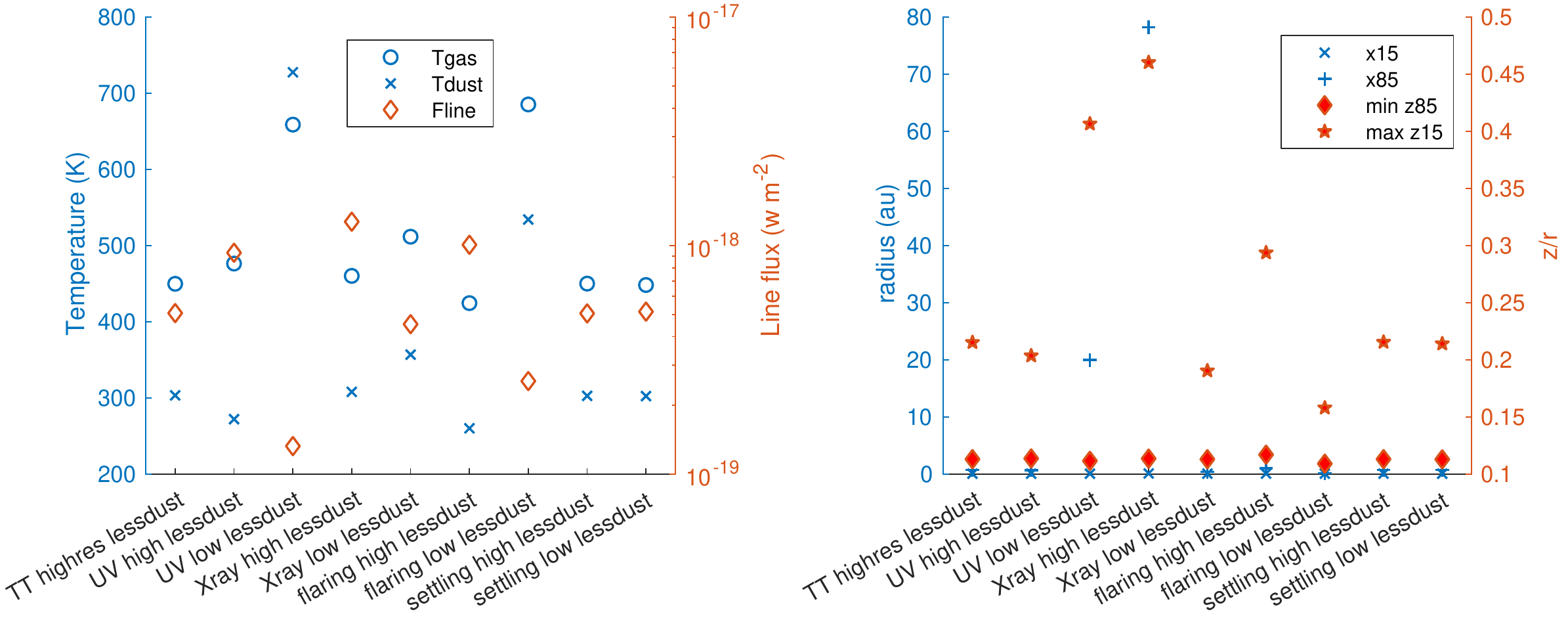}
			\caption{Properties of the line-emitting area of the \cem{OH} line at $20.1151~\micron$, calculated using the escape probability method for our series of models with a gas-to-dust ratio of $1000{:}1$. The description of each sub-figure is the same as  \cref{fig:CO2_lineemission}.}
			\label{fig:OH_lessdust_lineemission}		
		\end{figure*}

		\begin{figure*} \centering    
			\makebox[\textwidth][c]{\includegraphics[width=0.47\textwidth]{./figures/LER_plots/TT_highres_C2H2_H_132039_LER.pdf} \hspace{0.005\textwidth} 
				\includegraphics[width=0.47\textwidth]{./figures/LER_plots/TT_highres_HCN_H_140393_LER.pdf}}
			\makebox[\textwidth][c]{
				\includegraphics[width=0.47\textwidth]{./figures/LER_plots/TT_highres_CO2_H_149830_LER.pdf} \hspace{0.005\textwidth}
				\includegraphics[width=0.47\textwidth]{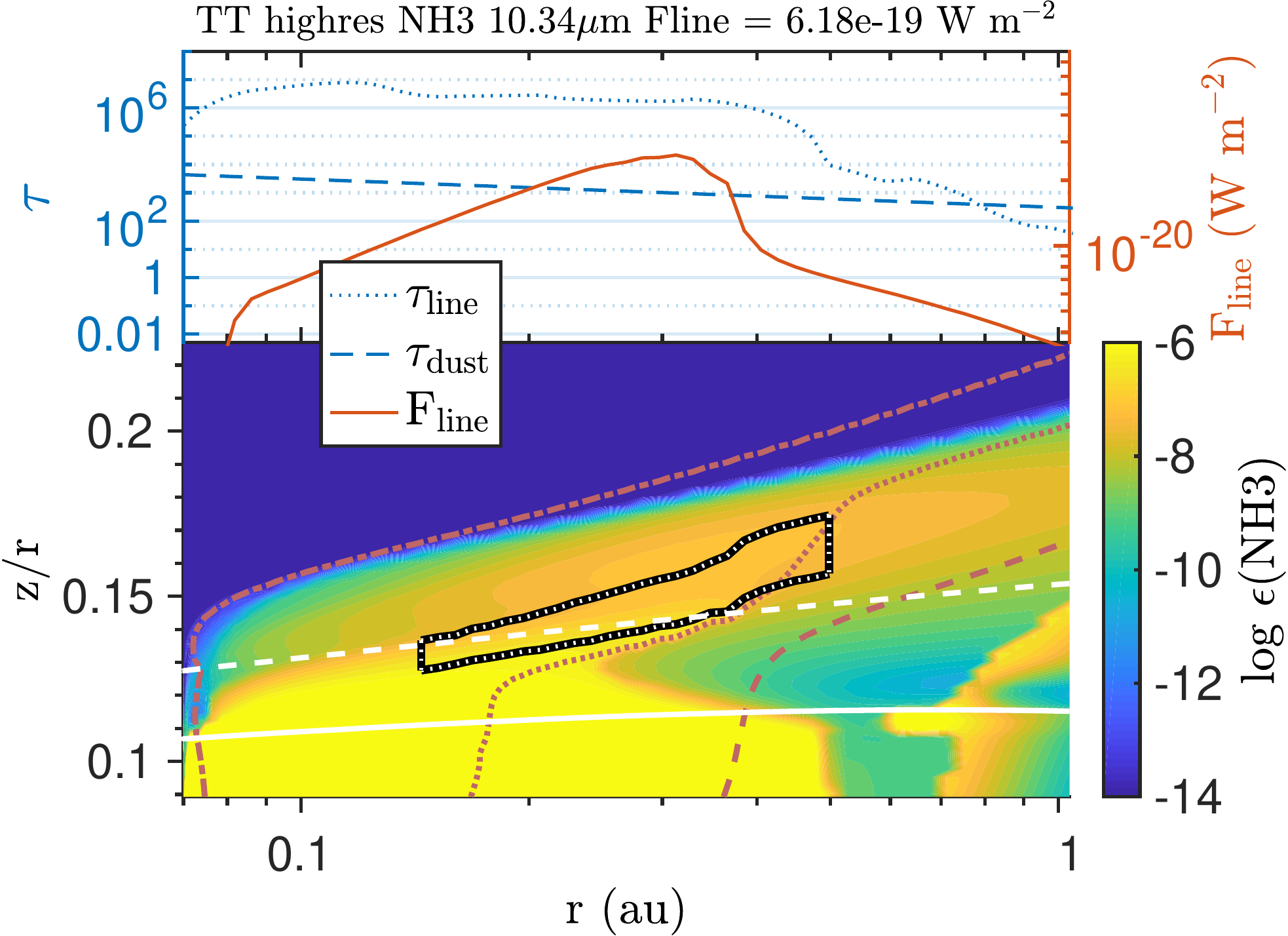}}
			\makebox[\textwidth][c]{
				\includegraphics[width=0.47\textwidth]{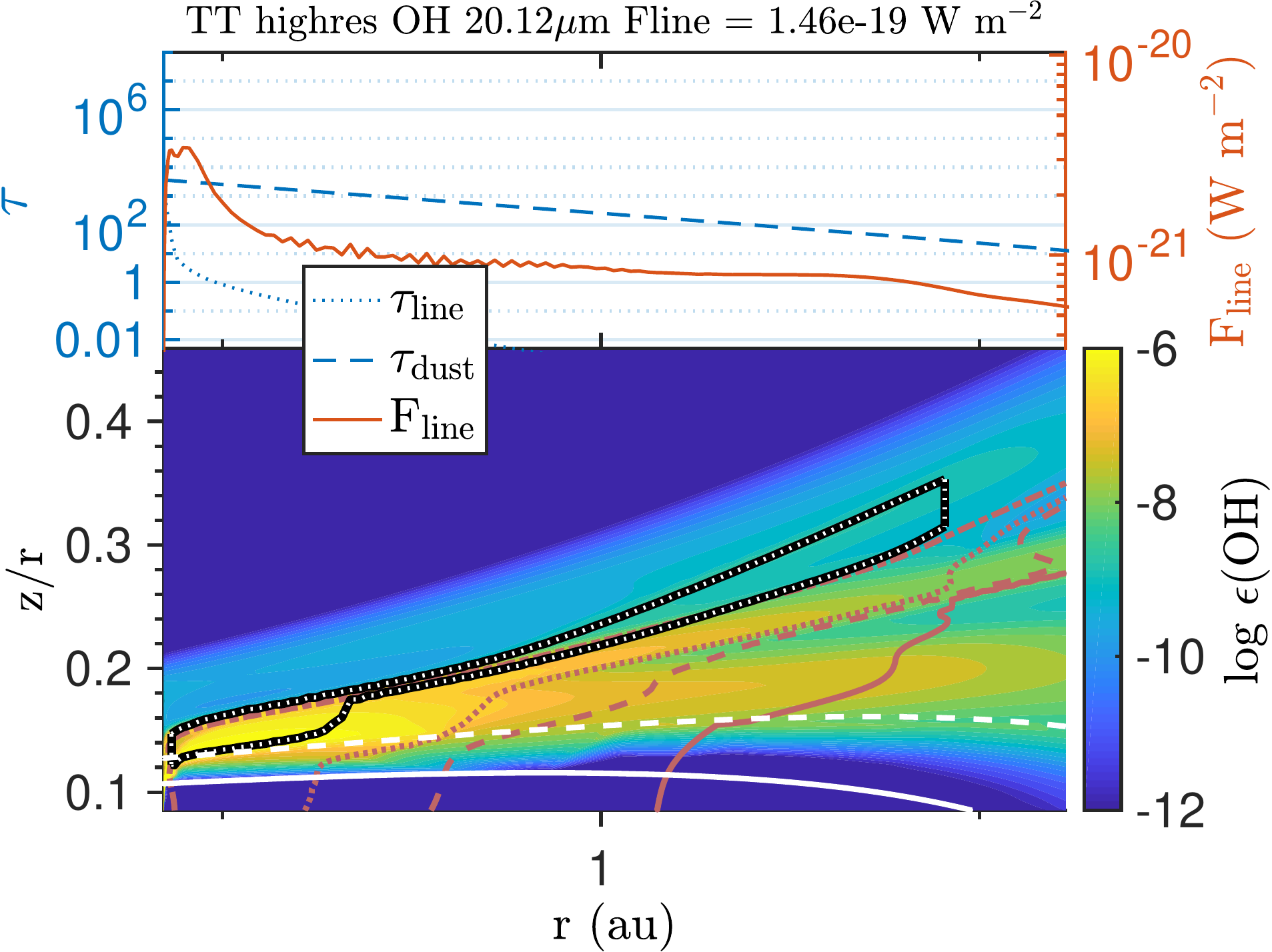} \hspace{0.005\textwidth}  
				\includegraphics[width=0.47\textwidth]{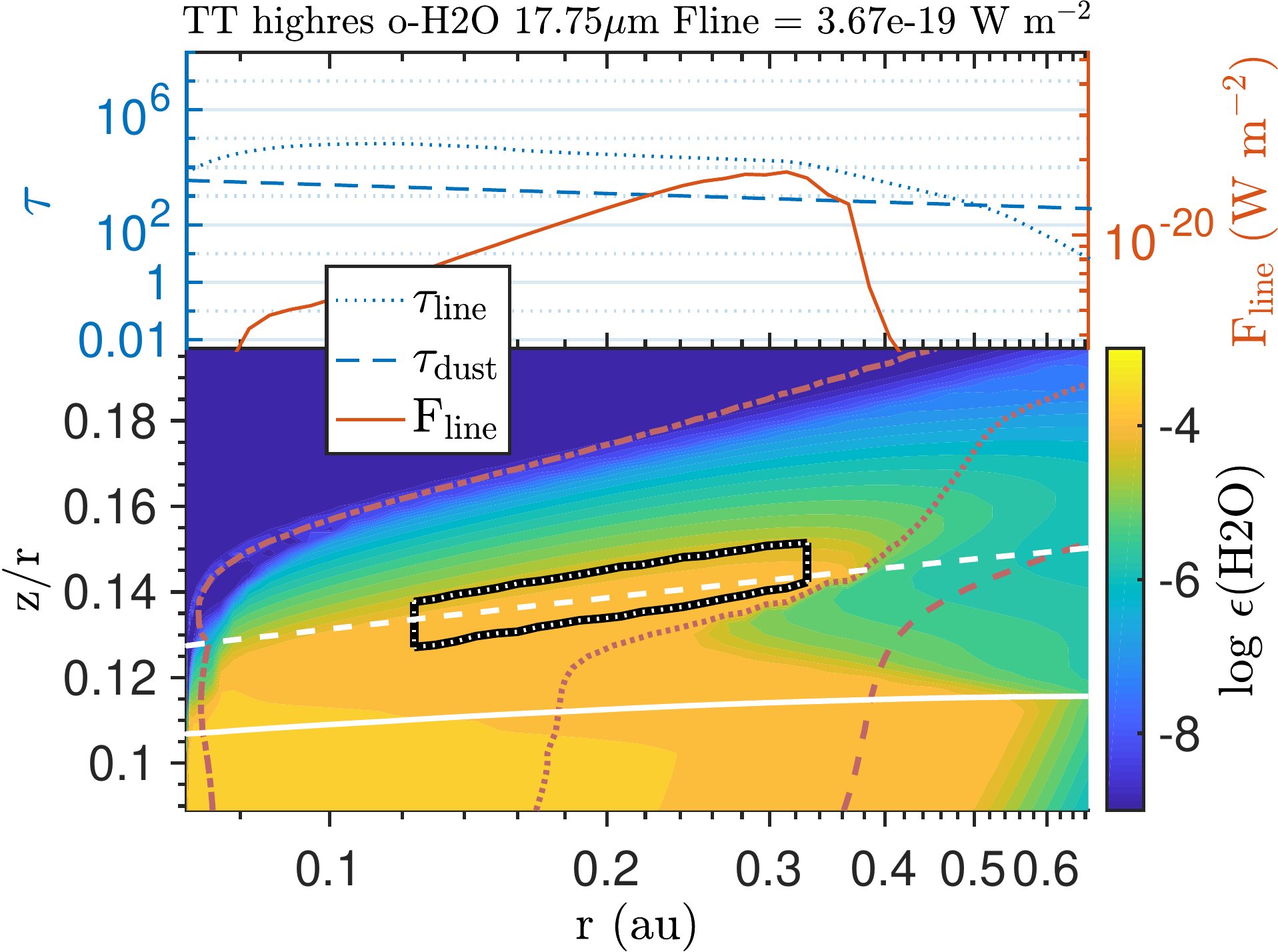}}
			\caption{Line-emitting regions for the model TT highres for \cem{C2H2}, \cem{HCN}, \cem{CO2}, \cem{NH3}, \cem{OH}, and \cem{o-H2O}. On the upper panel, the $20~\micron$ dust continuum and gas line optical depths and the line flux are plotted. The optical depths and line fluxes are calculated using the vertical escape probability. On the lower panel, the plotted colour map is the molecular abundance (relative to the total elemental hydrogen abundance, across all species), with contour increments every $0.5$ dex. The over-plotted contours are the main line-emitting area (black-and-white line), the visual extinction (dashed white line at $\av = 1$, and solid white line at $\av = 1$), and gas temperature. The gas temperature is plotted in red at $100$, $200$, $300$, and $1000~\kelvin$ (solid, dashed, dotted, and dash-dotted lines respectively). } 
			\label{fig:LER_TT_highres}  
		\end{figure*}
		
		\begin{figure*} \centering
			\makebox[\textwidth][c]{\includegraphics[width=0.47\textwidth]{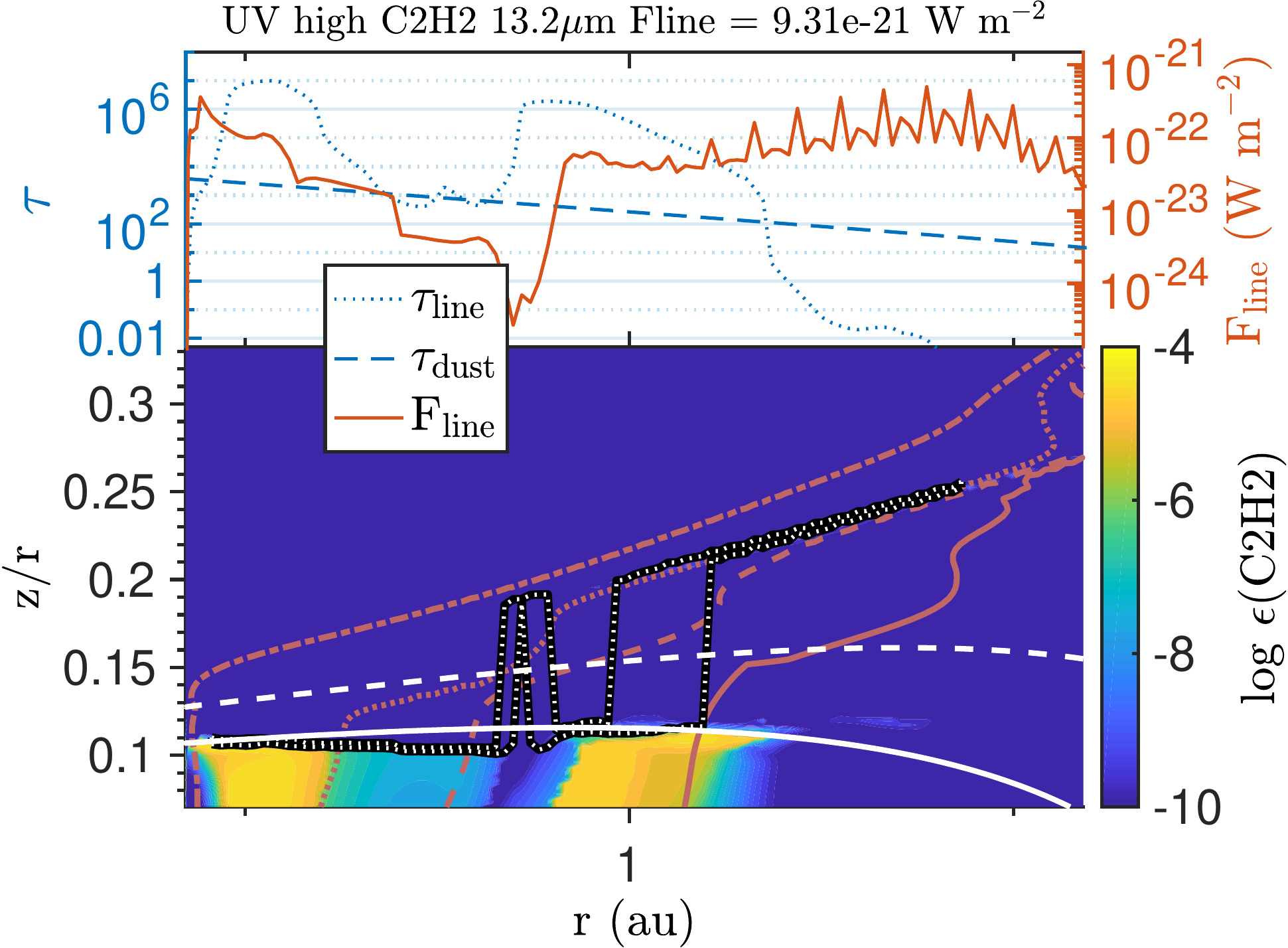} \hspace{0.005\textwidth}   
				\includegraphics[width=0.47\textwidth]{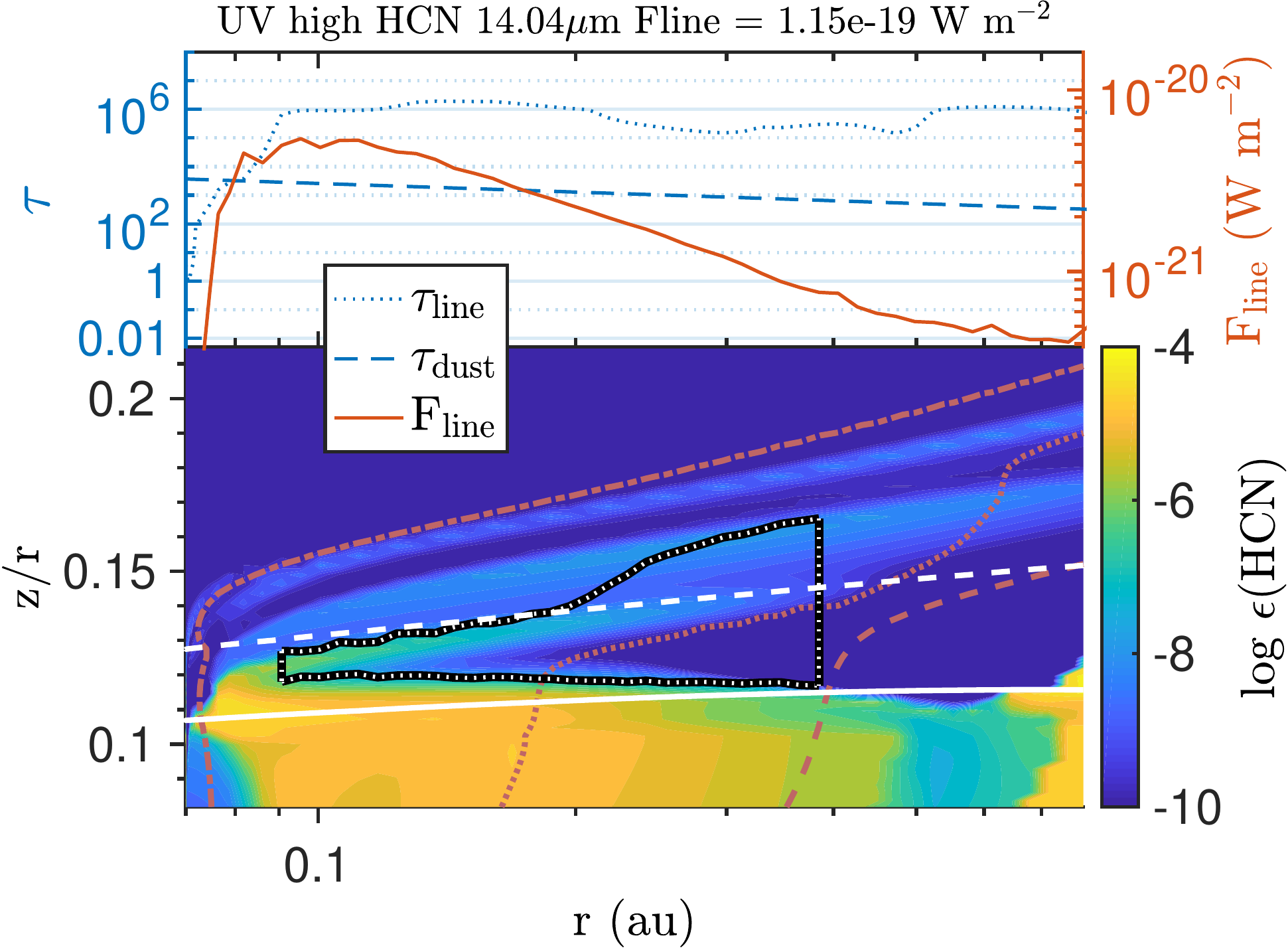}}      
			\makebox[\textwidth][c]{\includegraphics[width=0.47\textwidth]{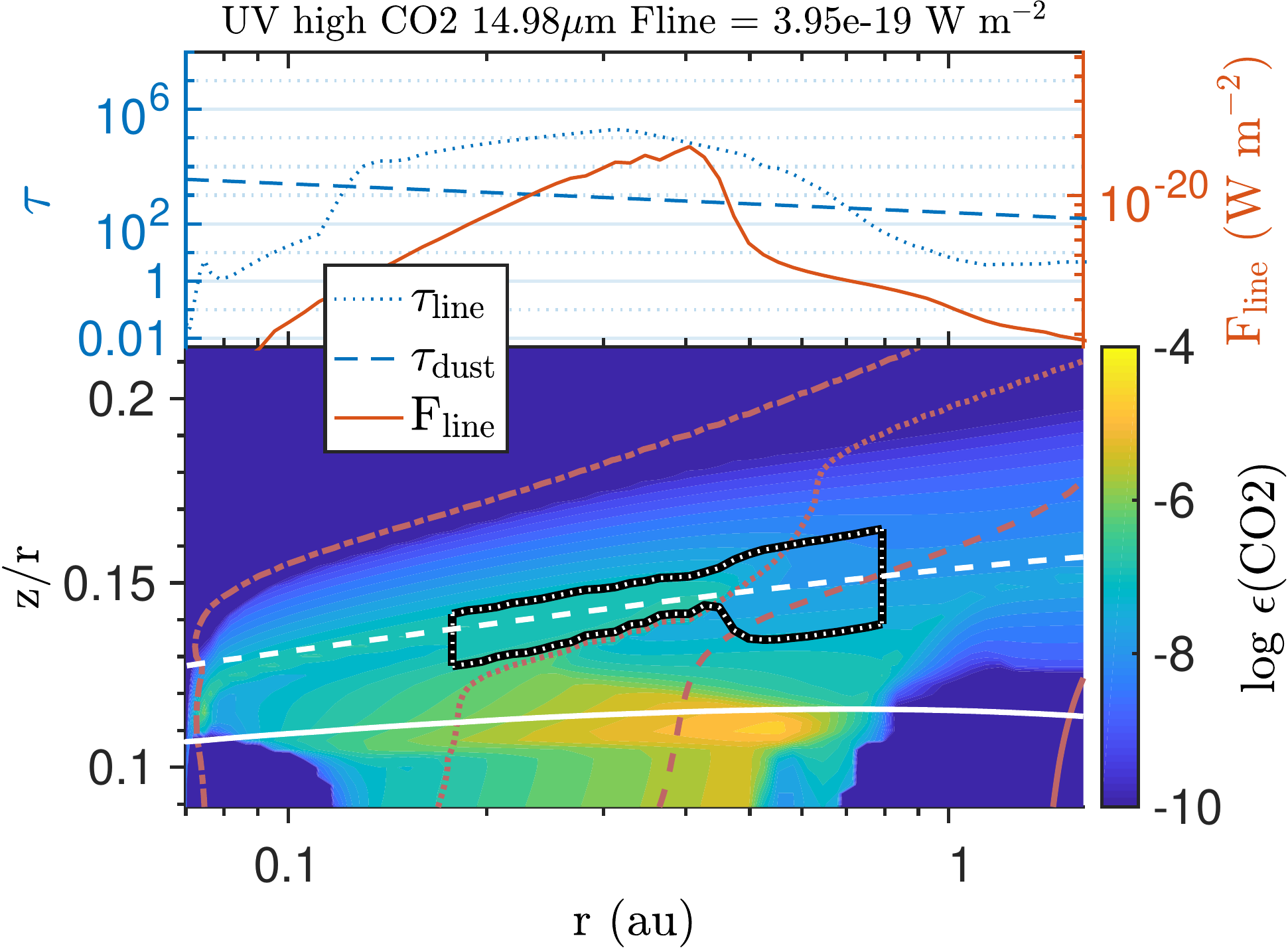} \hspace{0.005\textwidth}
				\includegraphics[width=0.47\textwidth]{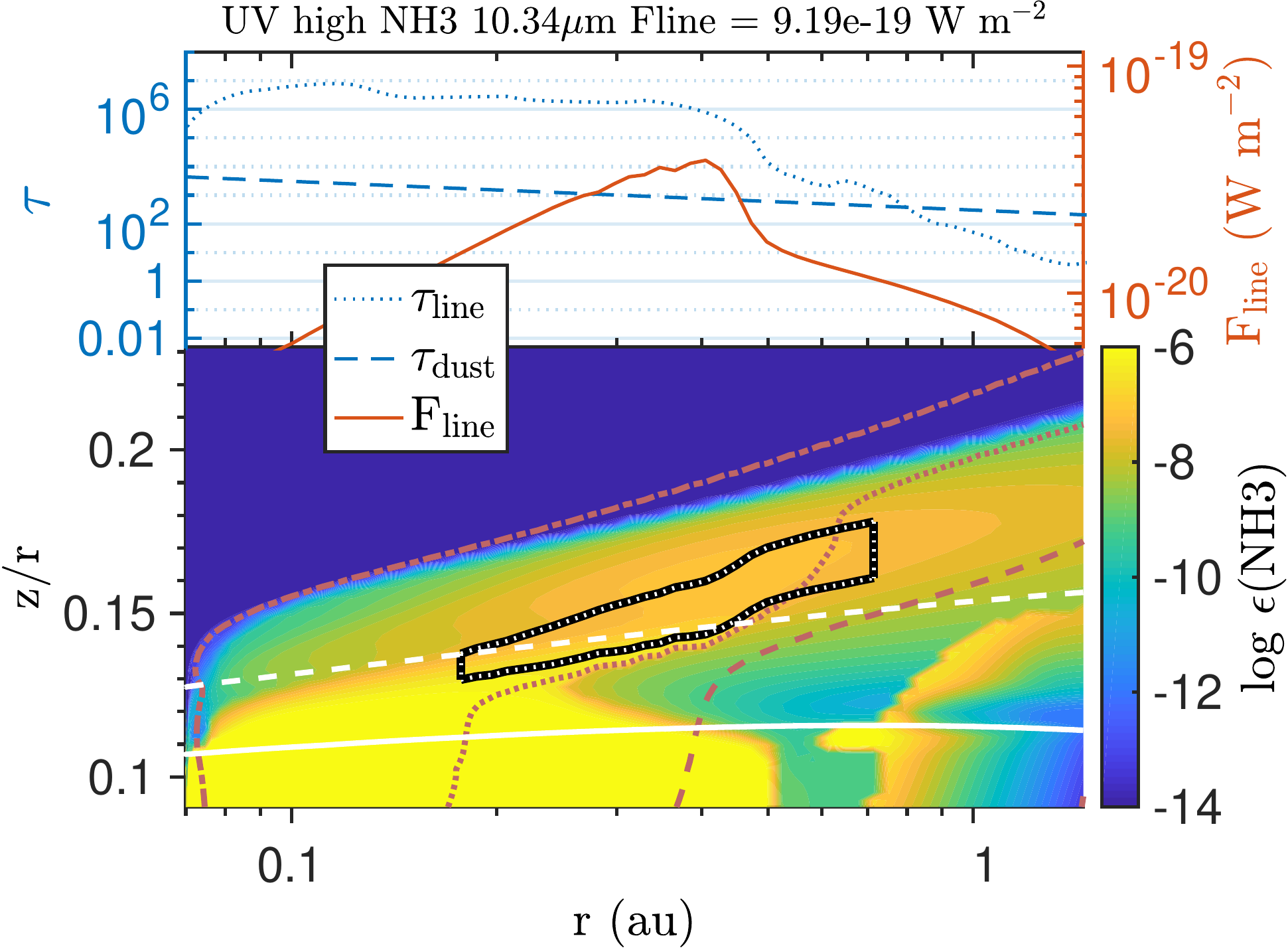}}     
			\makebox[\textwidth][c]{\includegraphics[width=0.47\textwidth]{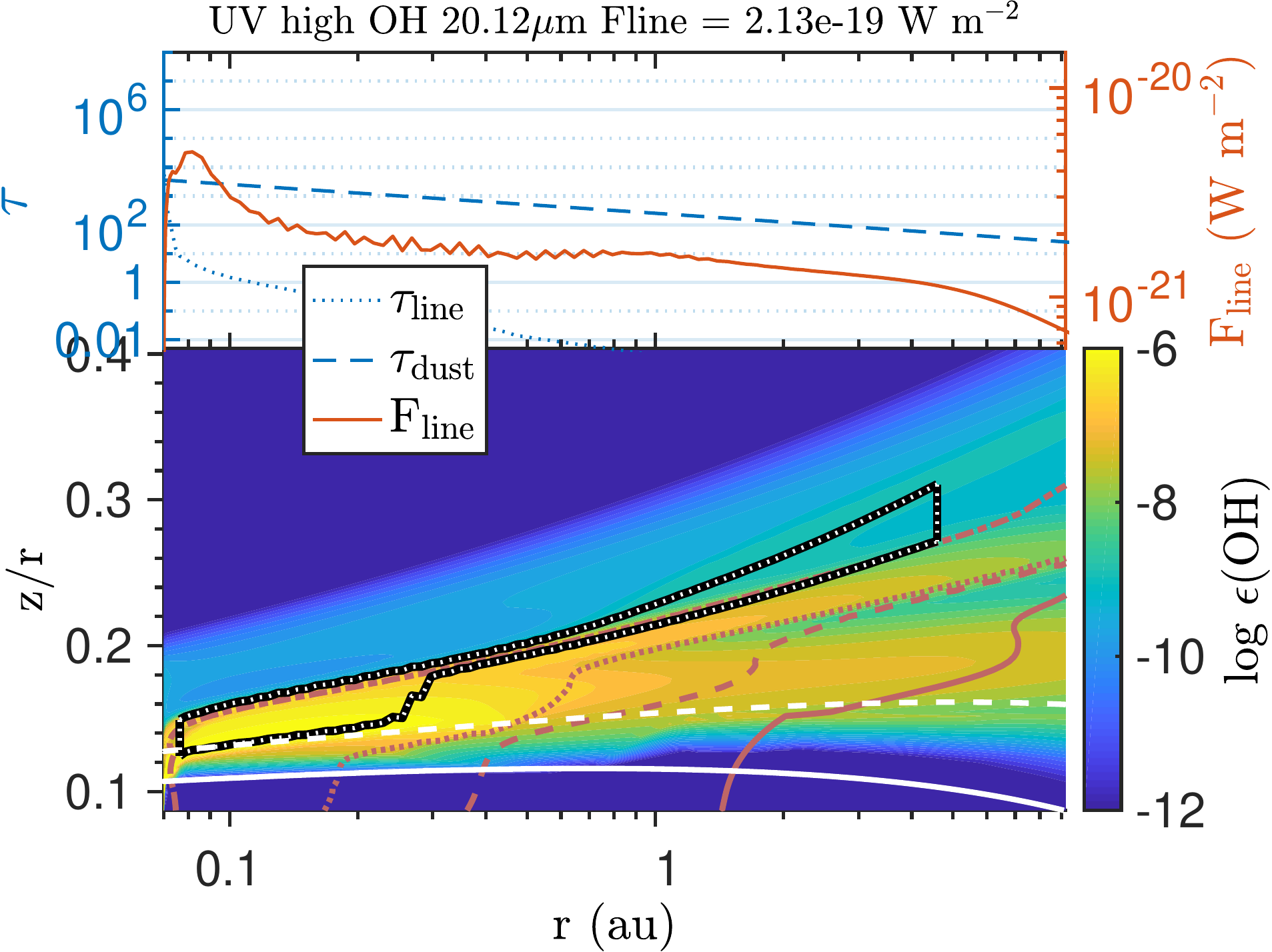} \hspace{0.005\textwidth}     
				\includegraphics[width=0.47\textwidth]{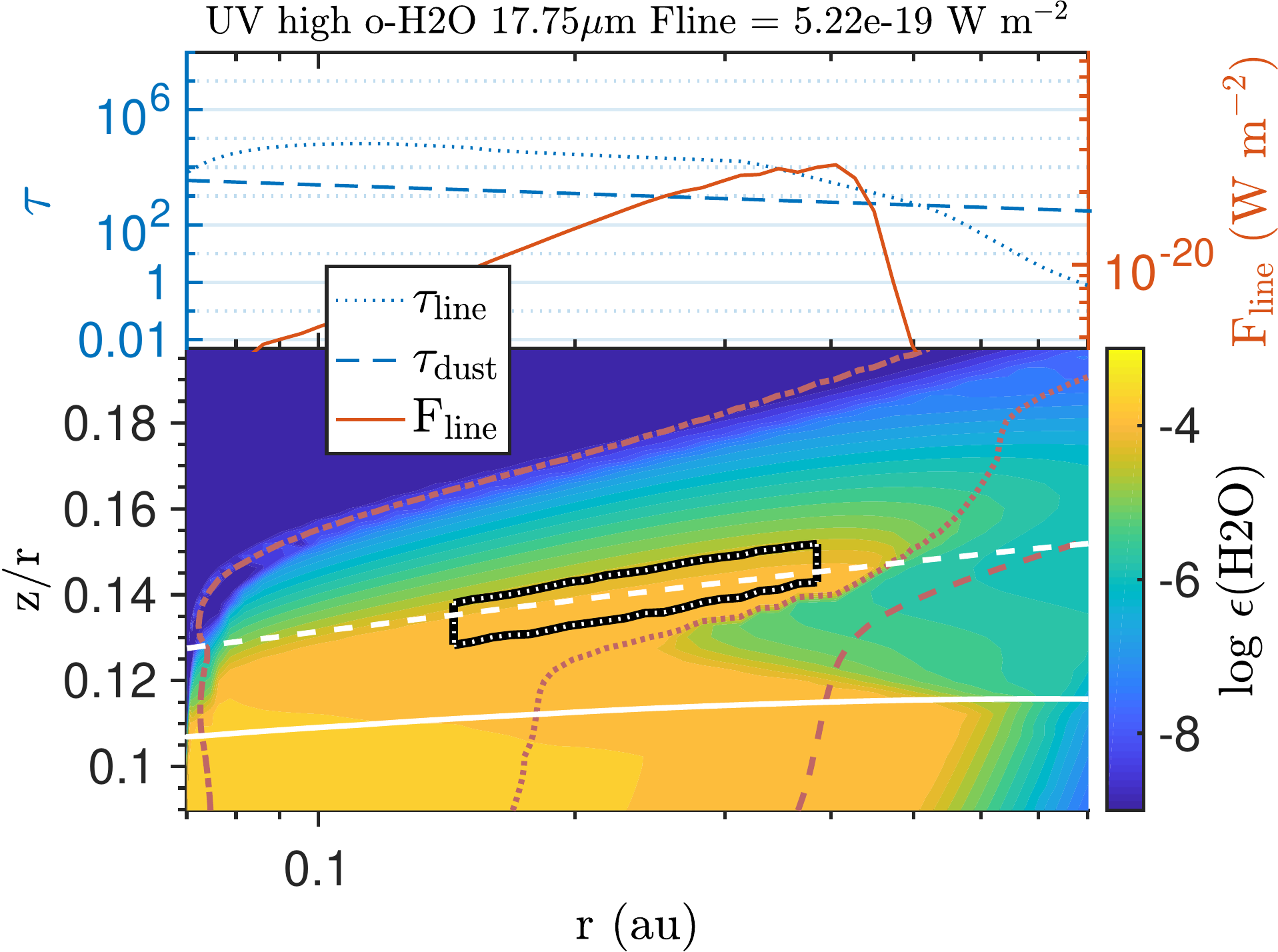}}     
			\caption{Line-emitting regions for the model UV high\. The plotted lines are \cem{C2H2}, \cem{HCN}, \cem{CO2}, \cem{NH3}, \cem{OH}, and \cem{o-H2O}. The rest of the figure is as described in  \cref{fig:LER_TT_highres}.  
			}\label{fig:LER_UV_high}     
		\end{figure*}
		
		\begin{figure*} \centering    
			\makebox[\textwidth][c]{\includegraphics[width=0.47\textwidth]{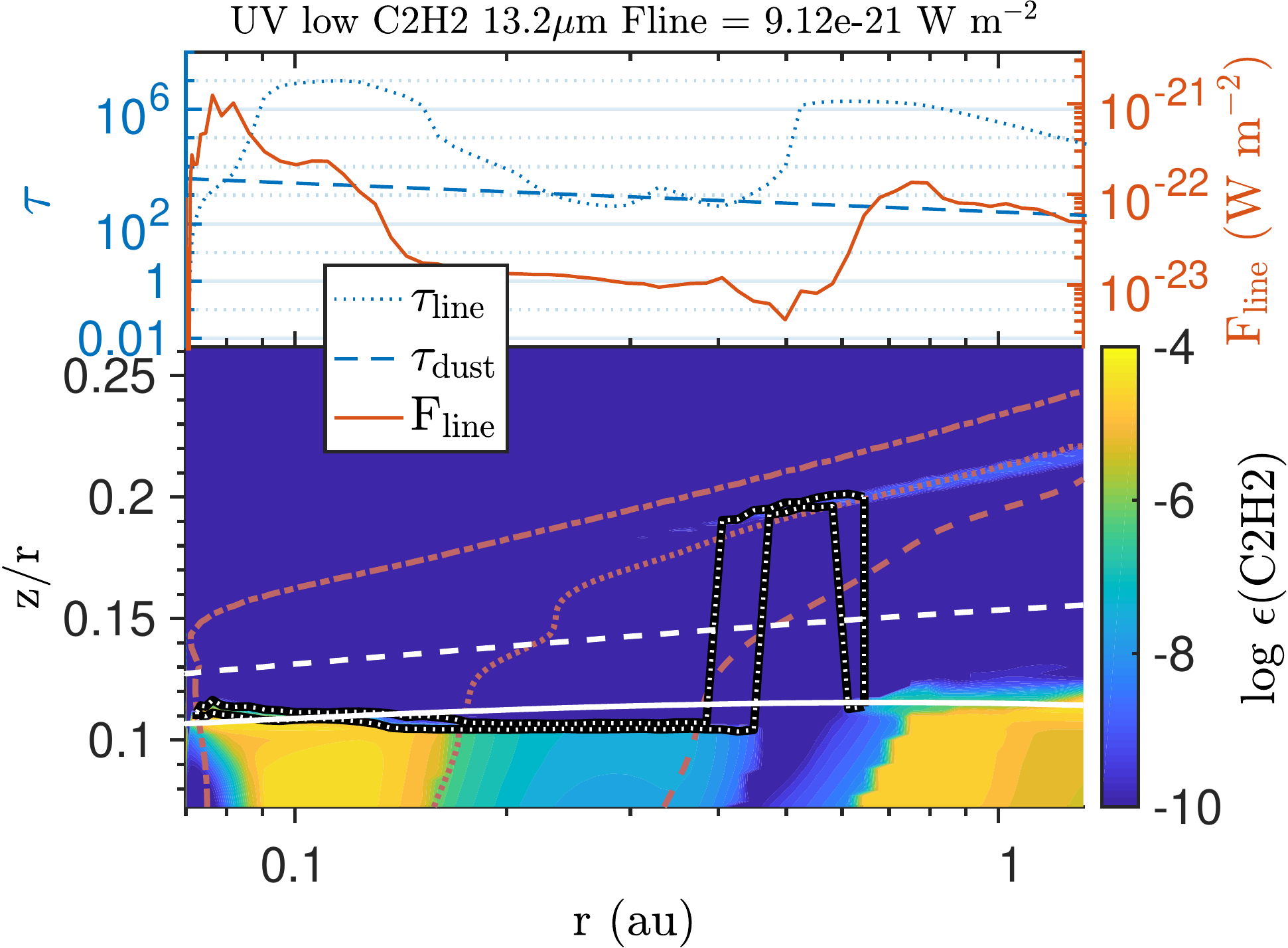} \hspace{0.005\textwidth}    
				\includegraphics[width=0.47\textwidth]{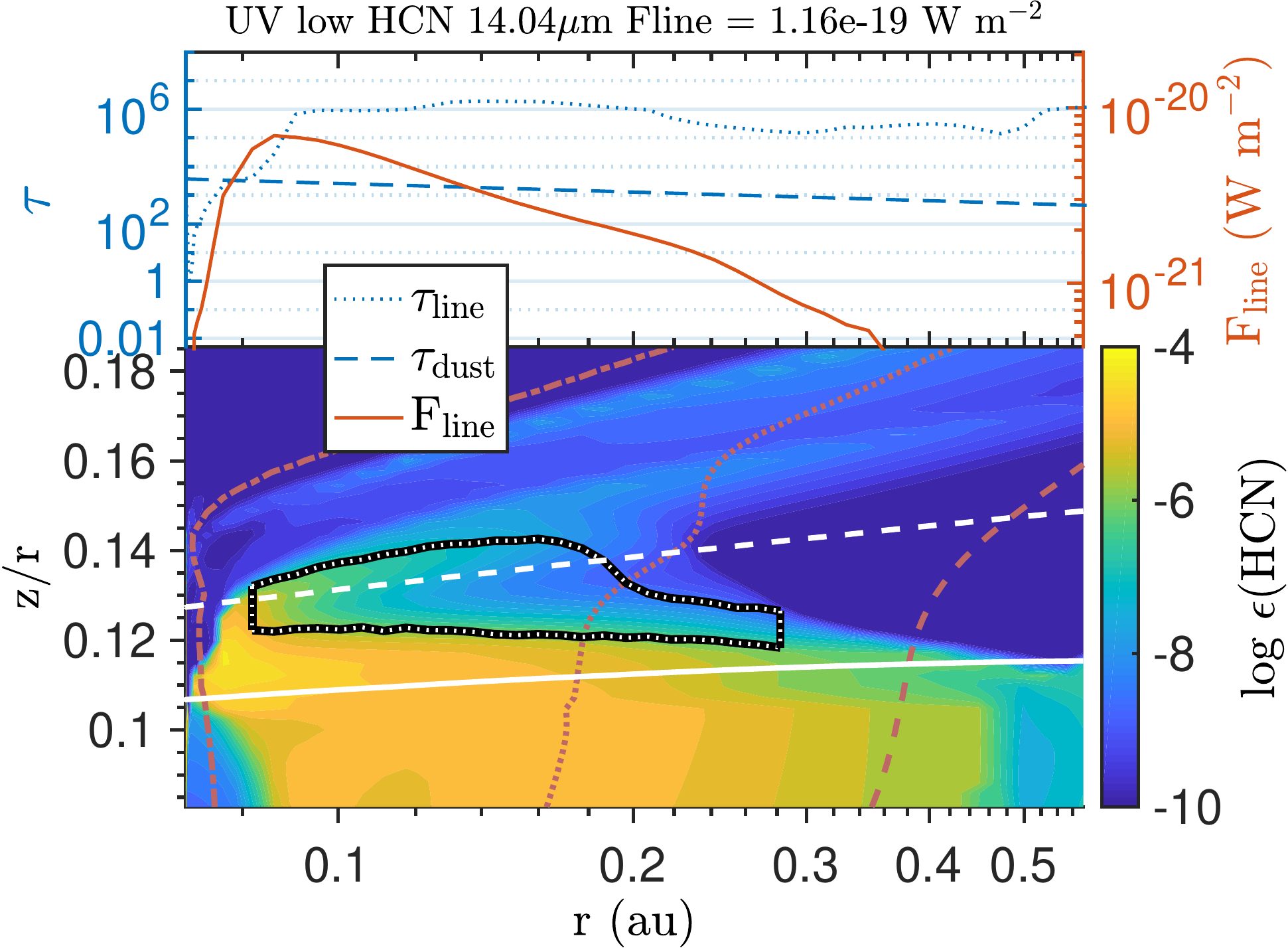}} 
			\makebox[\textwidth][c]{\includegraphics[width=0.47\textwidth]{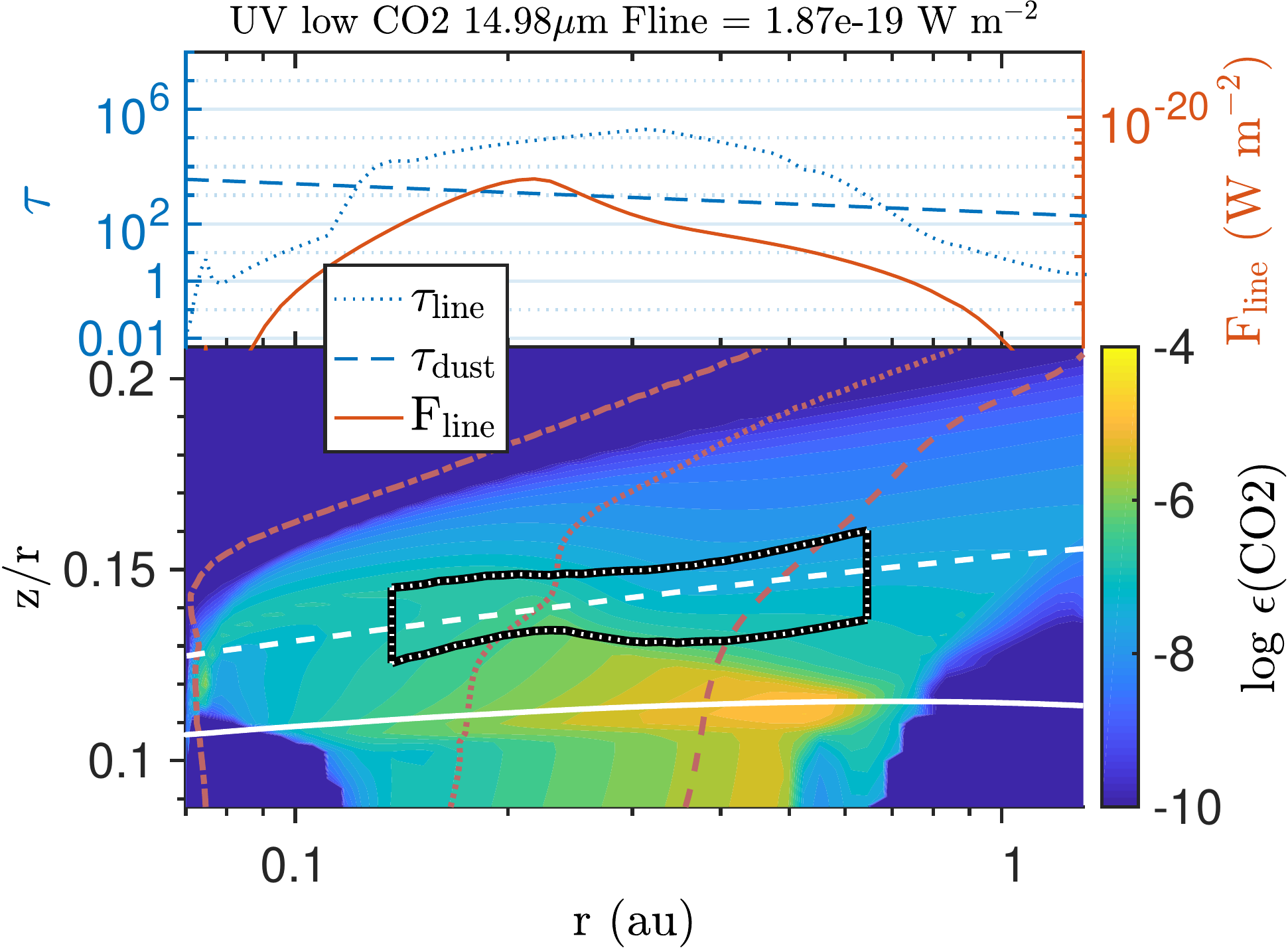} \hspace{0.005\textwidth}
				\includegraphics[width=0.47\textwidth]{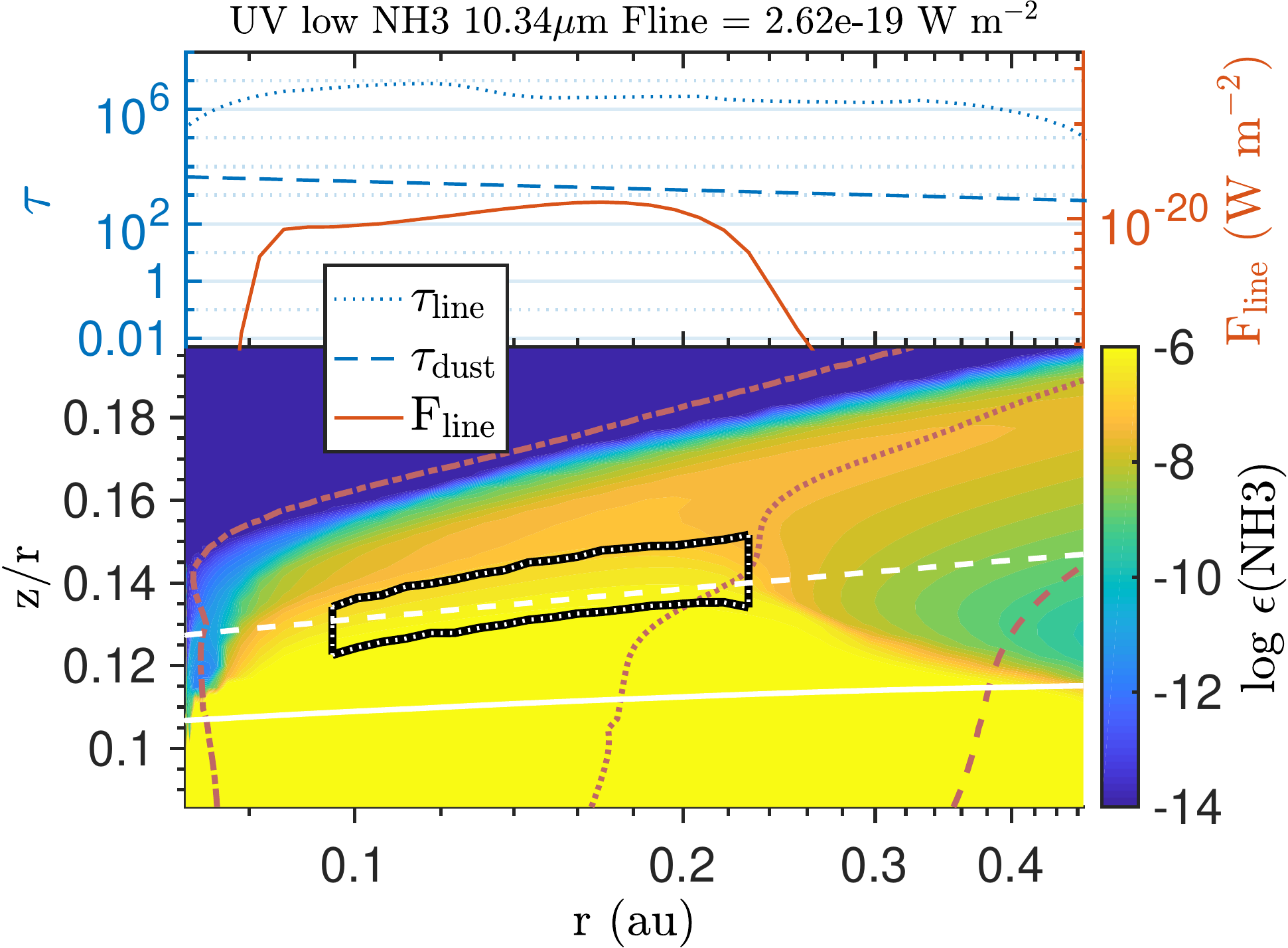}}      
			\makebox[\textwidth][c]{\includegraphics[width=0.47\textwidth]{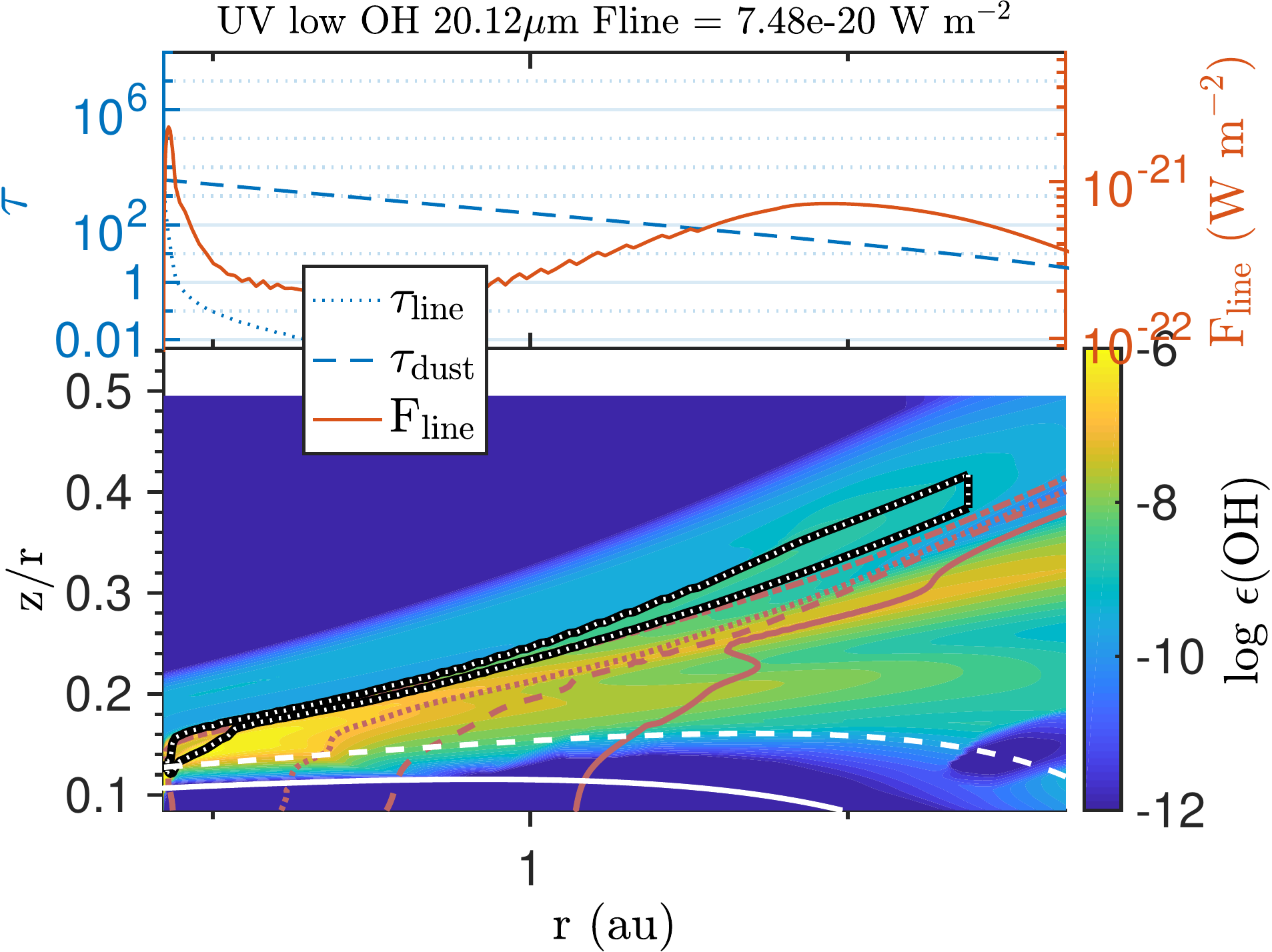} \hspace{0.005\textwidth}
				\includegraphics[width=0.47\textwidth]{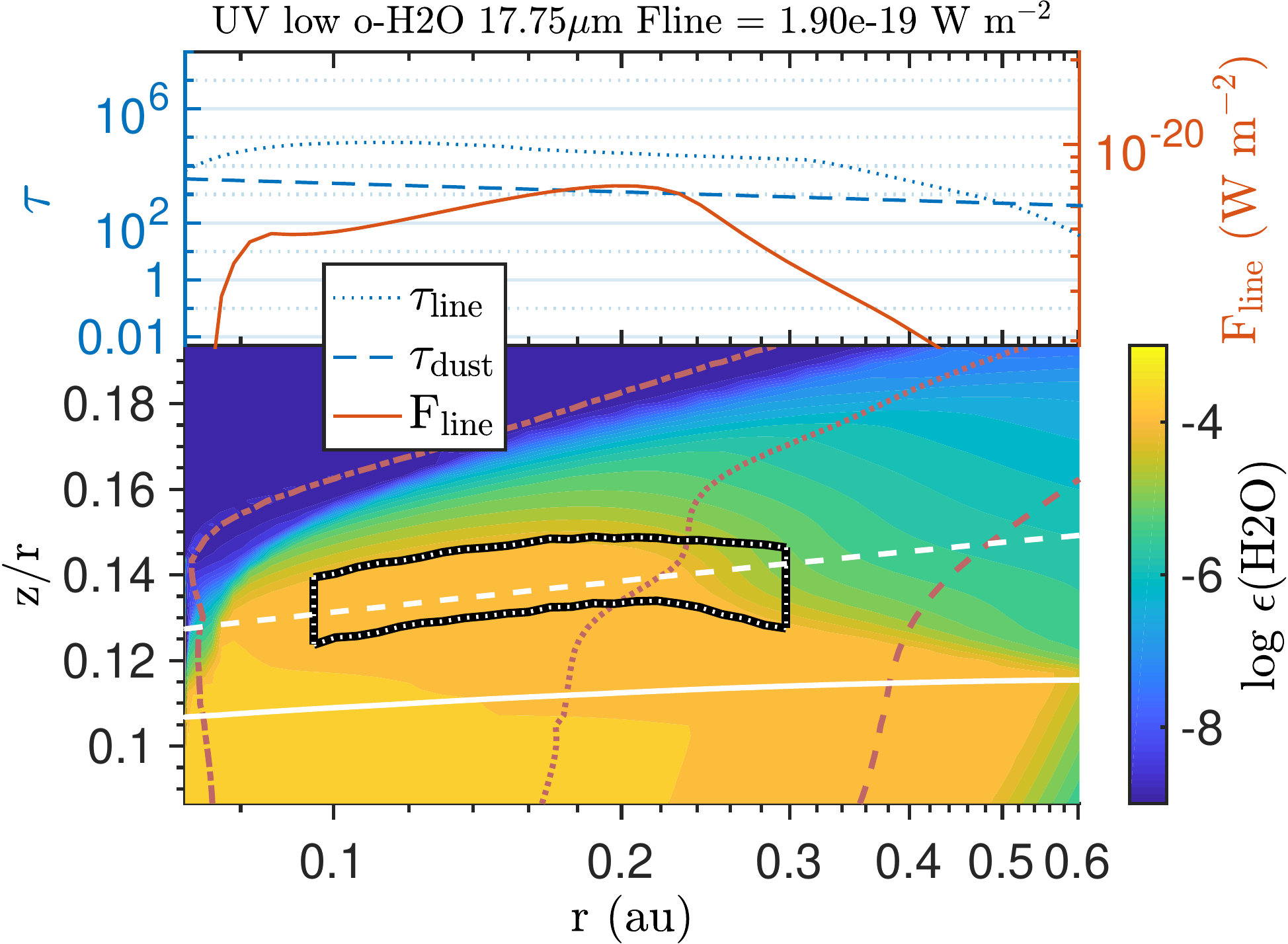}}
			\caption{Line-emitting regions for the model UV low. The plotted lines are \cem{C2H2}, \cem{HCN}, \cem{CO2}, \cem{NH3}, \cem{OH}, and \cem{o-H2O}. The rest of the figure is as described in \cref{fig:LER_TT_highres}.   
			}\label{fig:LER_UV_low}
		\end{figure*}
		
		\begin{figure*} \centering    
			\makebox[\textwidth][c]{\includegraphics[width=0.47\textwidth]{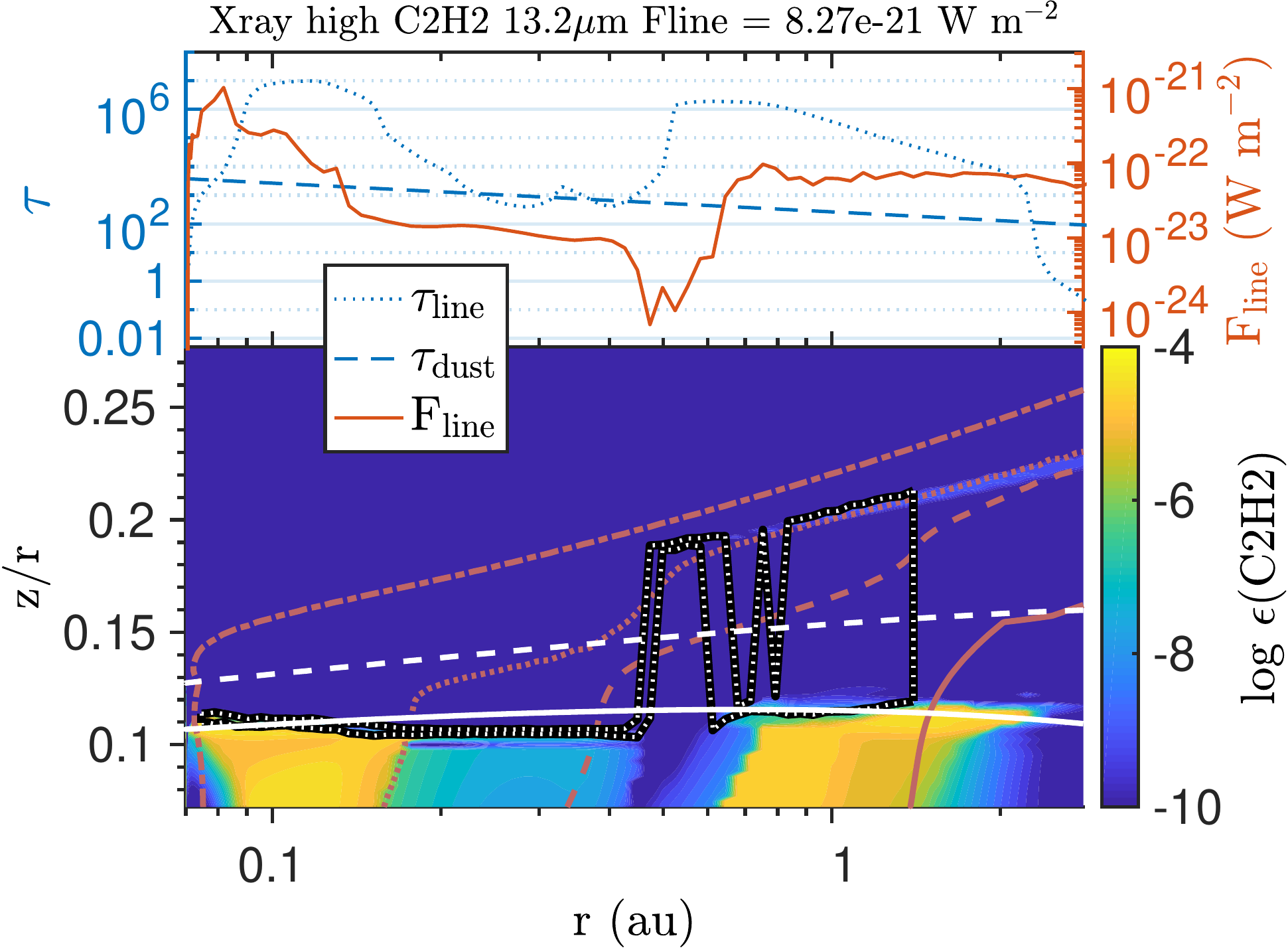} \hspace{0.005\textwidth} 
				\includegraphics[width=0.47\textwidth]{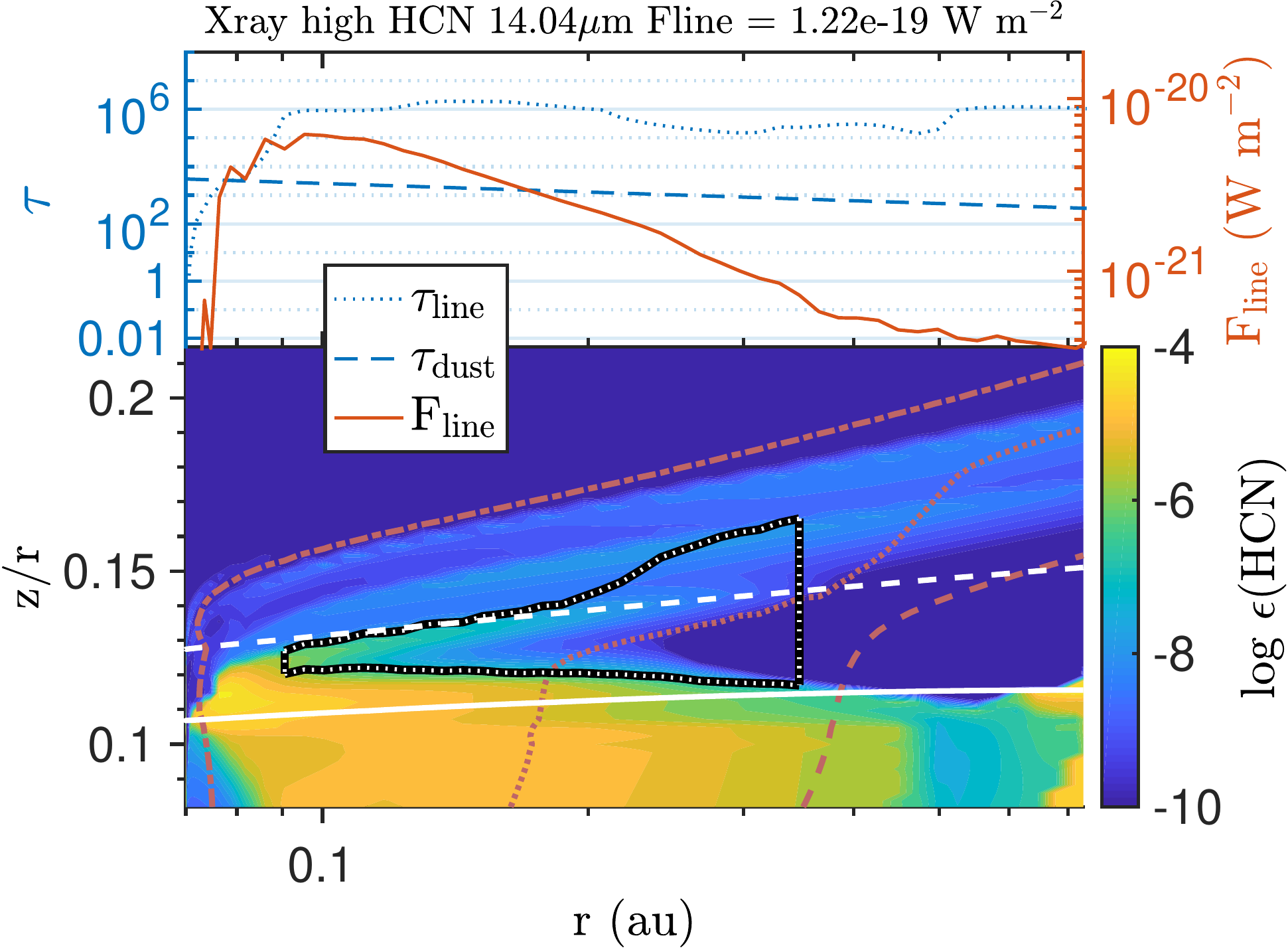}}    
			\makebox[\textwidth][c]{\includegraphics[width=0.47\textwidth]{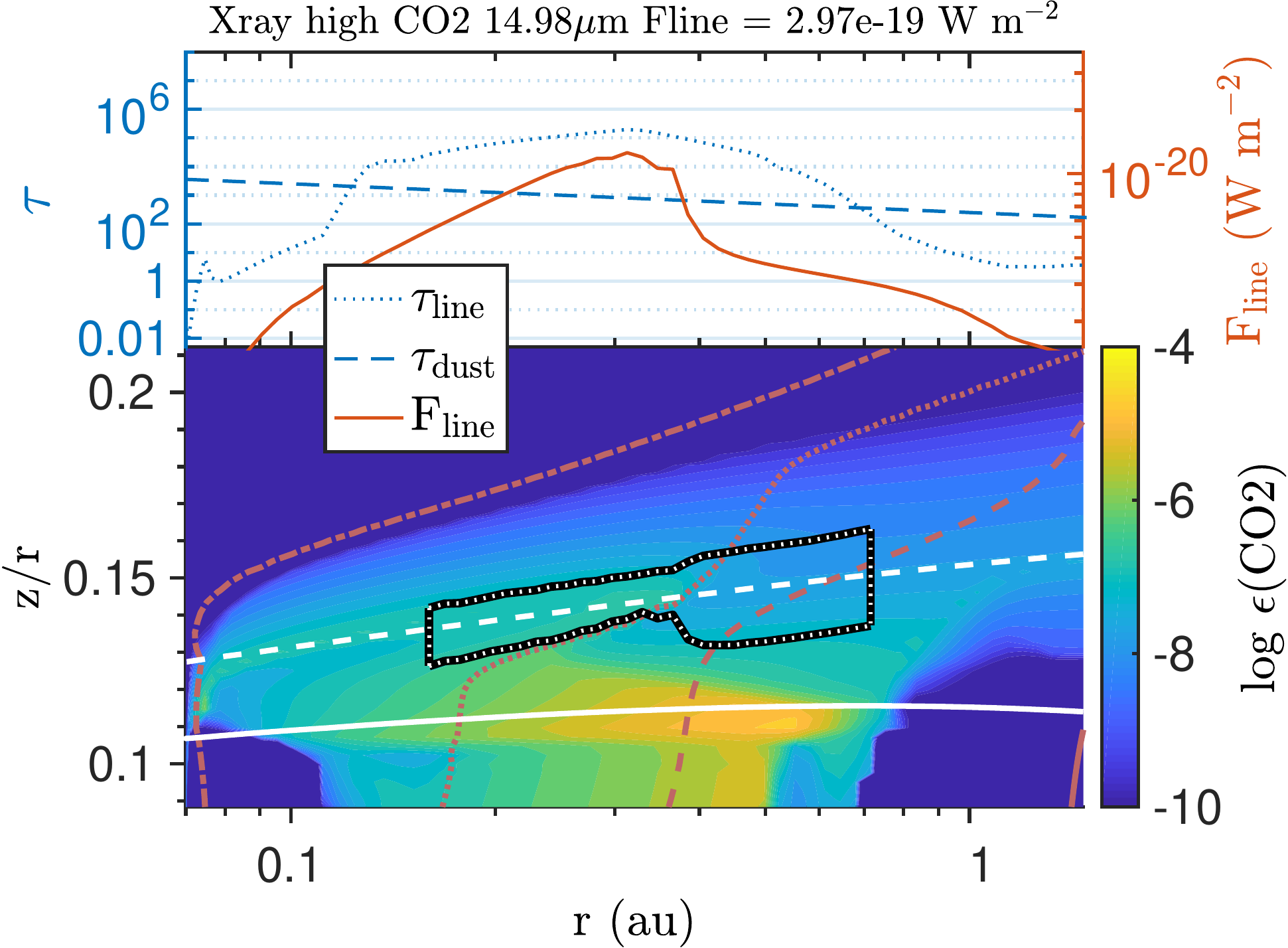} \hspace{0.005\textwidth}
				\includegraphics[width=0.47\textwidth]{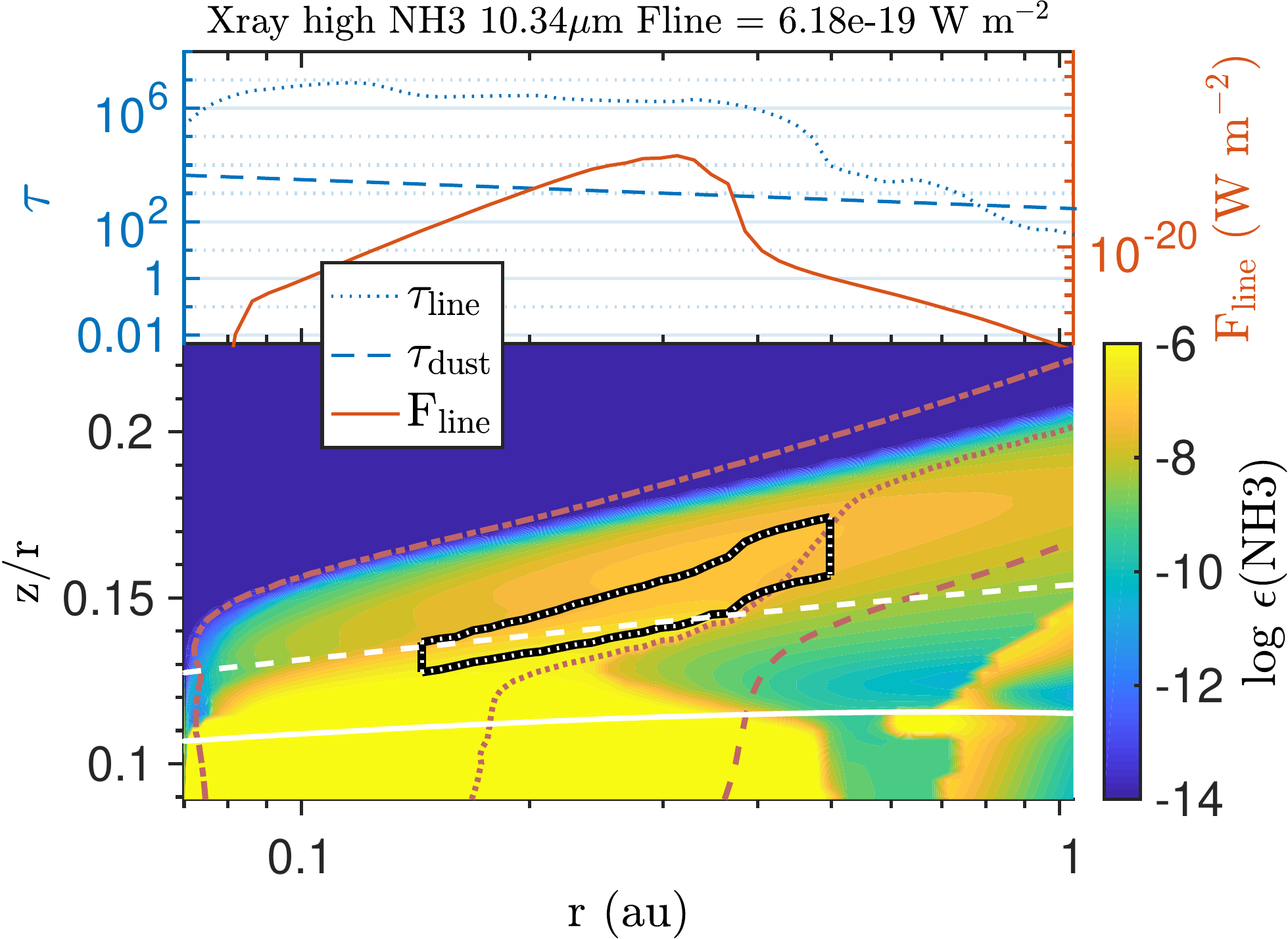}}   
			\makebox[\textwidth][c]{\includegraphics[width=0.47\textwidth]{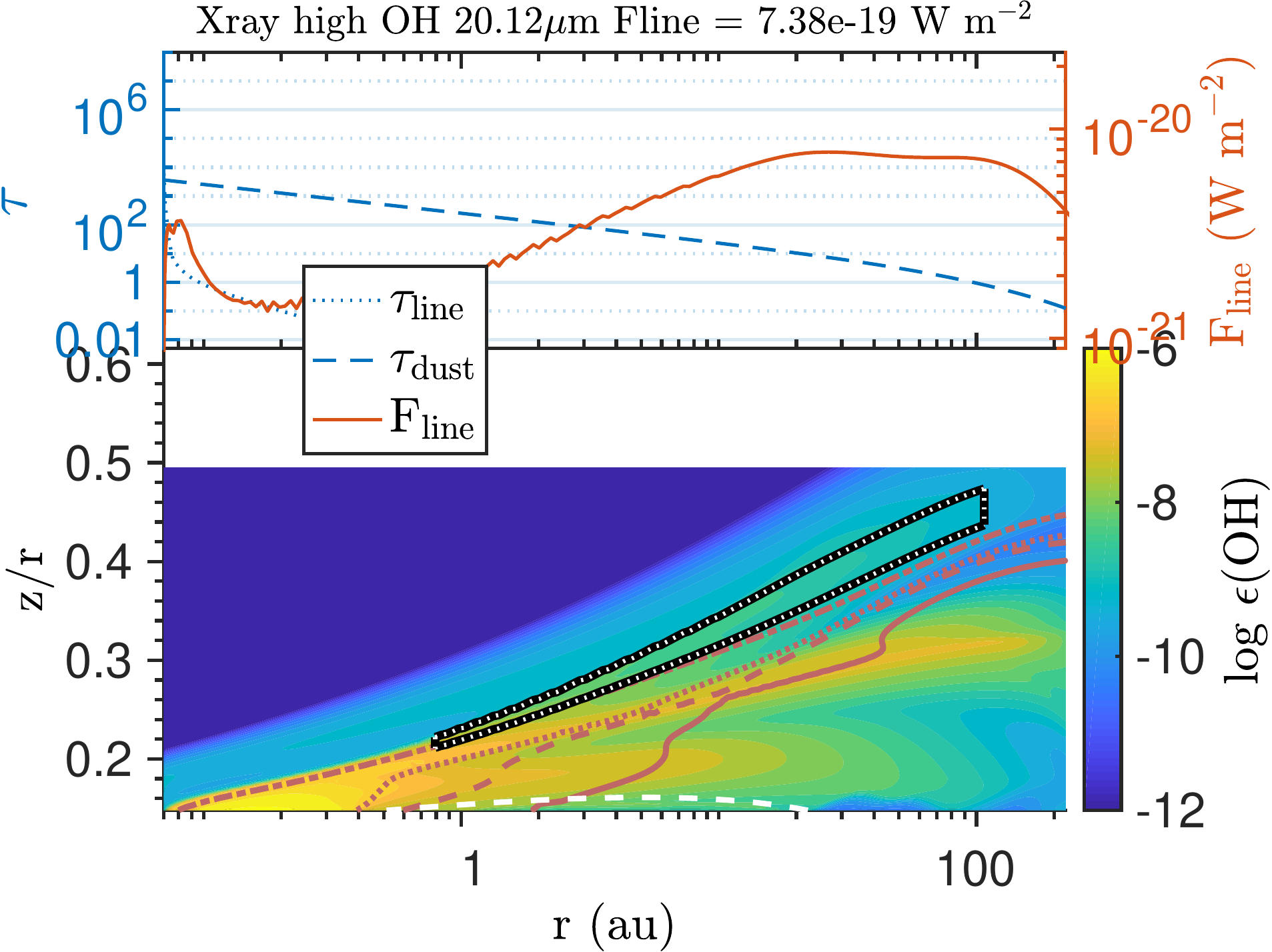} \hspace{0.005\textwidth}   
				\includegraphics[width=0.47\textwidth]{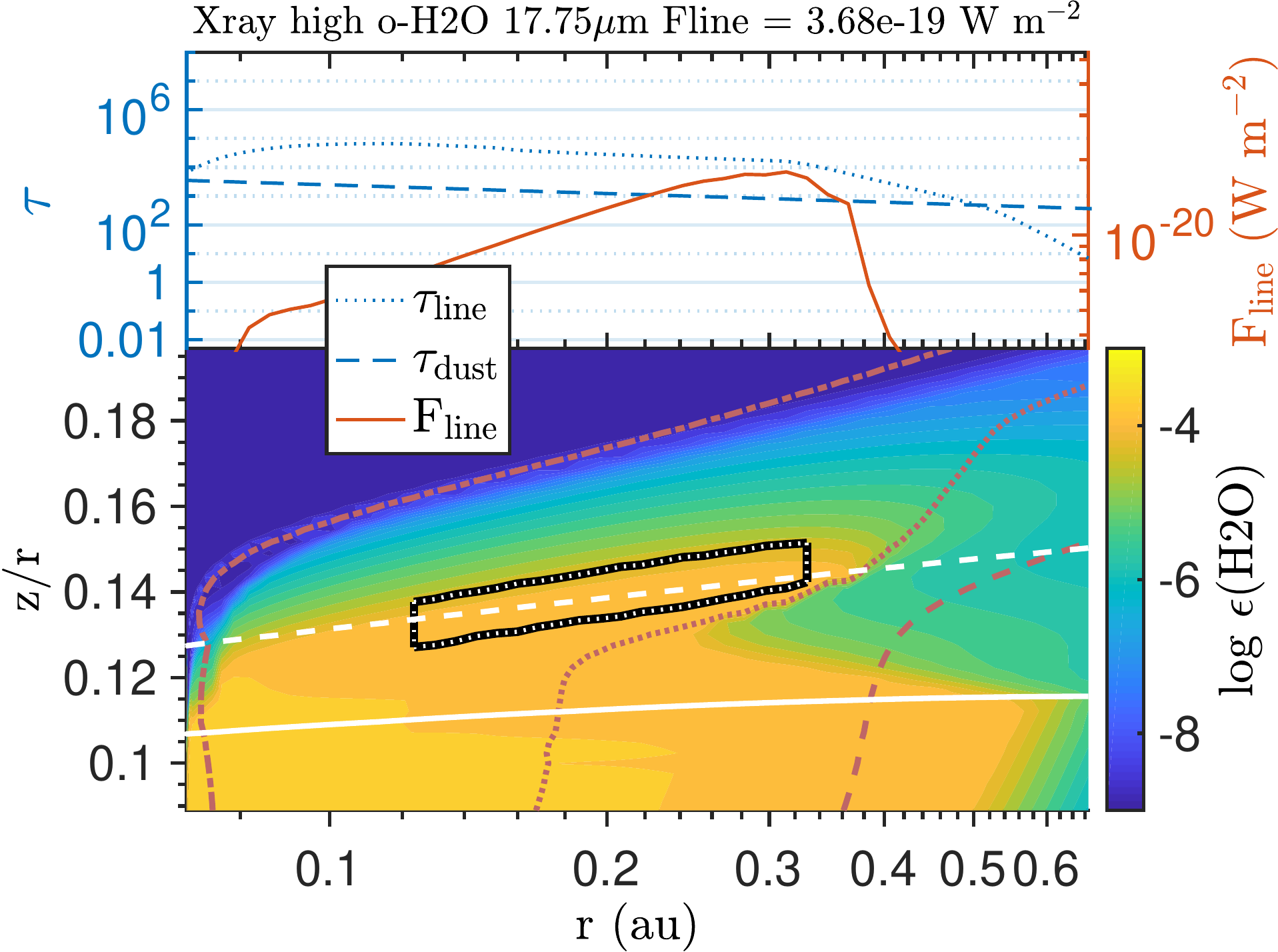}}  
			\caption{Line-emitting regions for the model Xray high. The plotted lines are \cem{C2H2}, \cem{HCN}, \cem{CO2}, \cem{NH3}, \cem{OH}, and \cem{o-H2O}. The rest of the figure is as described in \cref{fig:LER_TT_highres}.
			}\label{fig:LER_Xray_high}   
		\end{figure*}
		
		\begin{figure*} \centering    
			\makebox[\textwidth][c]{\includegraphics[width=0.47\textwidth]{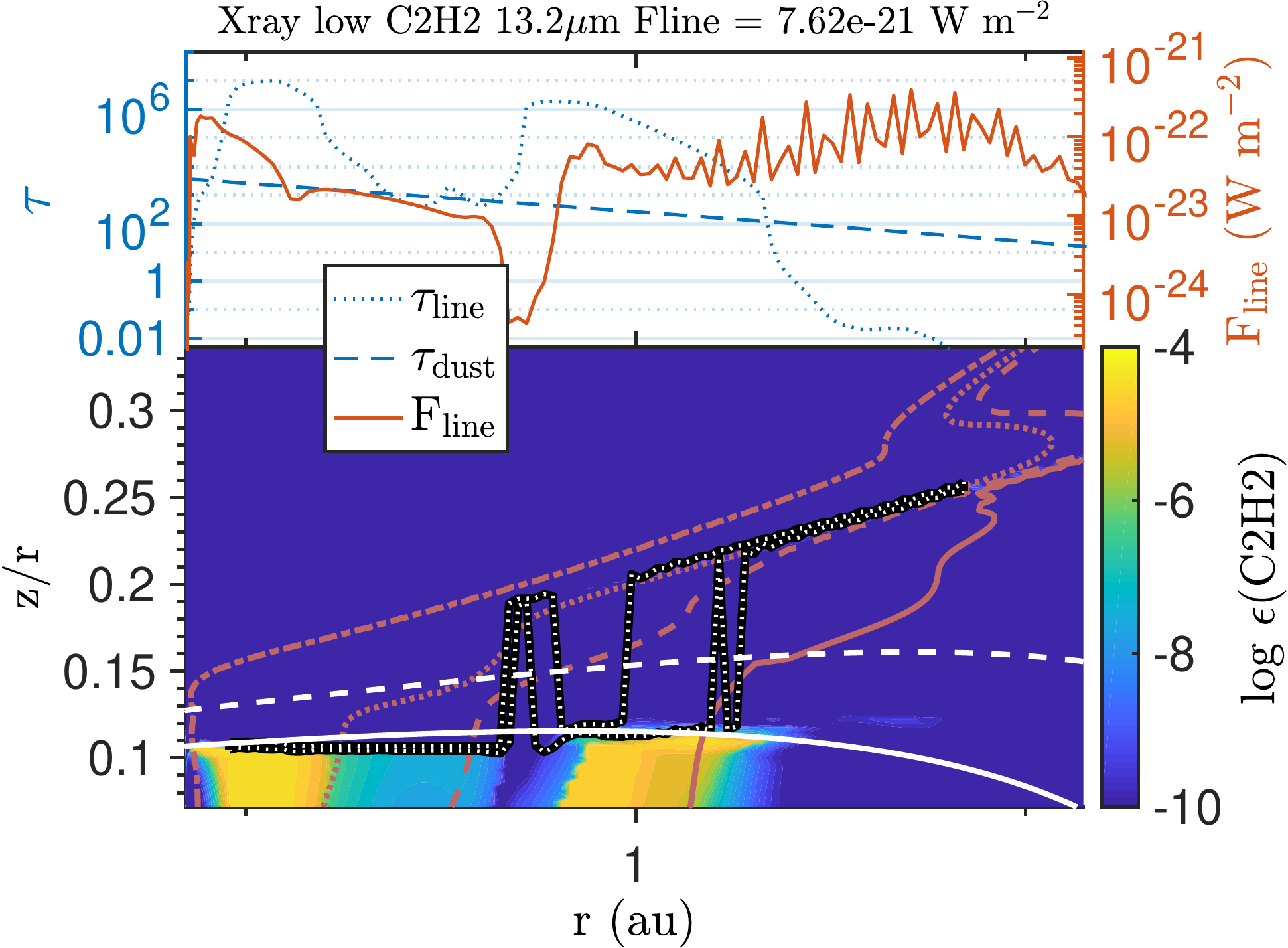} \hspace{0.005\textwidth}  
				\includegraphics[width=0.47\textwidth]{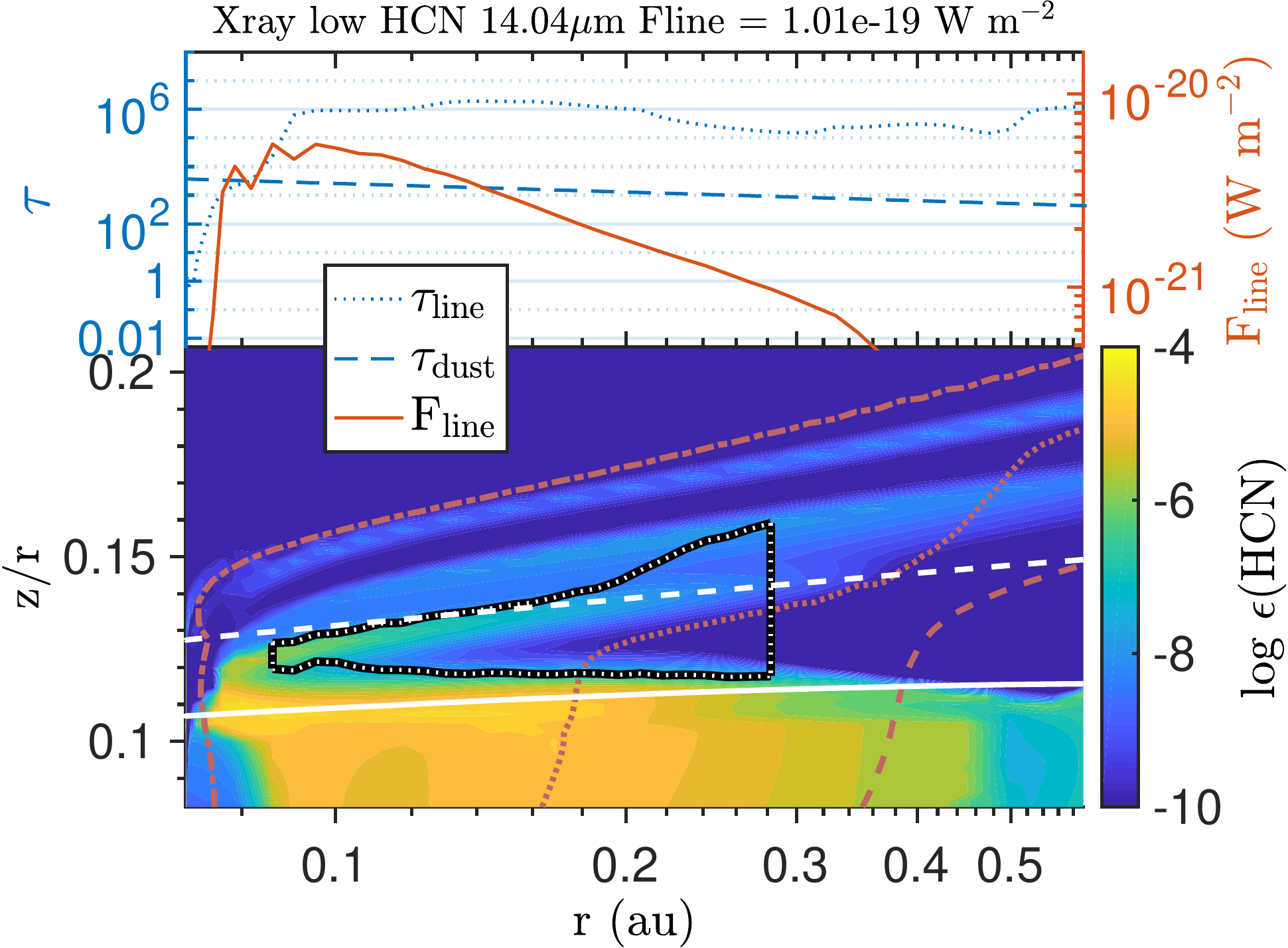}}     
			\makebox[\textwidth][c]{\includegraphics[width=0.47\textwidth]{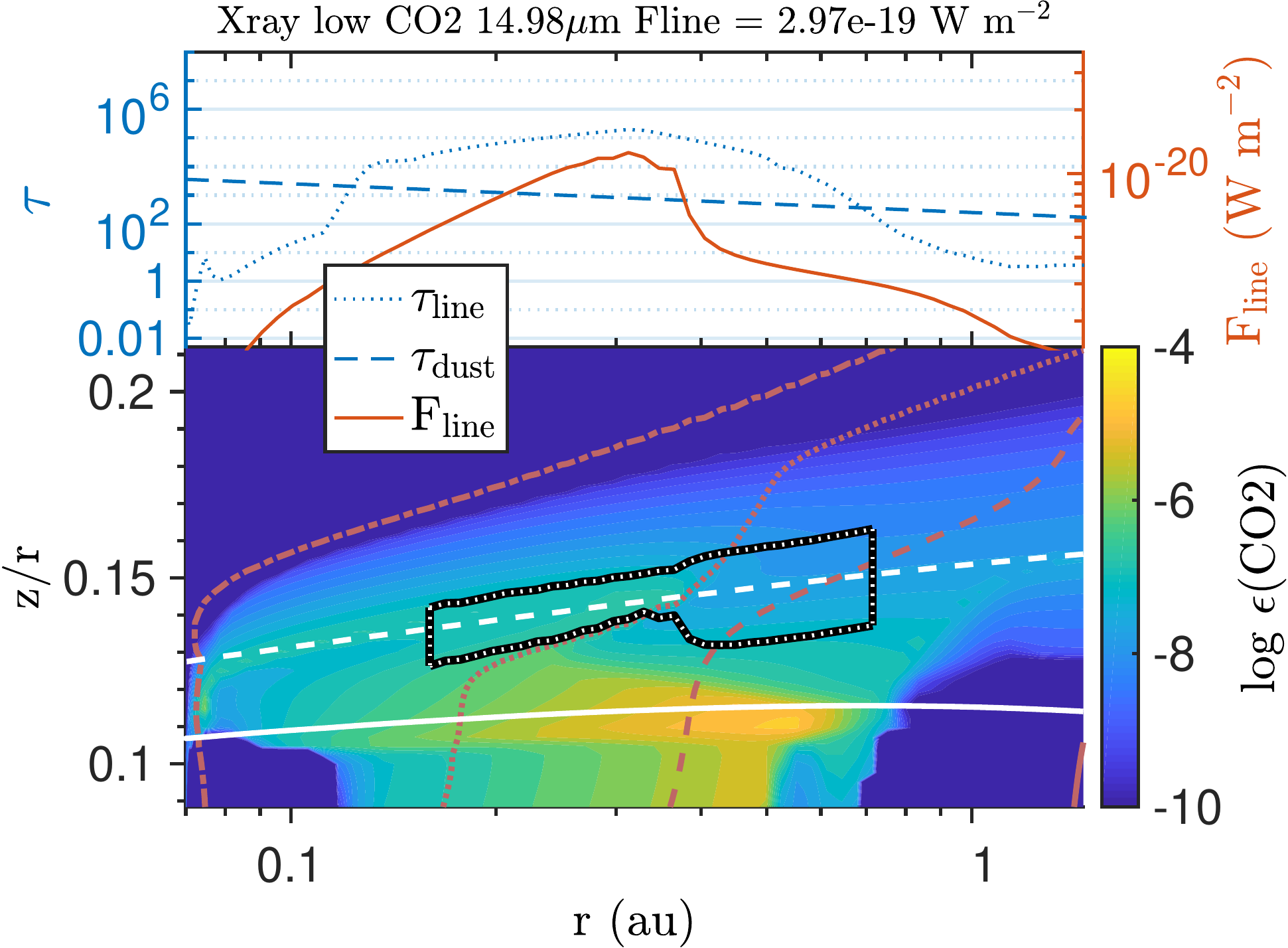} \hspace{0.005\textwidth}
				\includegraphics[width=0.47\textwidth]{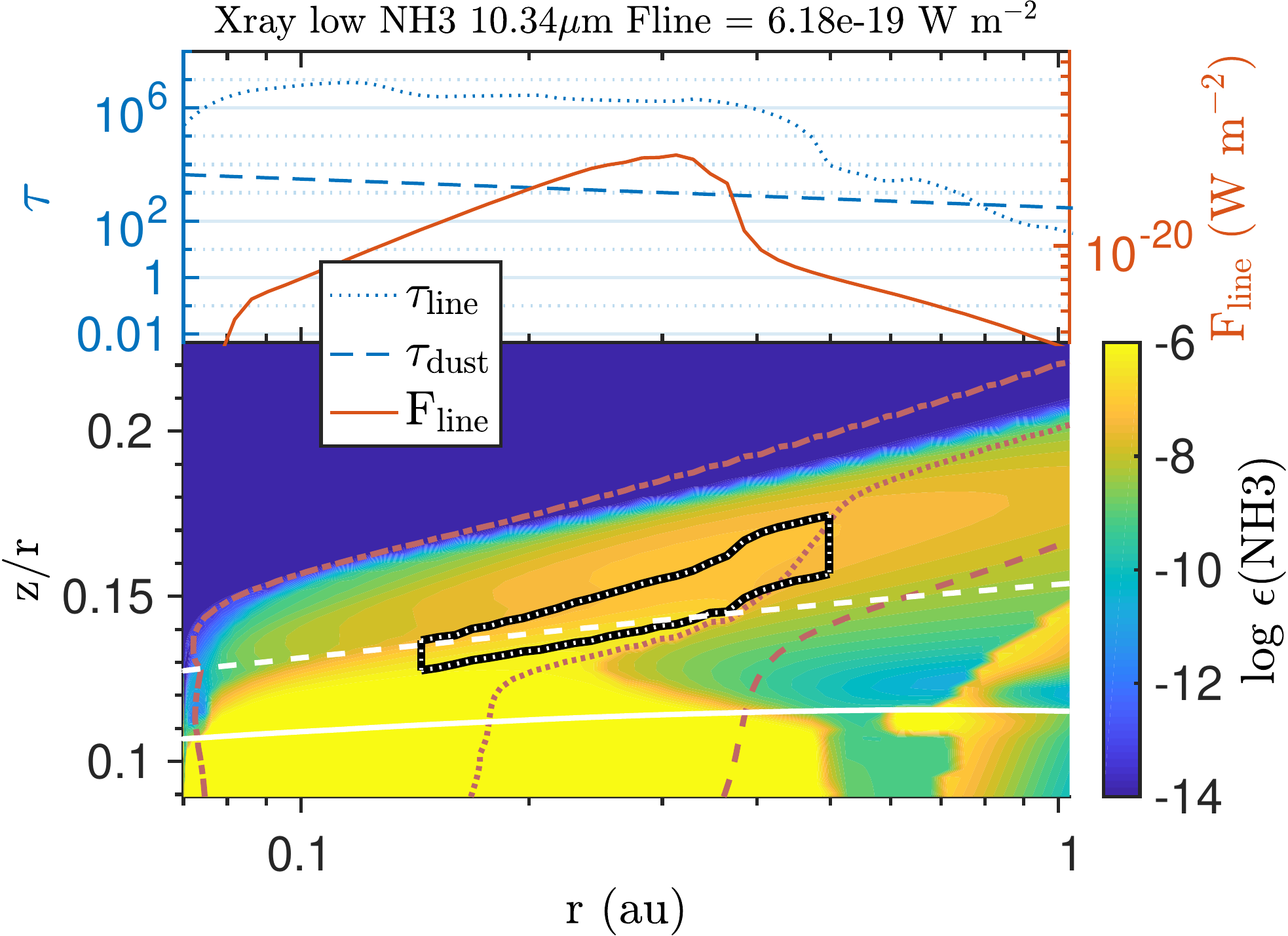}}    
			\makebox[\textwidth][c]{\includegraphics[width=0.47\textwidth]{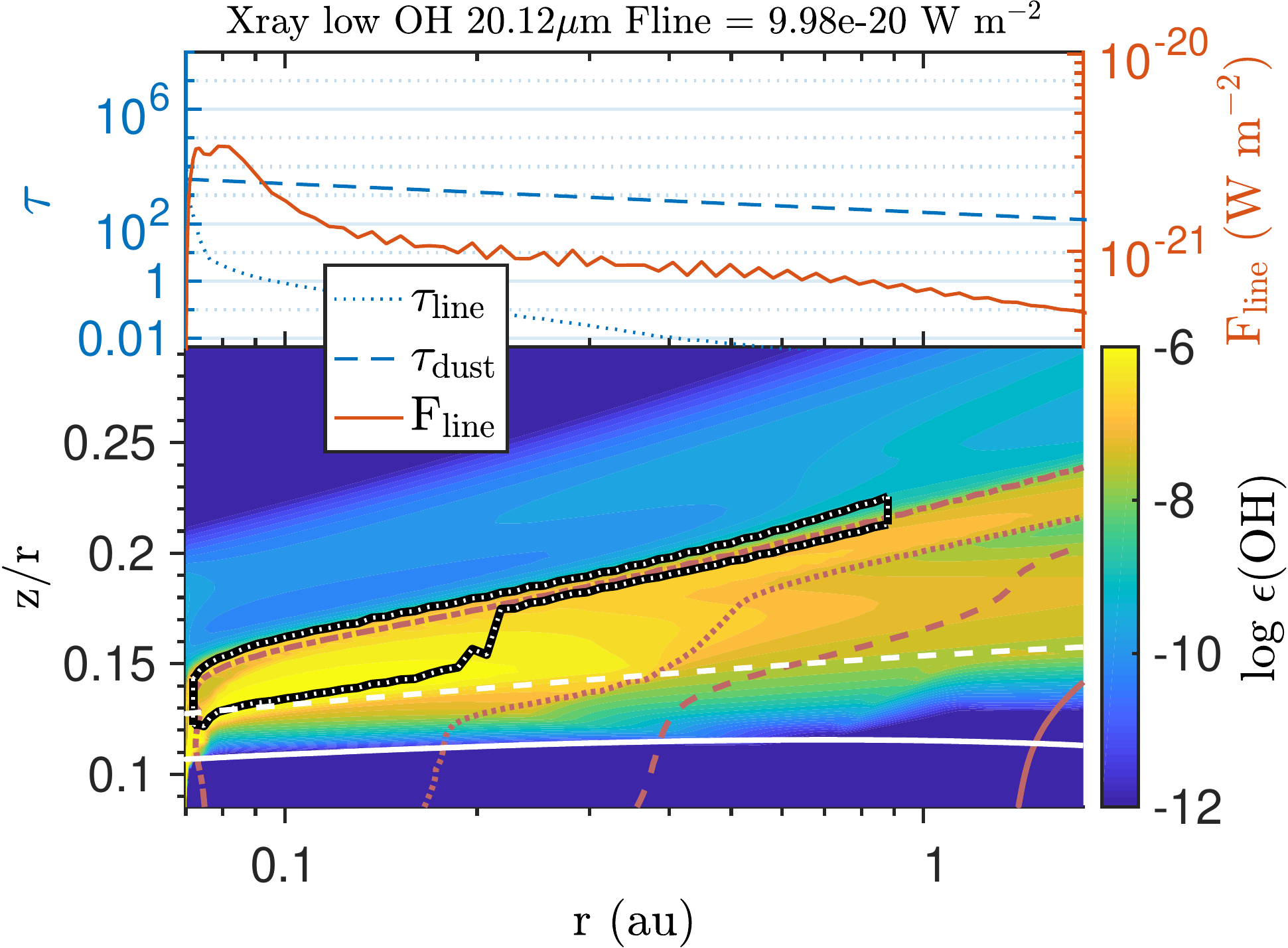} \hspace{0.005\textwidth}    
				\includegraphics[width=0.47\textwidth]{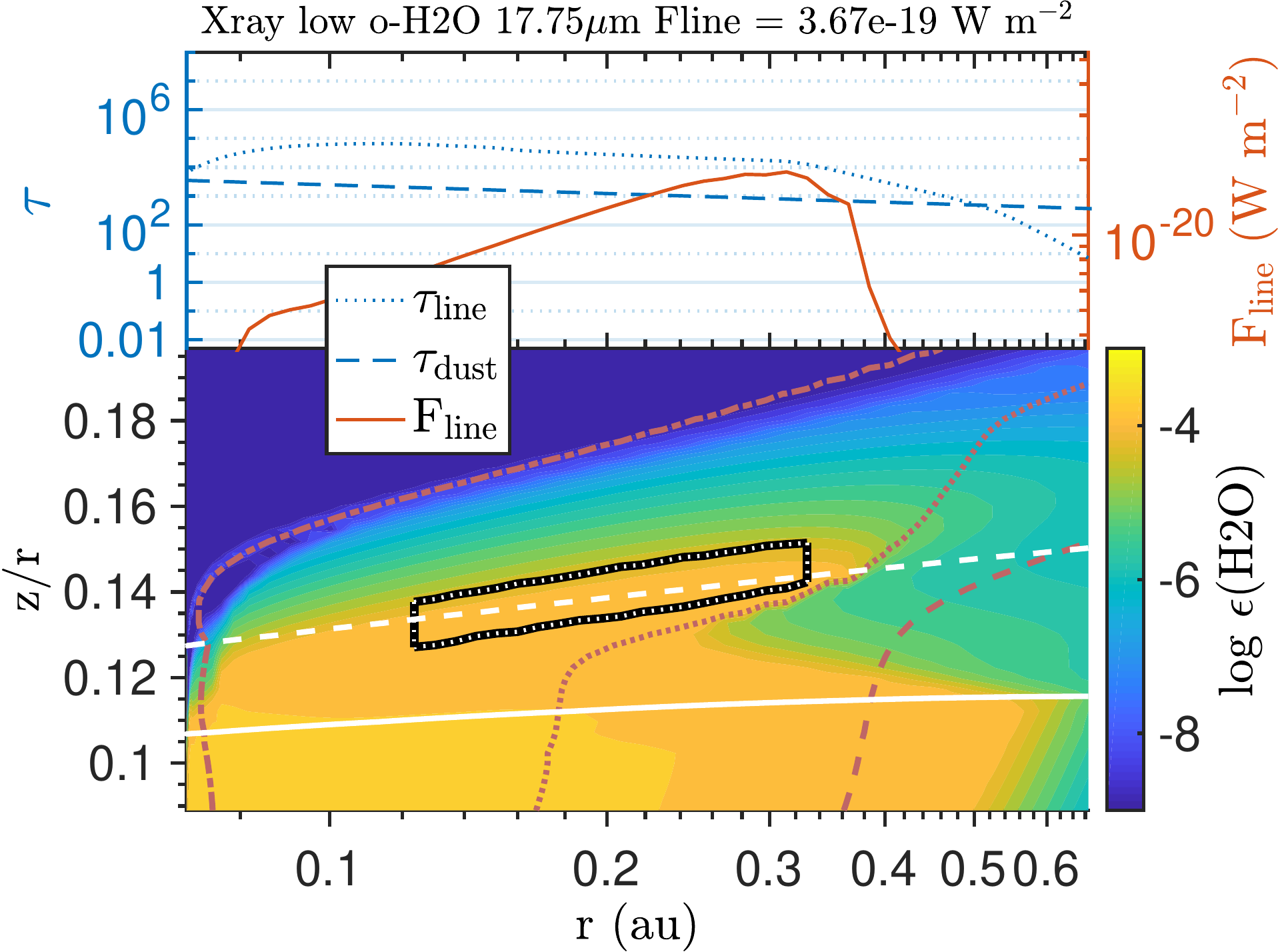}}  
			\caption{Line-emitting regions for the model Xray low. The plotted lines are \cem{C2H2}, \cem{HCN}, \cem{CO2}, \cem{NH3}, \cem{OH}, and \cem{o-H2O}. The rest of the figure is as described in \cref{fig:LER_TT_highres}. 
			}\label{fig:LER_Xray_low}    
		\end{figure*}
		
		\begin{figure*} \centering    
			\makebox[\textwidth][c]{\includegraphics[width=0.47\textwidth]{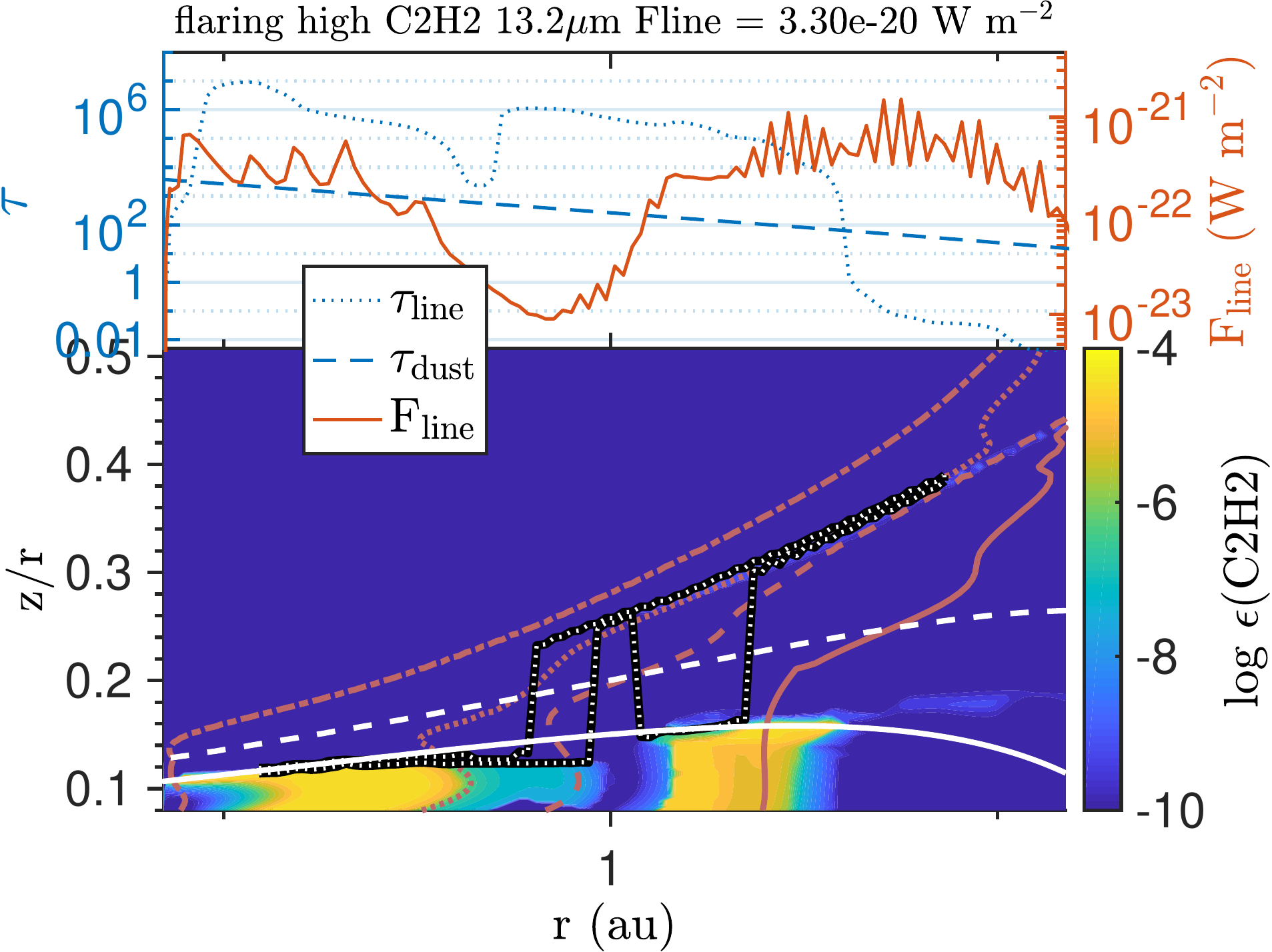} \hspace{0.005\textwidth}    
				\includegraphics[width=0.47\textwidth]{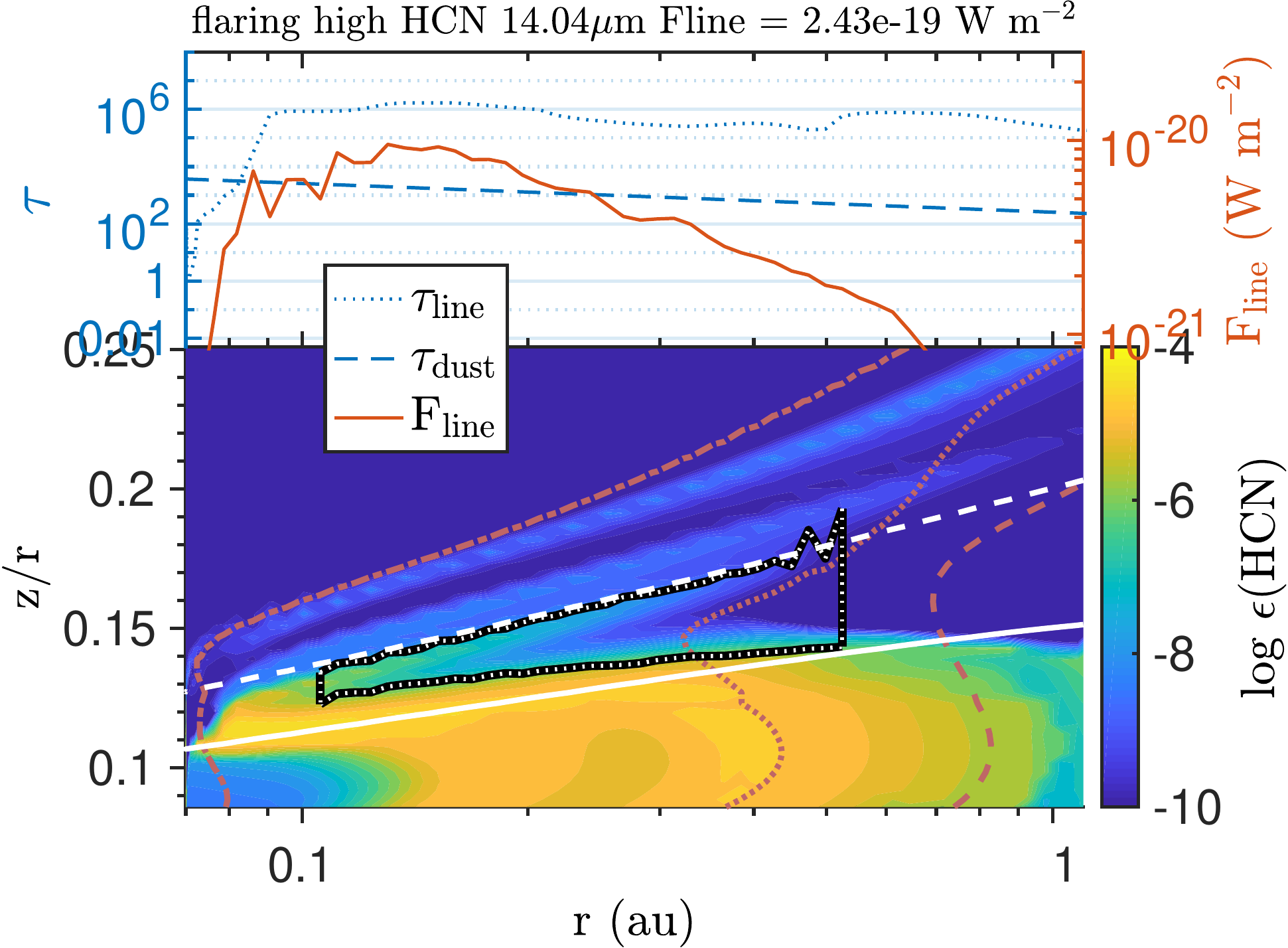}} 
			\makebox[\textwidth][c]{\includegraphics[width=0.47\textwidth]{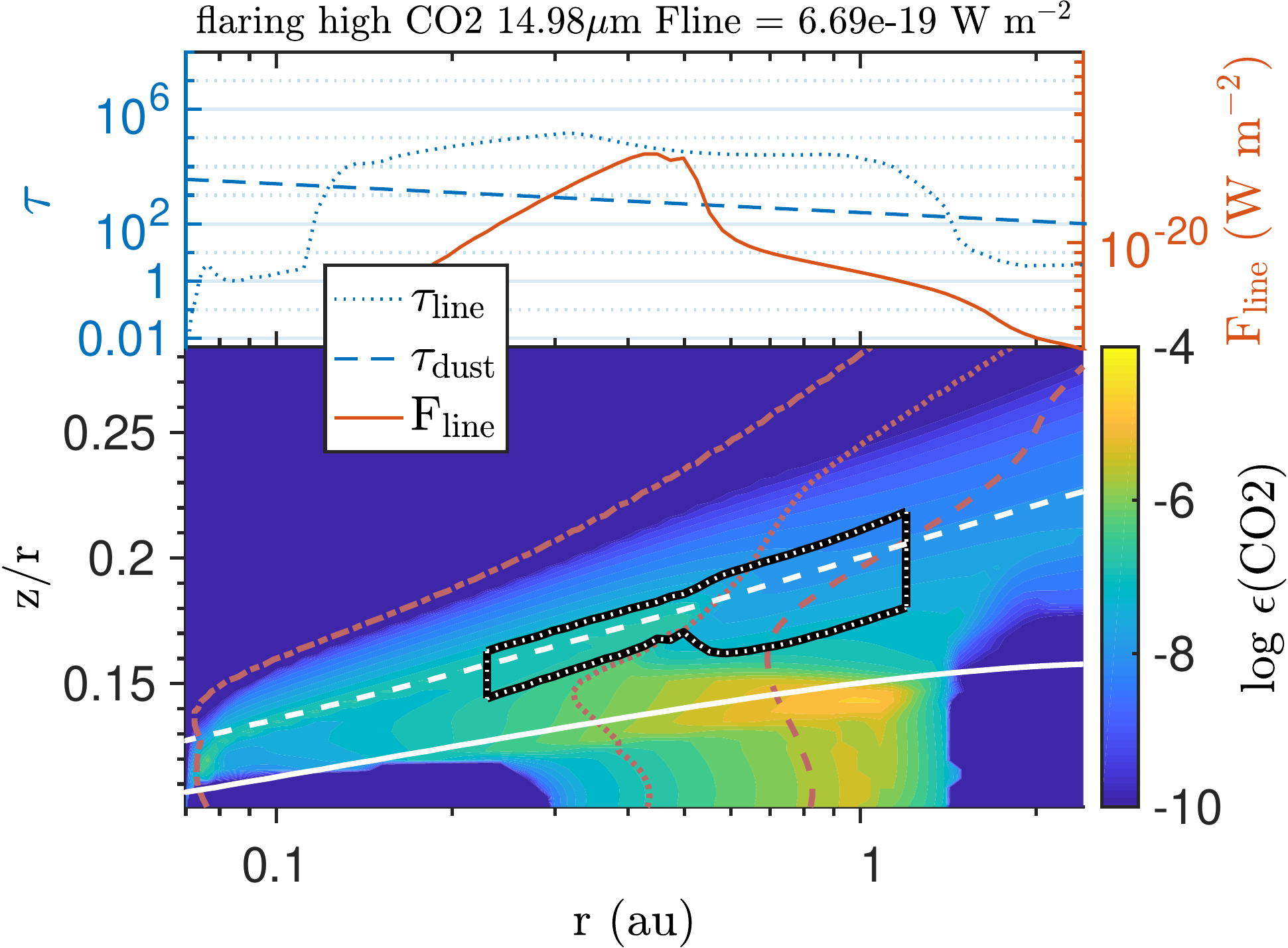} \hspace{0.005\textwidth}
				\includegraphics[width=0.47\textwidth]{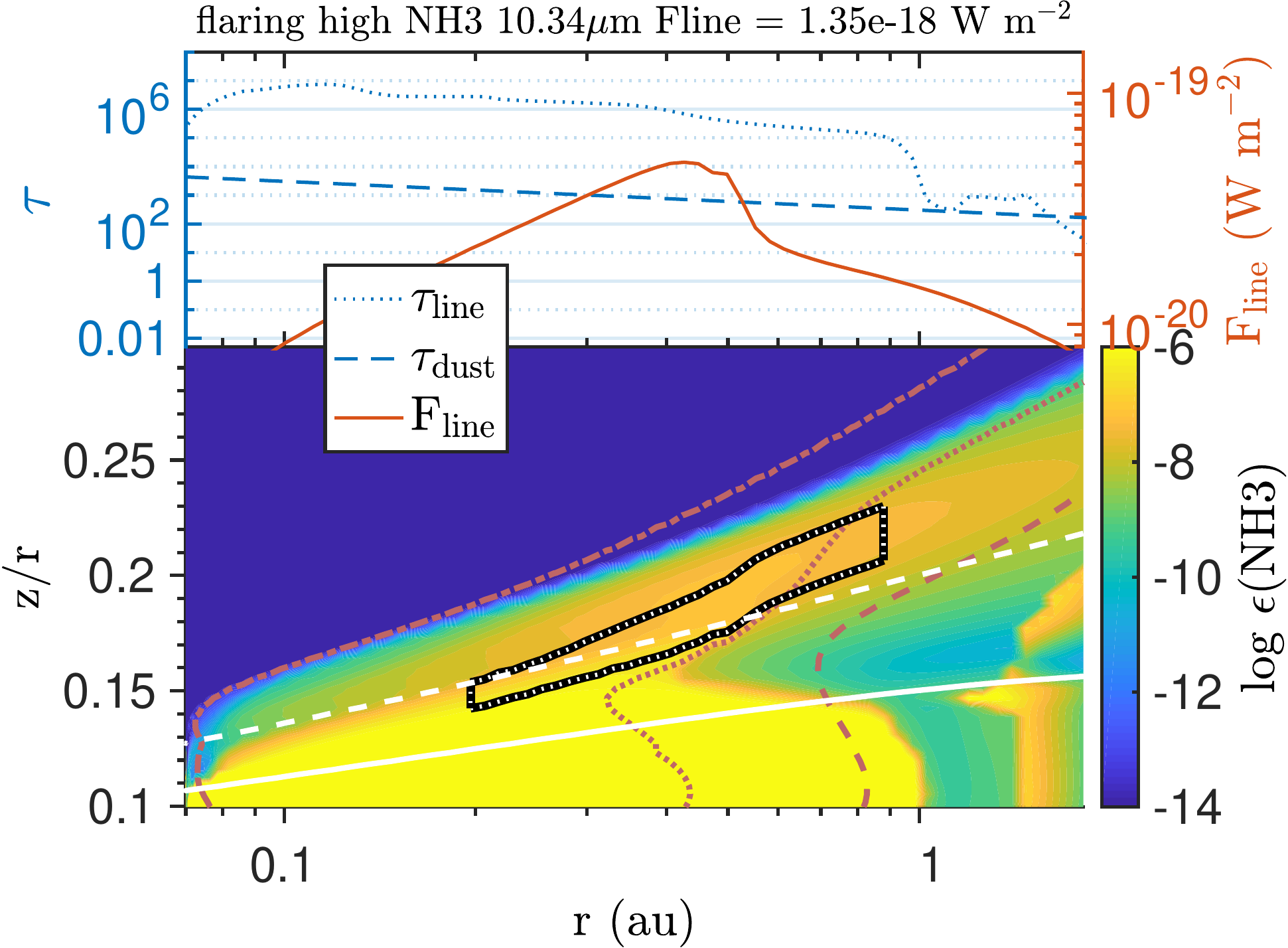}}      
			\makebox[\textwidth][c]{\includegraphics[width=0.47\textwidth]{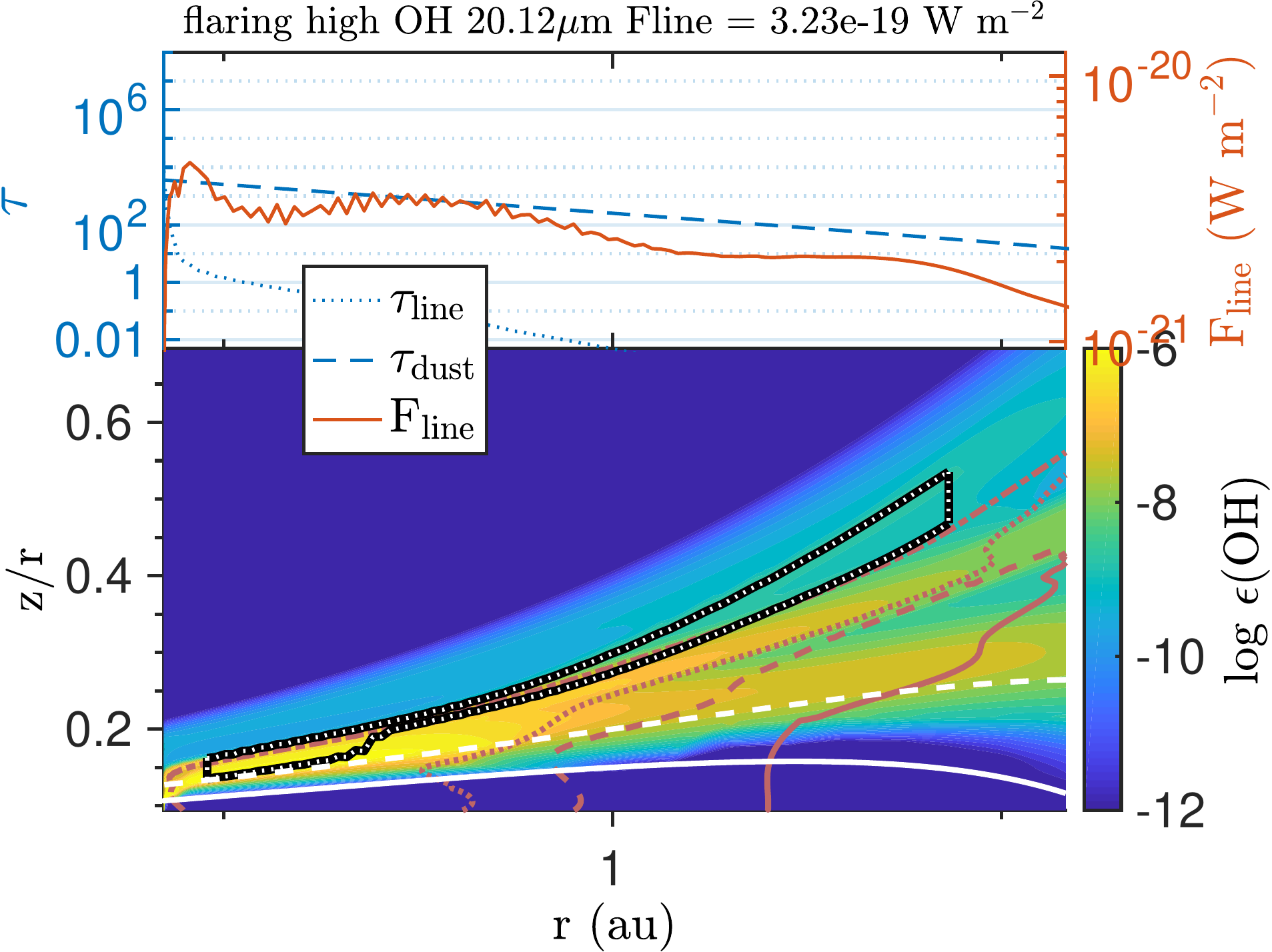} \hspace{0.005\textwidth}
				\includegraphics[width=0.47\textwidth]{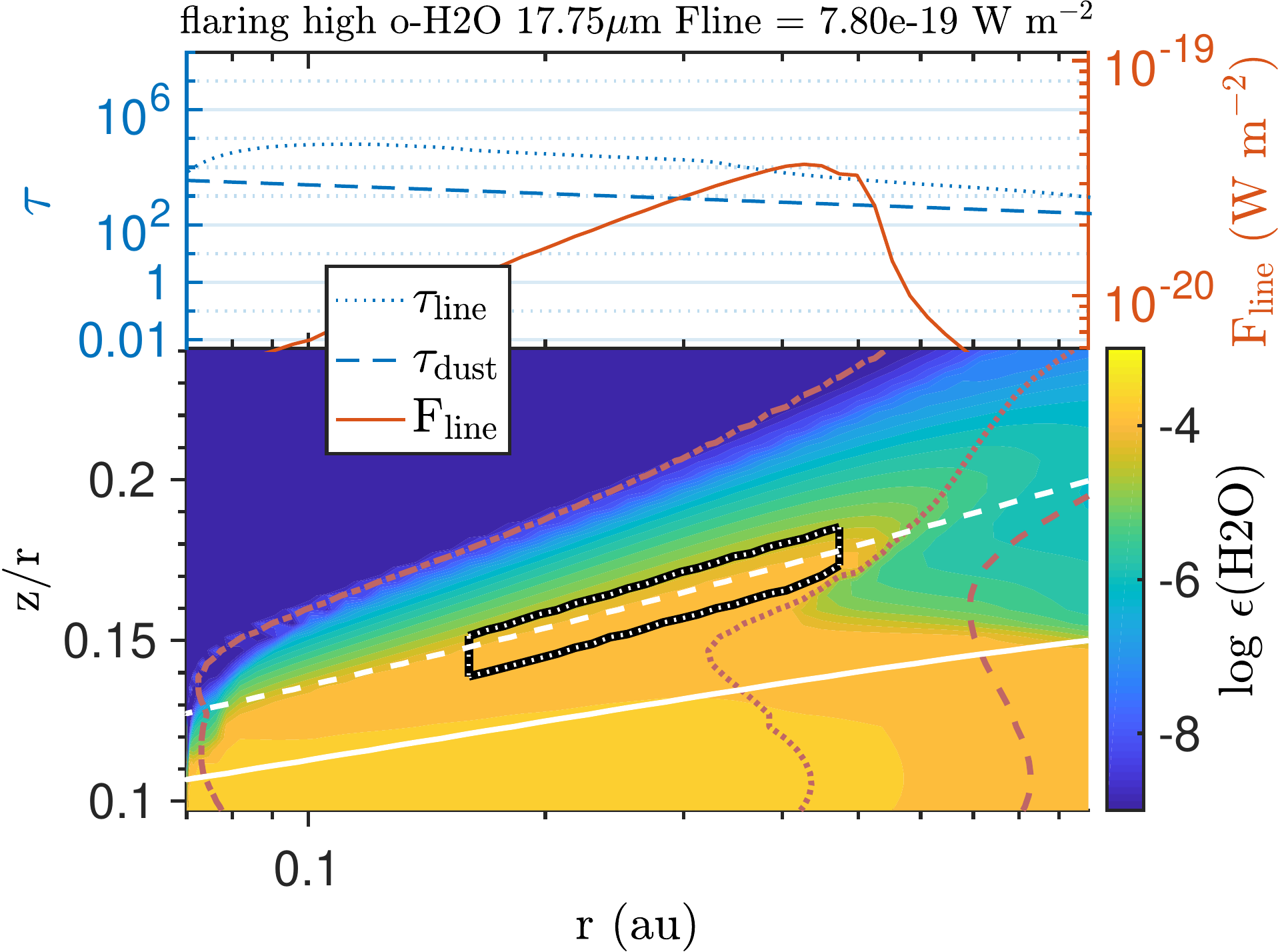}}    
			\caption{Line-emitting regions for the model flaring high. The plotted lines are \cem{C2H2}, \cem{HCN}, \cem{CO2}, \cem{NH3}, \cem{OH}, and \cem{o-H2O}. The rest of the figure is as described in \cref{fig:LER_TT_highres}.   
			}\label{fig:LER_flaring_high}
		\end{figure*}
		
		\begin{figure*} \centering    
			\makebox[\textwidth][c]{\includegraphics[width=0.47\textwidth]{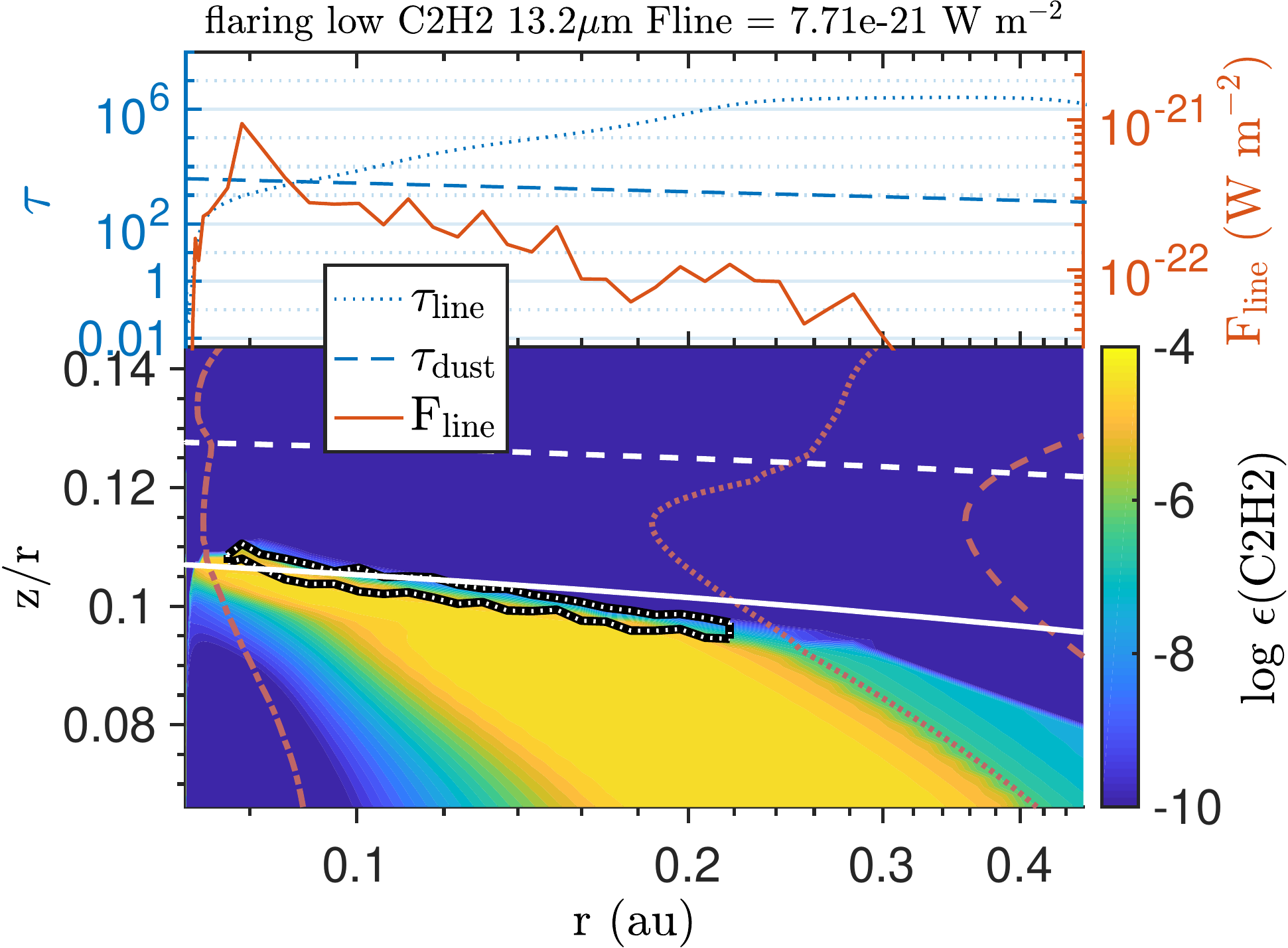} \hspace{0.005\textwidth}   
				\includegraphics[width=0.47\textwidth]{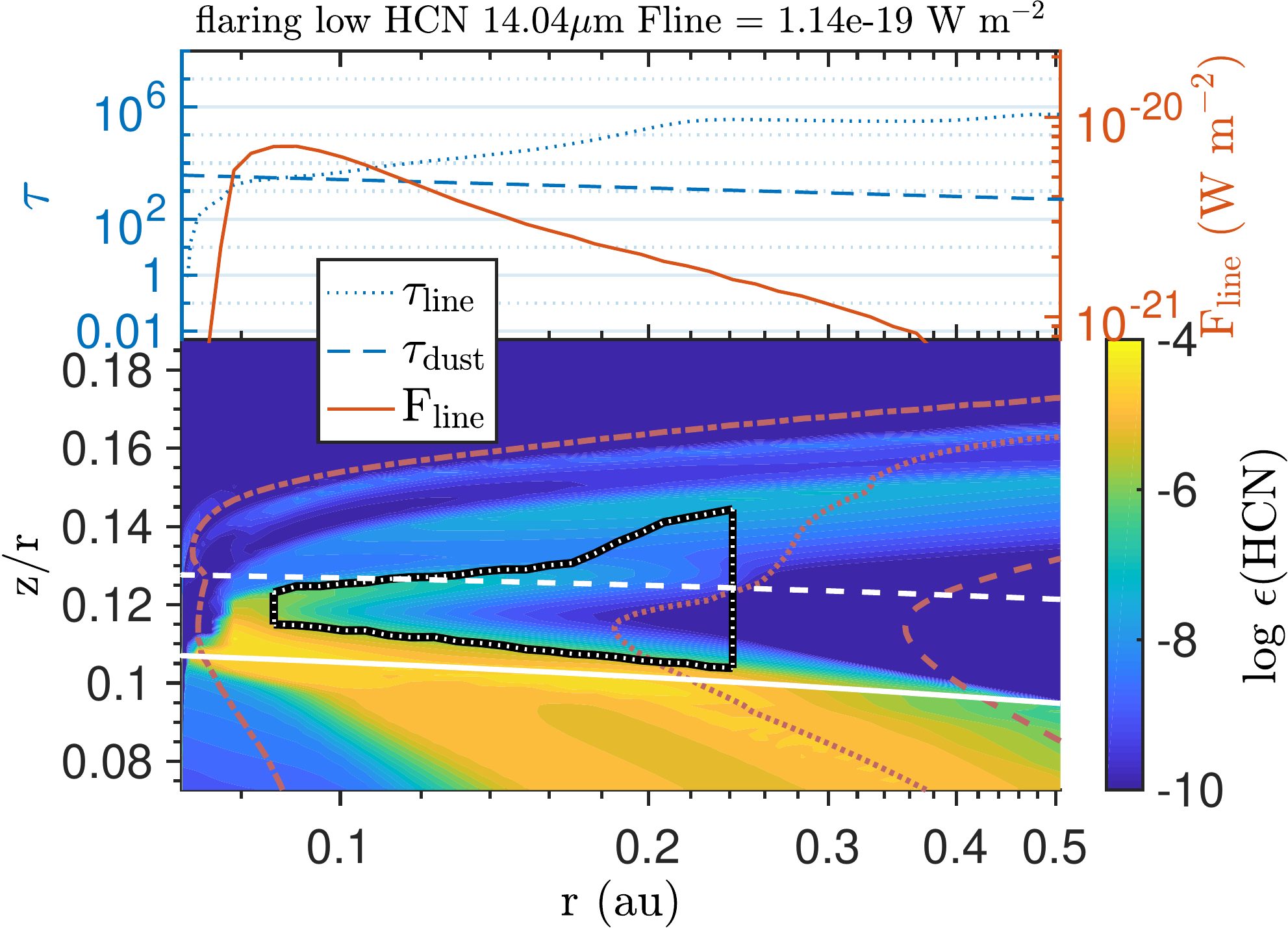}}      
			\makebox[\textwidth][c]{\includegraphics[width=0.47\textwidth]{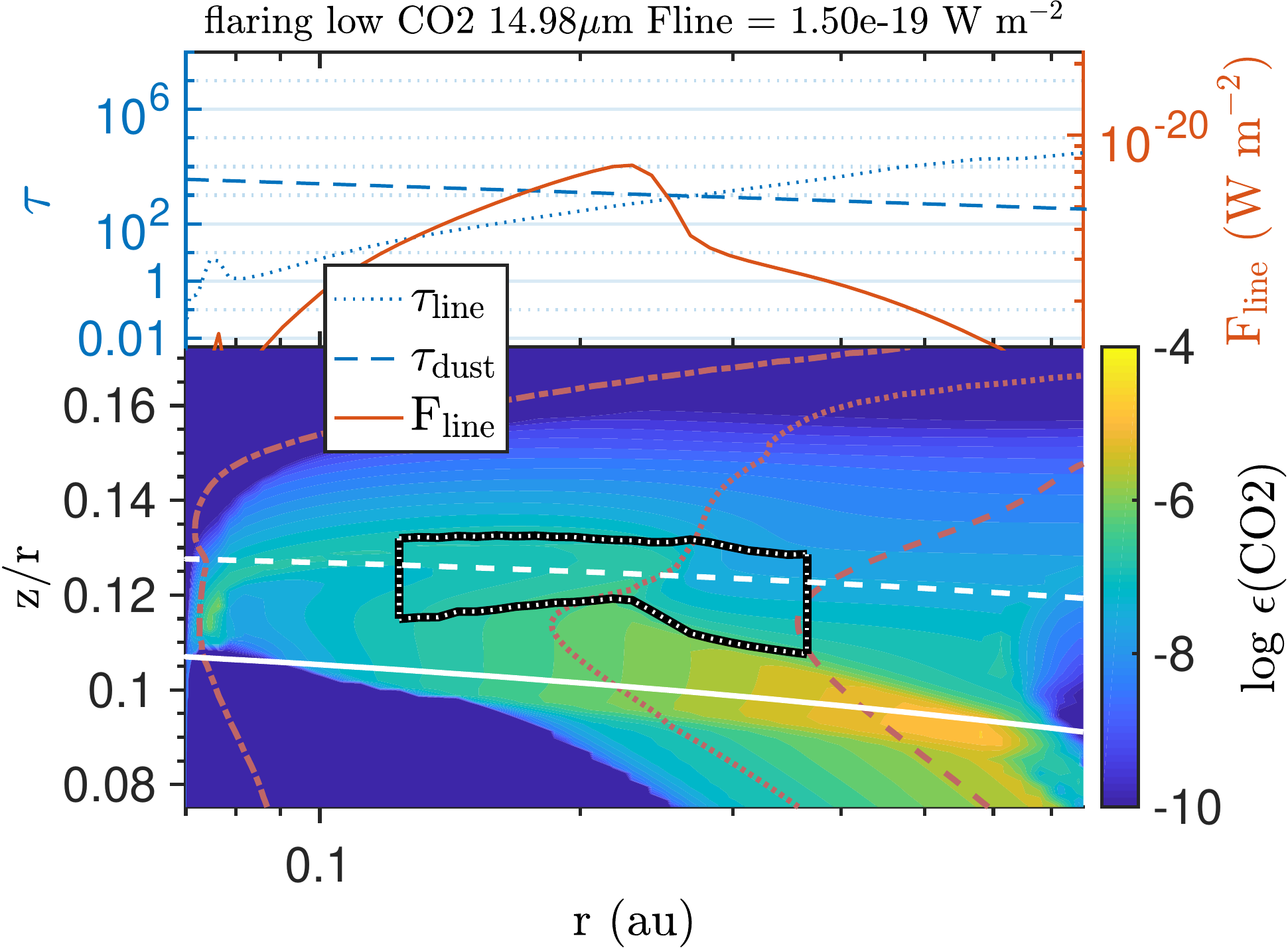} \hspace{0.005\textwidth}
				\includegraphics[width=0.47\textwidth]{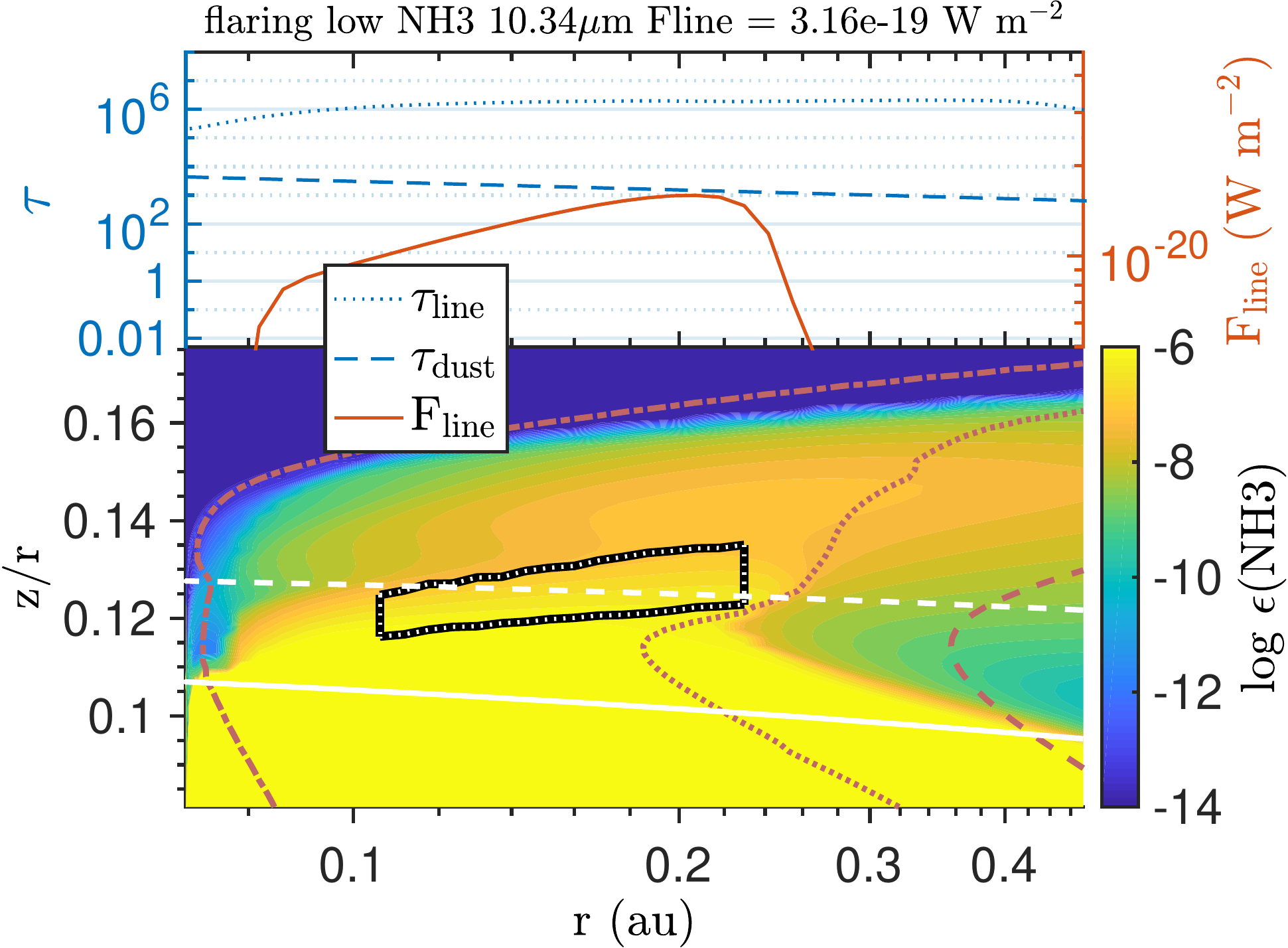}}     
			\makebox[\textwidth][c]{\includegraphics[width=0.47\textwidth]{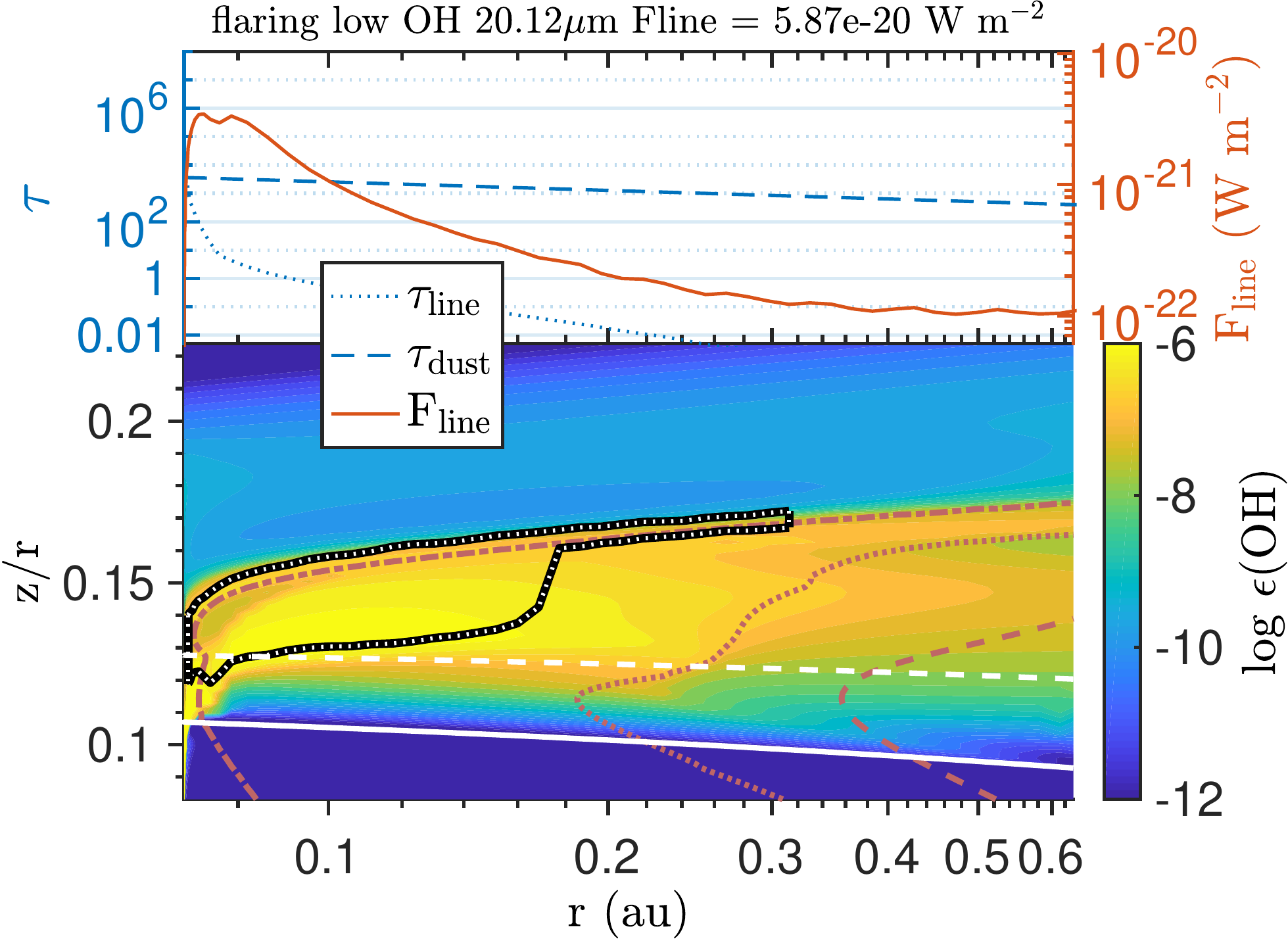} \hspace{0.005\textwidth}     
				\includegraphics[width=0.47\textwidth]{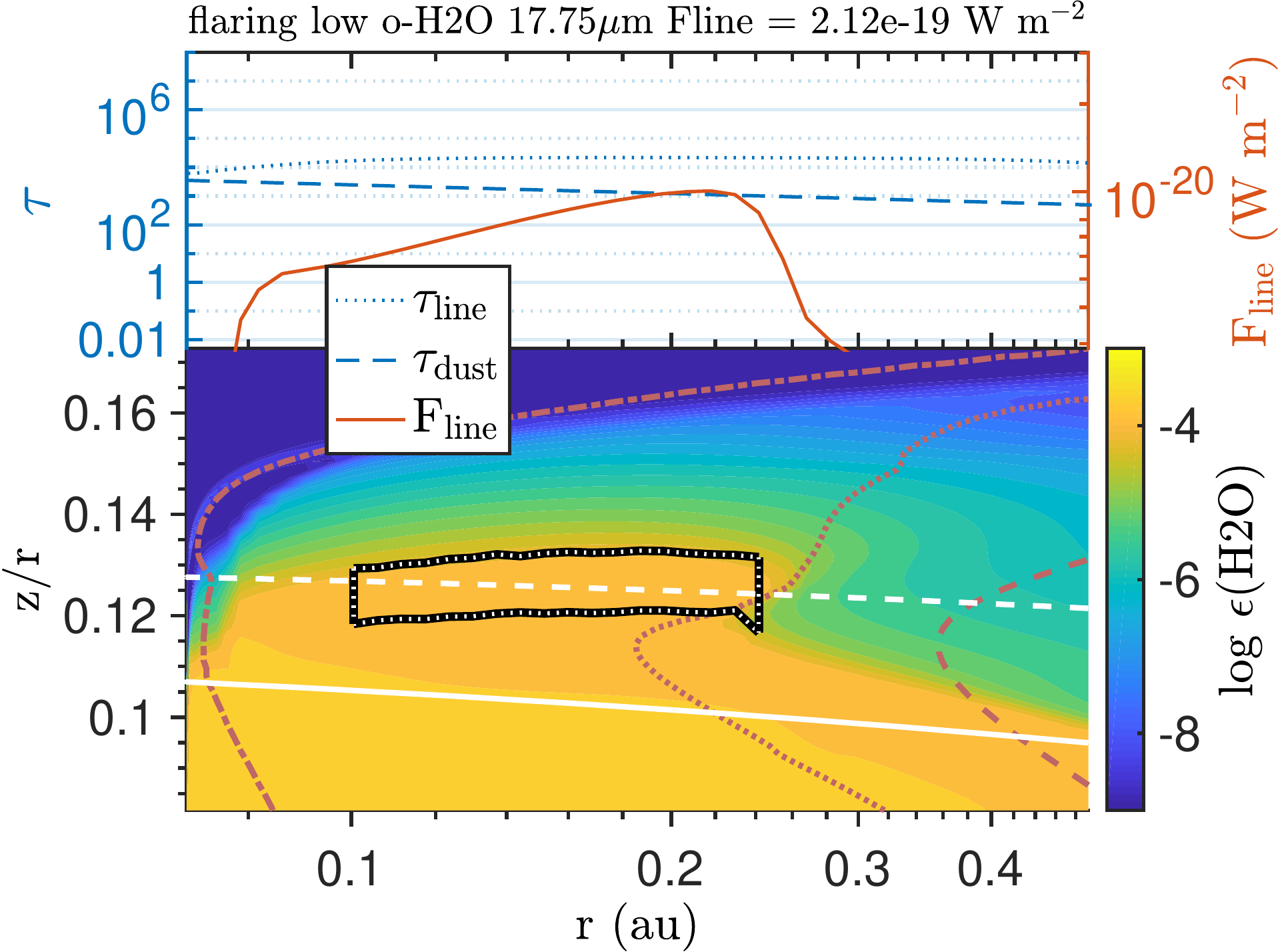}}    
			\caption{Line-emitting regions for the model flaring low. The plotted lines are \cem{C2H2}, \cem{HCN}, \cem{CO2}, \cem{NH3}, \cem{OH}, and \cem{o-H2O}. The rest of the figure is as described in \cref{fig:LER_TT_highres}.
			}\label{fig:LER_flaring_low_scaleheightfix_run5}     
		\end{figure*}
		
		\begin{figure*} \centering    
			\makebox[\textwidth][c]{\includegraphics[width=0.47\textwidth]{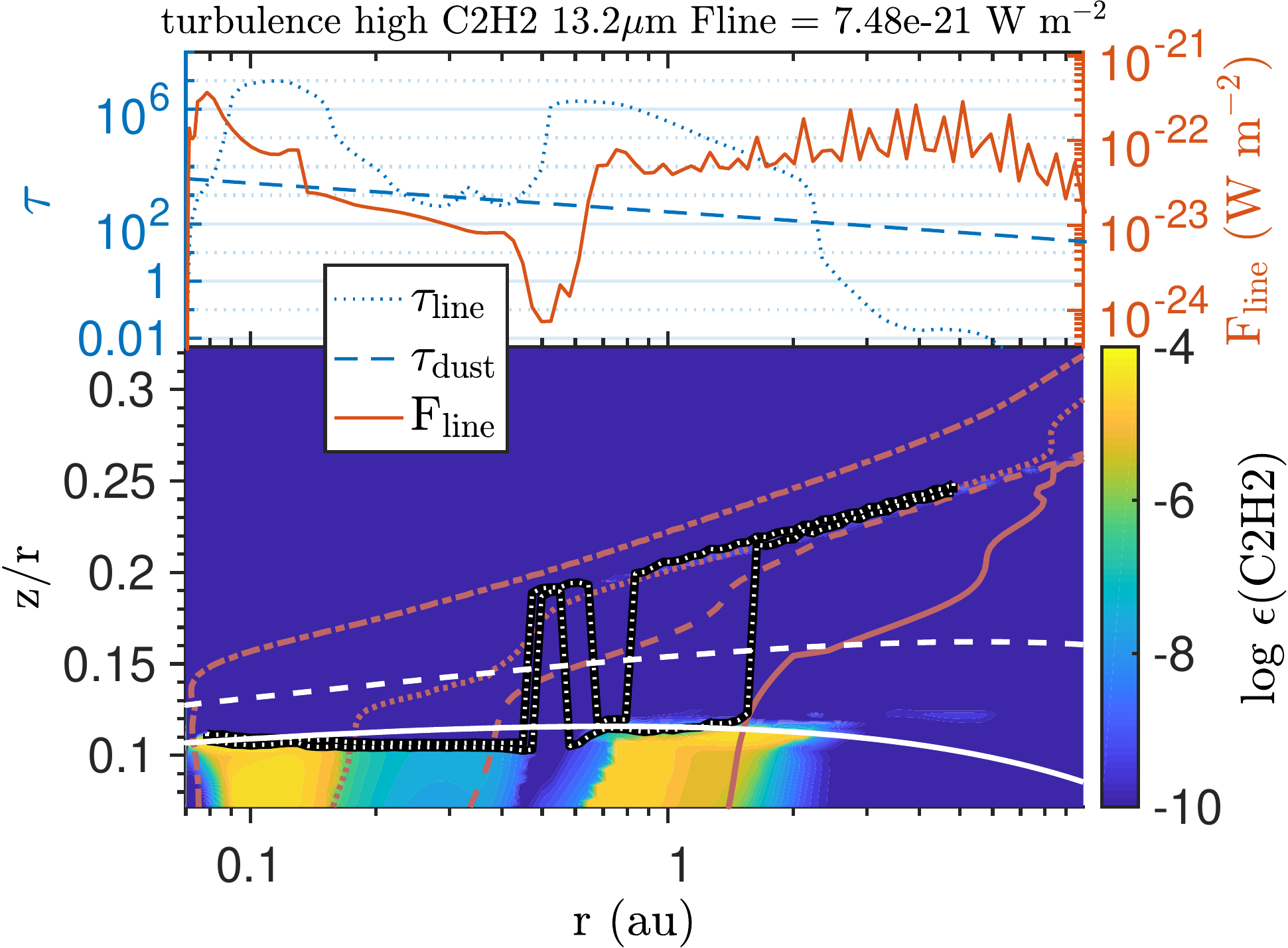} \hspace{0.005\textwidth}   
				\includegraphics[width=0.47\textwidth]{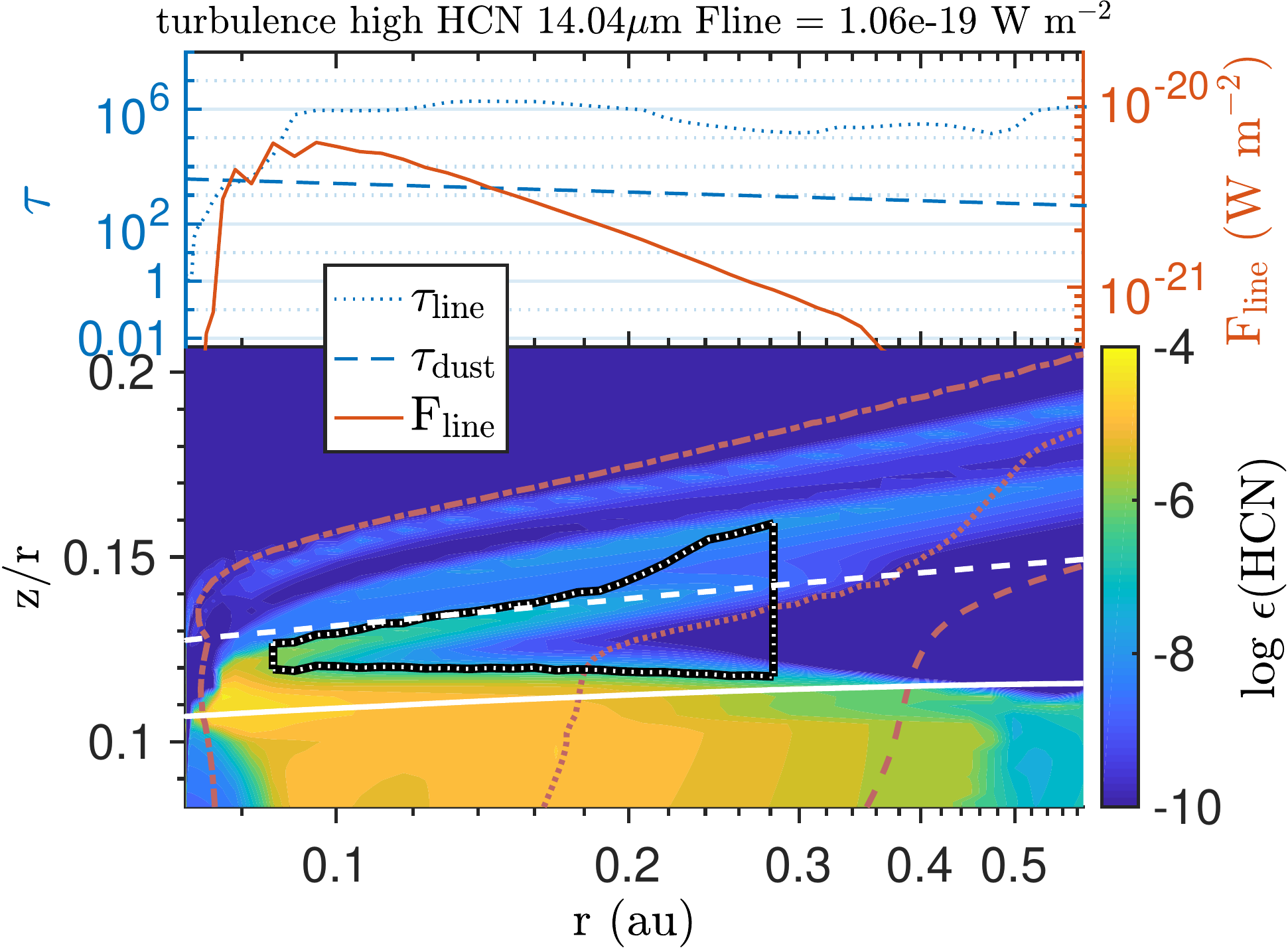}}      
			\makebox[\textwidth][c]{\includegraphics[width=0.47\textwidth]{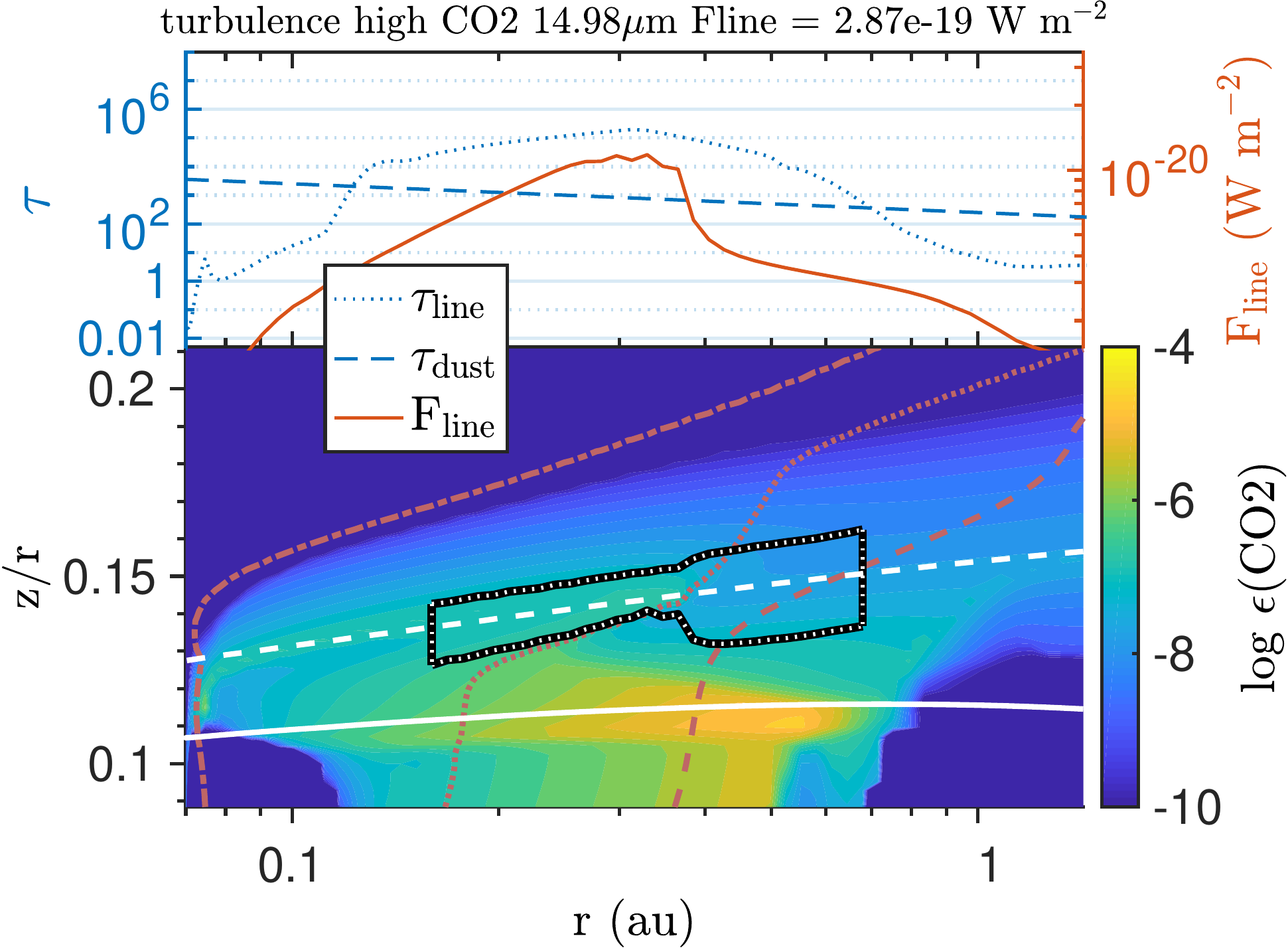} \hspace{0.005\textwidth}
				\includegraphics[width=0.47\textwidth]{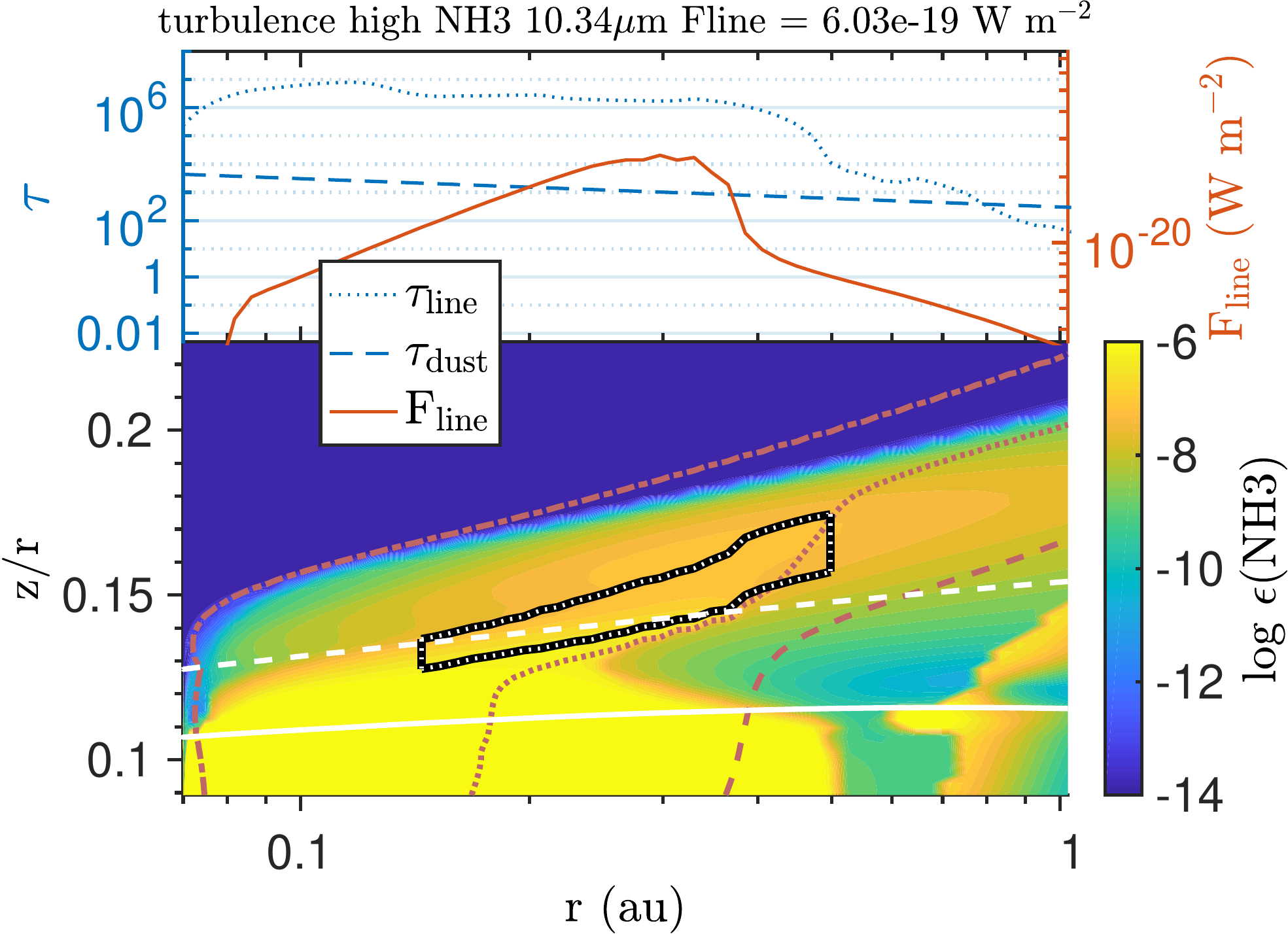}}     
			\makebox[\textwidth][c]{\includegraphics[width=0.47\textwidth]{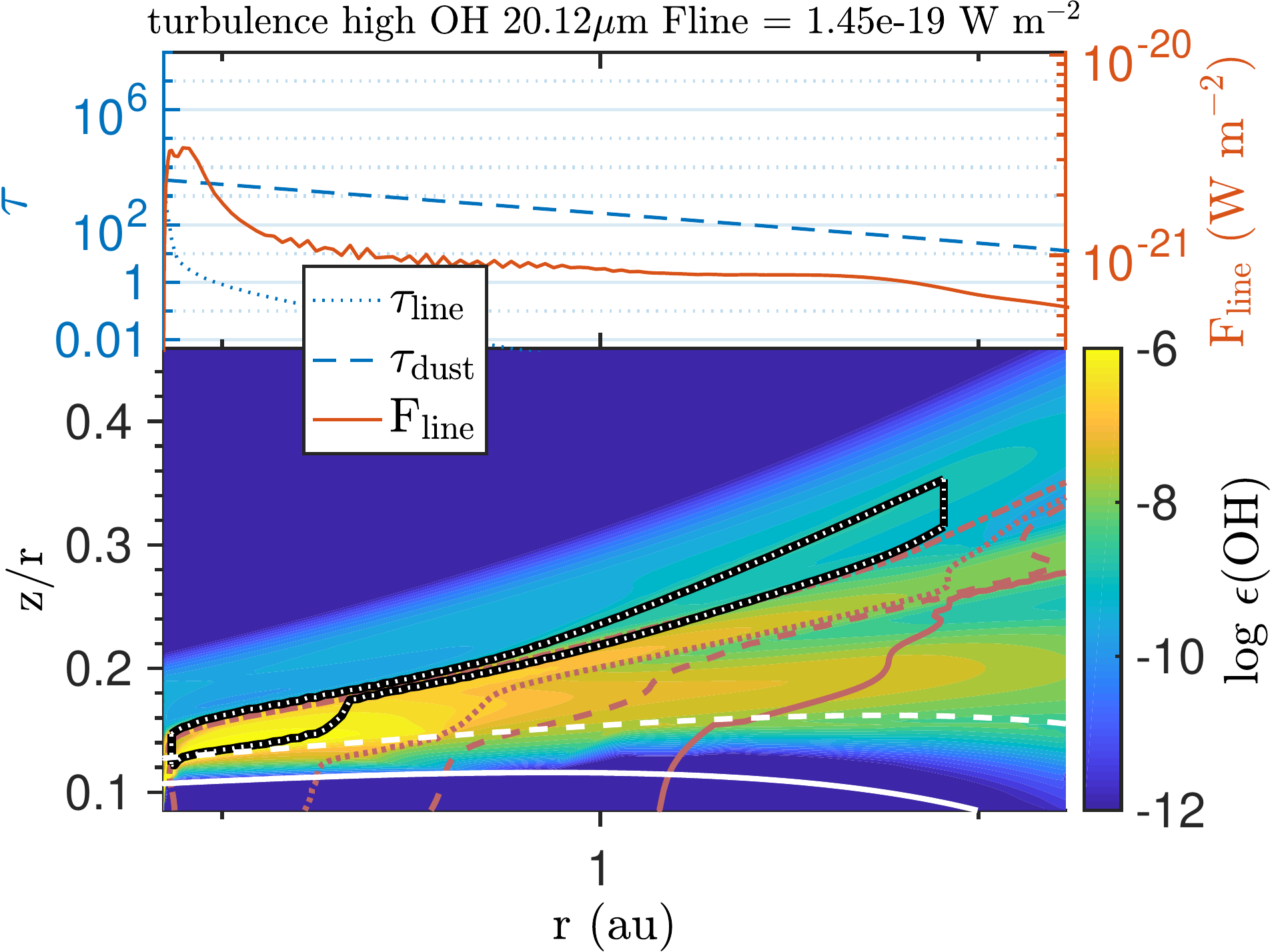} \hspace{0.005\textwidth}     
				\includegraphics[width=0.47\textwidth]{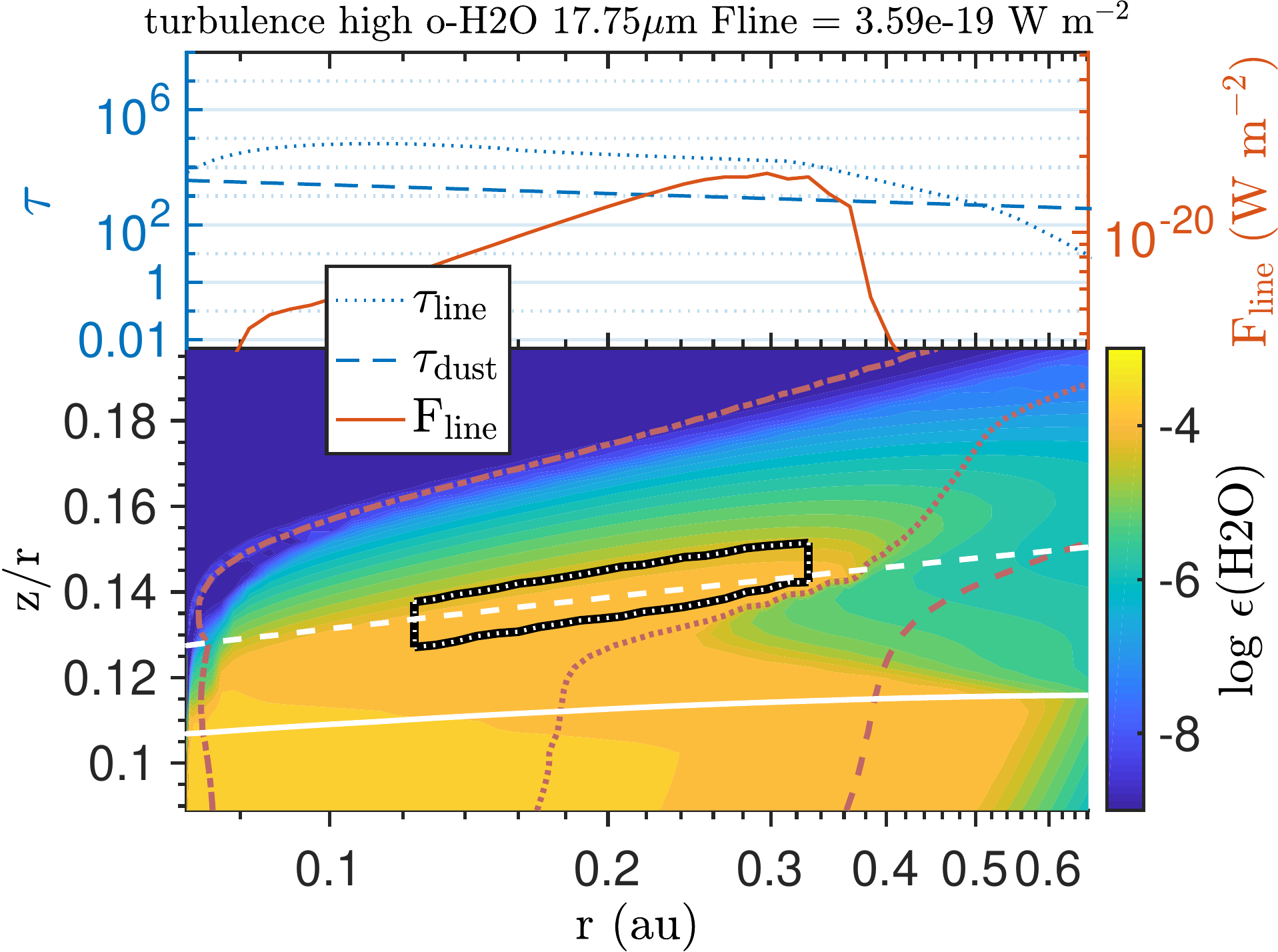}}    
			\caption{Line-emitting regions for the model turbulence high. The plotted lines are \cem{C2H2}, \cem{HCN}, \cem{CO2}, \cem{NH3}, \cem{OH}, and \cem{o-H2O}. The rest of the figure is as described in \cref{fig:LER_TT_highres}.  
			}\label{fig:LER_settling_high}     
		\end{figure*}
		
		\begin{figure*} \centering    
			\makebox[\textwidth][c]{\includegraphics[width=0.47\textwidth]{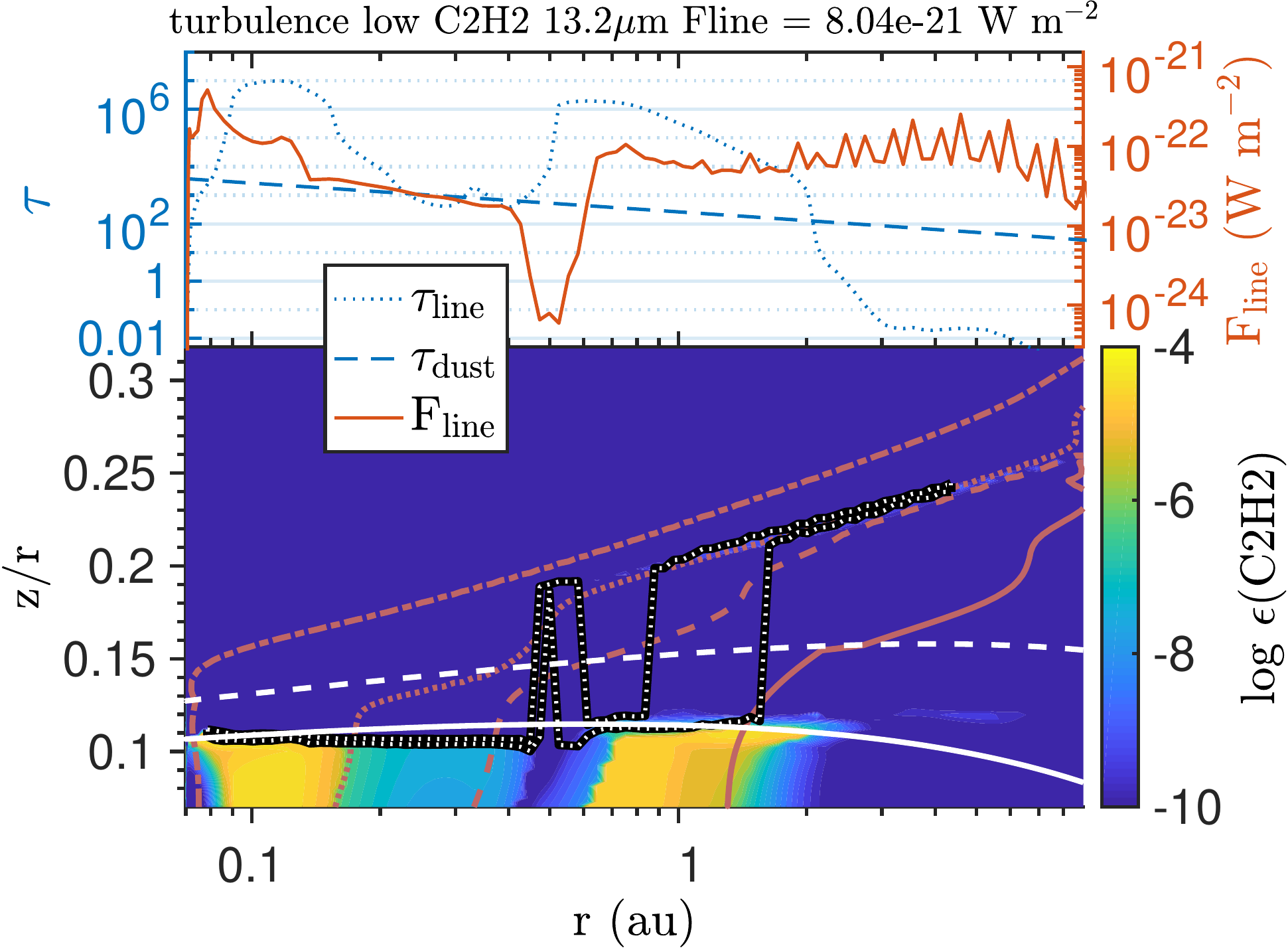} \hspace{0.005\textwidth}    
				\includegraphics[width=0.47\textwidth]{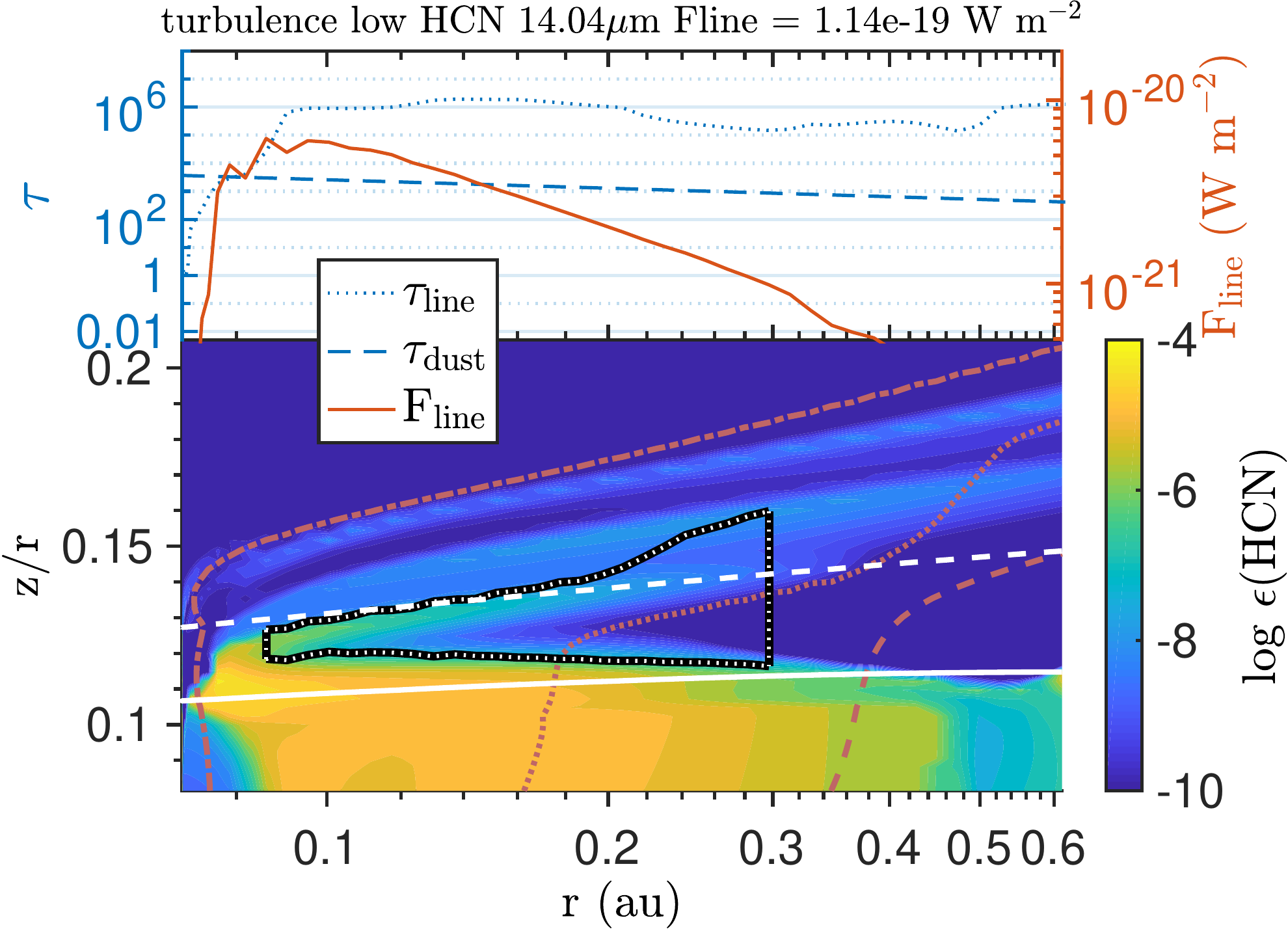}} 
			\makebox[\textwidth][c]{\includegraphics[width=0.47\textwidth]{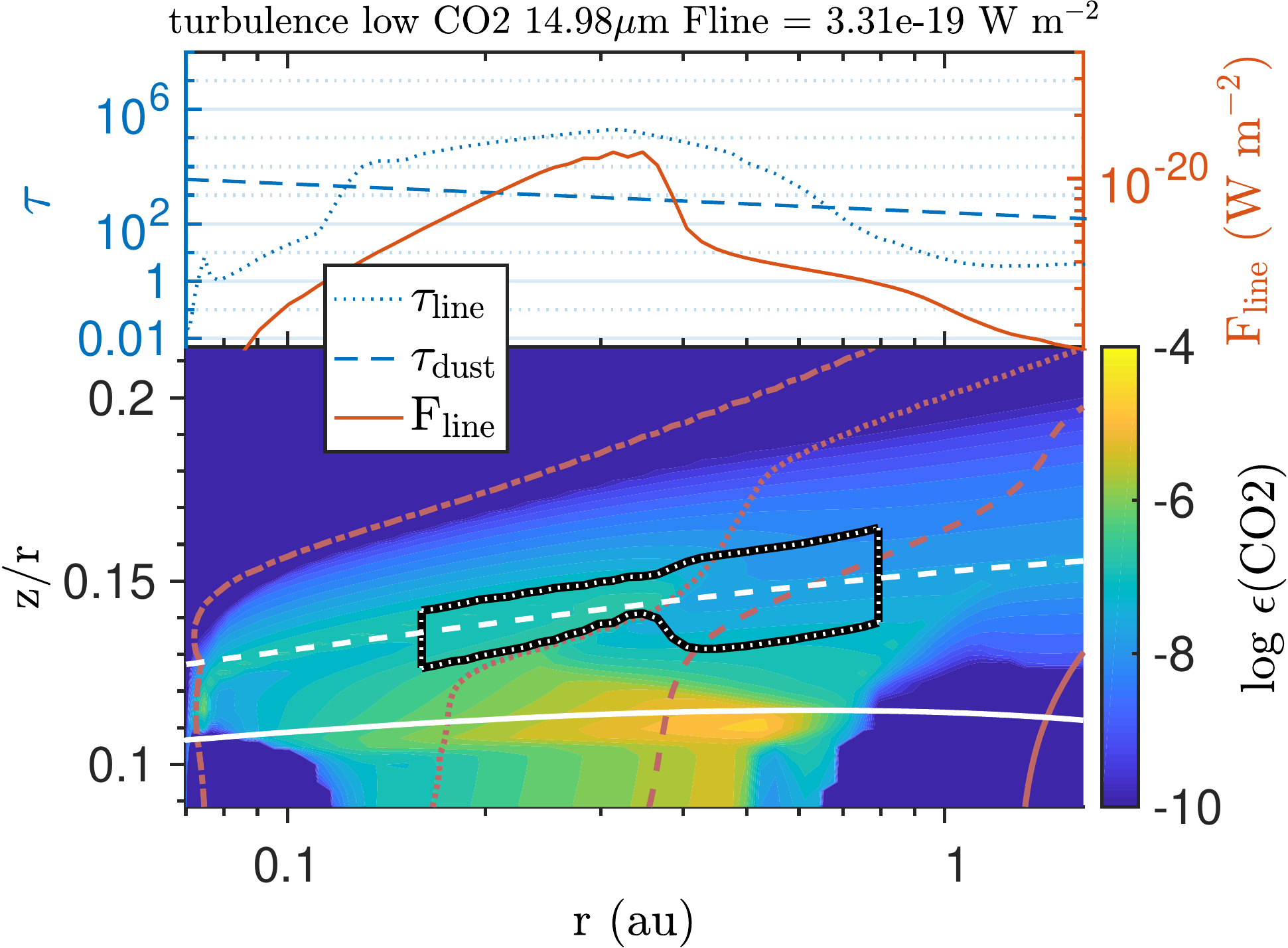} \hspace{0.005\textwidth}
				\includegraphics[width=0.47\textwidth]{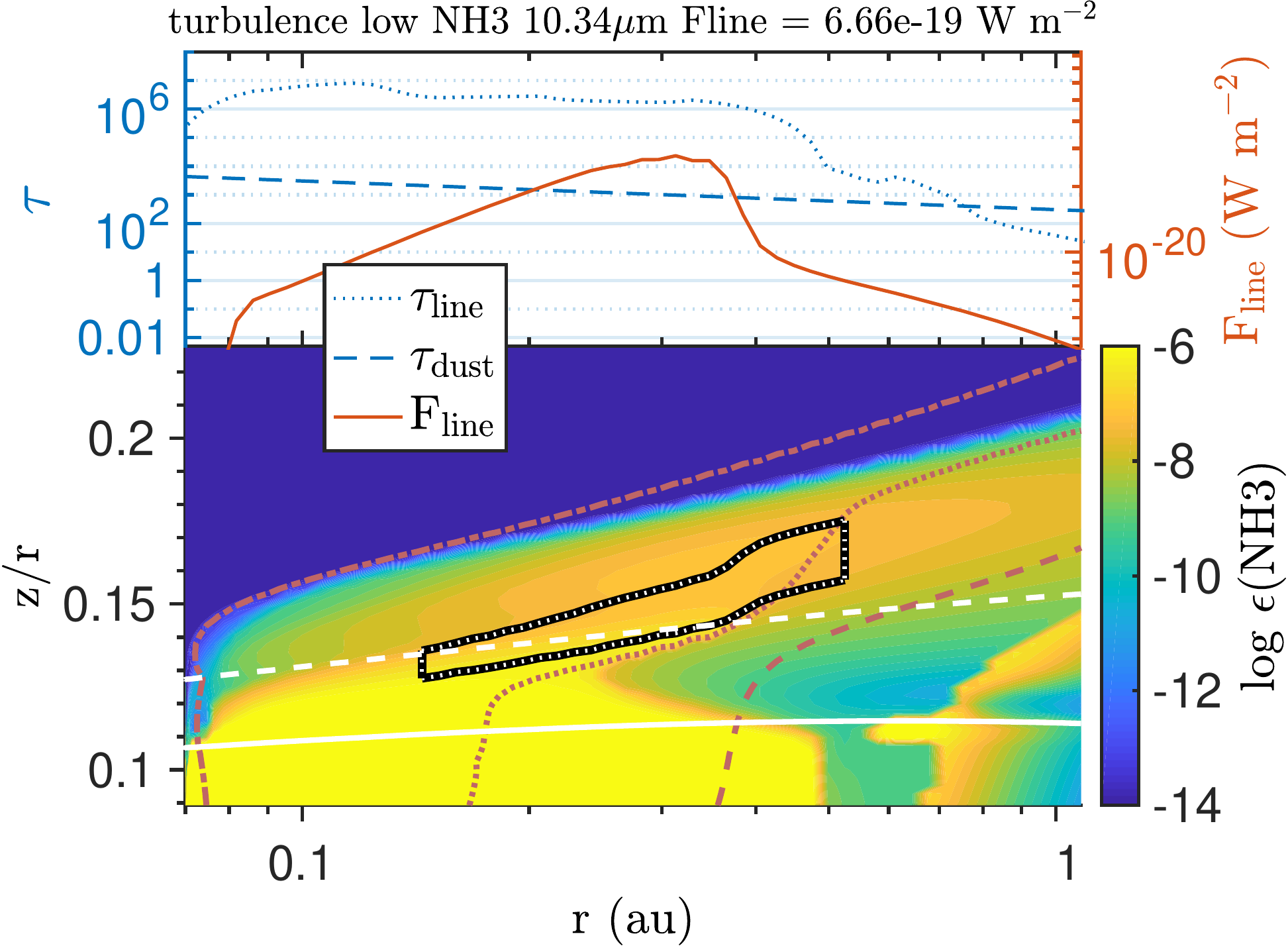}}      
			\makebox[\textwidth][c]{\includegraphics[width=0.47\textwidth]{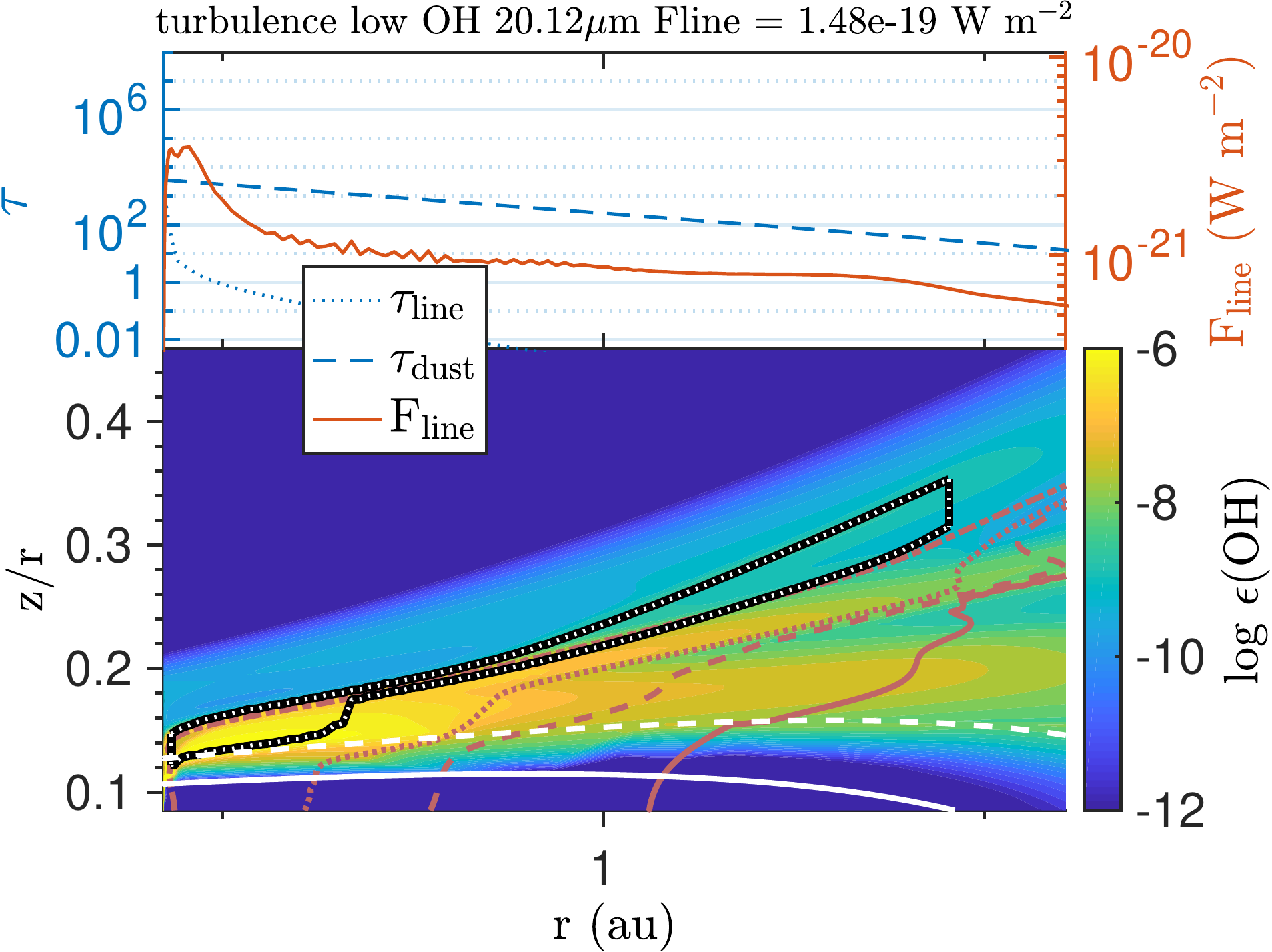} \hspace{0.005\textwidth}
				\includegraphics[width=0.47\textwidth]{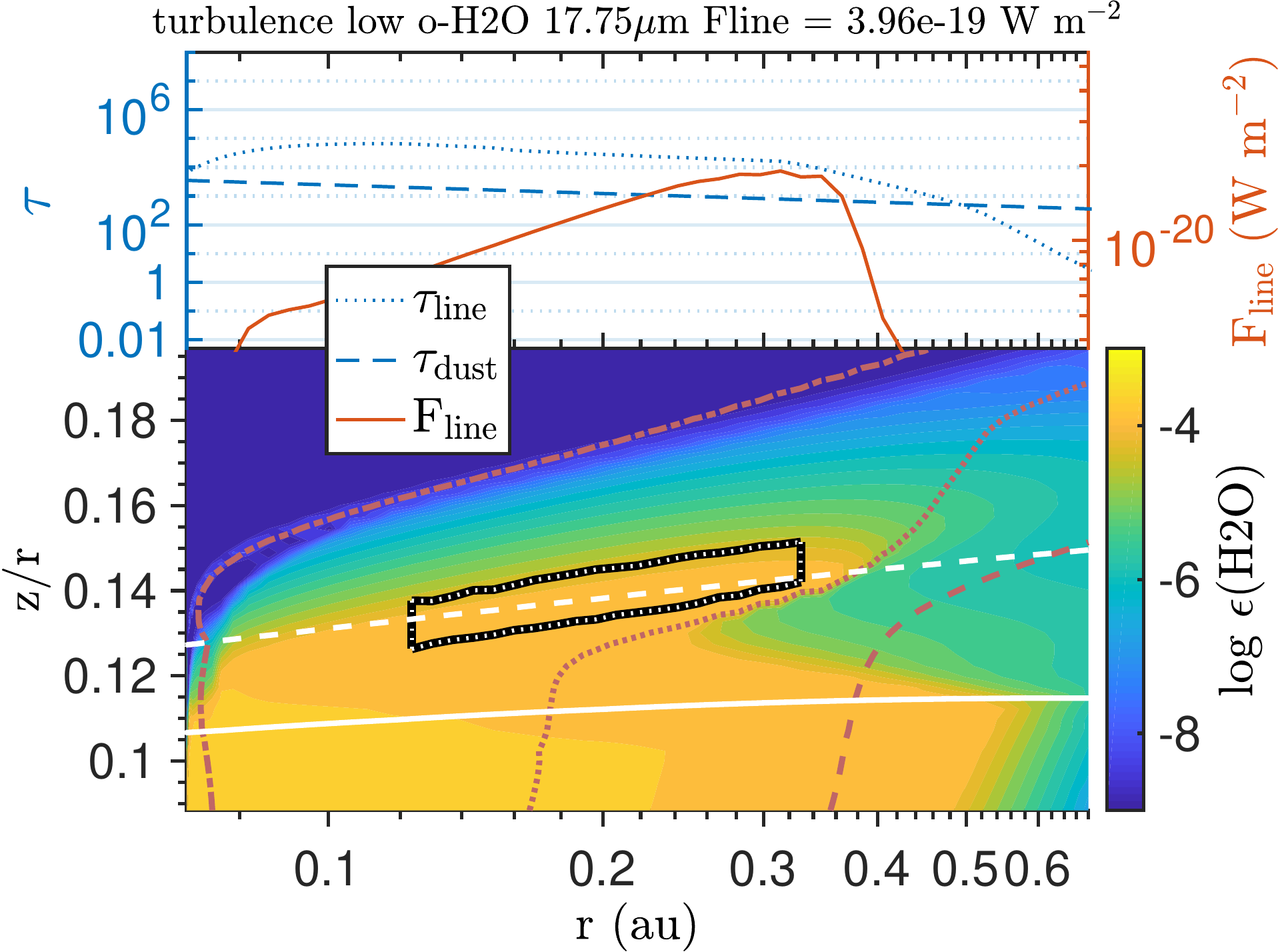}}
			\caption{Line-emitting regions for the model turbulence low. The plotted lines are \cem{C2H2}, \cem{HCN}, \cem{CO2}, \cem{NH3}, \cem{OH}, and \cem{o-H2O}. The rest of the figure is as described in \cref{fig:LER_TT_highres}.
			}\label{fig:LER_settling_low}   
		\end{figure*}
		
		\begin{figure*} \centering    
			\makebox[\textwidth][c]{\includegraphics[width=0.47\textwidth]{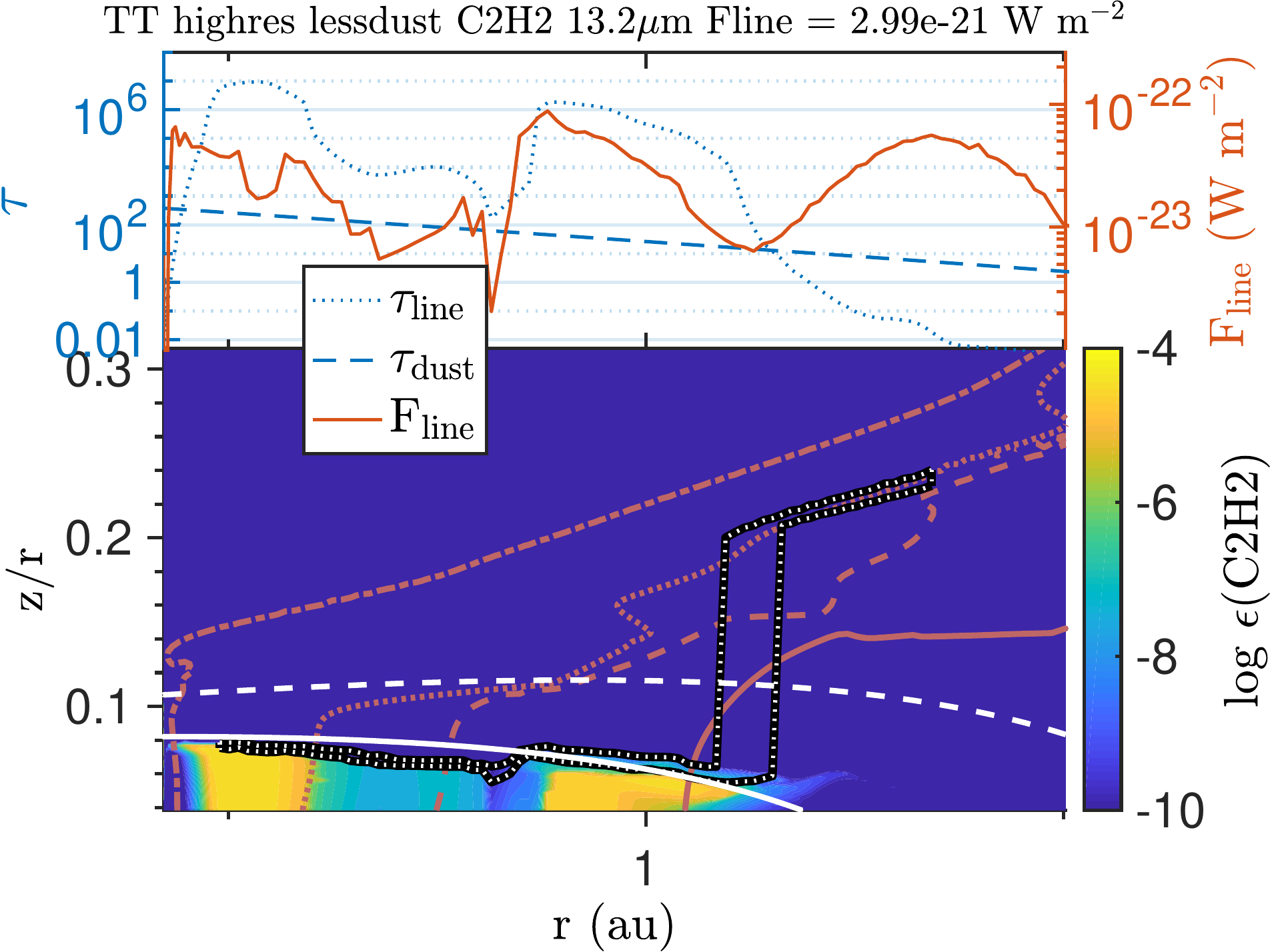} \hspace{0.005\textwidth}
				\includegraphics[width=0.47\textwidth]{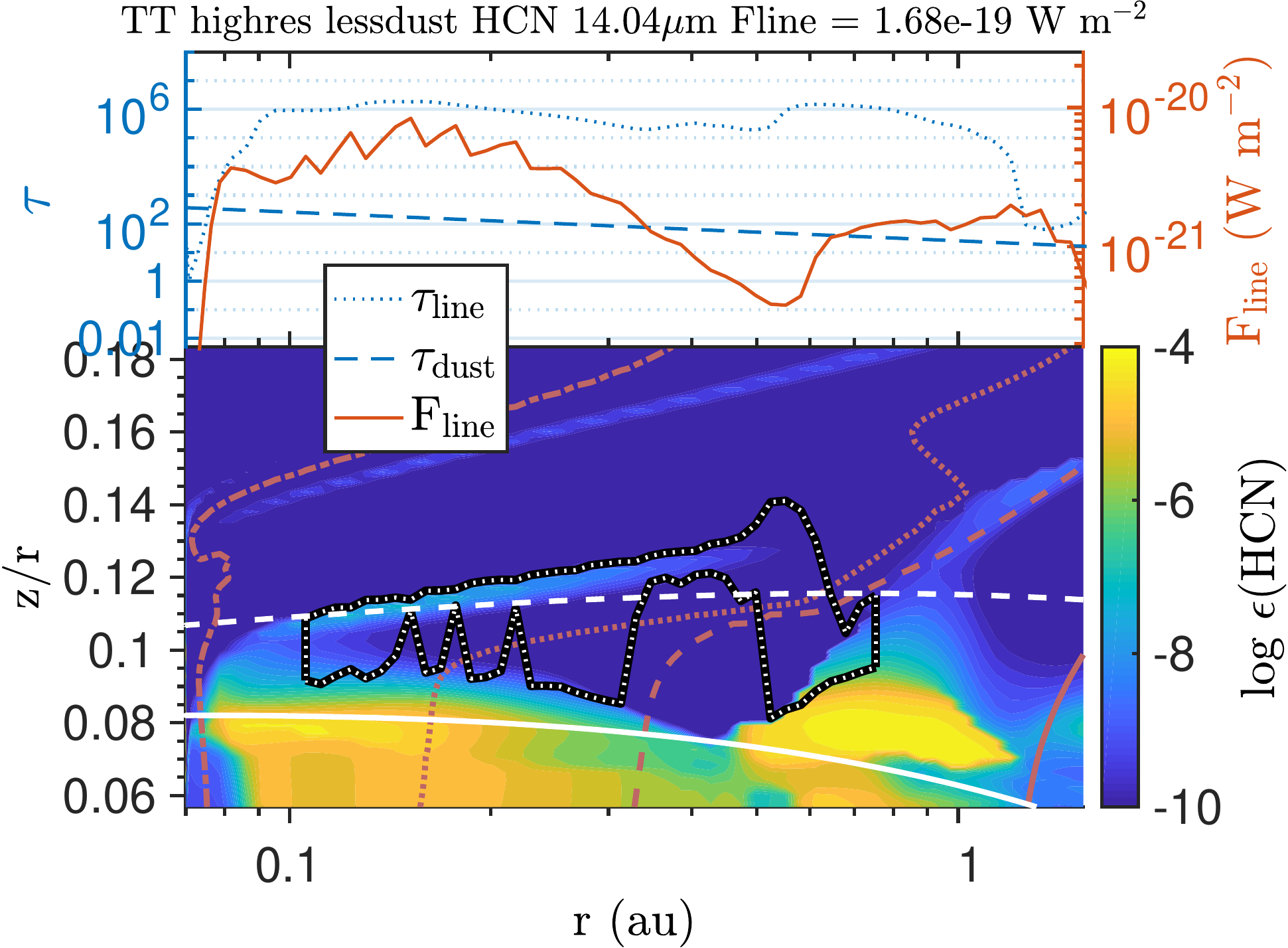}}   
			\makebox[\textwidth][c]{\includegraphics[width=0.47\textwidth]{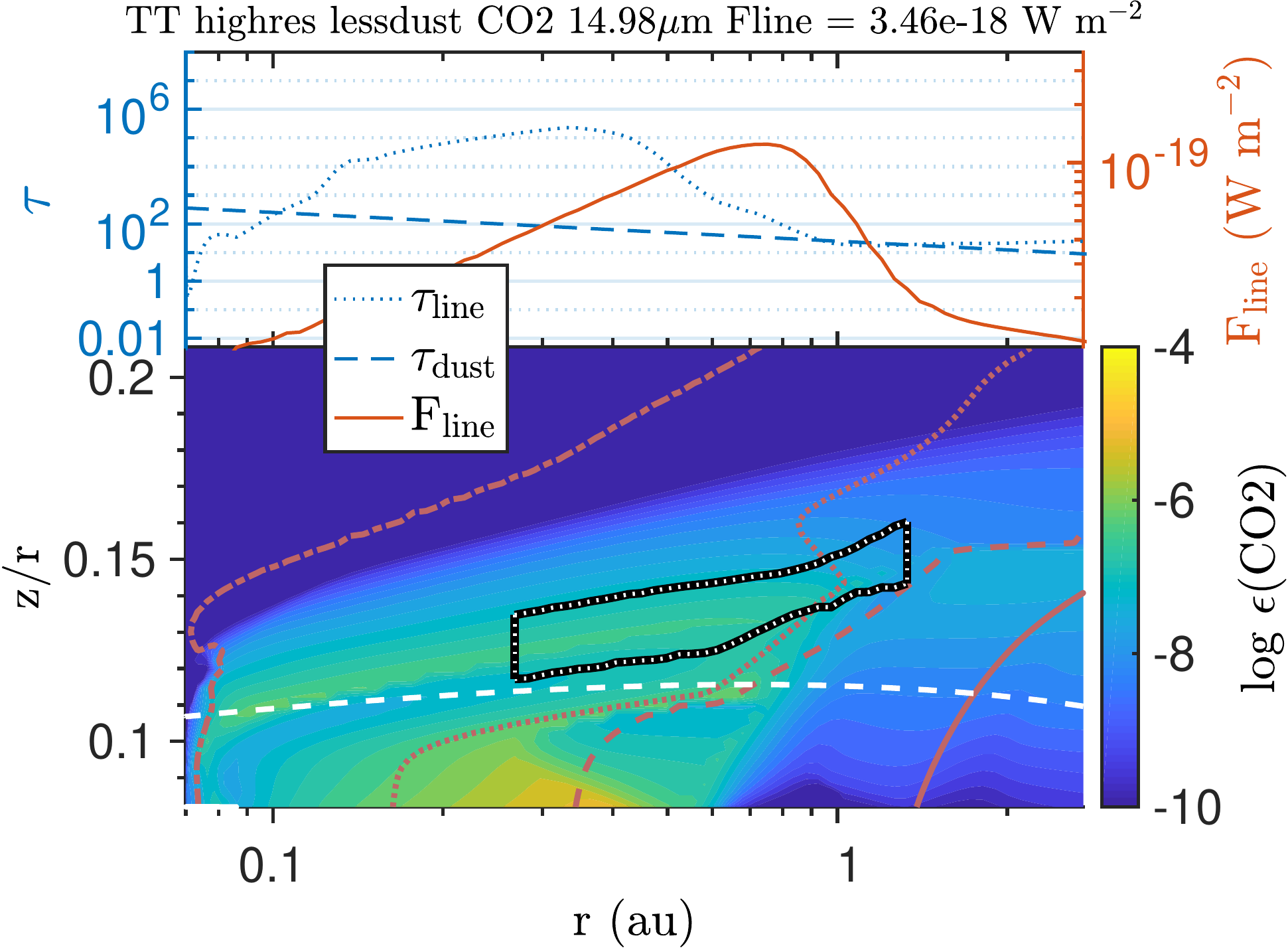} \hspace{0.005\textwidth}
				\includegraphics[width=0.47\textwidth]{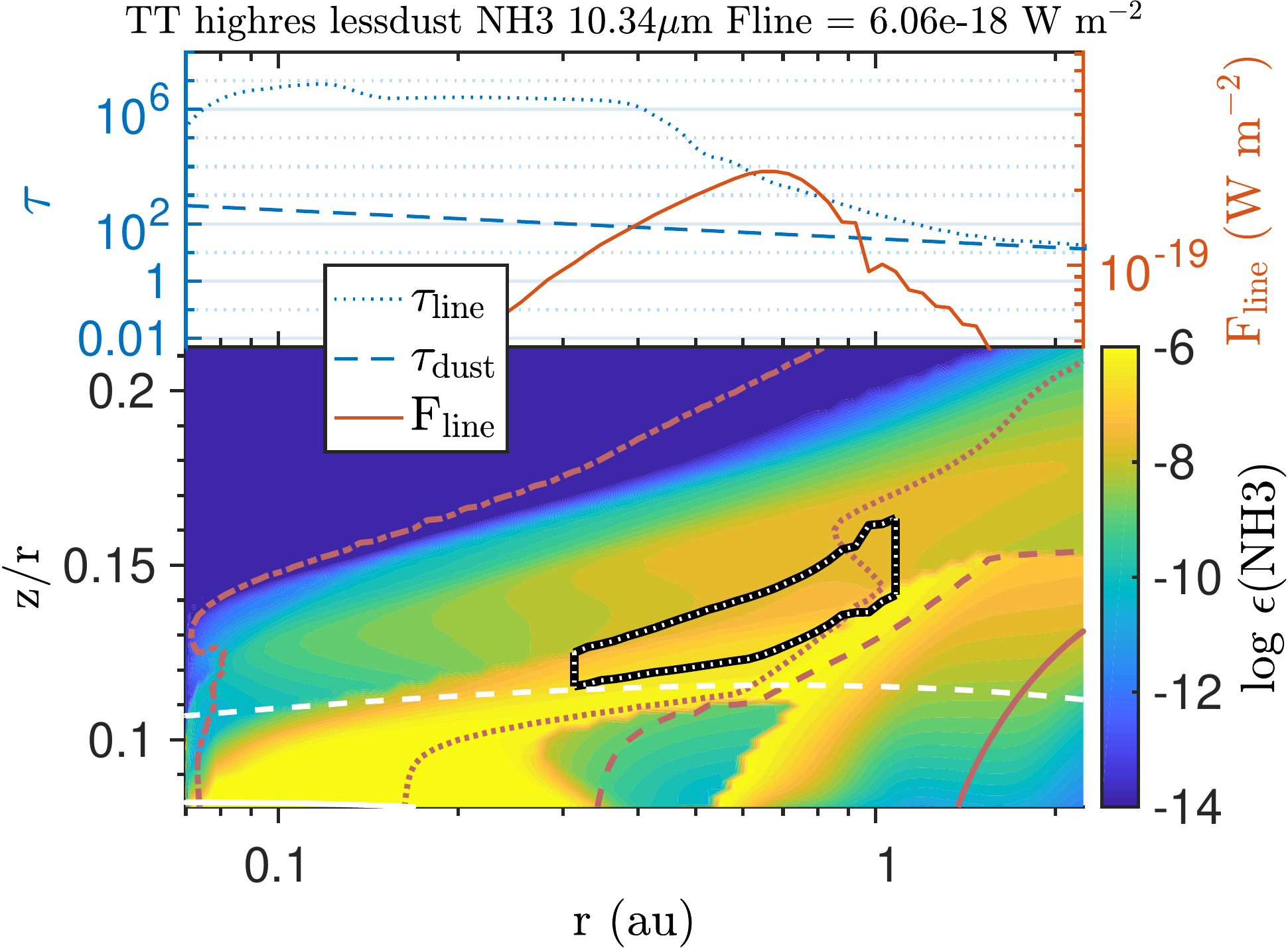}}  
			\makebox[\textwidth][c]{\includegraphics[width=0.47\textwidth]{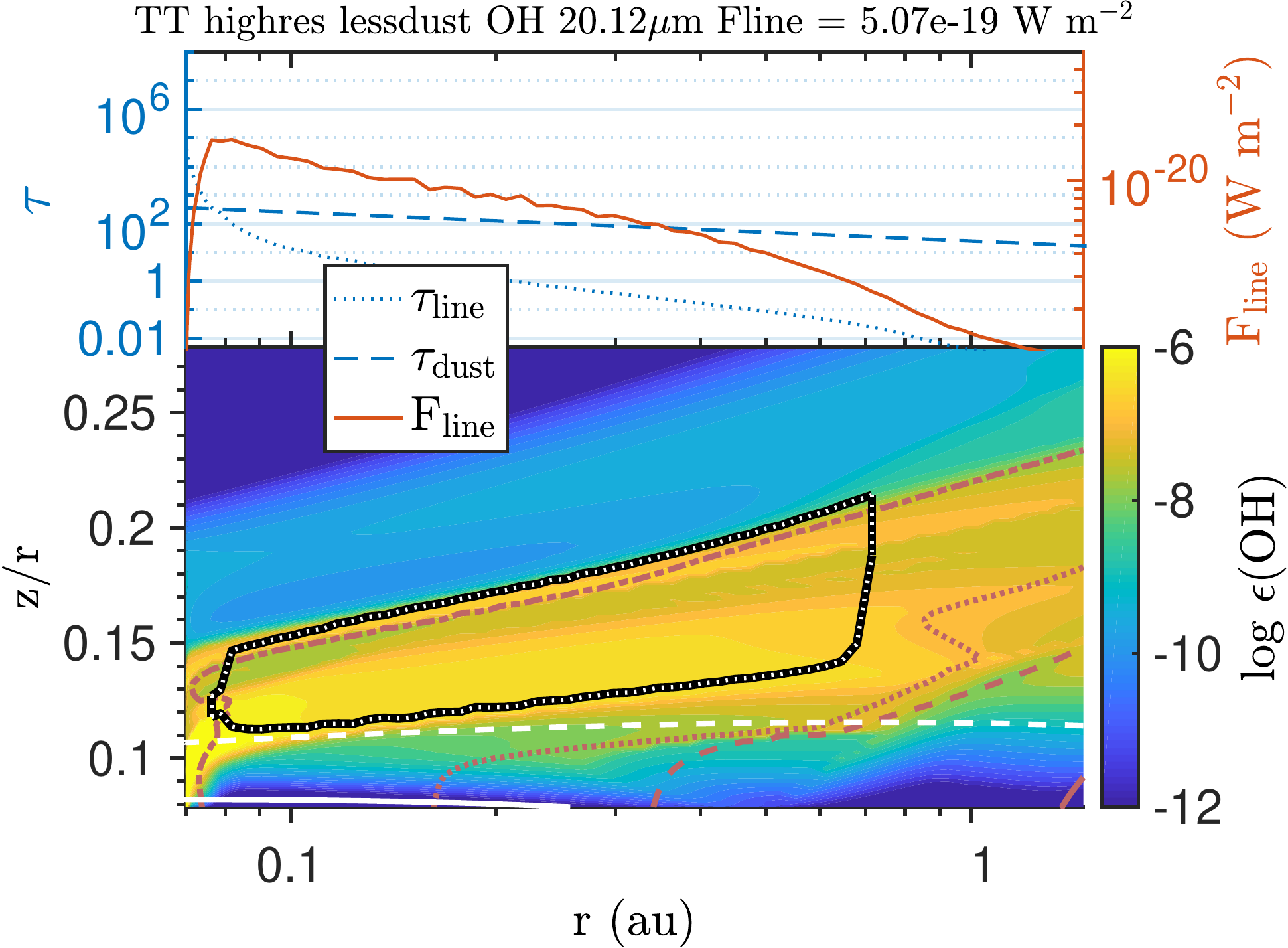} \hspace{0.005\textwidth}  
				\includegraphics[width=0.47\textwidth]{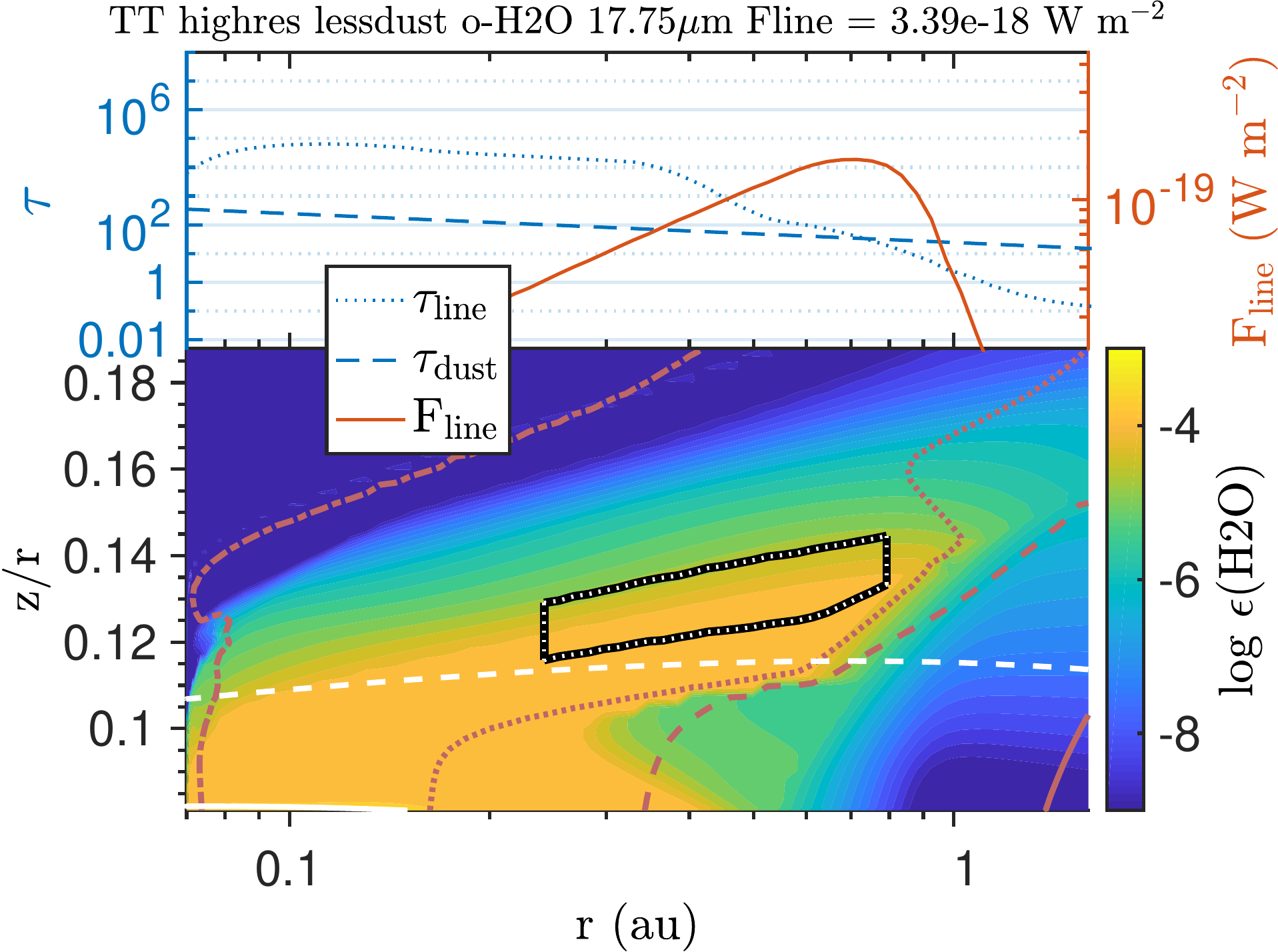}} 
			\caption{Line-emitting regions for the model TT highres lessdust. The plotted lines are \cem{C2H2}, \cem{HCN}, \cem{CO2}, \cem{NH3}, \cem{OH}, and \cem{o-H2O}. The rest of the figure is as described in \cref{fig:LER_TT_highres}.     
			}\label{fig:LER_TT_highres_lessdust}  
		\end{figure*}
		
		\begin{figure*} \centering    
			\makebox[\textwidth][c]{\includegraphics[width=0.47\textwidth]{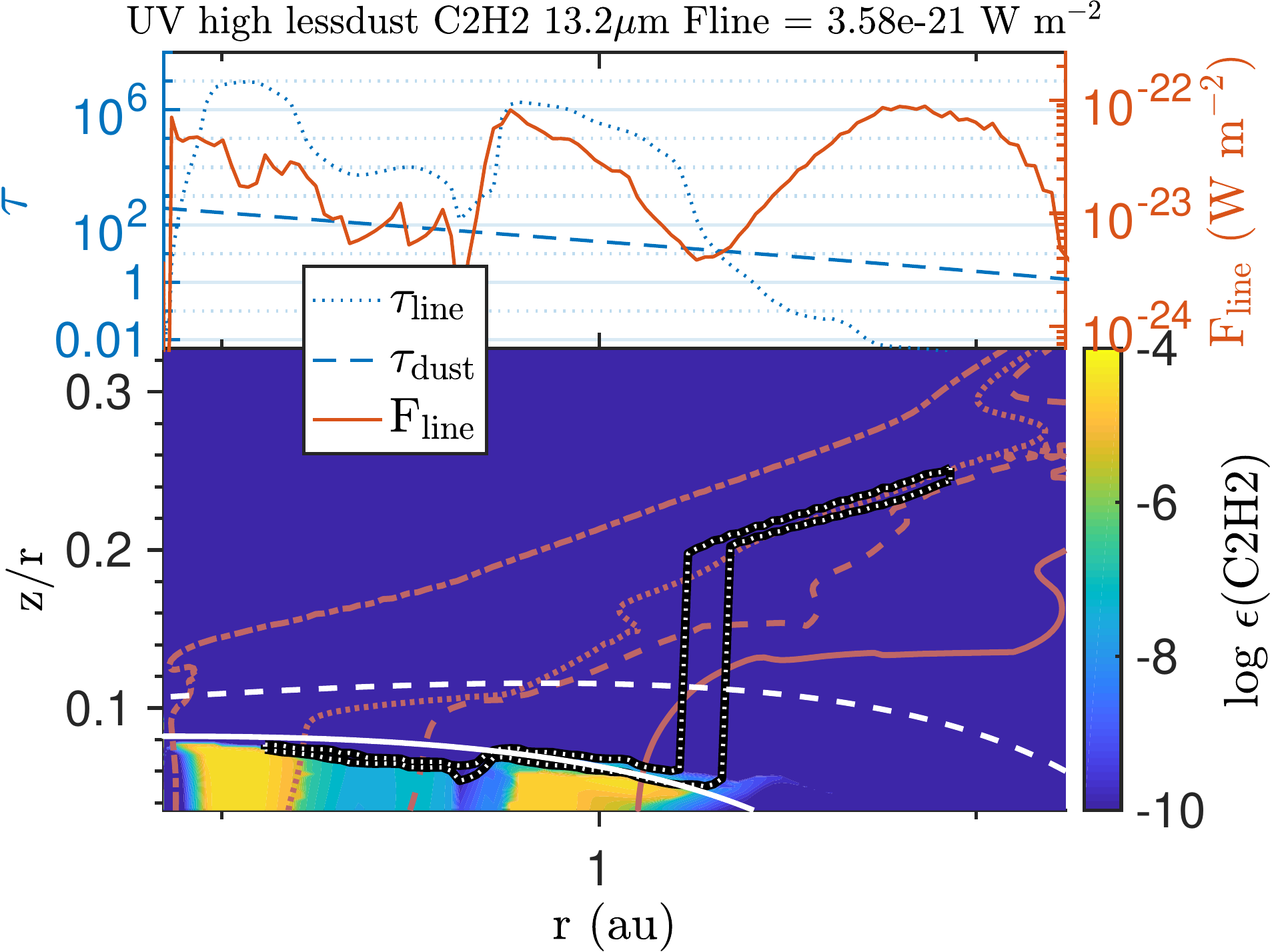} \hspace{0.005\textwidth}   
				\includegraphics[width=0.47\textwidth]{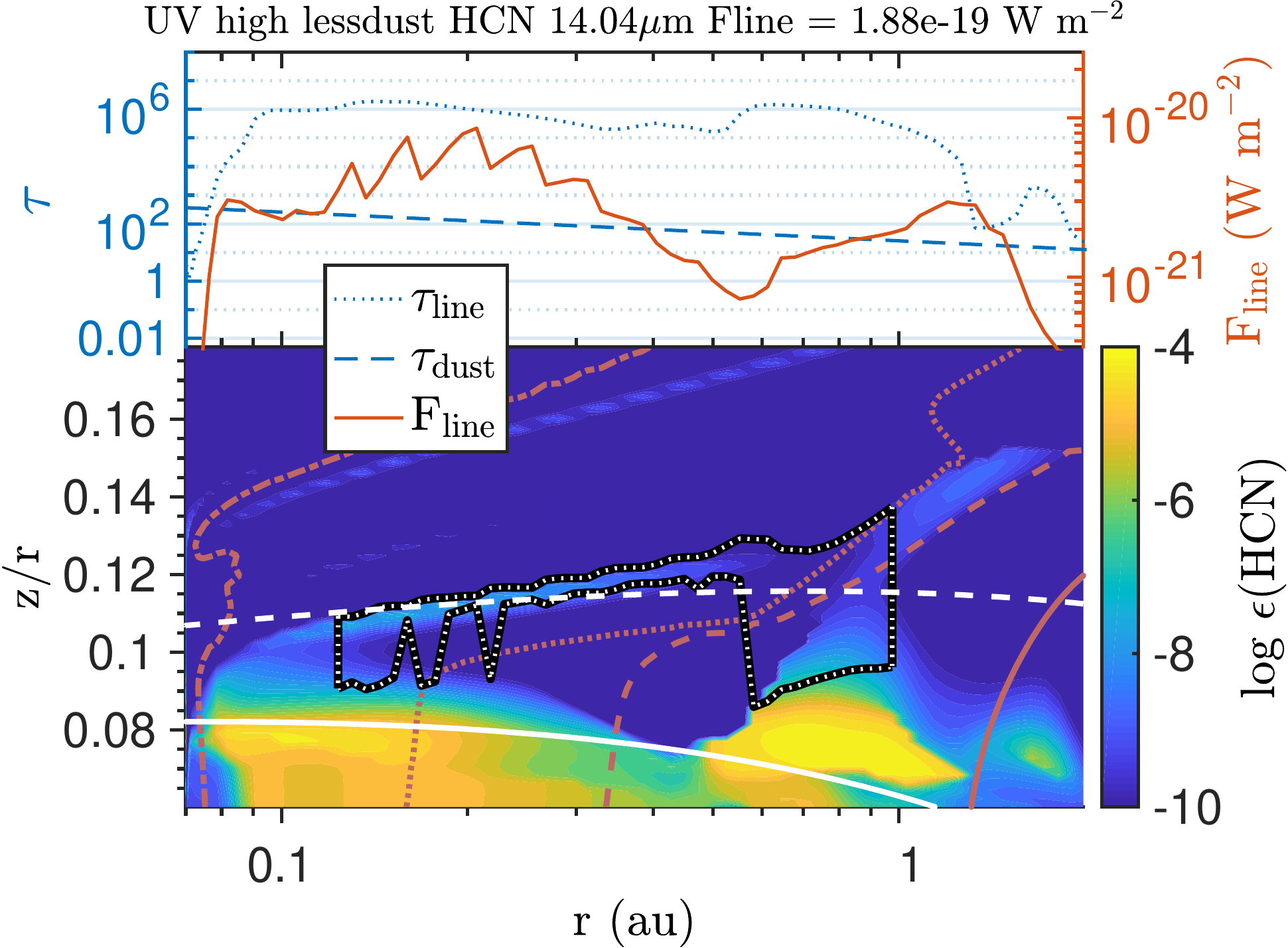}}      
			\makebox[\textwidth][c]{\includegraphics[width=0.47\textwidth]{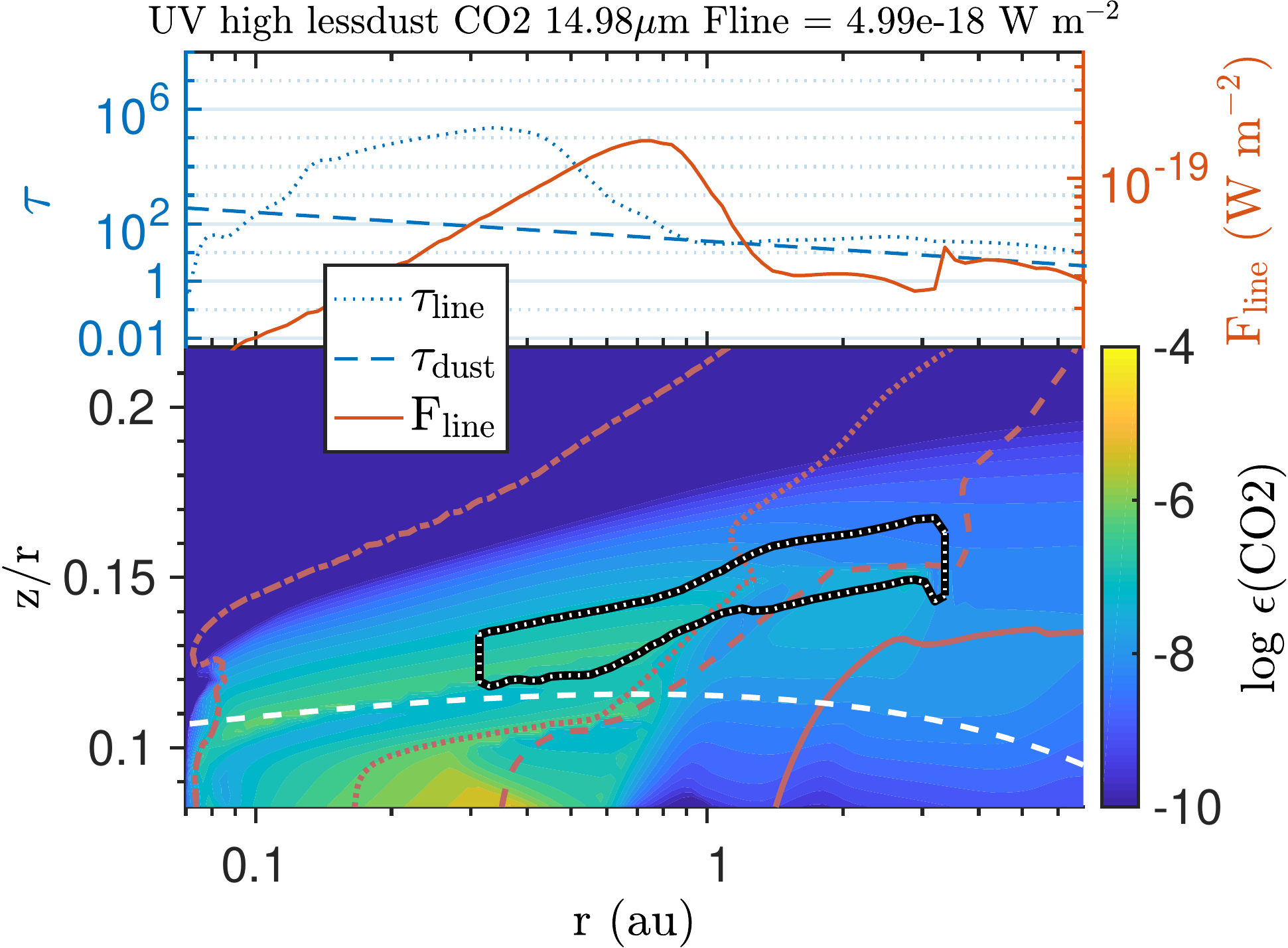} \hspace{0.005\textwidth}
				\includegraphics[width=0.47\textwidth]{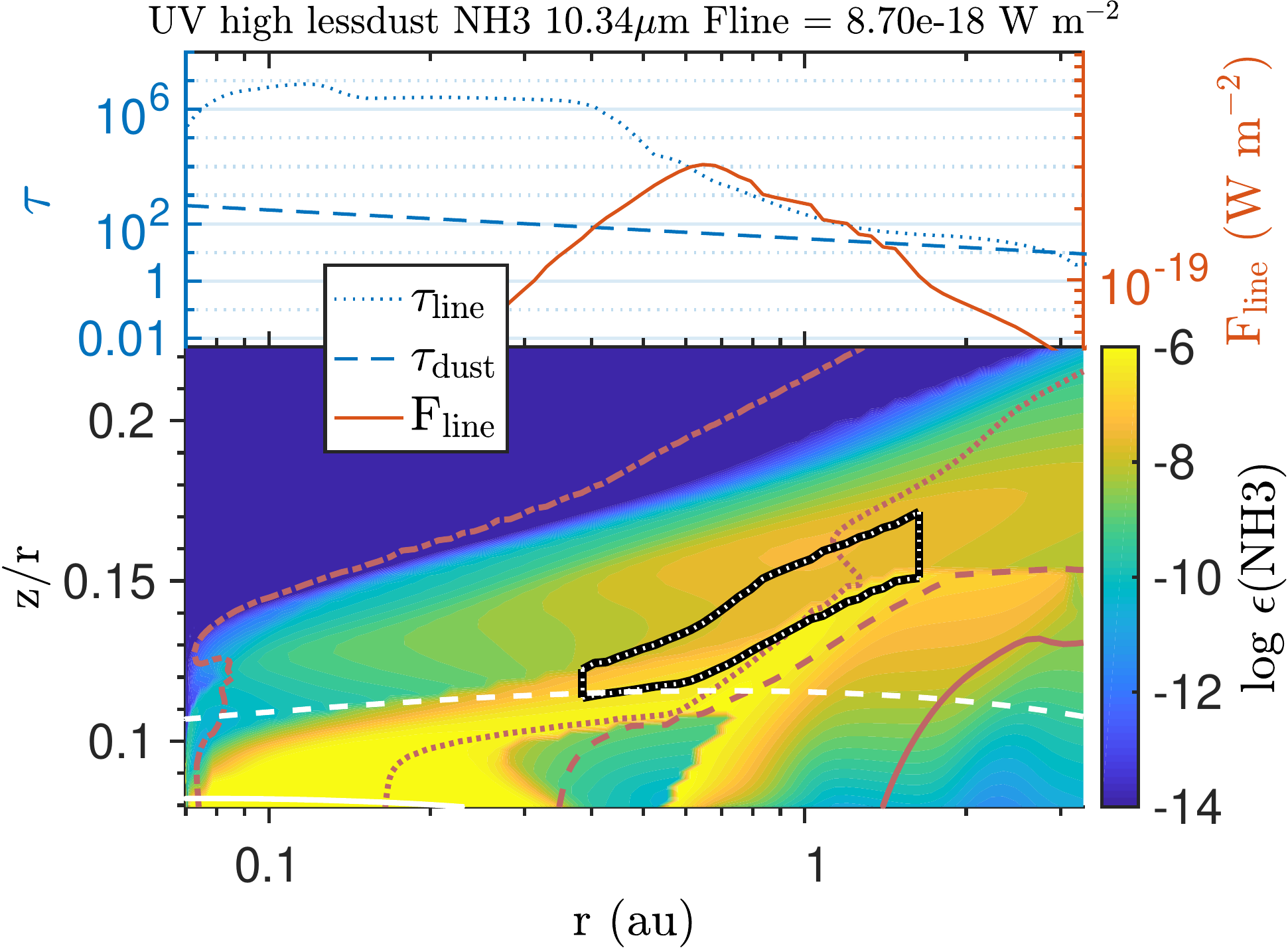}}     
			\makebox[\textwidth][c]{\includegraphics[width=0.47\textwidth]{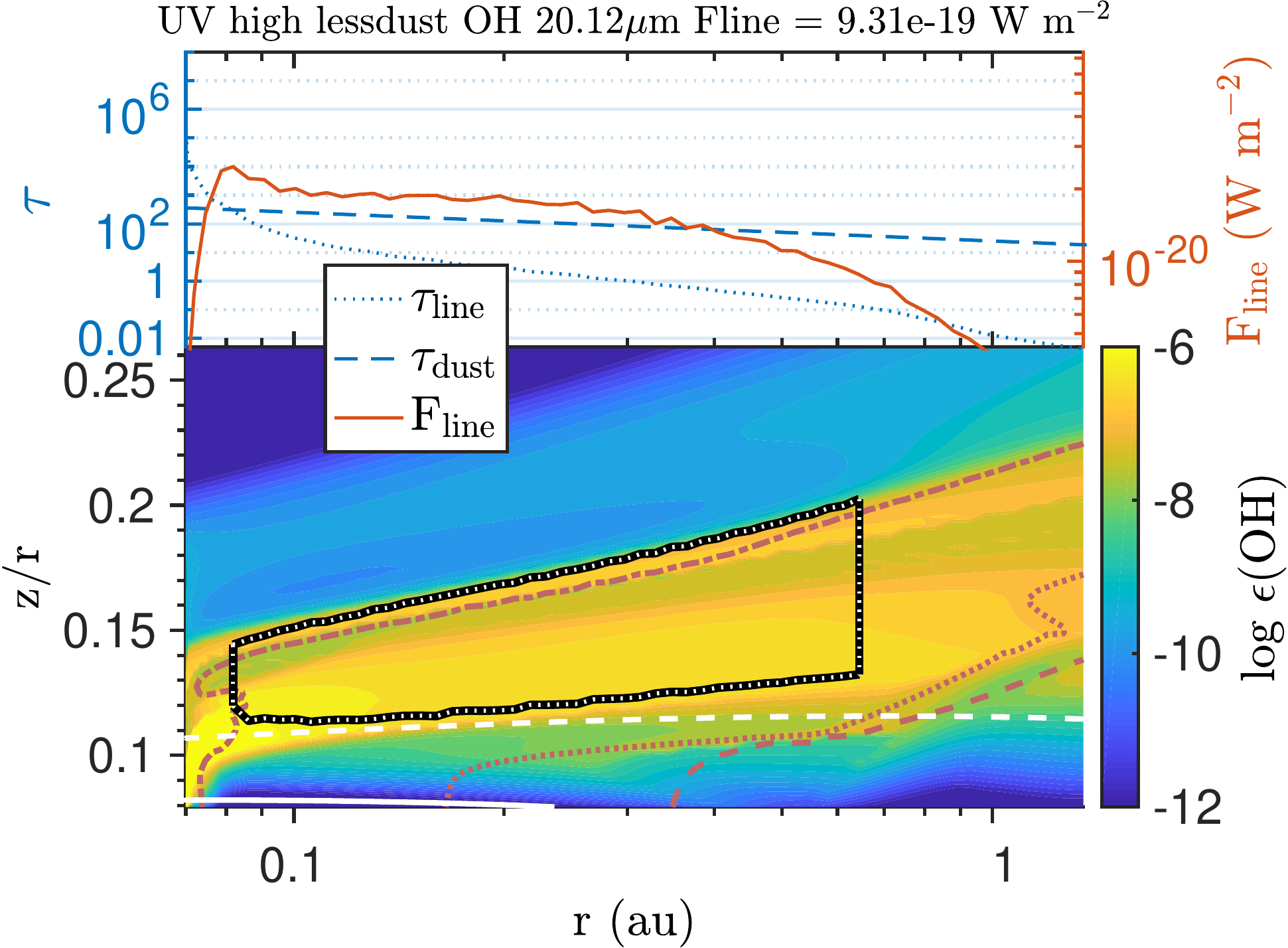} \hspace{0.005\textwidth}     
				\includegraphics[width=0.47\textwidth]{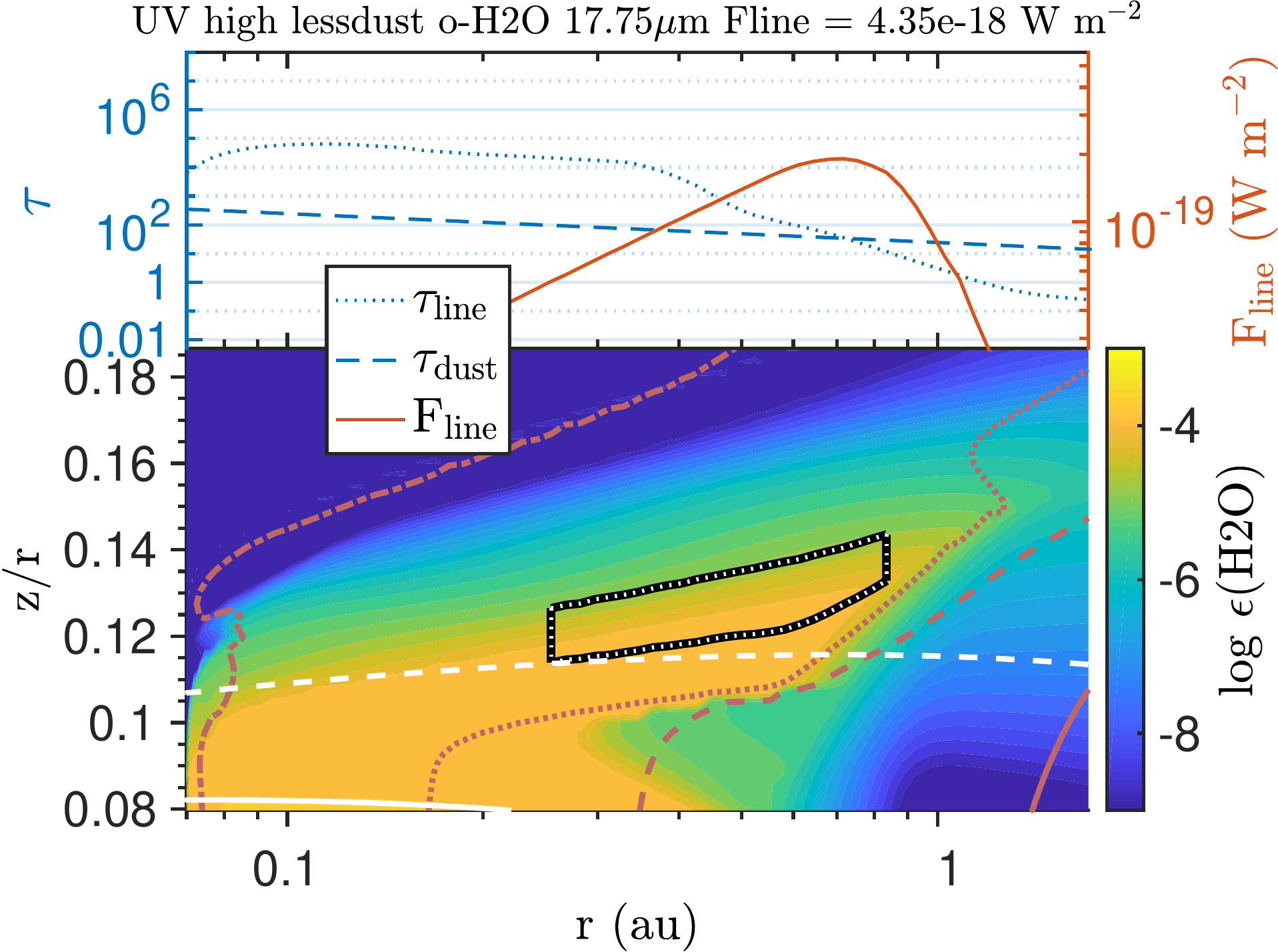}}      
			\caption{Line-emitting regions for the model UV high lessdust. The plotted lines are \cem{C2H2}, \cem{HCN}, \cem{CO2}, \cem{NH3}, \cem{OH}, and \cem{o-H2O}. The rest of the figure is as described in \cref{fig:LER_TT_highres}.  
			}\label{fig:LER_UV_high_lessdust}     
		\end{figure*}
		
		\begin{figure*} \centering    
			\makebox[\textwidth][c]{\includegraphics[width=0.47\textwidth]{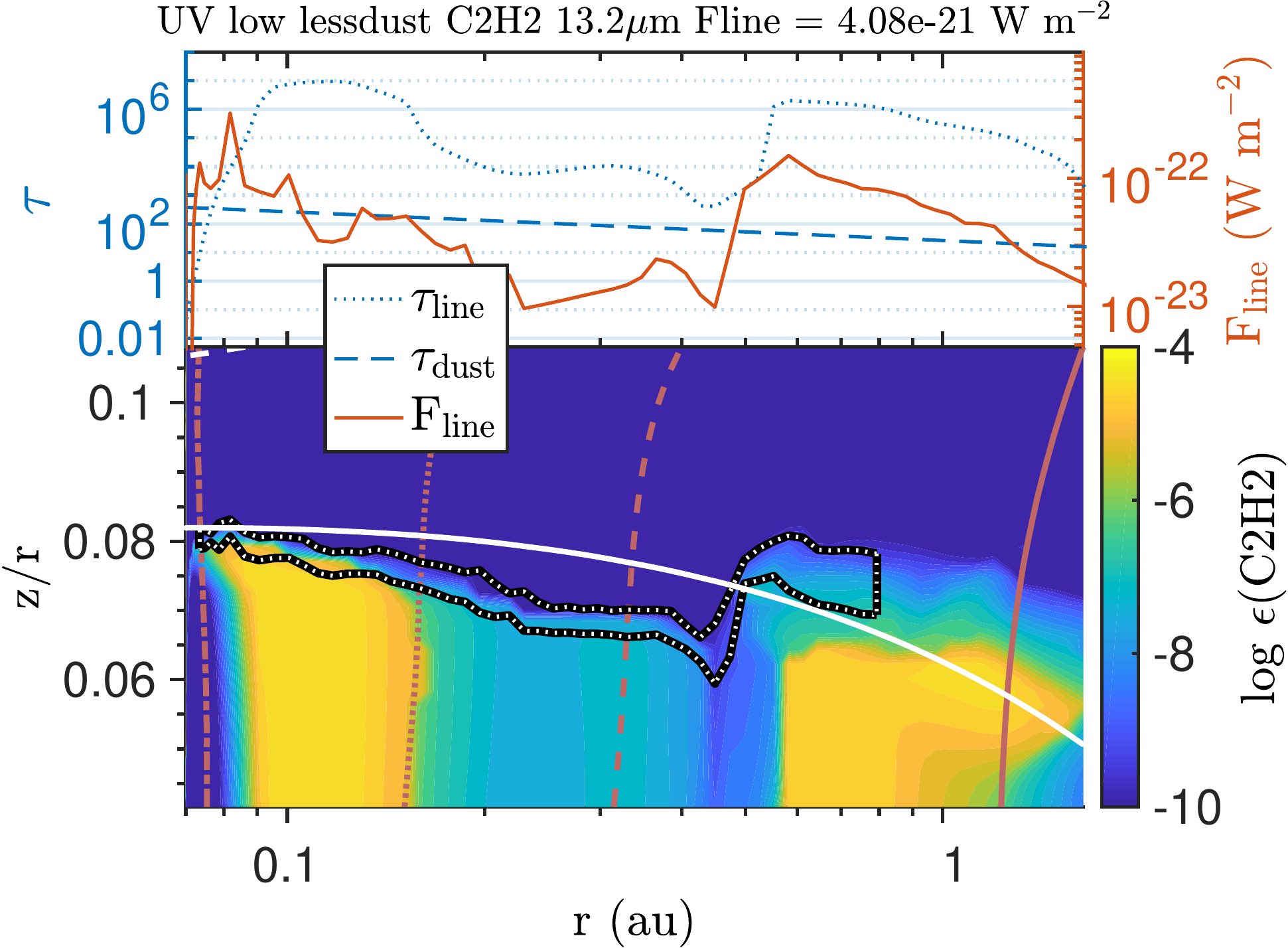} \hspace{0.005\textwidth}    
				\includegraphics[width=0.47\textwidth]{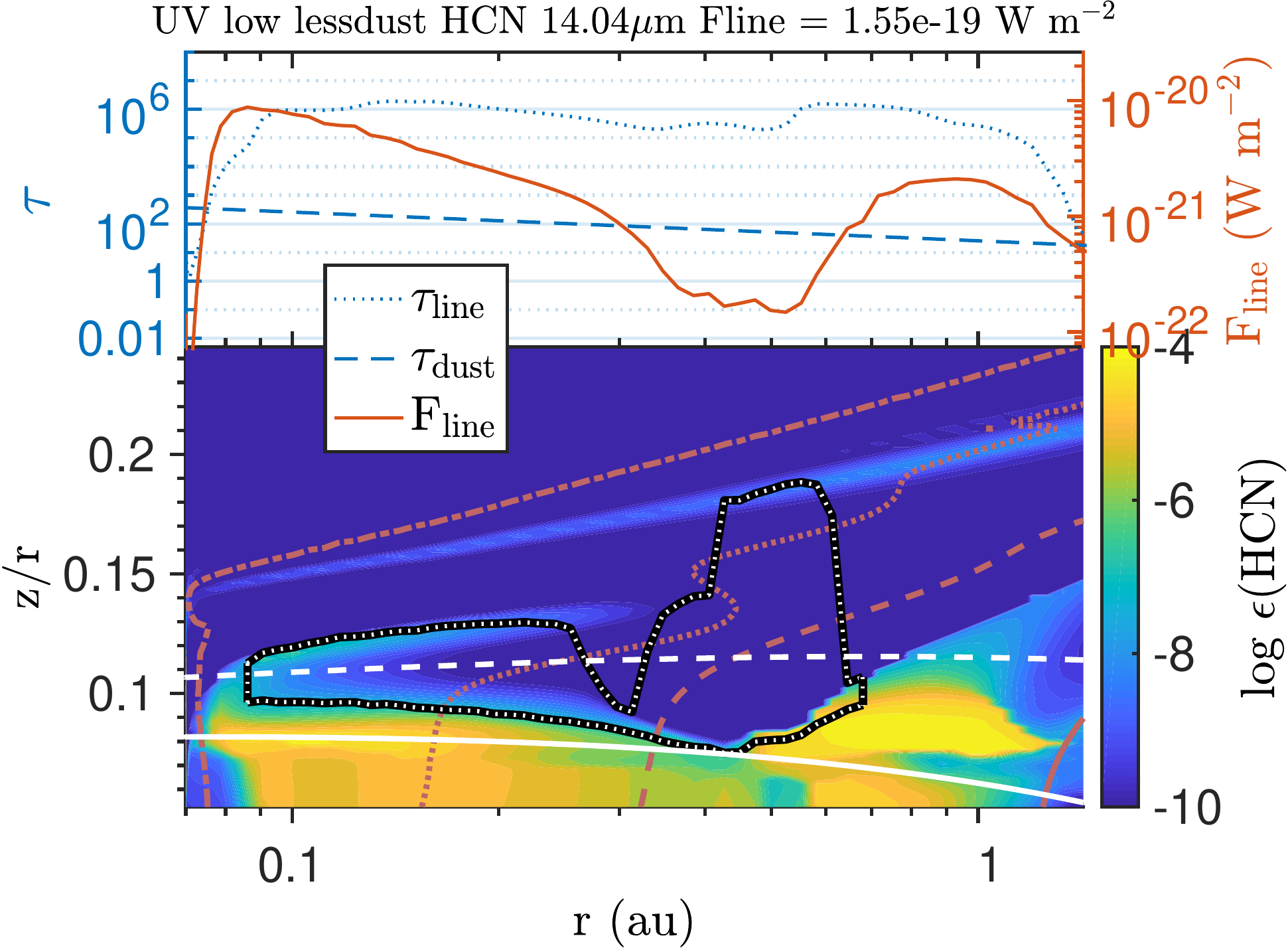}} 
			\makebox[\textwidth][c]{\includegraphics[width=0.47\textwidth]{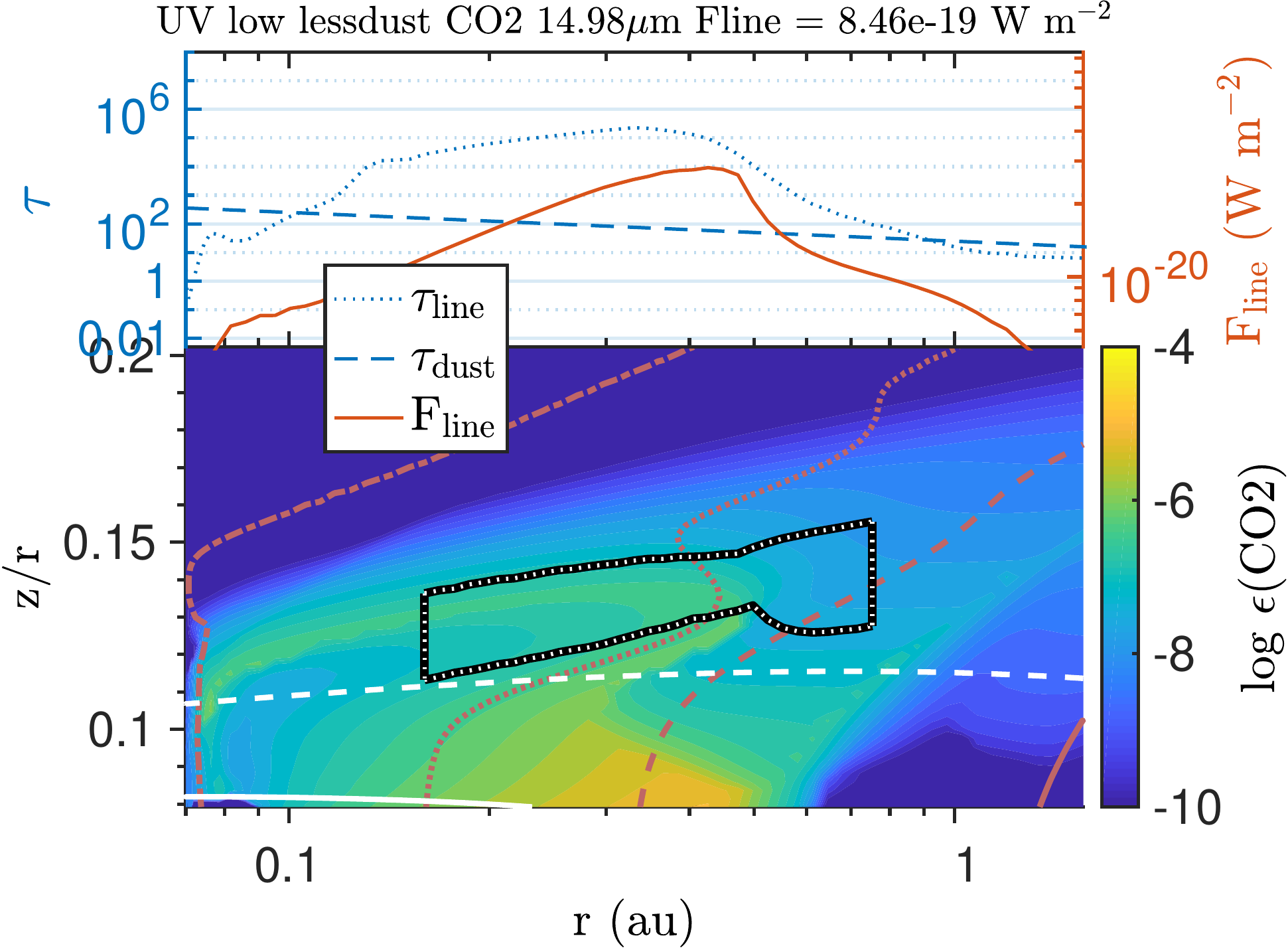} \hspace{0.005\textwidth}
				\includegraphics[width=0.47\textwidth]{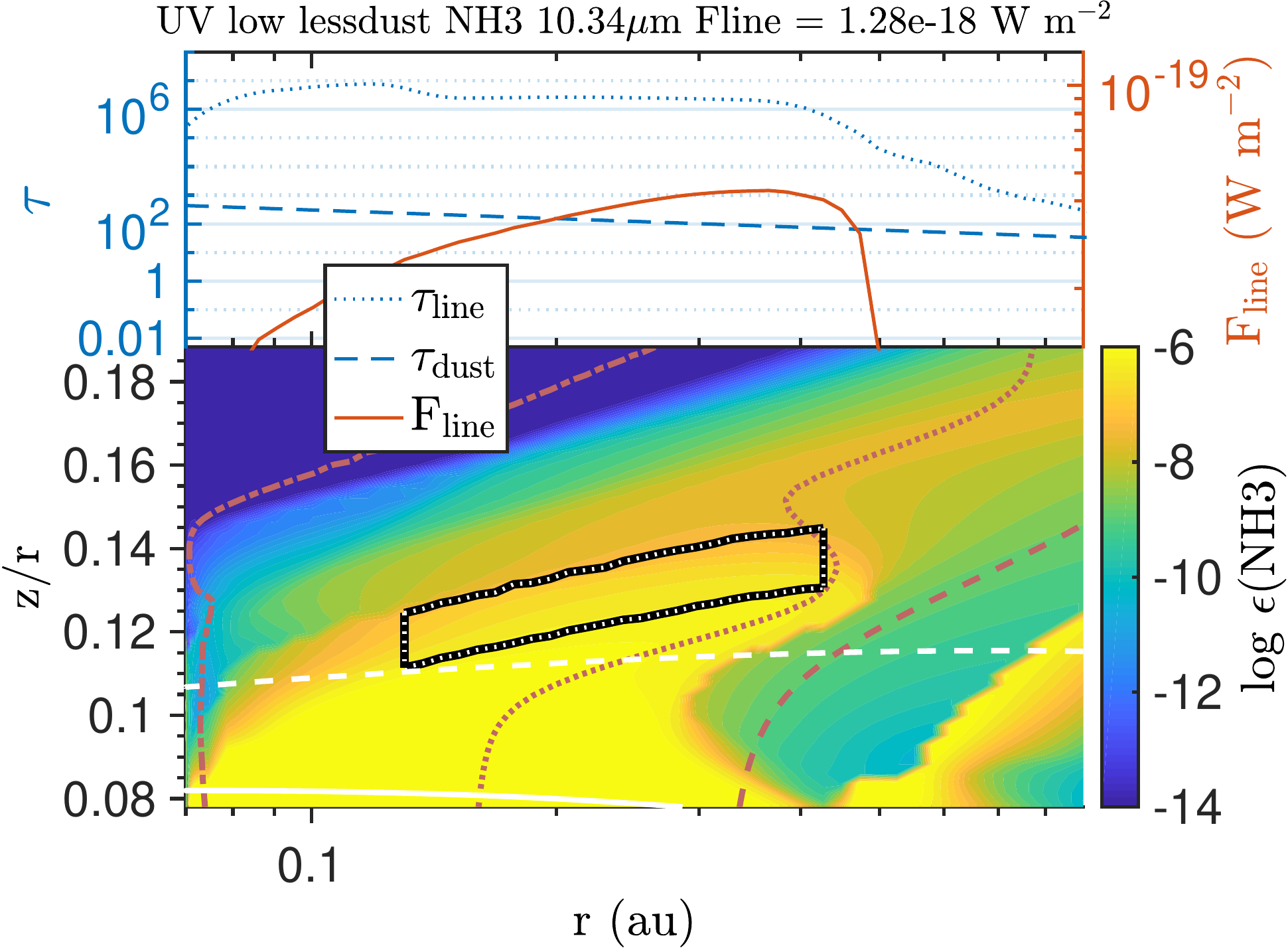}}      
			\makebox[\textwidth][c]{\includegraphics[width=0.47\textwidth]{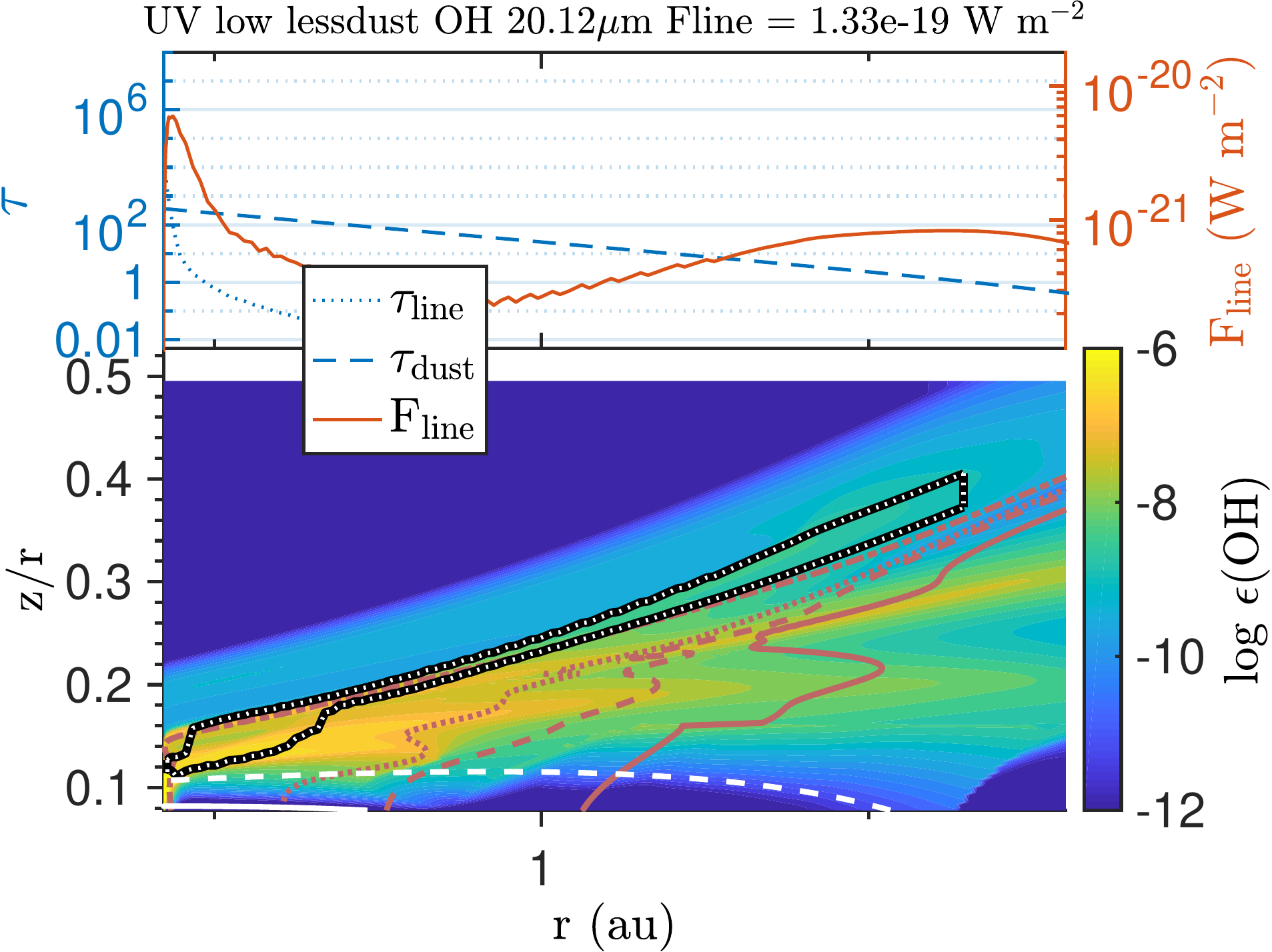} \hspace{0.005\textwidth}
				\includegraphics[width=0.47\textwidth]{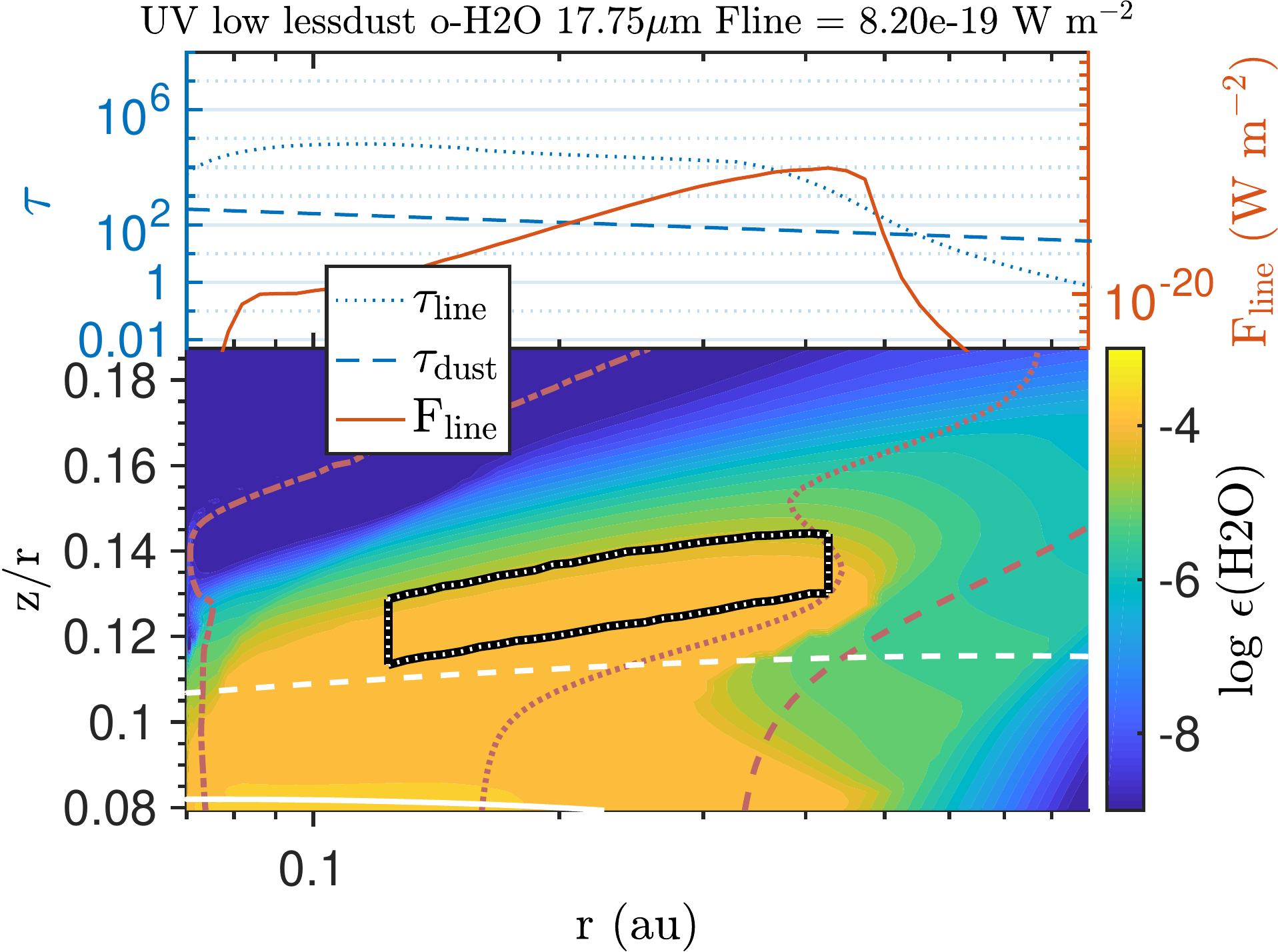}}
			\caption{Line-emitting regions for the model UV low lessdust. The plotted lines are \cem{C2H2}, \cem{HCN}, \cem{CO2}, \cem{NH3}, \cem{OH}, and \cem{o-H2O}. The rest of the figure is as described in \cref{fig:LER_TT_highres}.   
			}\label{fig:LER_UV_low_lessdust}
		\end{figure*}
		
		\begin{figure*} \centering    
			\makebox[\textwidth][c]{\includegraphics[width=0.47\textwidth]{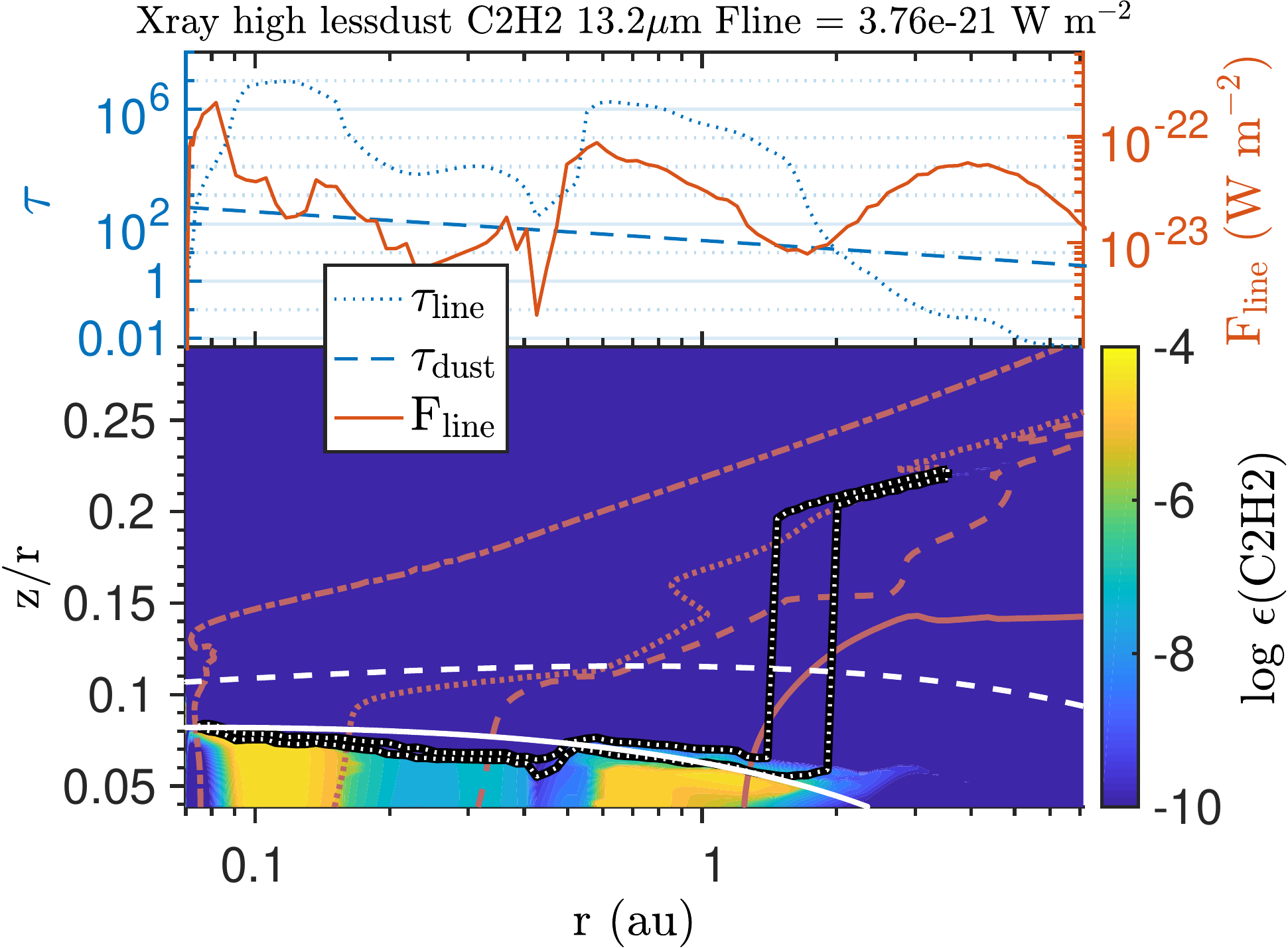} \hspace{0.005\textwidth} 
				\includegraphics[width=0.47\textwidth]{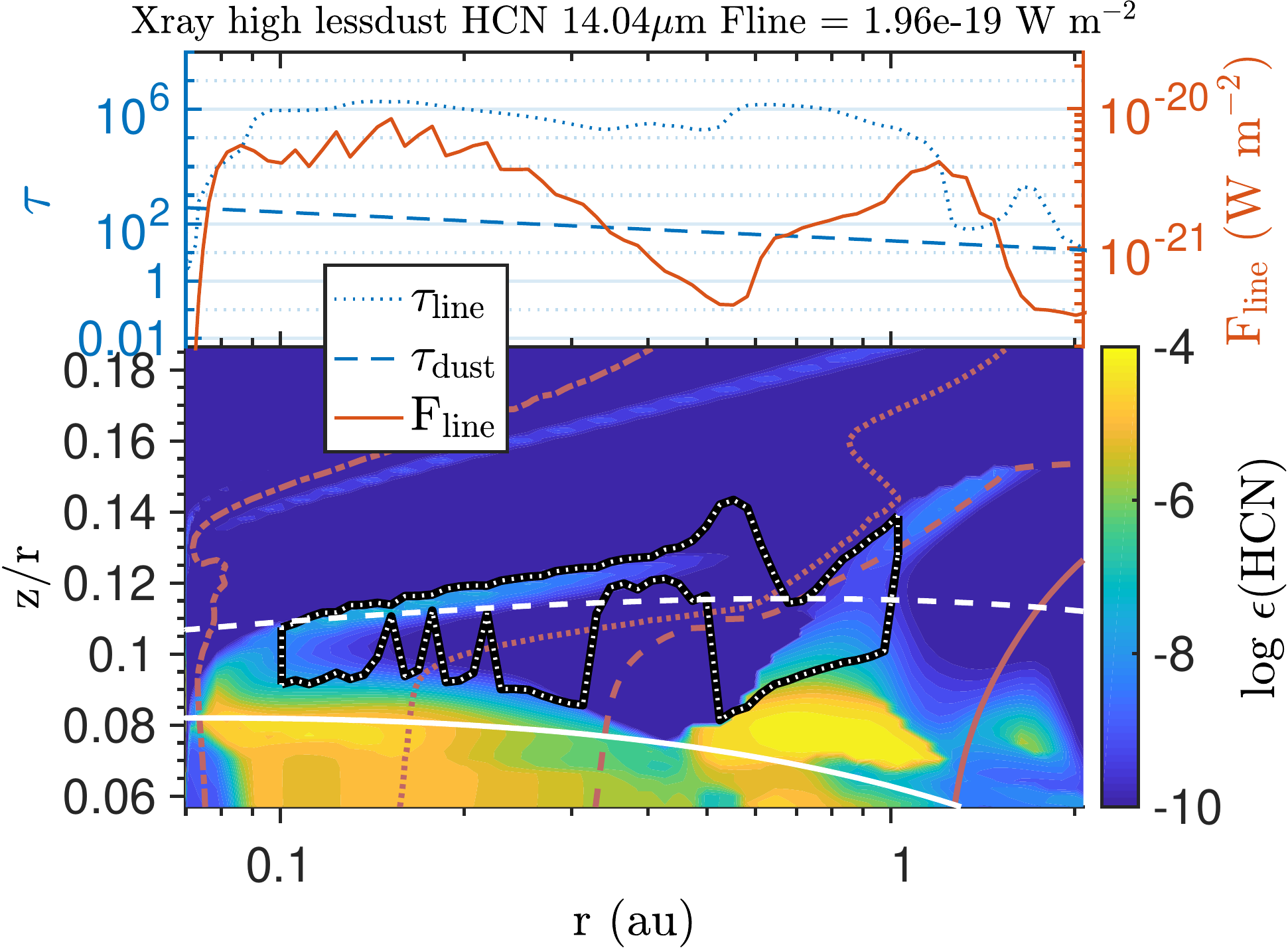}}    
			\makebox[\textwidth][c]{\includegraphics[width=0.47\textwidth]{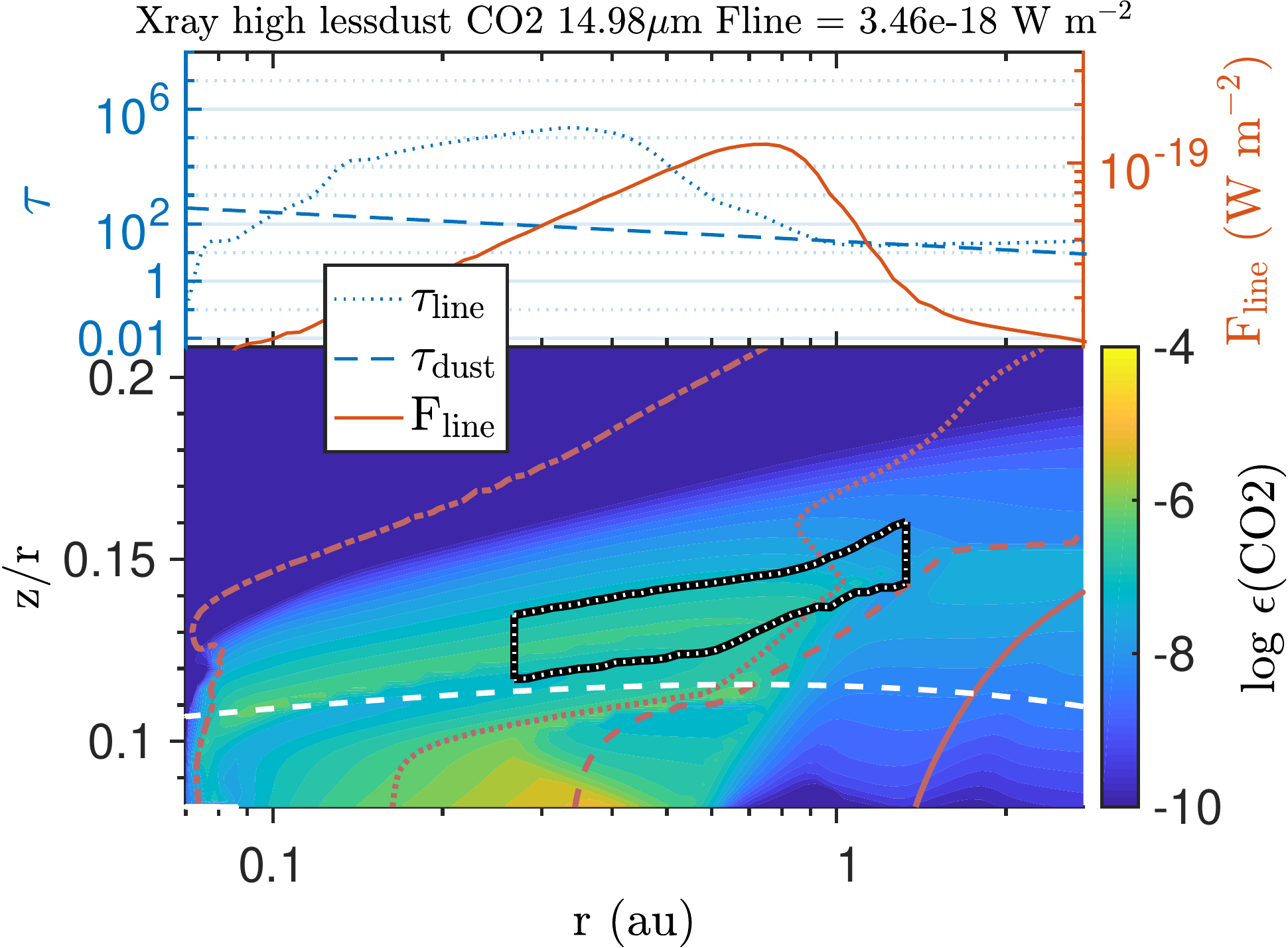} \hspace{0.005\textwidth}
				\includegraphics[width=0.47\textwidth]{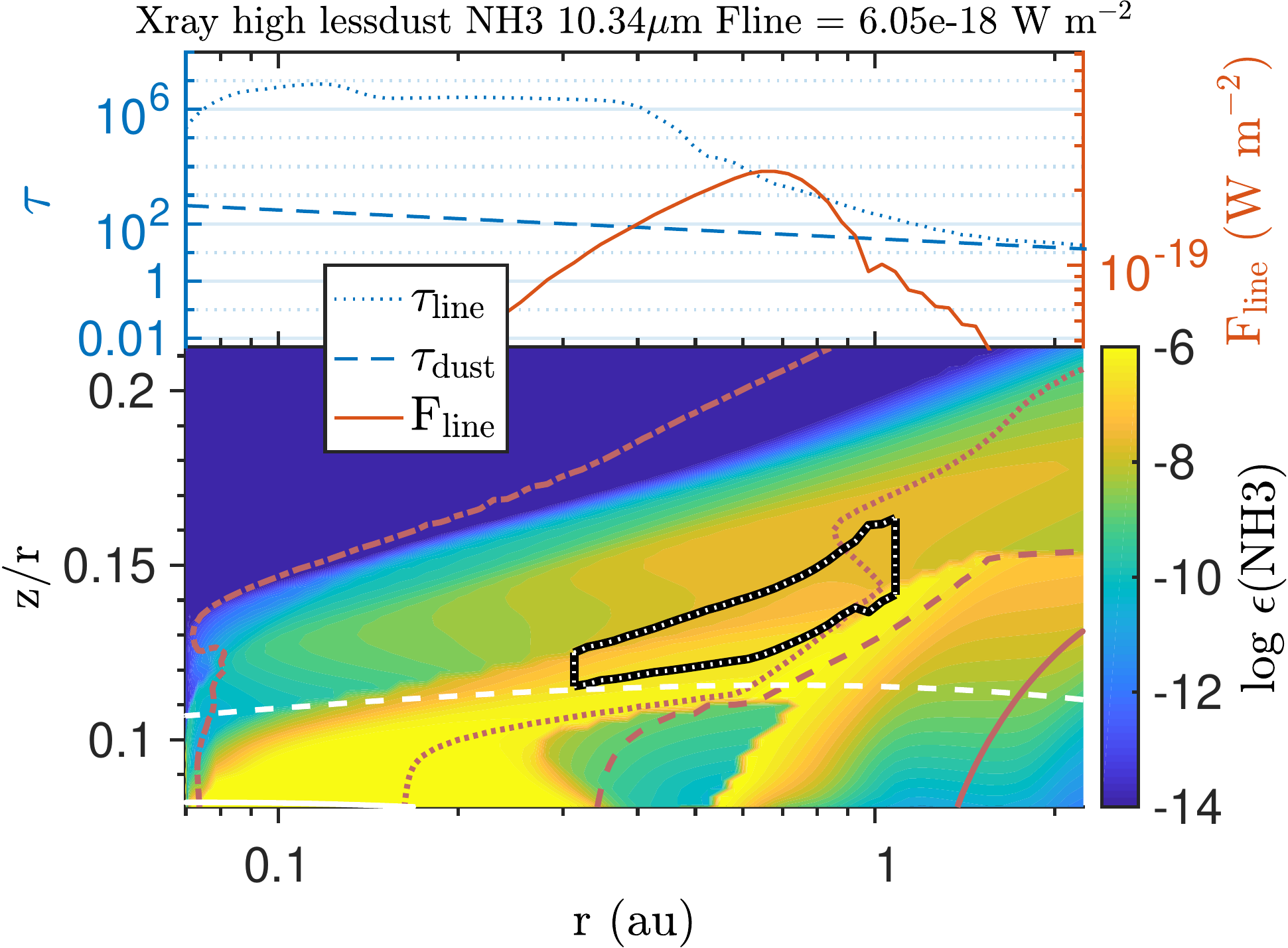}}   
			\makebox[\textwidth][c]{\includegraphics[width=0.47\textwidth]{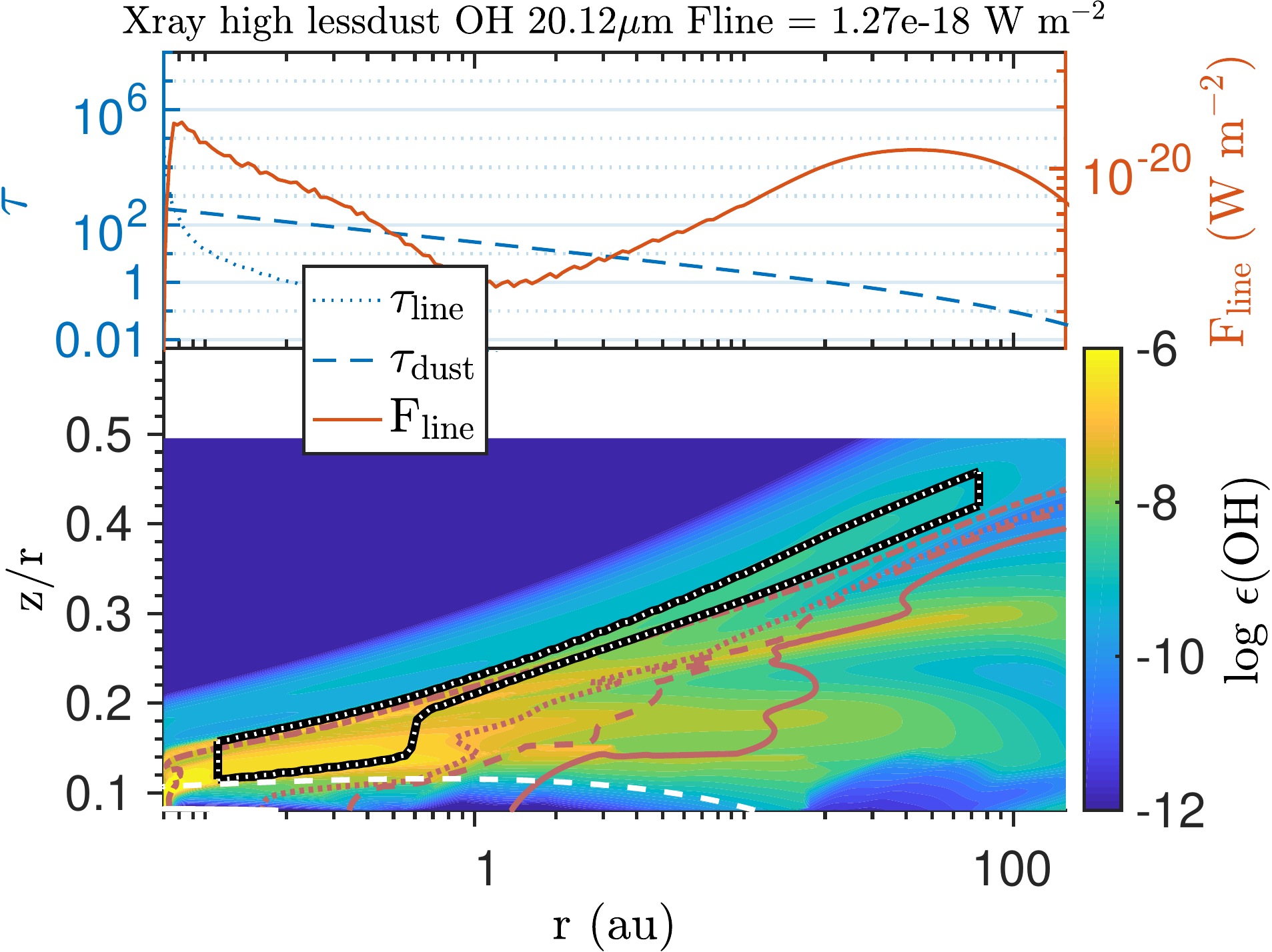} \hspace{0.005\textwidth}   
				\includegraphics[width=0.47\textwidth]{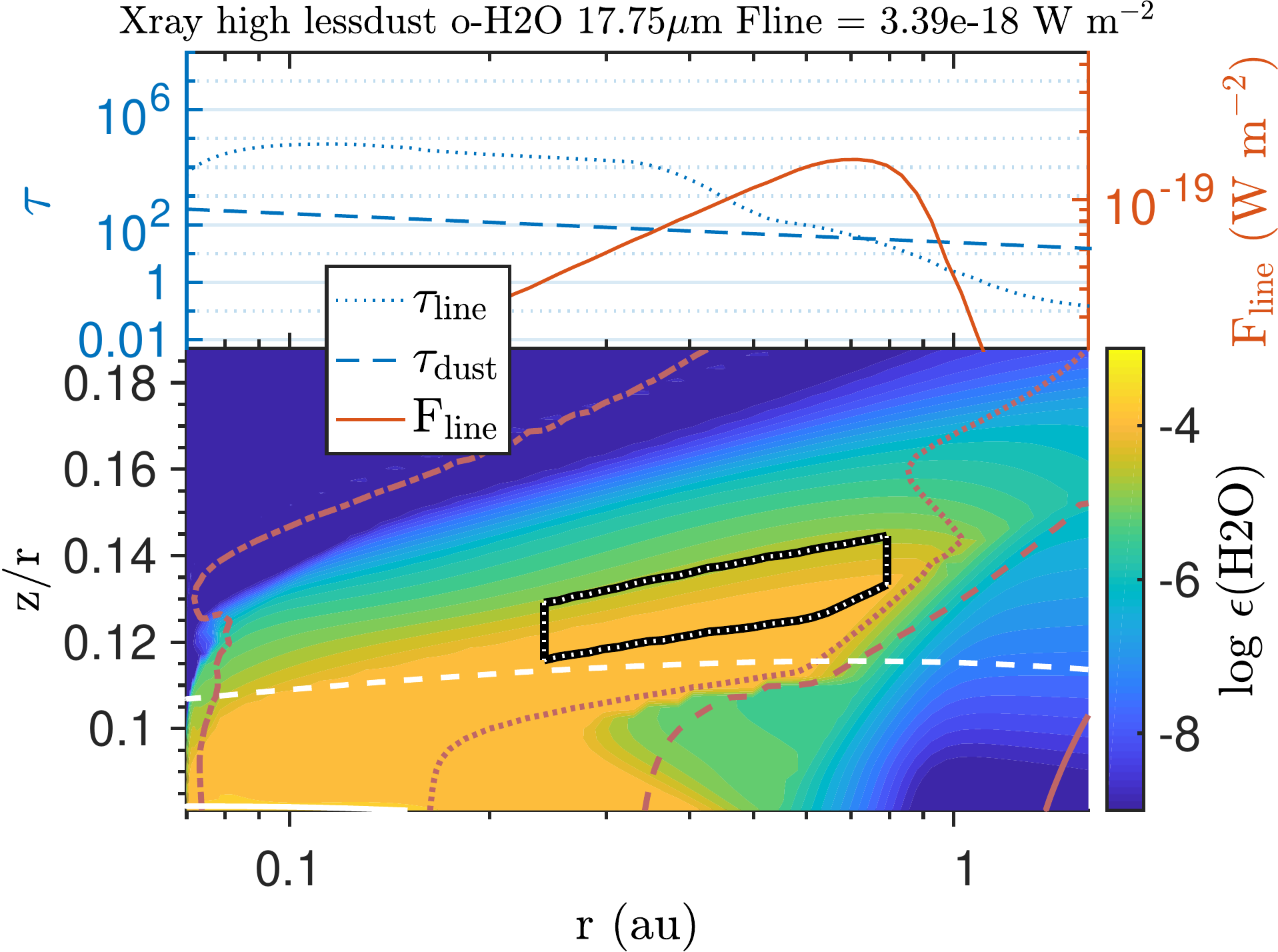}}  
			\caption{Line-emitting regions for the model Xray high lessdust. The plotted lines are \cem{C2H2}, \cem{HCN}, \cem{CO2}, \cem{NH3}, \cem{OH}, and \cem{o-H2O}. The rest of the figure is as described in \cref{fig:LER_TT_highres}.
			}\label{fig:LER_Xray_high_lessdust}   
		\end{figure*}
		
		\begin{figure*} \centering    
			\makebox[\textwidth][c]{\includegraphics[width=0.47\textwidth]{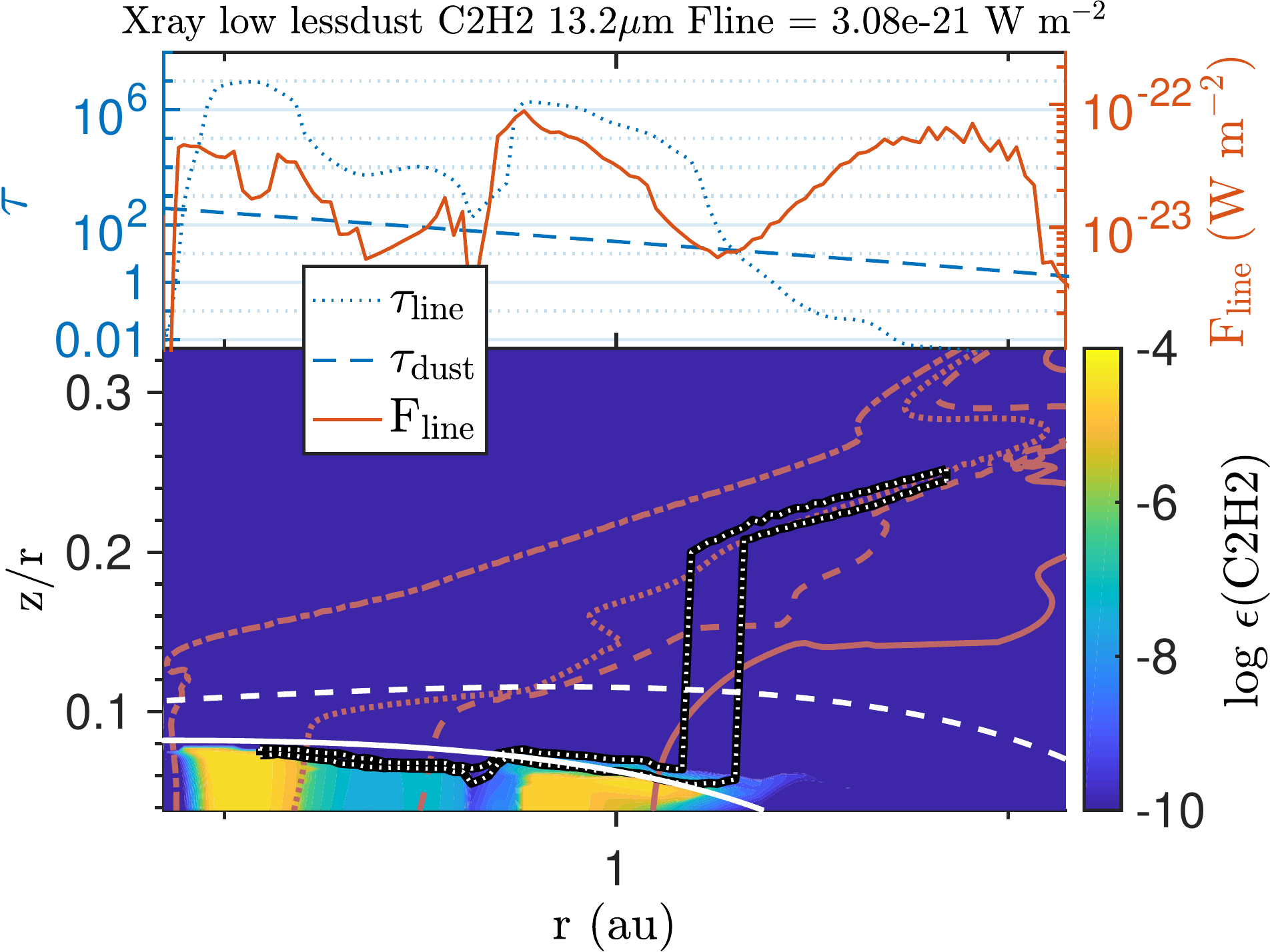} \hspace{0.005\textwidth}  
				\includegraphics[width=0.47\textwidth]{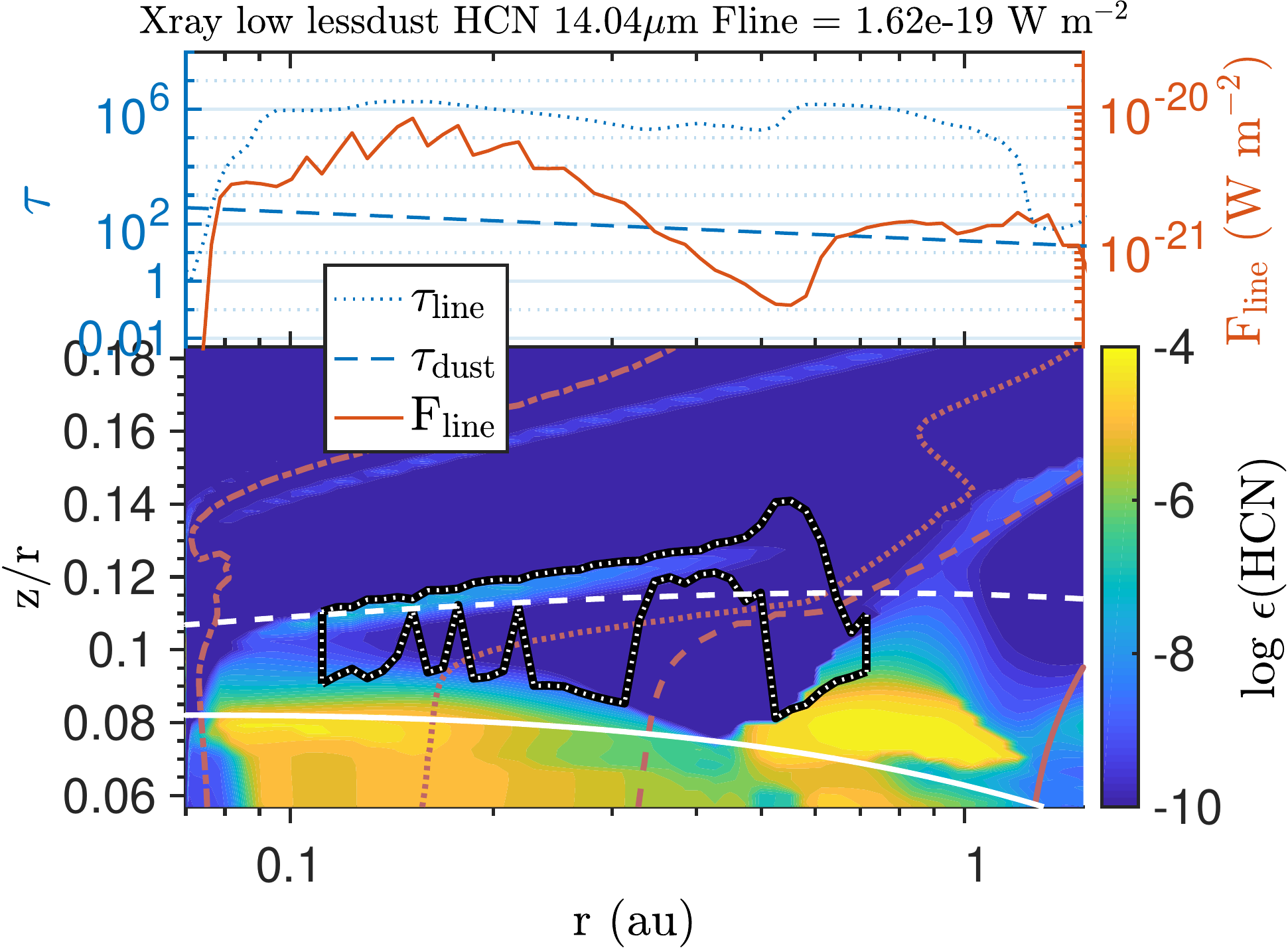}}     
			\makebox[\textwidth][c]{\includegraphics[width=0.47\textwidth]{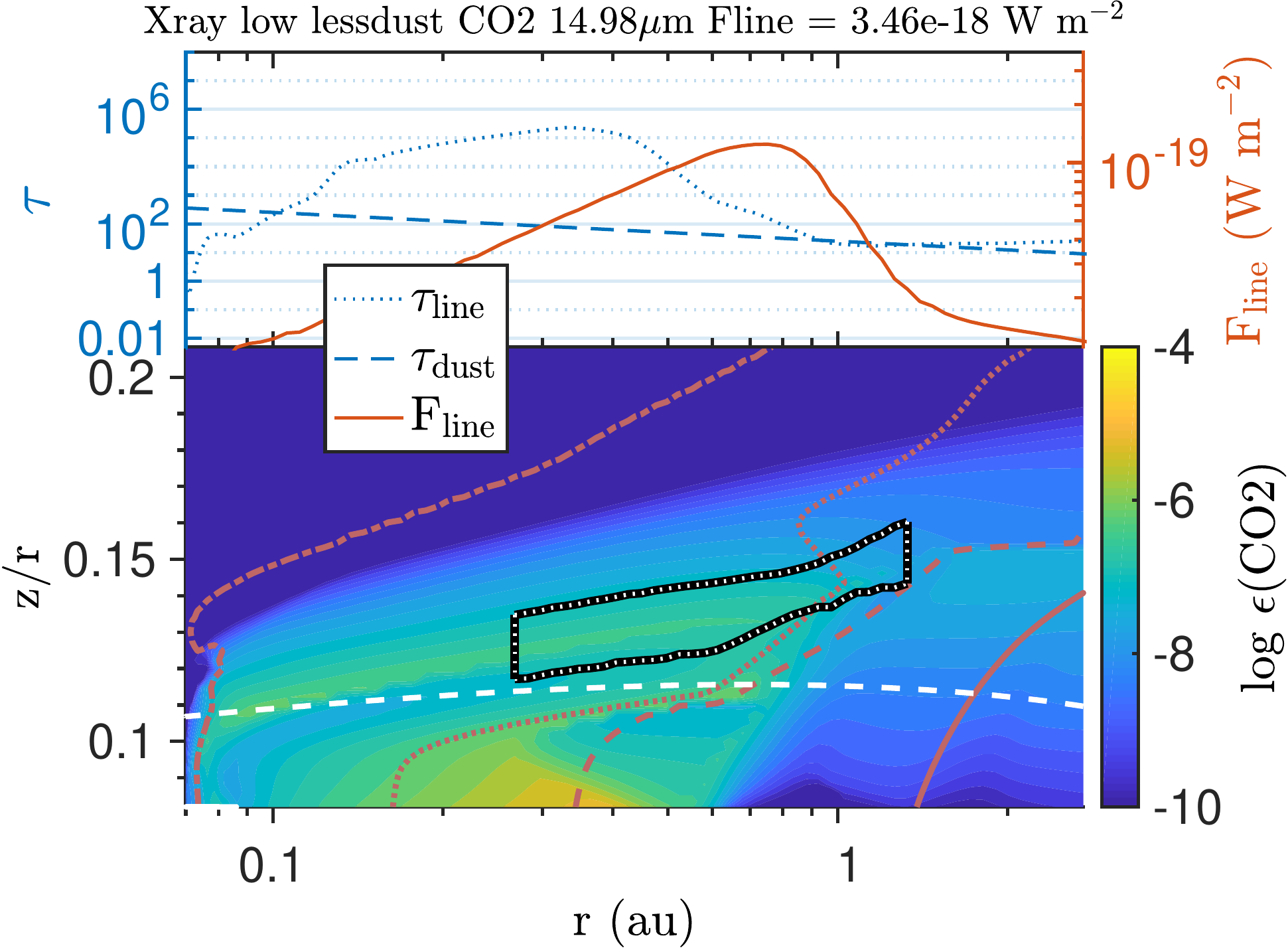} \hspace{0.005\textwidth}
				\includegraphics[width=0.47\textwidth]{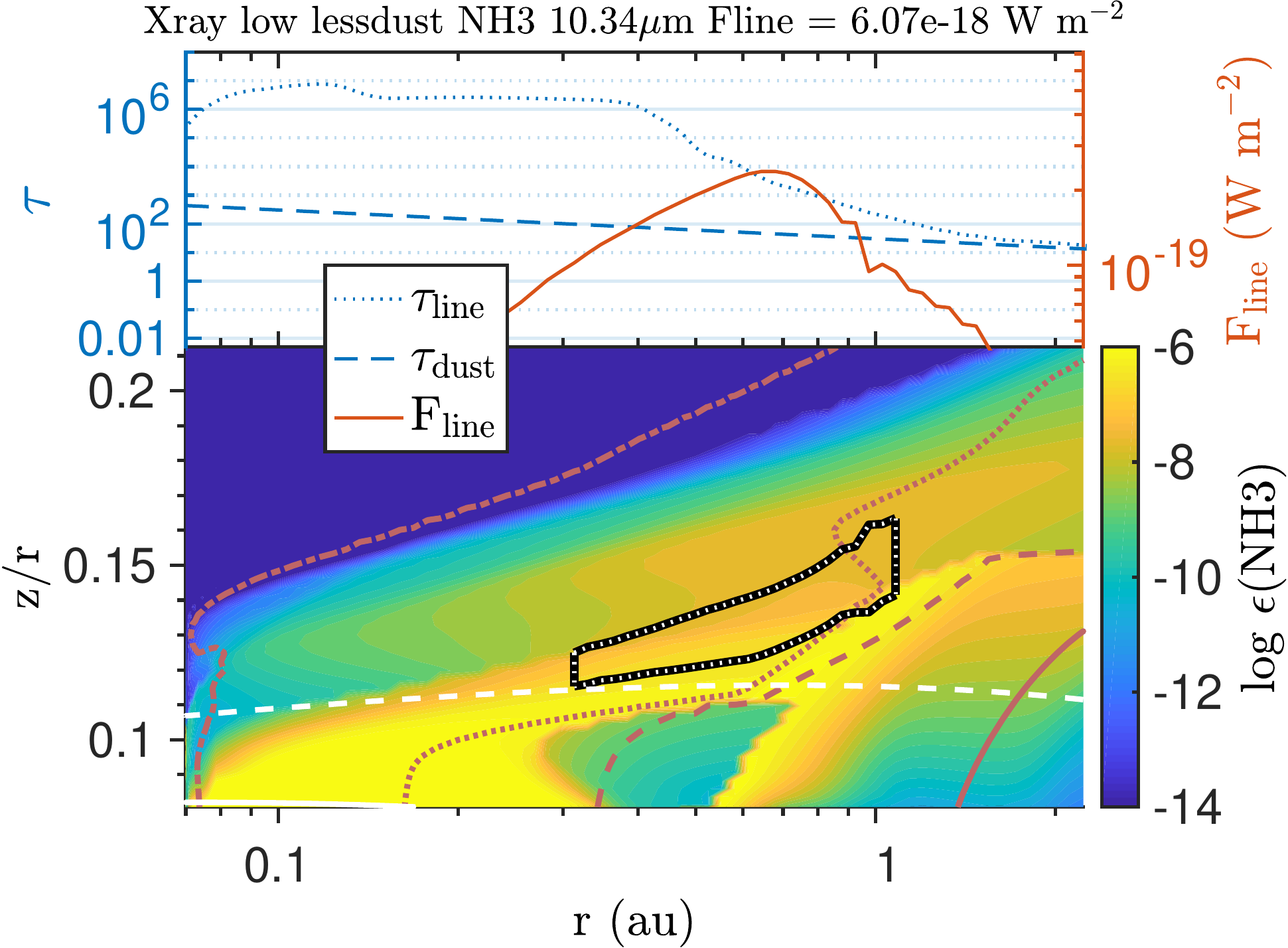}}    
			\makebox[\textwidth][c]{\includegraphics[width=0.47\textwidth]{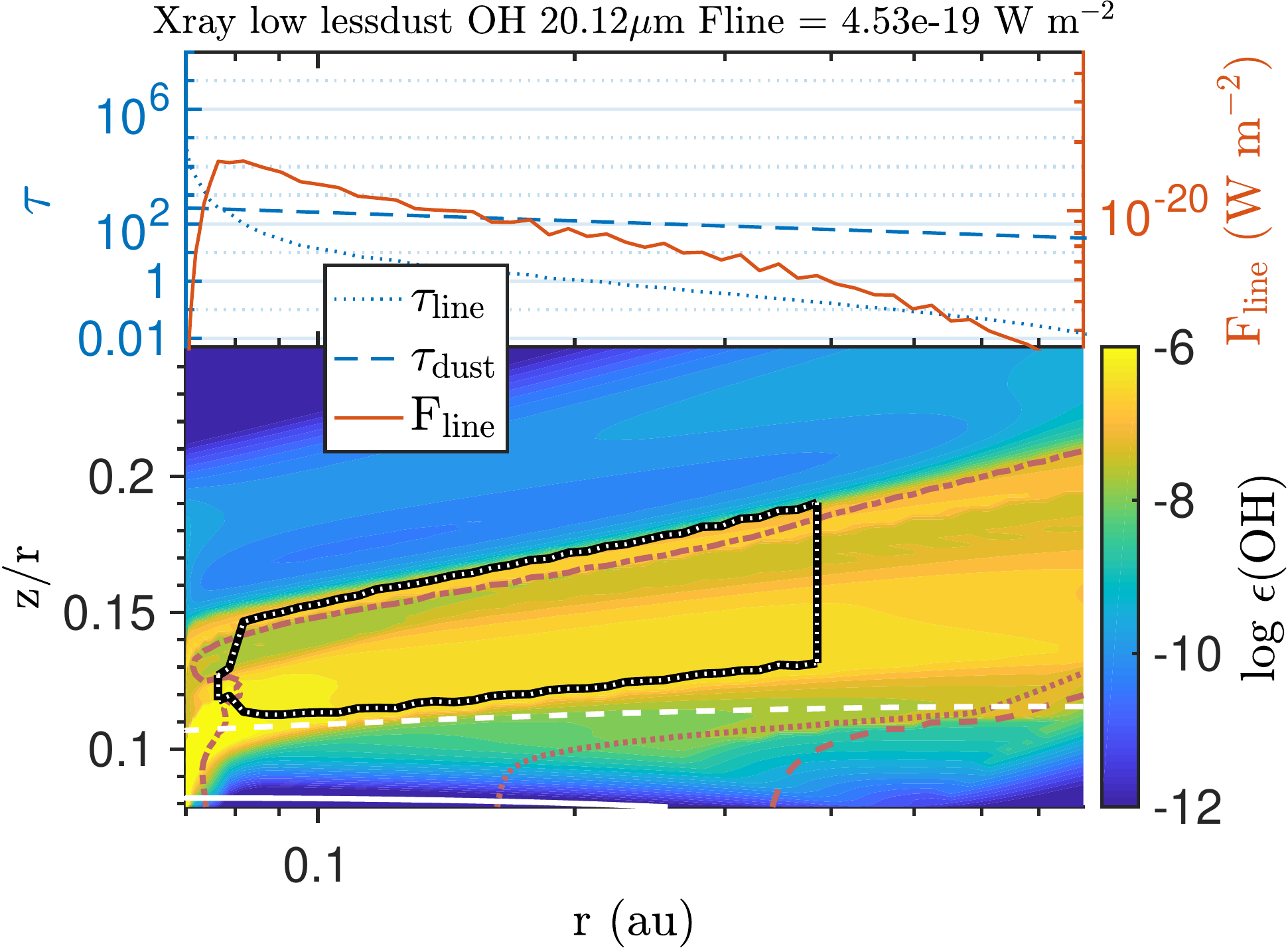} \hspace{0.005\textwidth}    
				\includegraphics[width=0.47\textwidth]{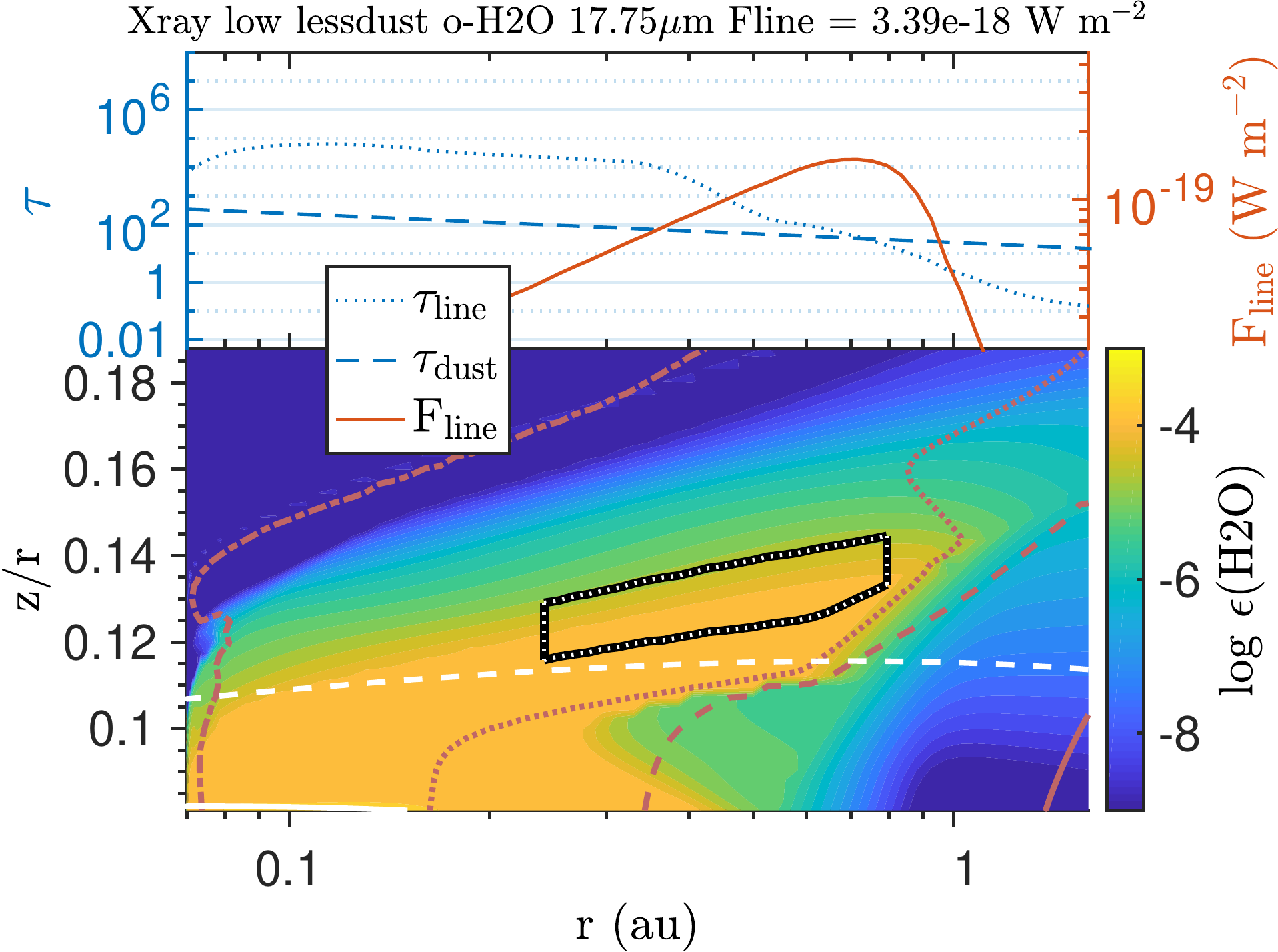}}  
			\caption{Line-emitting regions for the model Xray low lessdust. The plotted lines are \cem{C2H2}, \cem{HCN}, \cem{CO2}, \cem{NH3}, \cem{OH}, and \cem{o-H2O}. The rest of the figure is as described in \cref{fig:LER_TT_highres}. 
			}\label{fig:LER_Xray_low_lessdust}    
		\end{figure*}
		
		\begin{figure*} \centering    
			\makebox[\textwidth][c]{\includegraphics[width=0.47\textwidth]{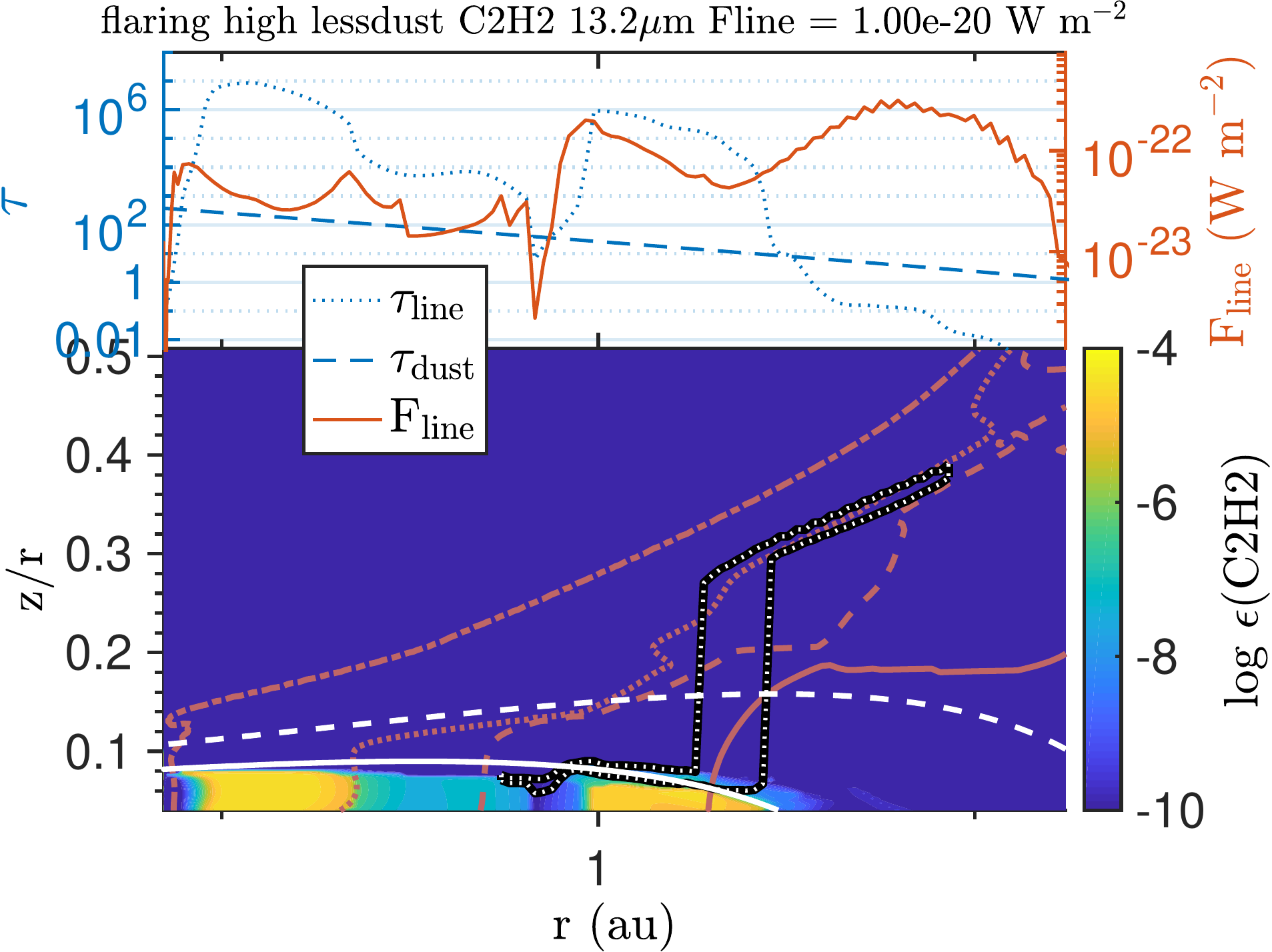} \hspace{0.005\textwidth}    
				\includegraphics[width=0.47\textwidth]{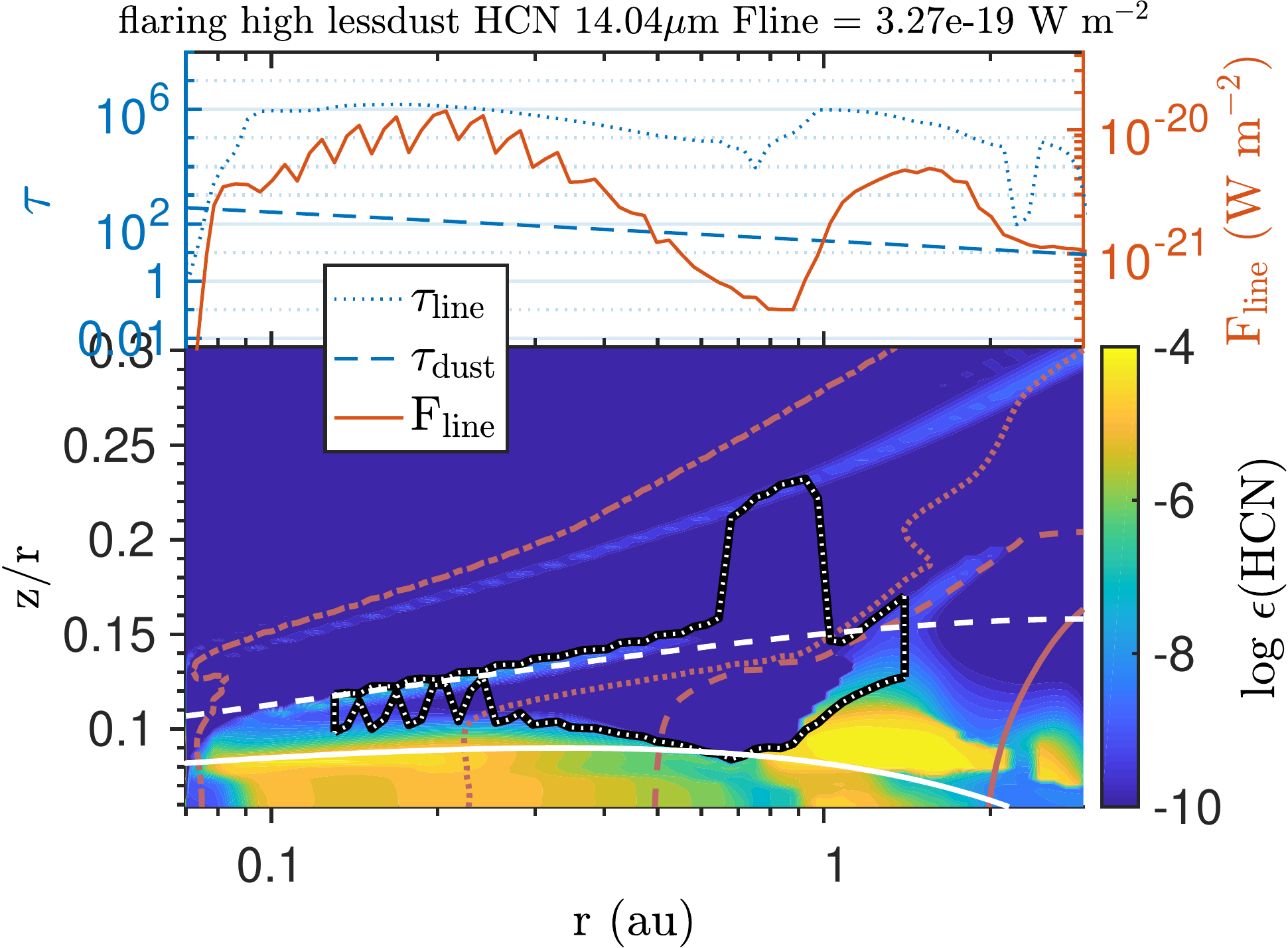}} 
			\makebox[\textwidth][c]{\includegraphics[width=0.47\textwidth]{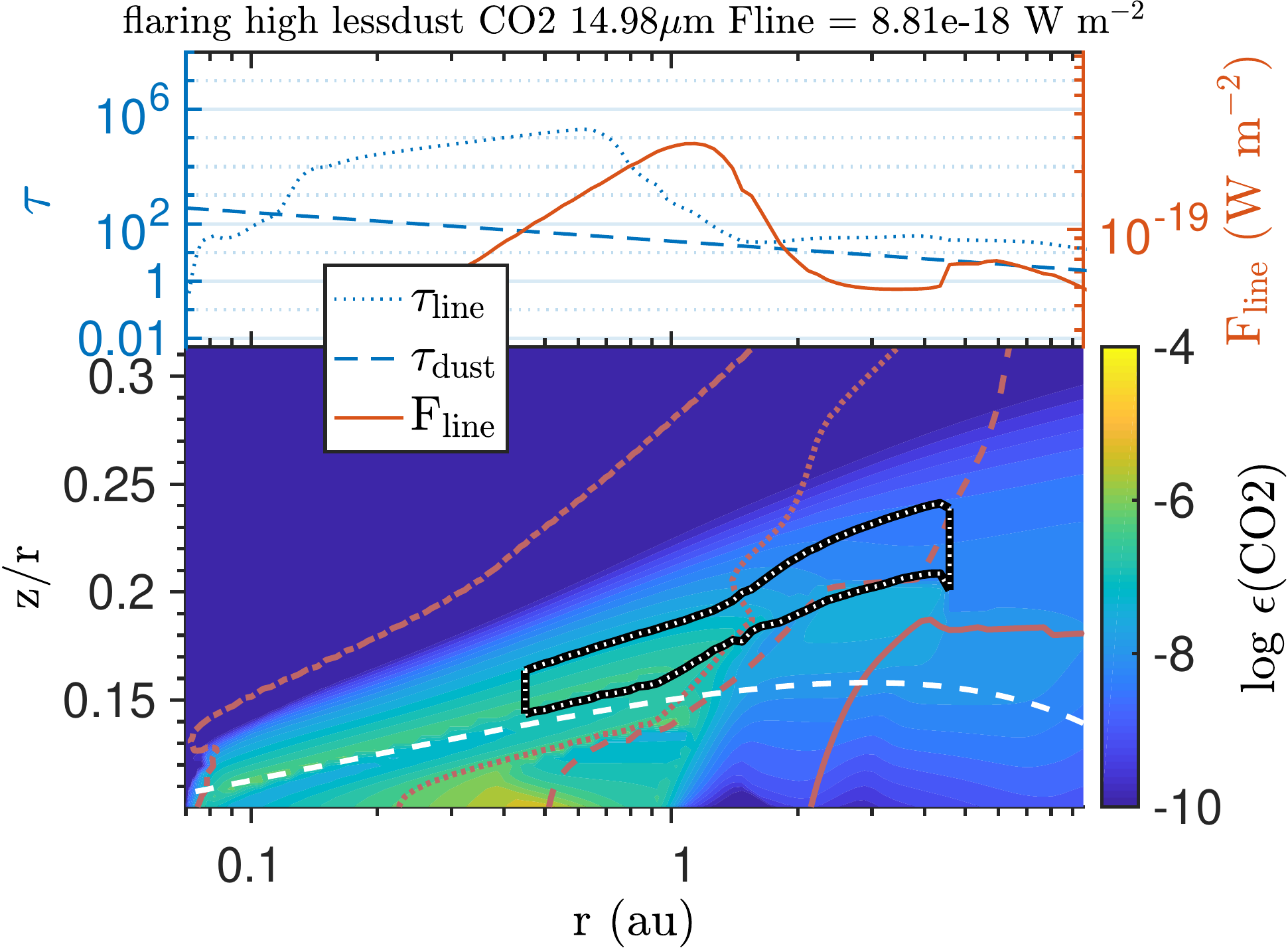} \hspace{0.005\textwidth}
				\includegraphics[width=0.47\textwidth]{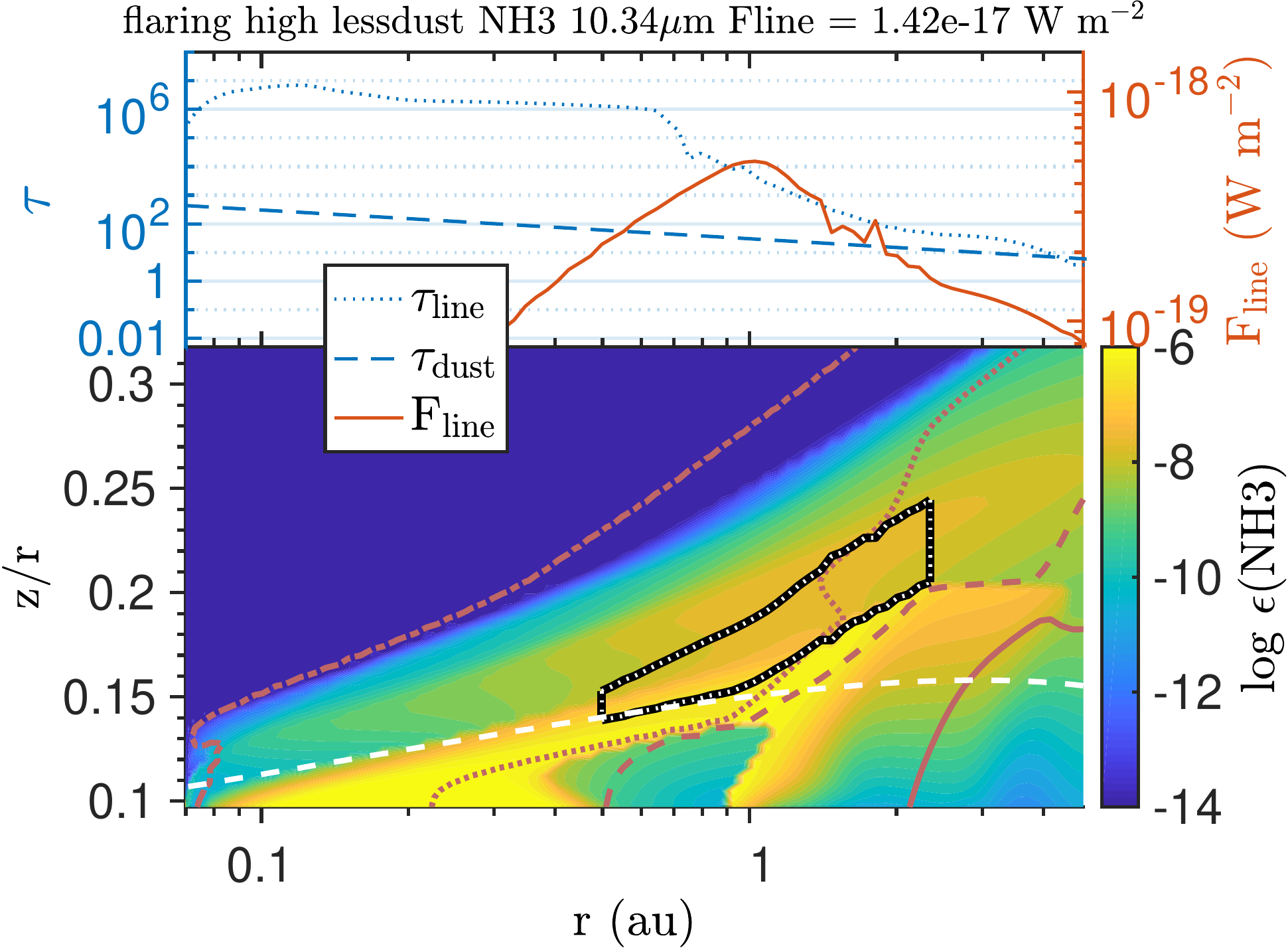}}      
			\makebox[\textwidth][c]{\includegraphics[width=0.47\textwidth]{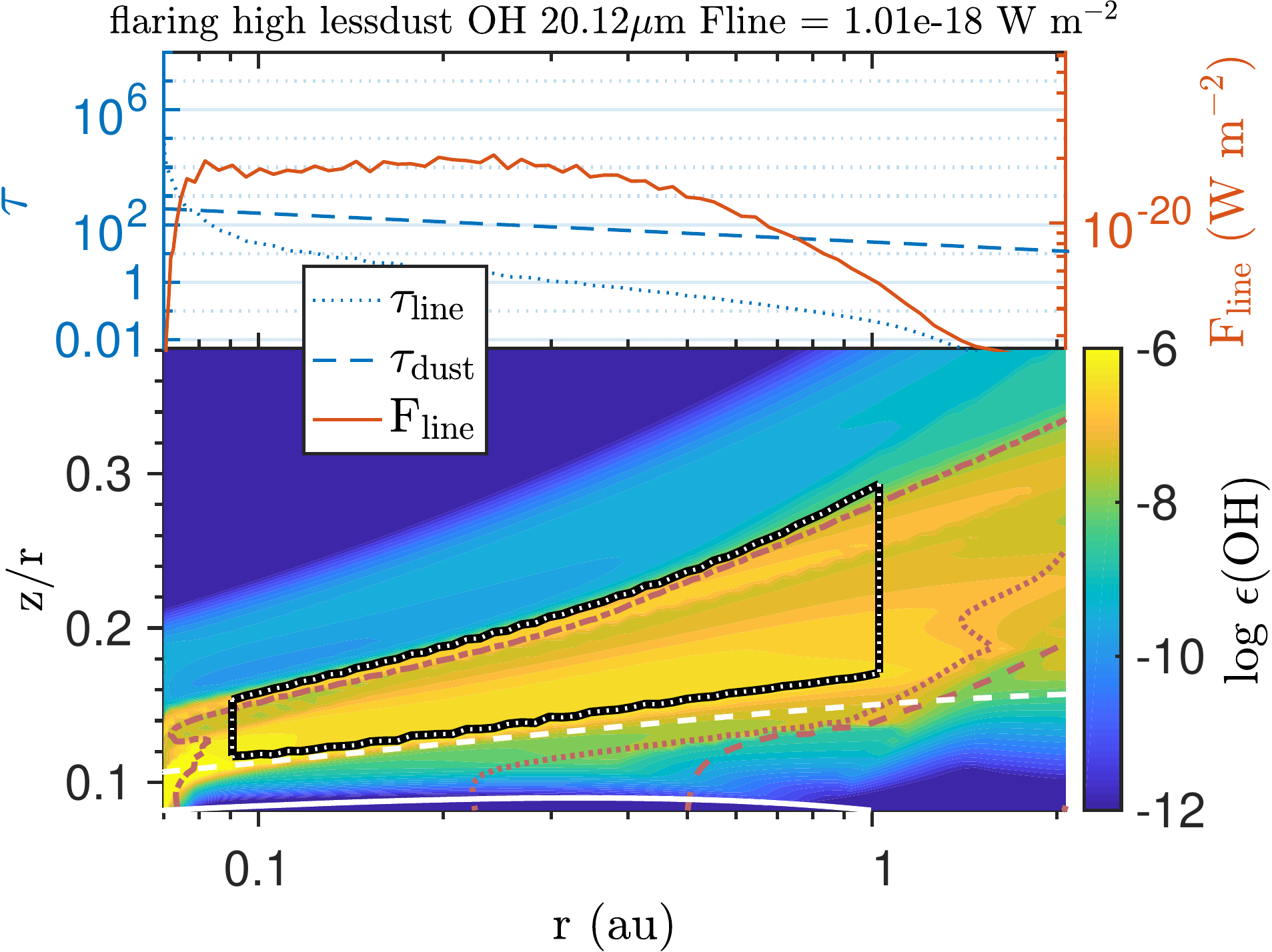} \hspace{0.005\textwidth}
				\includegraphics[width=0.47\textwidth]{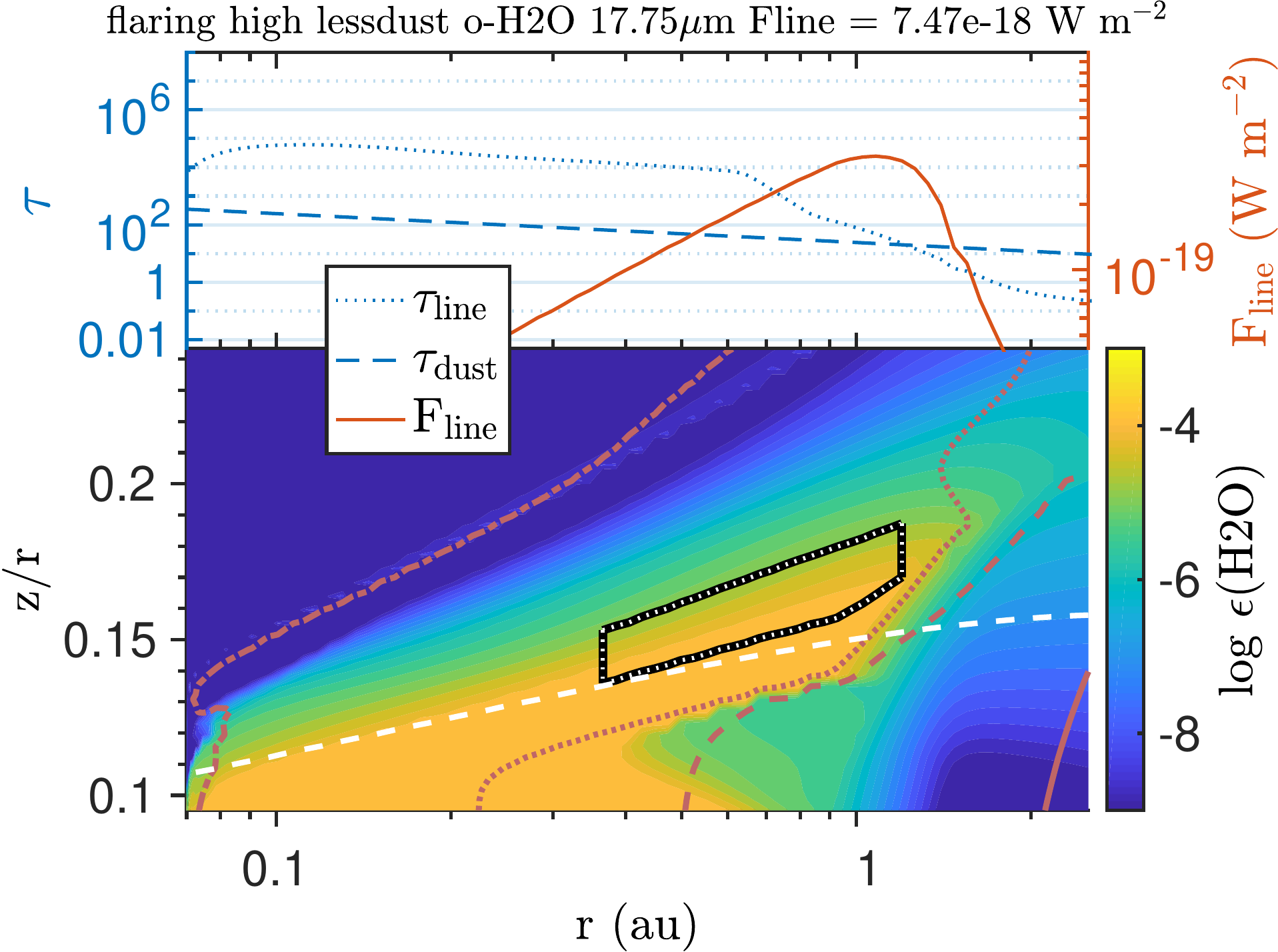}}
			\caption{Line-emitting regions for the model flaring high lessdust. The plotted lines are \cem{C2H2}, \cem{HCN}, \cem{CO2}, \cem{NH3}, \cem{OH}, and \cem{o-H2O}. The rest of the figure is as described in \cref{fig:LER_TT_highres}.   
			}\label{fig:LER_flaring_high_lessdust}
		\end{figure*}
		
		\begin{figure*} \centering    
			\makebox[\textwidth][c]{\includegraphics[width=0.47\textwidth]{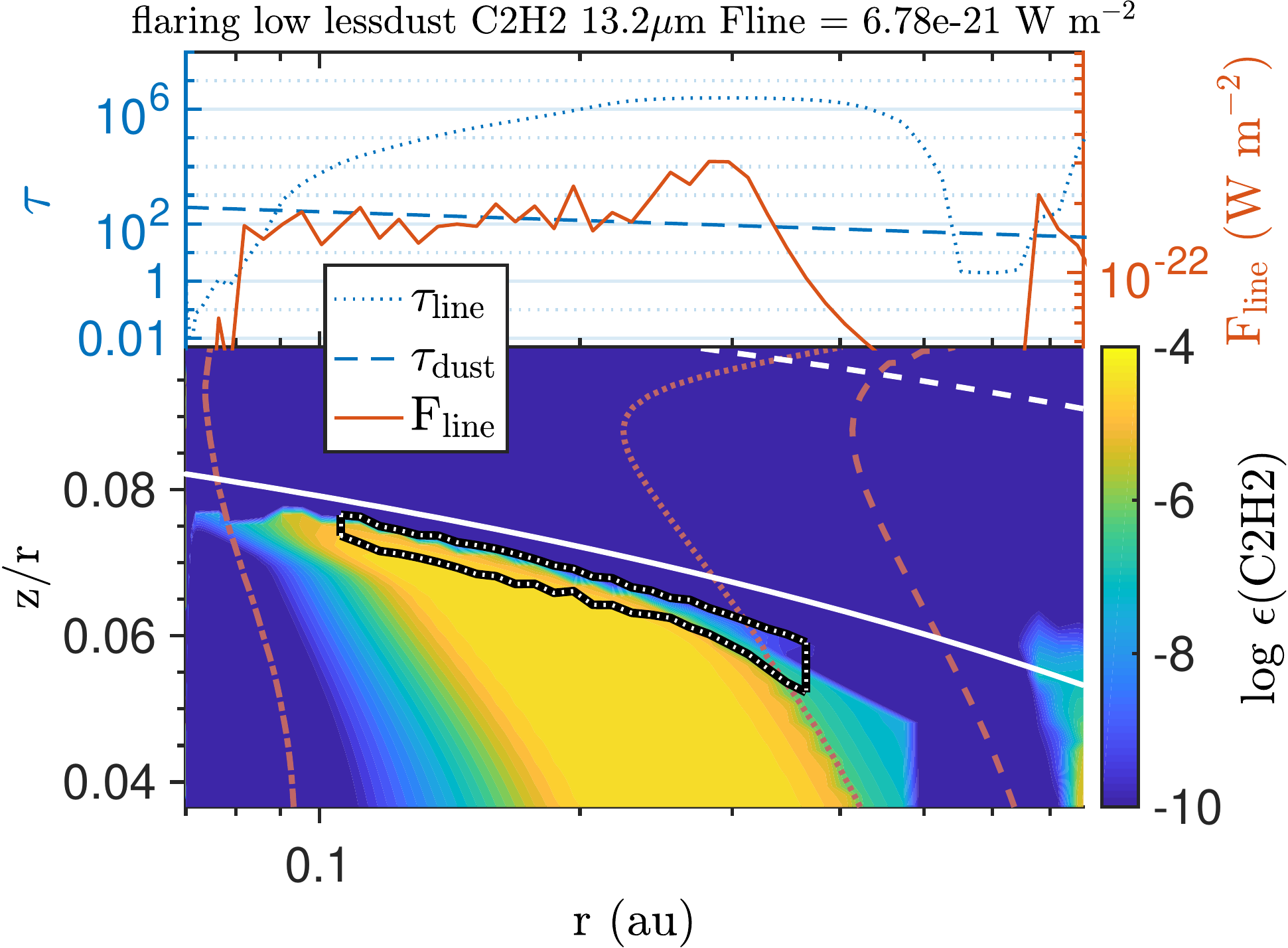} \hspace{0.005\textwidth}   
				\includegraphics[width=0.47\textwidth]{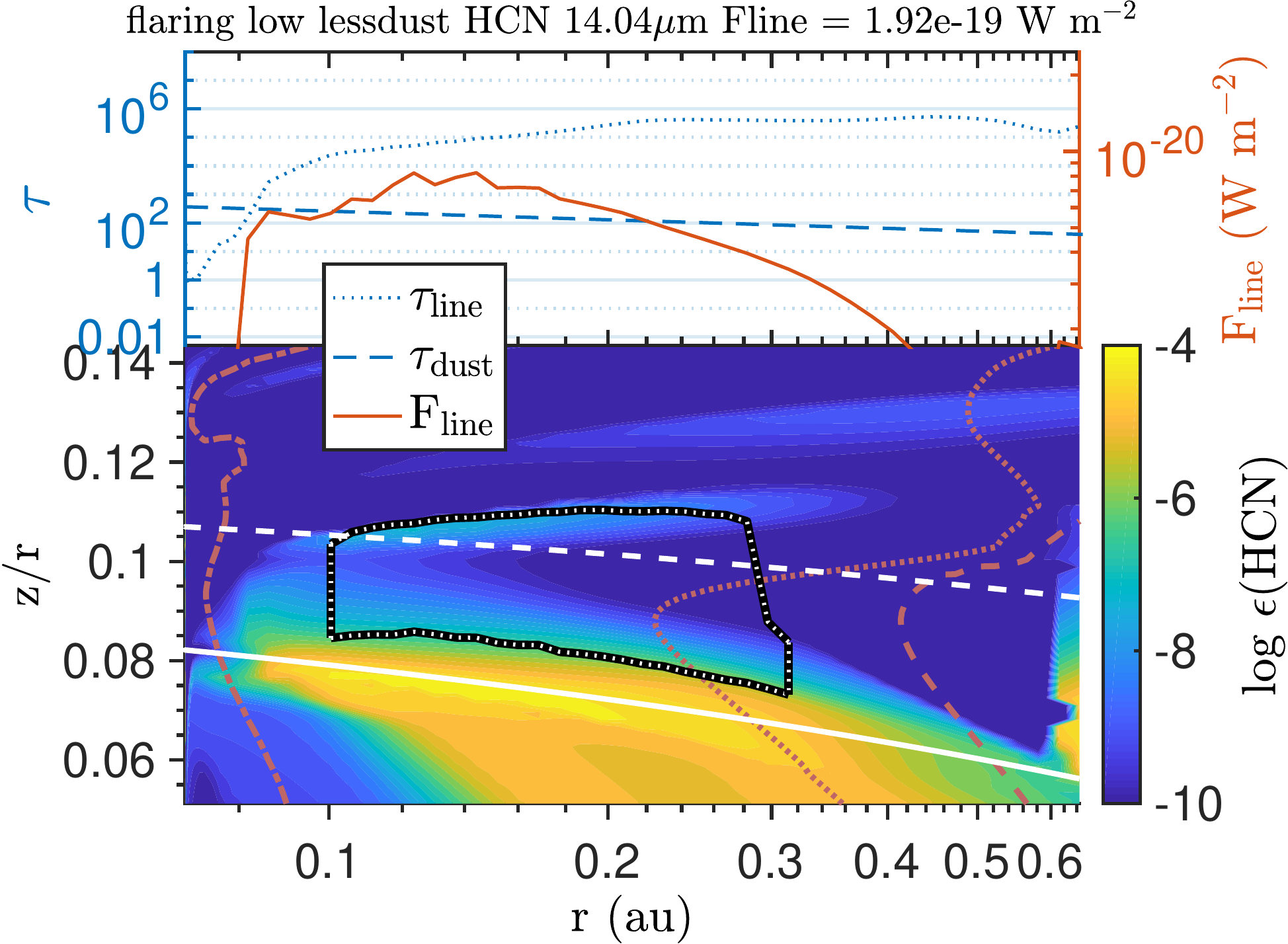}}      
			\makebox[\textwidth][c]{\includegraphics[width=0.47\textwidth]{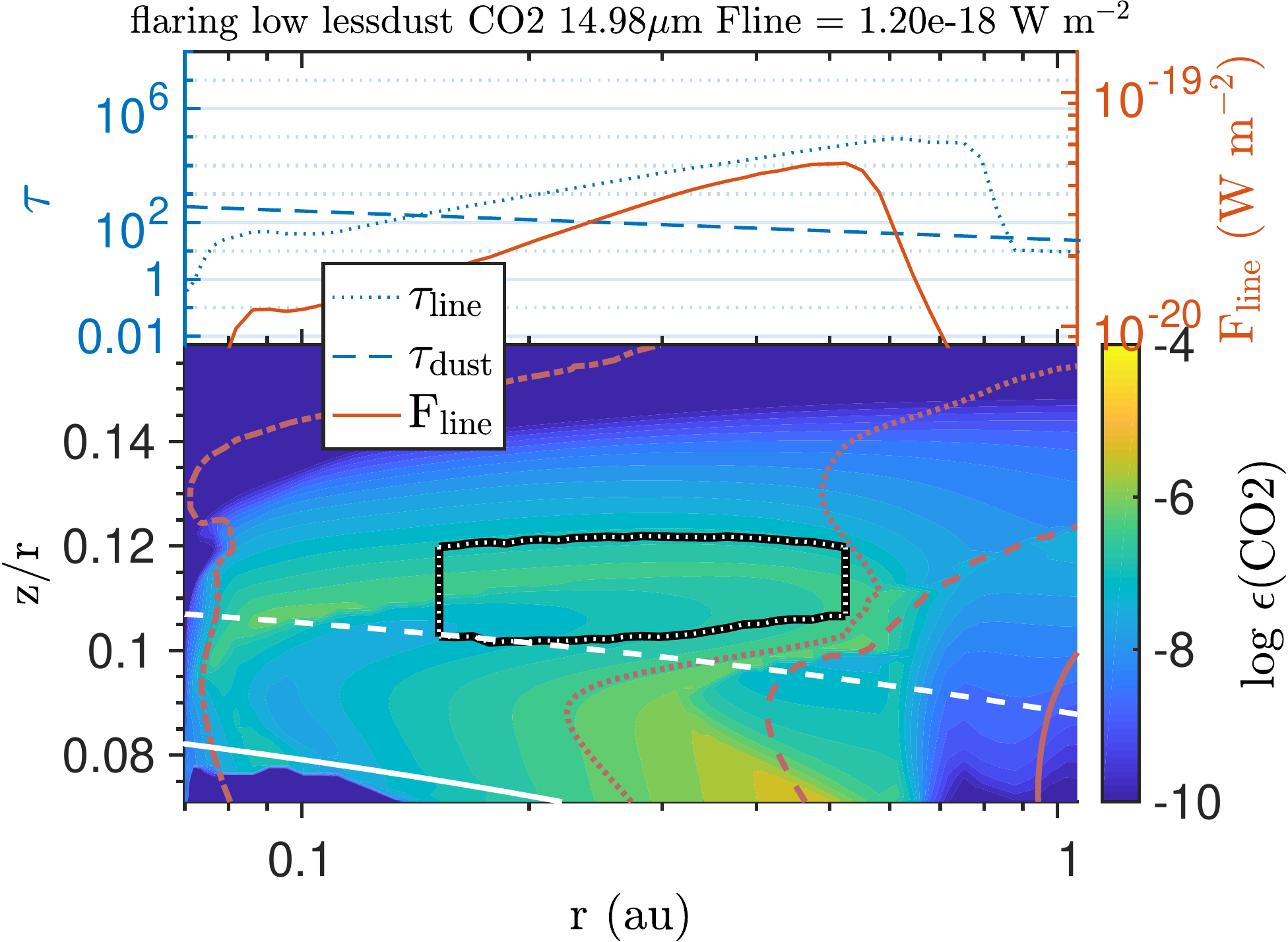} \hspace{0.005\textwidth}
				\includegraphics[width=0.47\textwidth]{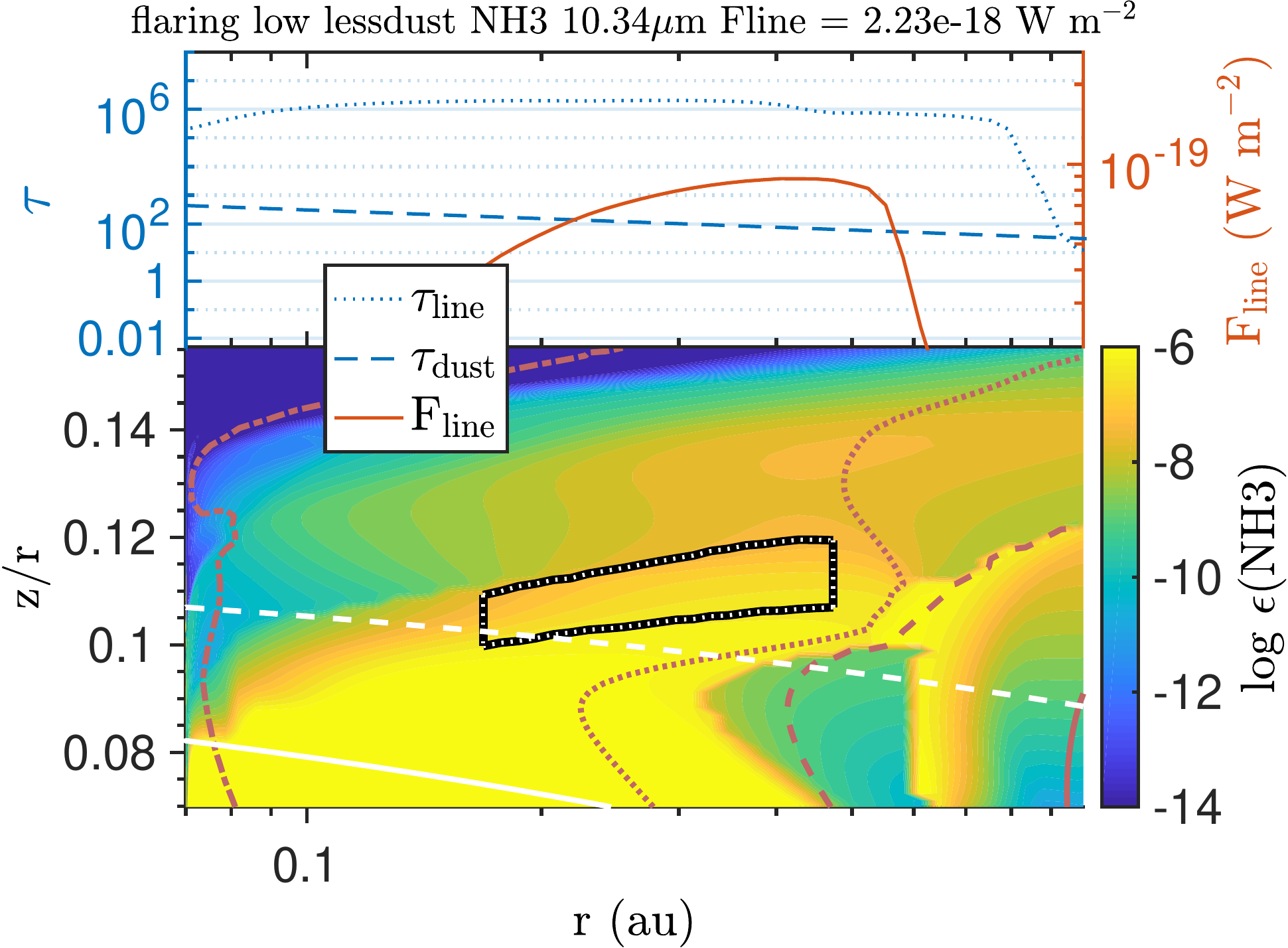}}     
			\makebox[\textwidth][c]{\includegraphics[width=0.47\textwidth]{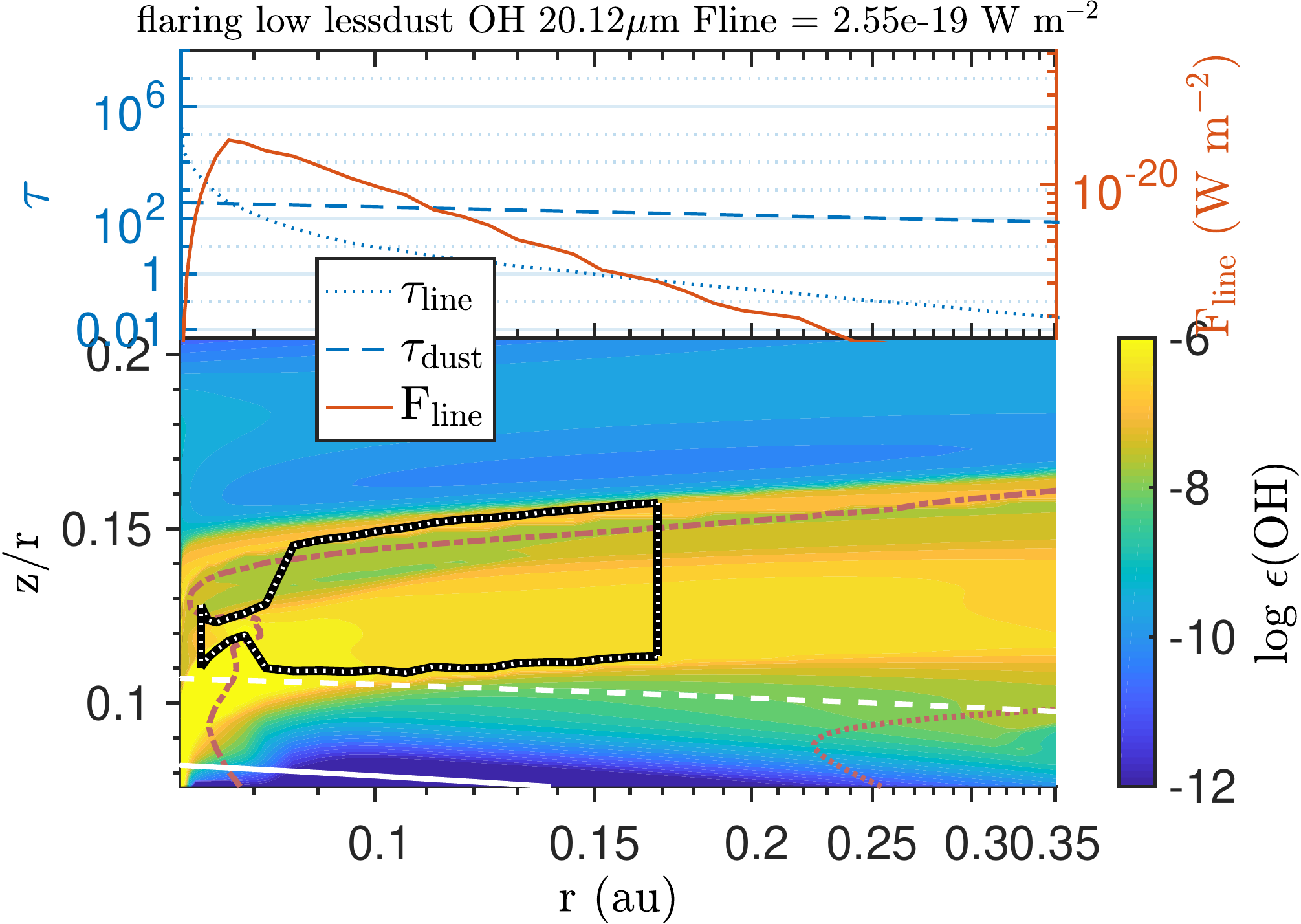} \hspace{0.005\textwidth}     
				\includegraphics[width=0.47\textwidth]{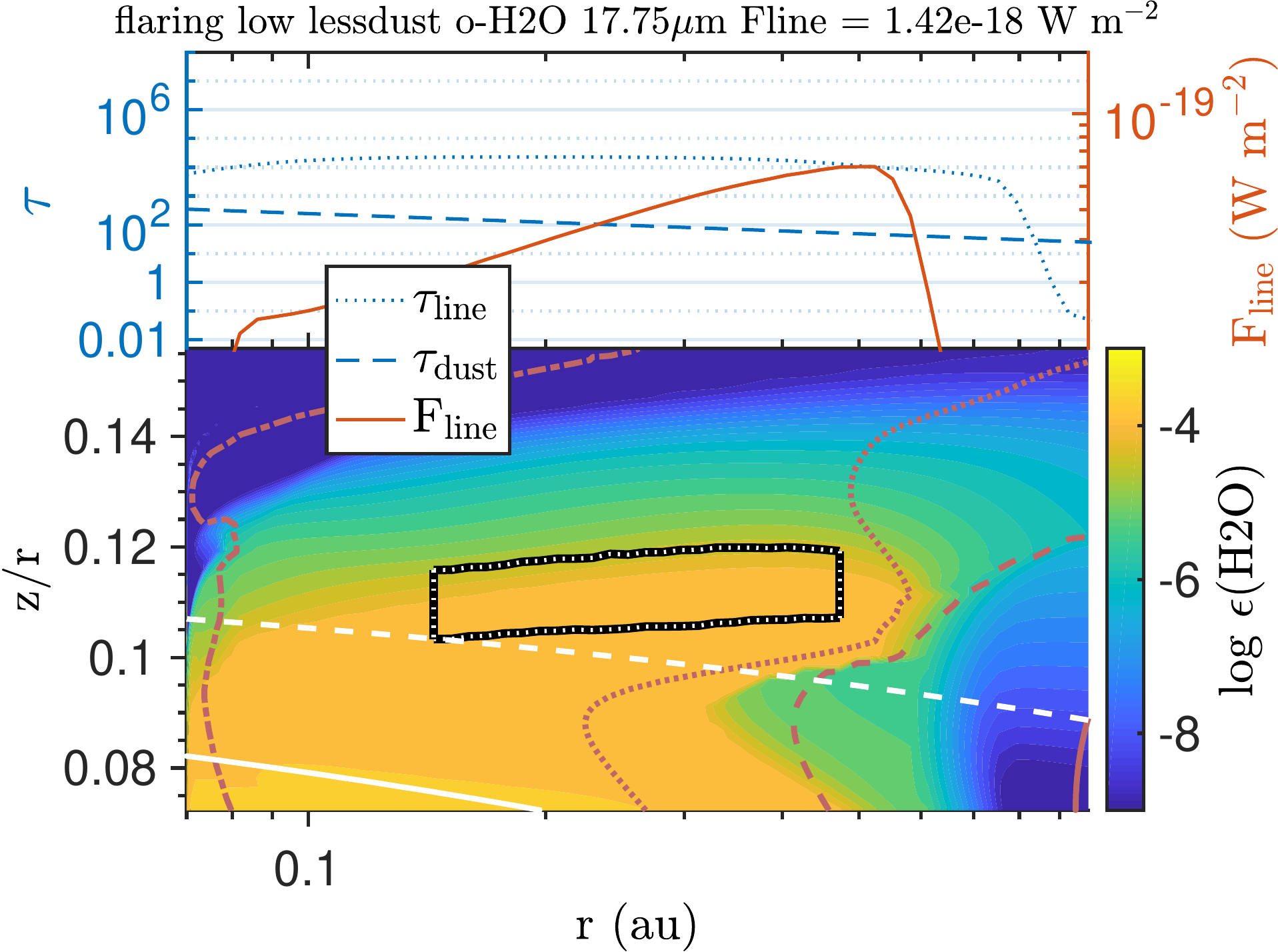}}     
			\caption{Line-emitting regions for the model flaring low lessdust. The plotted lines are \cem{C2H2}, \cem{HCN}, \cem{CO2}, \cem{NH3}, \cem{OH}, and \cem{o-H2O}. The rest of the figure is as described in \cref{fig:LER_TT_highres}.
			}\label{fig:LER_flaring_low_scaleheightfix_run5_lessdust}     
		\end{figure*}
		
		\begin{figure*} \centering    
			\makebox[\textwidth][c]{\includegraphics[width=0.47\textwidth]{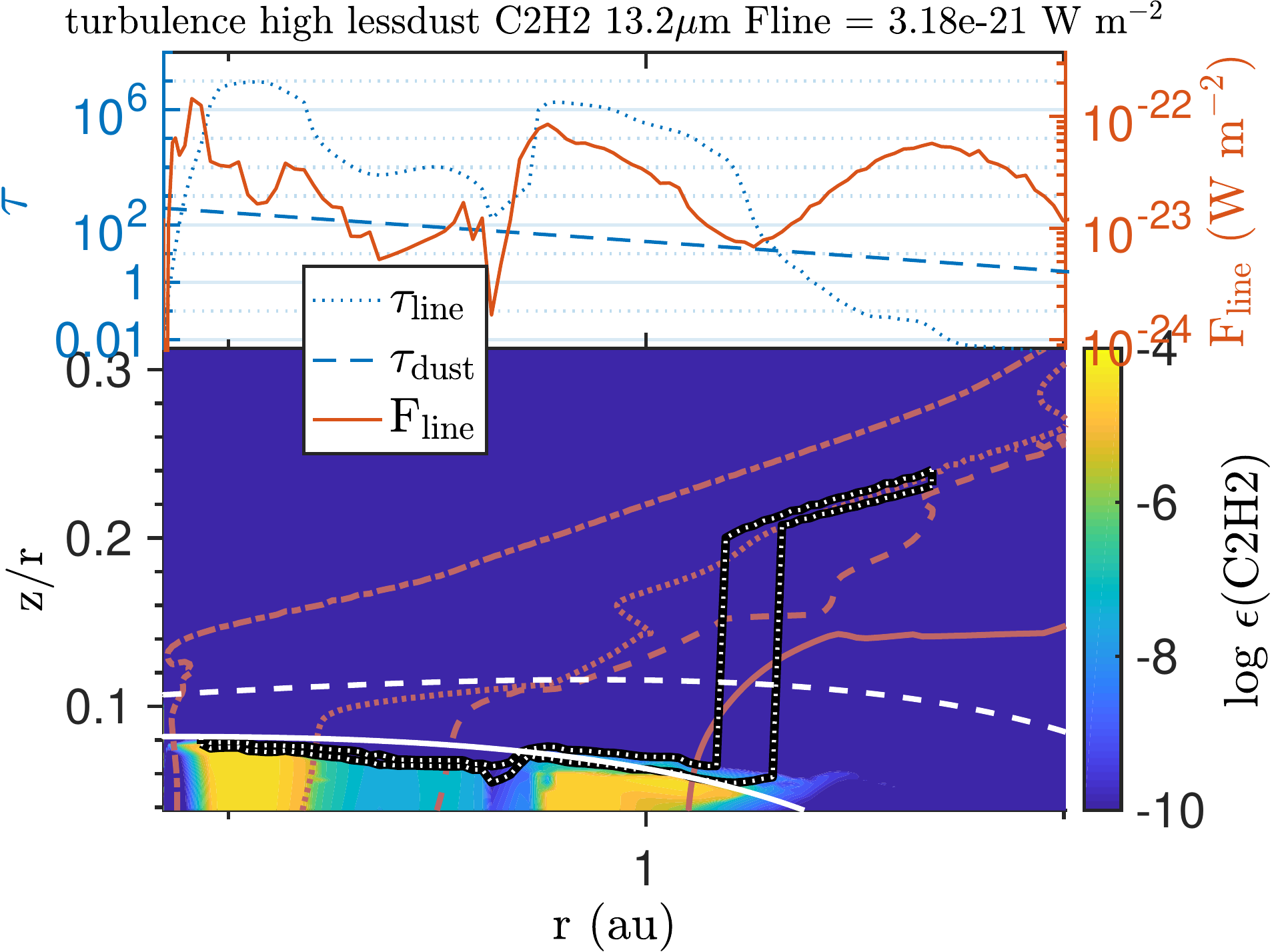} \hspace{0.005\textwidth}   
				\includegraphics[width=0.47\textwidth]{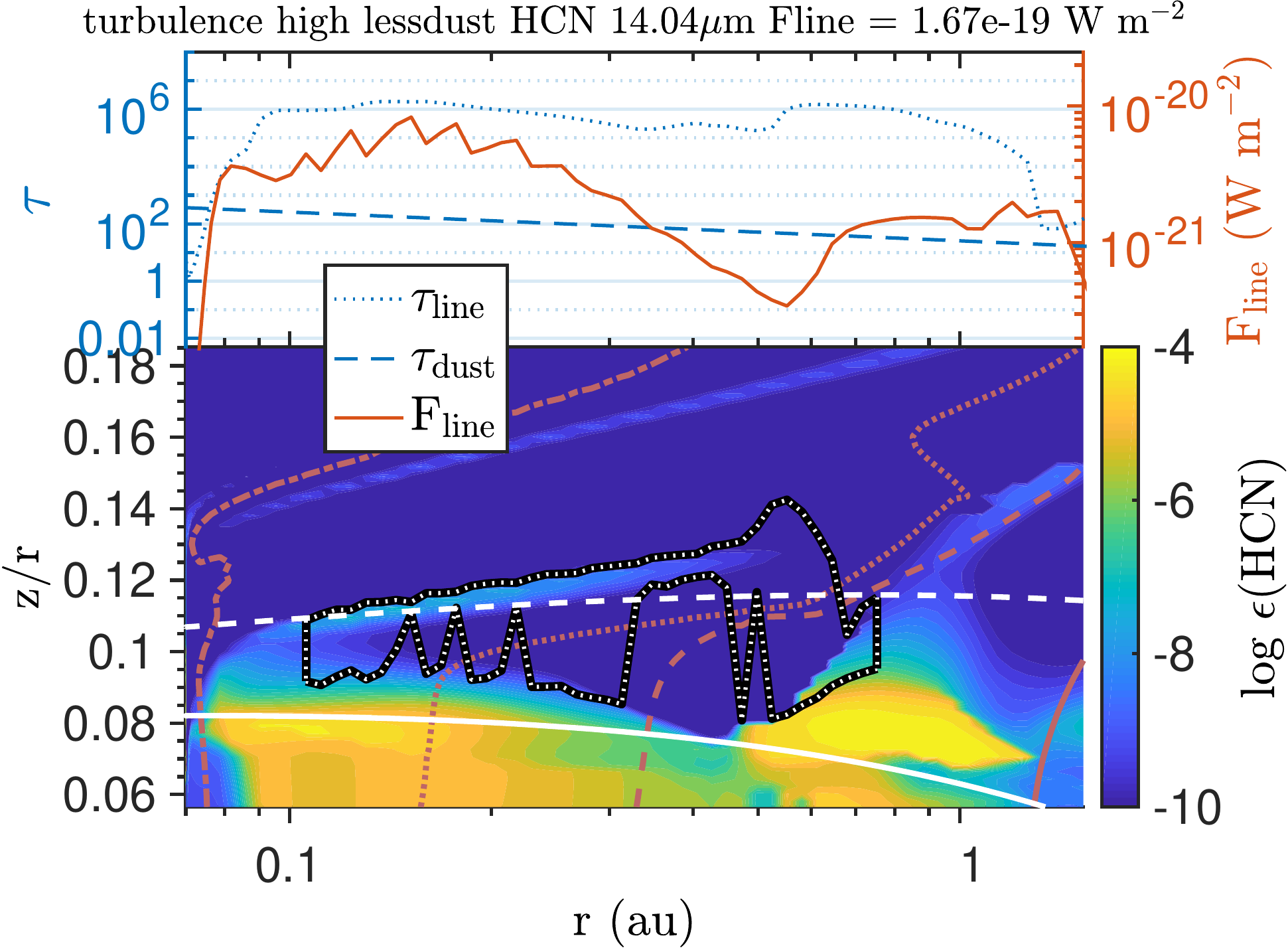}}      
			\makebox[\textwidth][c]{\includegraphics[width=0.47\textwidth]{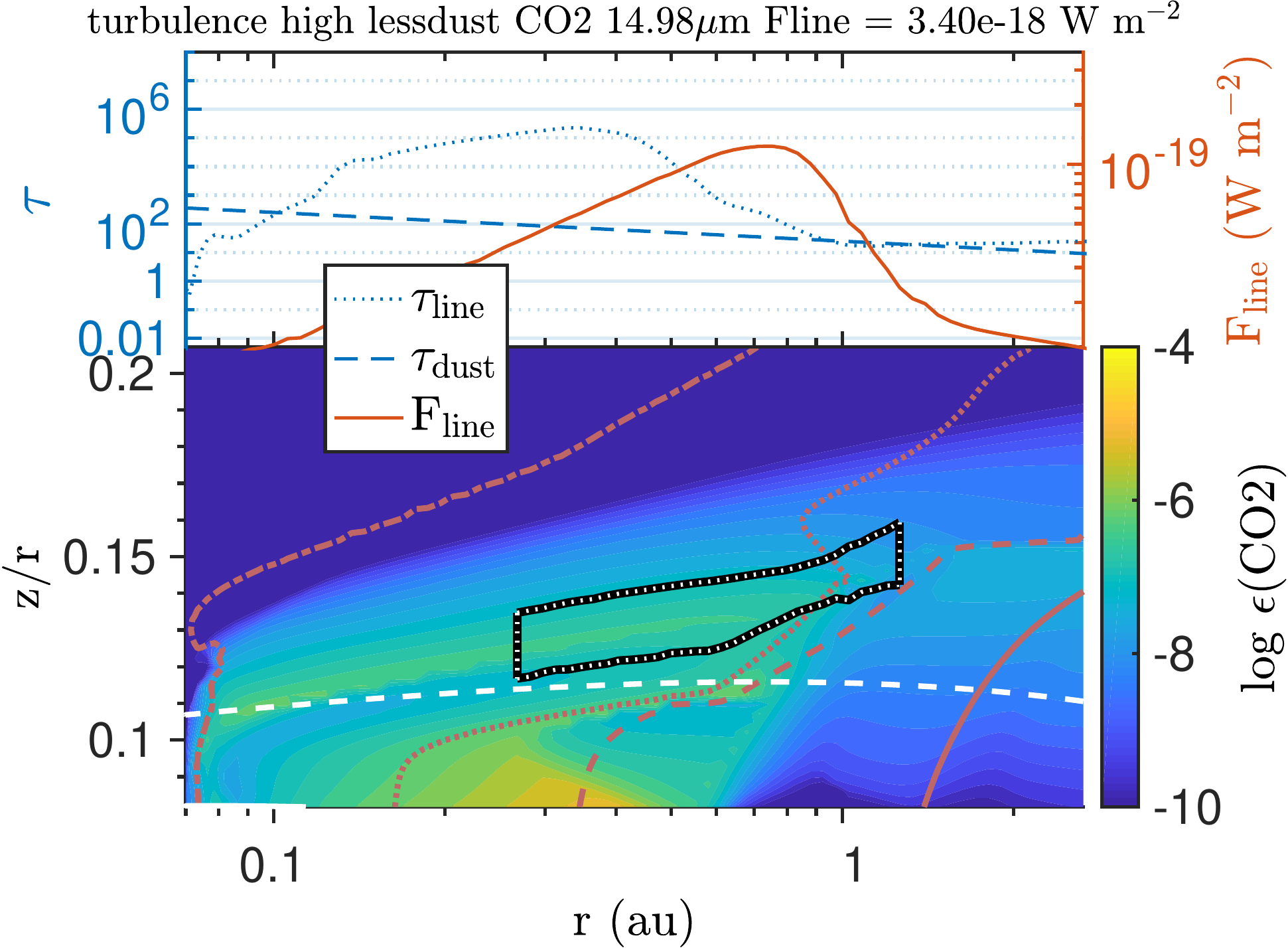} \hspace{0.005\textwidth}
				\includegraphics[width=0.47\textwidth]{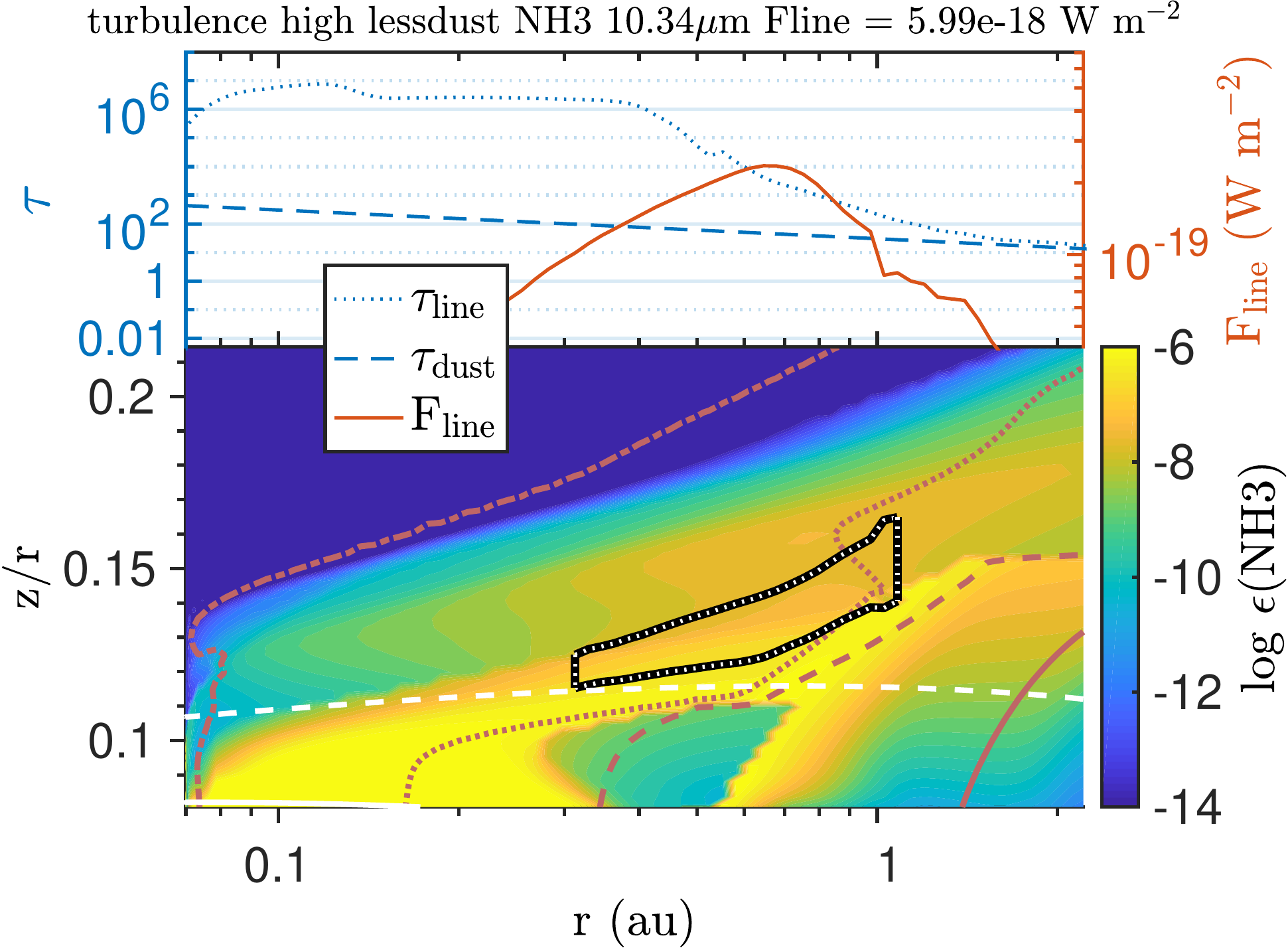}}     
			\makebox[\textwidth][c]{\includegraphics[width=0.47\textwidth]{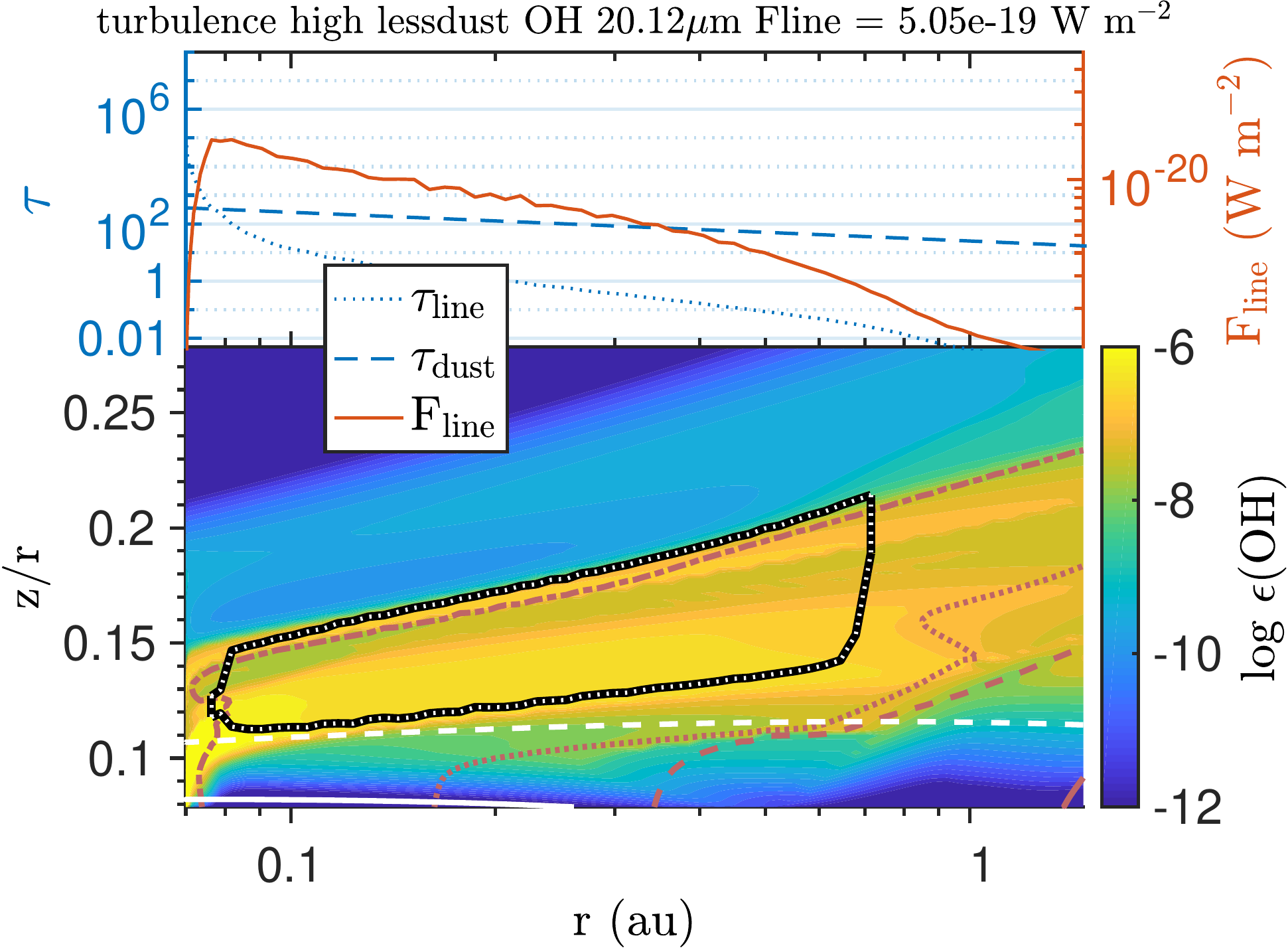} \hspace{0.005\textwidth}     
				\includegraphics[width=0.47\textwidth]{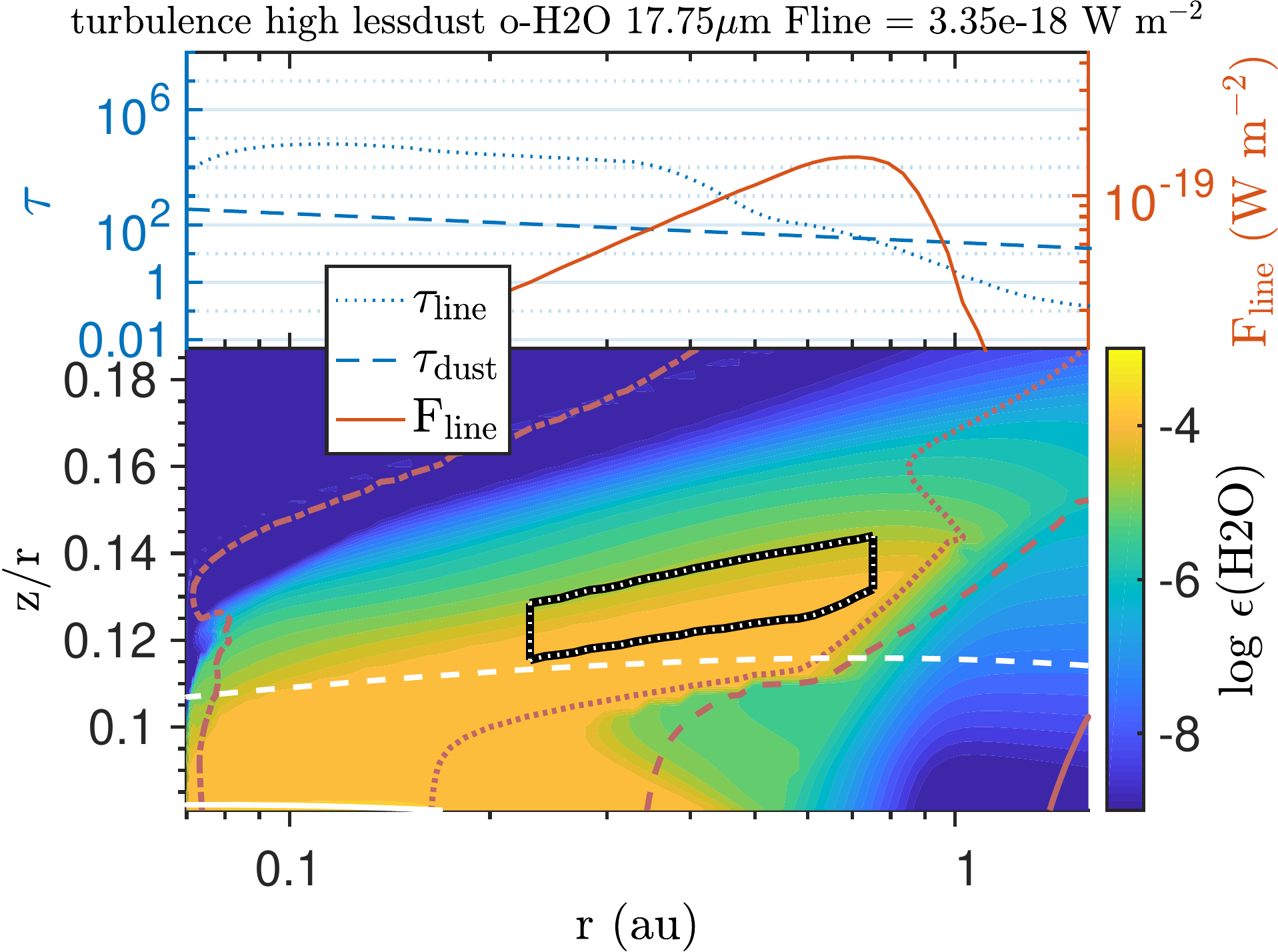}}     
			\caption{Line-emitting regions for the model turbulence high lessdust. The plotted lines are \cem{C2H2}, \cem{HCN}, \cem{CO2}, \cem{NH3}, \cem{OH}, and \cem{o-H2O}. The rest of the figure is as described in \cref{fig:LER_TT_highres}.  
			}\label{fig:LER_settling_high_lessdust}     
		\end{figure*}
		
		\begin{figure*} \centering    
			\makebox[\textwidth][c]{\includegraphics[width=0.47\textwidth]{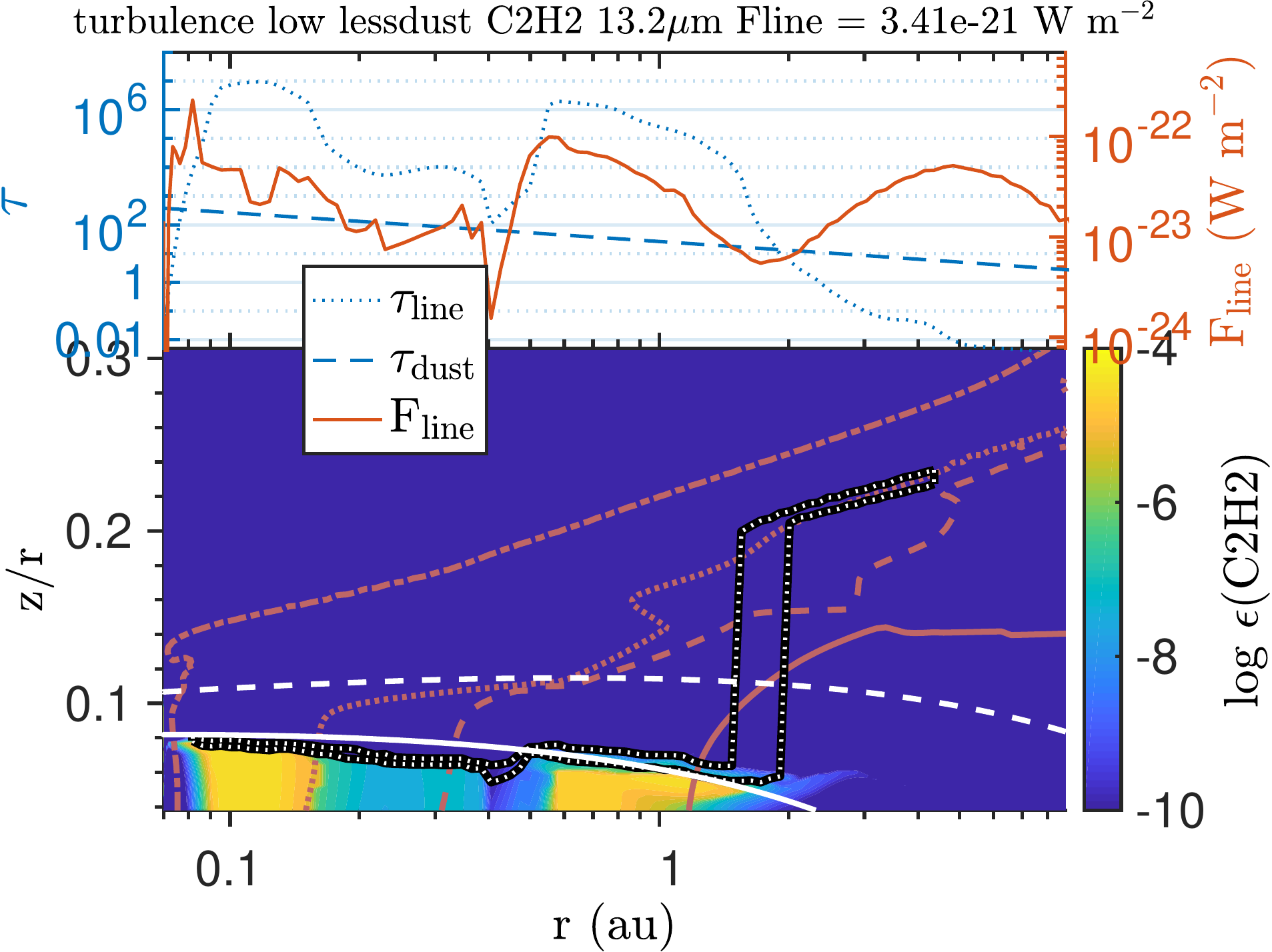} \hspace{0.005\textwidth}    
				\includegraphics[width=0.47\textwidth]{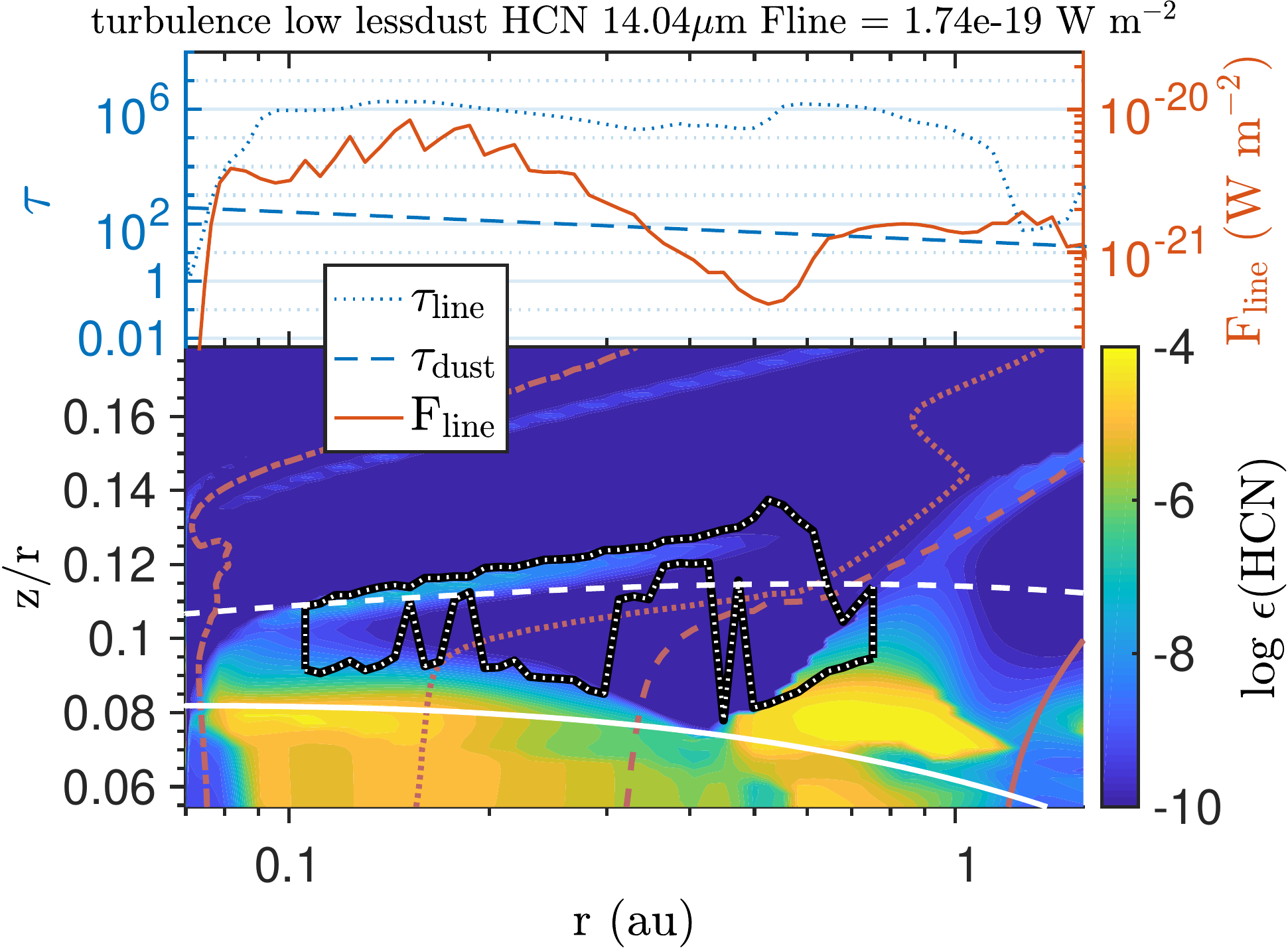}} 
			\makebox[\textwidth][c]{\includegraphics[width=0.47\textwidth]{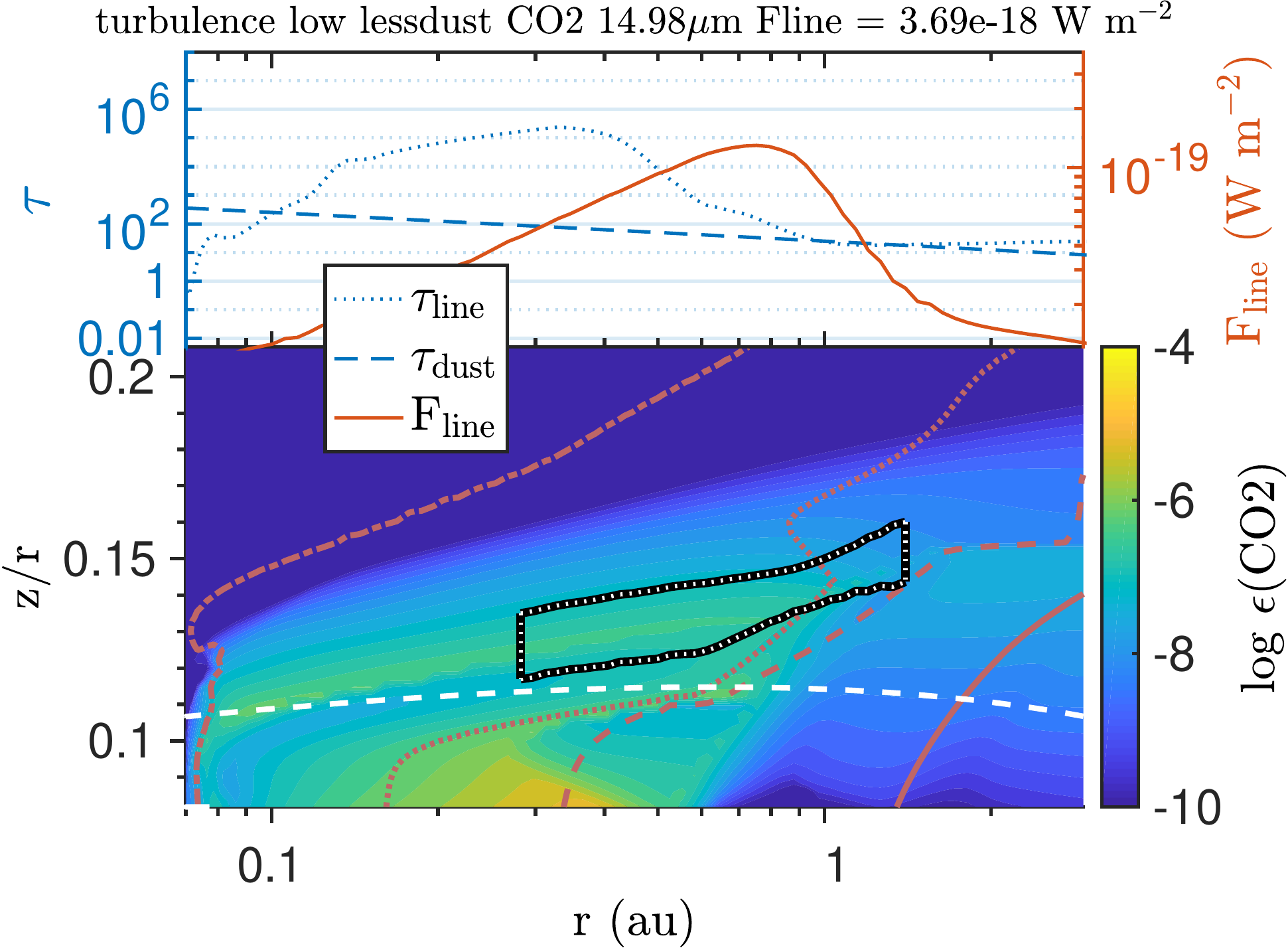} \hspace{0.005\textwidth}
				\includegraphics[width=0.47\textwidth]{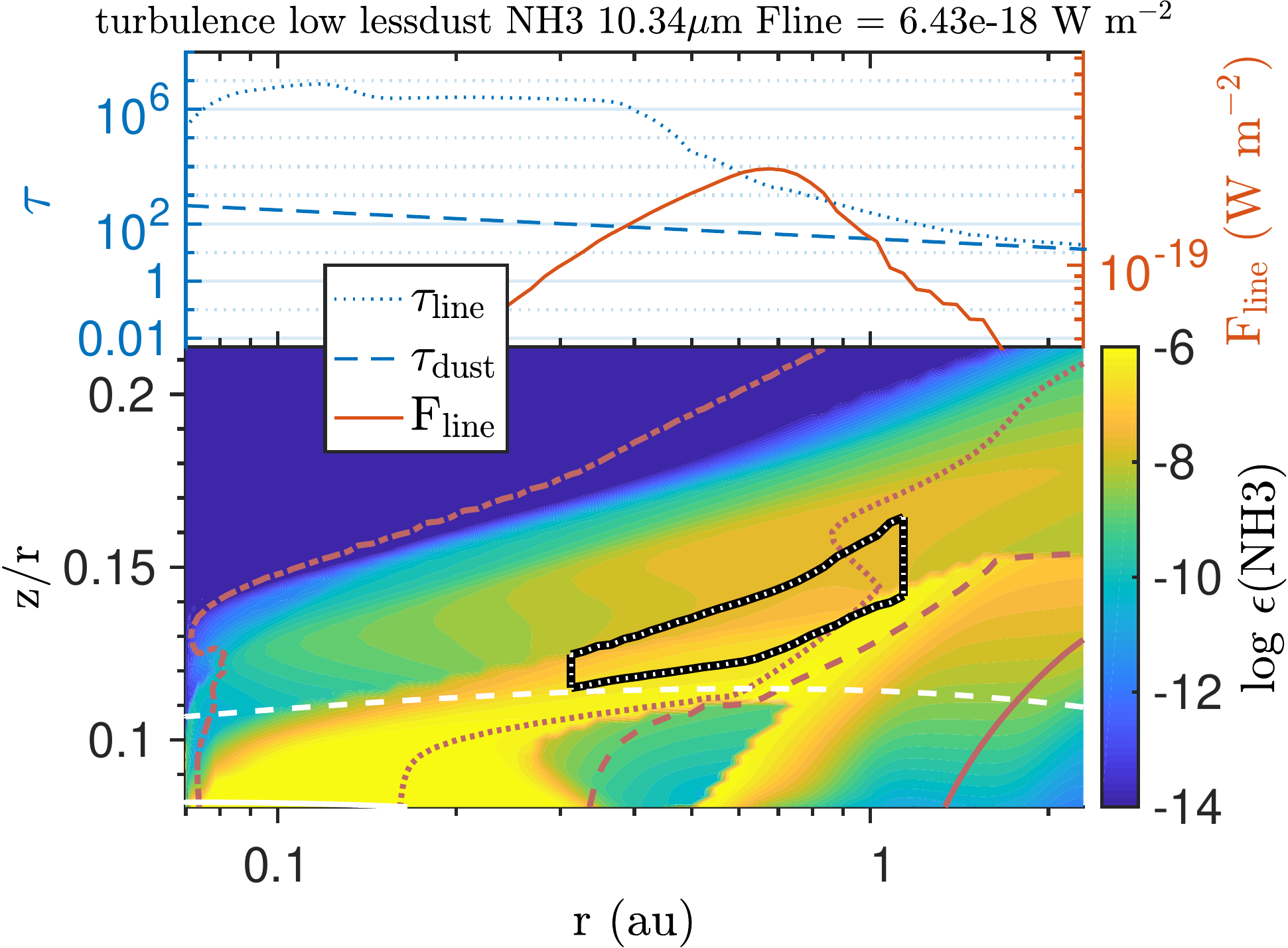}}      
			\makebox[\textwidth][c]{\includegraphics[width=0.47\textwidth]{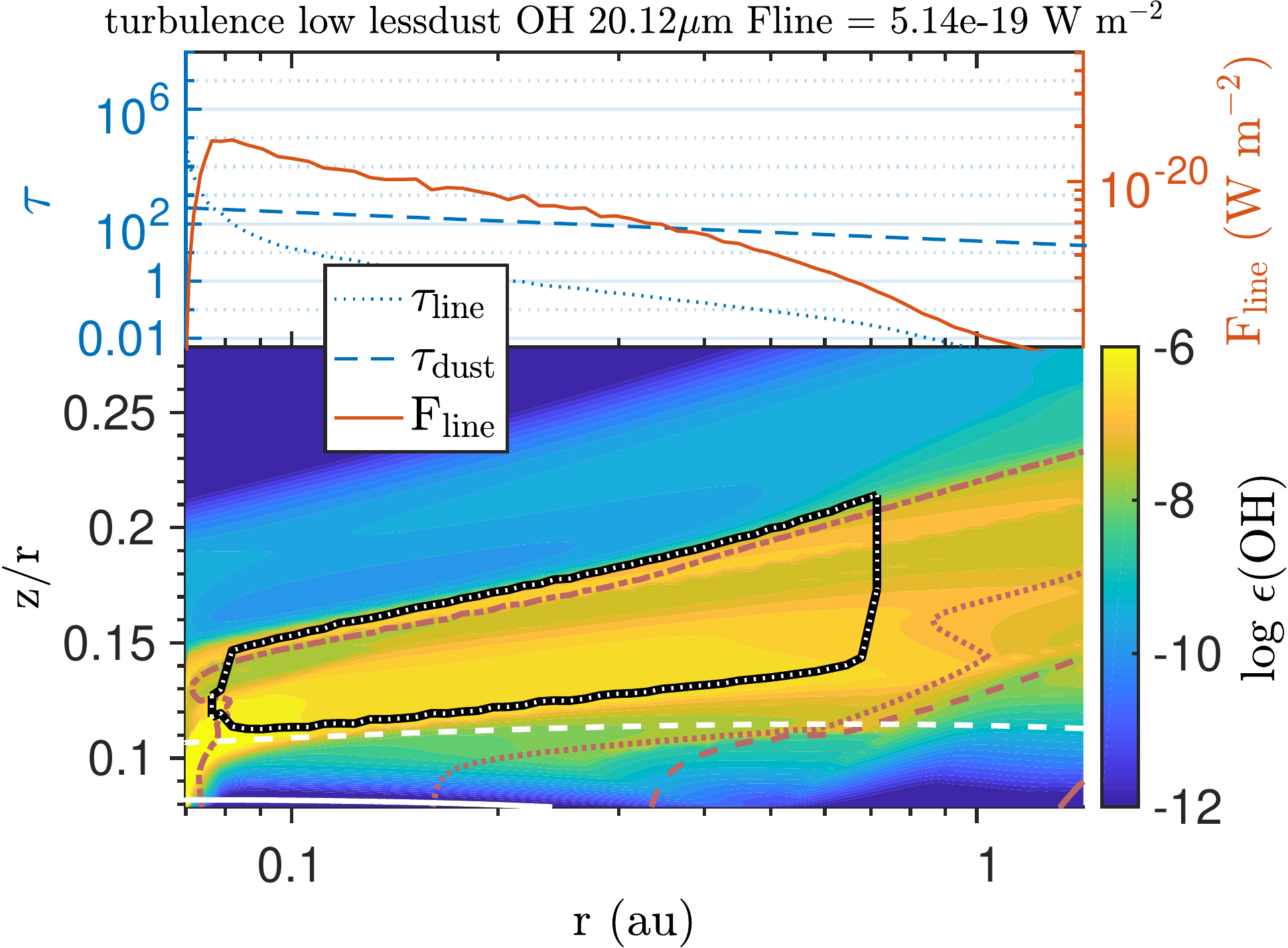} \hspace{0.005\textwidth}
				\includegraphics[width=0.47\textwidth]{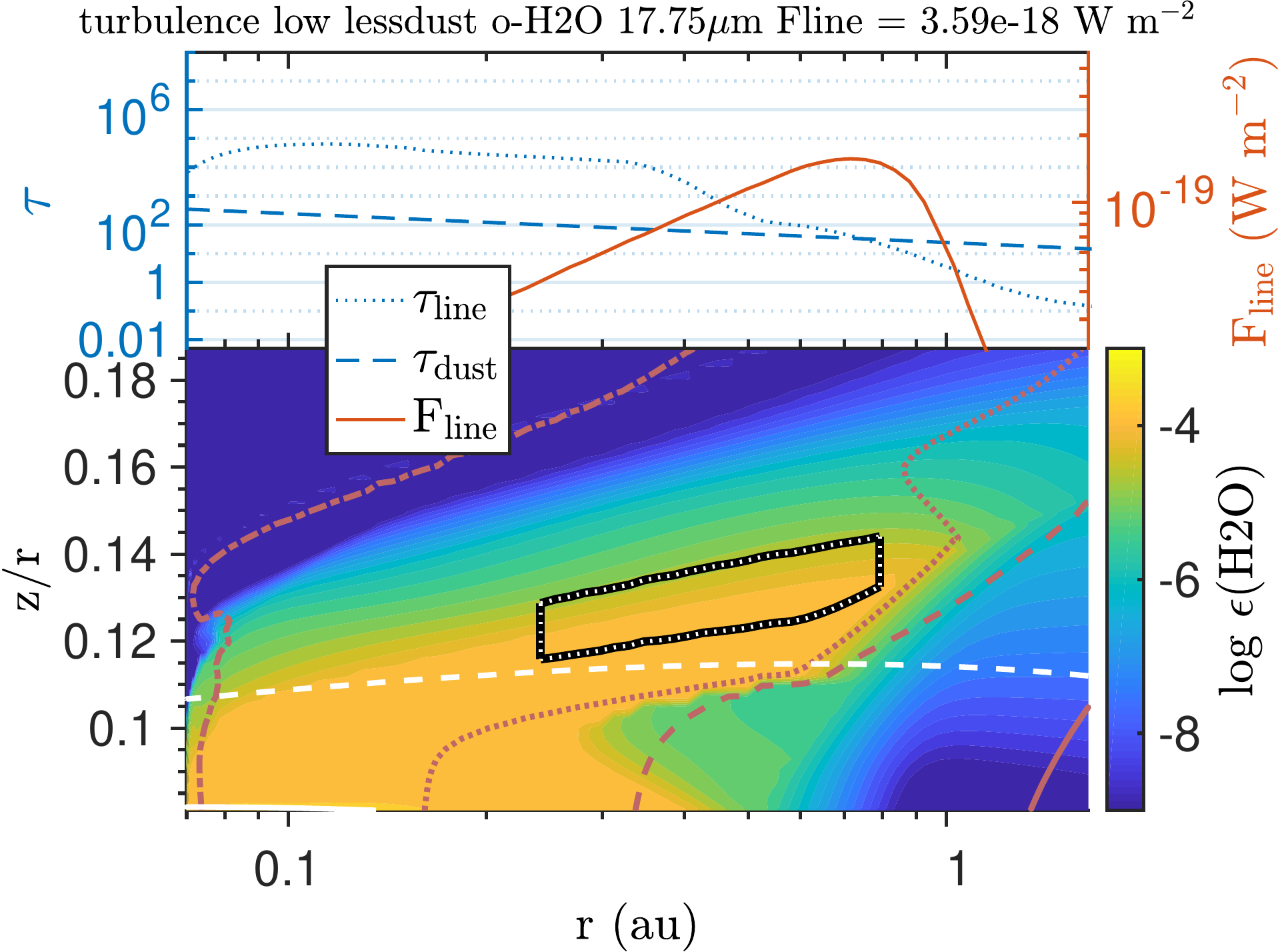}}
			\caption{Line-emitting regions for the model turbulence low lessdust. The plotted lines are \cem{C2H2}, \cem{HCN}, \cem{CO2}, \cem{NH3}, \cem{OH}, and \cem{o-H2O}. The rest of the figure is as described in \cref{fig:LER_TT_highres}.
			}\label{fig:LER_settling_low_lessdust}    
		\end{figure*}

	\end{document}